\documentclass[prd,aps,showpacs,preprintnumbers,amsmath,amssymb,superscriptaddress,twocolumn,floatfix]{revtex4}
\usepackage{alltt}
\usepackage{epsfig}
\usepackage{calc,rotating,xspace}
\usepackage{pifont}

\newcommand{\TeV}{\ensuremath{\mathrm{Te\kern -0.1em V}}}
\newcommand{\GeV}{\ensuremath{\mathrm{Ge\kern -0.1em V}}}
\newcommand{\MeV}{\ensuremath{\mathrm{Me\kern -0.1em V}}}
\newcommand{\KeV}{\ensuremath{\mathrm{Ke\kern -0.1em V}}}
\newcommand{\eV}{\ensuremath{\mathrm{e\kern -0.1em V}}}

\newcommand{\GeVC}{\ensuremath{\GeV\!/c}}
\newcommand{\MeVC}{\ensuremath{\MeV\!/c}}

\newcommand{\GeVCSq}{\ensuremath{\GeV\!/c^2}}

\newcommand{\Tesla}{\ensuremath{\mathrm{T}}}
\newcommand{\um}{\mu \mathrm{m}}
\newcommand{\mm}{\mathrm{mm}}
\newcommand{\cm}{\mathrm{cm}}
\newcommand{\m}{\mathrm{m}}

\newcommand{\ns}{\mathrm{ns}}
\newcommand{\us}{\mu \mathrm{s}}

\newcommand{\s}{\mathrm{s}}
\newcommand{\MHz}{\mathrm{MHz}}
\newcommand{\KHz}{\mathrm{kHz}}
\newcommand{\Hz}{\mathrm{Hz}}

\newcommand{\pbinv}{\ensuremath{\mathrm{pb}^{-1}}}

\newcommand{\pb}{\ensuremath{\mathrm{pb}}}
\newcommand{\nb}{\ensuremath{\mathrm{nb}}}

\newcommand{\mb}{\ensuremath{\mathrm{mb}}}

\newcommand{\W}{\ensuremath{W}}
\newcommand{\Z}{\ensuremath{Z}}
\newcommand{\zg}{\gamma^{*}/Z}
\newcommand{\zll}{Z \rightarrow \ell \ell}
\newcommand{\zgll}{\gamma^{*}/Z \rightarrow \ell \ell}
\newcommand{\wlnu}{W \rightarrow \ell \nu} 
\newcommand{\zgee}{\gamma^{*}/Z \rightarrow e e}
\newcommand{\zee}{Z \rightarrow e e}
\newcommand{\zmm}{Z \rightarrow \mu \mu}
\newcommand{\zgmm}{\gamma^{*}/Z \rightarrow \mu \mu}
\newcommand{\ztt}{Z \rightarrow \tau \tau}

\newcommand{\zgtt}{\gamma^{*}/Z \rightarrow \tau \tau}
\newcommand{\wenu}{W \rightarrow e \nu} 
\newcommand{\wmnu}{W \rightarrow \mu \nu} 
\newcommand{\wtnu}{W \rightarrow \tau \nu}

\newcommand{\sigmawen}{\sigma_{W}}
\newcommand{\sigmazee}{\sigma_{Z}}
\newcommand{\sigmaw}{\sigma_{W}}
\newcommand{\sigmaz}{\sigma_{Z}}
\newcommand{\sigmappw}{\sigma (\ppbar \rightarrow W)}
\newcommand{\sigmappz}{\sigma (\ppbar \rightarrow Z)}
\newcommand{\sigmabrppwln}{\sigma_{W} \times Br (\ppbar \rightarrow W \rightarrow \ell \nu)}

\newcommand{\sigmabrppzgll}{\sigma_{\gamma^{*}/Z} \times Br (\ppbar \rightarrow \gamma^{*}/Z \rightarrow \ell \ell)}
\newcommand{\sigmappzg}{\sigma (\ppbar \rightarrow \gamma^{*}/Z)}
\newcommand{\sigmabrwen}{\sigma_{W} \cdot Br(\wenu)}

\newcommand{\sigmabrzgee}{\sigma_{\gamma^{*}/Z} \cdot Br(\zgee)}
\newcommand{\sigmabrwln}{\sigma_{W} \cdot Br(\wlnu)}
\newcommand{\sigmabrzll}{\sigma_{Z} \cdot Br(\zll)}
\newcommand{\sigmabrzgll}{\sigma_{\gamma^{*}/Z} \cdot Br(\zgll)}
\newcommand{\sigmabrwmn}{\sigma_{W} \cdot Br(\wmnu)}

\newcommand{\sigmabrzgmm}{\sigma_{\gamma^{*}/Z} \cdot Br(\zgmm)}

\newcommand{\gam}{\Gamma(W)}
\newcommand{\gamz}{\Gamma(Z)}
\newcommand{\vcs}{V_{cs}}

\newcommand{\et}{\ensuremath{E_{T}}}
\newcommand{\ex}{\ensuremath{E_{x}}}
\newcommand{\ey}{\ensuremath{E_{y}}}
\newcommand{\pt}{\ensuremath{p_{T}}}

\newcommand{\ppbar}{p\overline{p}}
\newcommand{\lum}{{\cal L}}

\newcommand {\absdeteta} {\mbox{$\mid \eta_{\mathrm{det}} \mid $}}

\newcommand{\met}{\mbox{${\hbox{$E$\kern-0.6em\lower-.1ex\hbox{/}}}_T\:$}}
\newcommand{\metx}{\mbox{${\hbox{$E$\kern-0.6em\lower-.1ex\hbox{/}}}_{T,x}\:$}}
\newcommand{\mety}{\mbox{${\hbox{$E$\kern-0.6em\lower-.1ex\hbox{/}}}_{T,y}\:$}}

\newcommand{\bit}{\begin{itemize}}
\newcommand{\eit}{\end{itemize}}
\newcommand{\bce}{\begin{center}}
\newcommand{\beq}{\begin{equation}}
\newcommand{\ece}{\end{center}}

\begin{document}
\clearpage

\title{\boldmath{Measurements of Inclusive $\W$ and $\Z$ Cross
Sections in $\ppbar$ Collisions at $\sqrt{s} =$~1.96~$\TeV$}}

\affiliation{Institute of Physics, Academia Sinica, Taipei, Taiwan 11529, Republic of China} 
\affiliation{Argonne National Laboratory, Argonne, Illinois 60439} 
\affiliation{Institut de Fisica d'Altes Energies, Universitat Autonoma de Barcelona, E-08193, Bellaterra (Barcelona), Spain} 
\affiliation{Baylor University, Waco, Texas  76798} 
\affiliation{Istituto Nazionale di Fisica Nucleare, University of Bologna, I-40127 Bologna, Italy} 
\affiliation{Brandeis University, Waltham, Massachusetts 02254} 
\affiliation{University of California, Davis, Davis, California  95616} 
\affiliation{University of California, Los Angeles, Los Angeles, California  90024} 
\affiliation{University of California, San Diego, La Jolla, California  92093} 
\affiliation{University of California, Santa Barbara, Santa Barbara, California 93106} 
\affiliation{Instituto de Fisica de Cantabria, CSIC-University of Cantabria, 39005 Santander, Spain} 
\affiliation{Carnegie Mellon University, Pittsburgh, PA  15213} 
\affiliation{Enrico Fermi Institute, University of Chicago, Chicago, Illinois 60637} 
\affiliation{Joint Institute for Nuclear Research, RU-141980 Dubna, Russia} 
\affiliation{Duke University, Durham, North Carolina  27708} 
\affiliation{Fermi National Accelerator Laboratory, Batavia, Illinois 60510} 
\affiliation{University of Florida, Gainesville, Florida  32611} 
\affiliation{Laboratori Nazionali di Frascati, Istituto Nazionale di Fisica Nucleare, I-00044 Frascati, Italy} 
\affiliation{University of Geneva, CH-1211 Geneva 4, Switzerland} 
\affiliation{Glasgow University, Glasgow G12 8QQ, United Kingdom} 
\affiliation{Harvard University, Cambridge, Massachusetts 02138} 
\affiliation{Division of High Energy Physics, Department of Physics, University of Helsinki and Helsinki Institute of Physics, FIN-00014, Helsinki, Finland} 
\affiliation{University of Illinois, Urbana, Illinois 61801} 
\affiliation{The Johns Hopkins University, Baltimore, Maryland 21218} 
\affiliation{Institut f\"{u}r Experimentelle Kernphysik, Universit\"{a}t Karlsruhe, 76128 Karlsruhe, Germany} 
\affiliation{High Energy Accelerator Research Organization (KEK), Tsukuba, Ibaraki 305, Japan} 
\affiliation{Center for High Energy Physics: Kyungpook National University, Taegu 702-701; Seoul National University, Seoul 151-742; and SungKyunKwan University, Suwon 440-746; Korea} 
\affiliation{Ernest Orlando Lawrence Berkeley National Laboratory, Berkeley, California 94720} 
\affiliation{University of Liverpool, Liverpool L69 7ZE, United Kingdom} 
\affiliation{University College London, London WC1E 6BT, United Kingdom} 
\affiliation{Massachusetts Institute of Technology, Cambridge, Massachusetts  02139} 
\affiliation{Institute of Particle Physics: McGill University, Montr\'{e}al, Canada H3A~2T8; and University of Toronto, Toronto, Canada M5S~1A7} 
\affiliation{University of Michigan, Ann Arbor, Michigan 48109} 
\affiliation{Michigan State University, East Lansing, Michigan  48824} 
\affiliation{Institution for Theoretical and Experimental Physics, ITEP, Moscow 117259, Russia} 
\affiliation{University of New Mexico, Albuquerque, New Mexico 87131} 
\affiliation{Northwestern University, Evanston, Illinois  60208} 
\affiliation{The Ohio State University, Columbus, Ohio  43210} 
\affiliation{Okayama University, Okayama 700-8530, Japan} 
\affiliation{Osaka City University, Osaka 588, Japan} 
\affiliation{University of Oxford, Oxford OX1 3RH, United Kingdom} 
\affiliation{University of Padova, Istituto Nazionale di Fisica Nucleare, Sezione di Padova-Trento, I-35131 Padova, Italy} 
\affiliation{University of Pennsylvania, Philadelphia, Pennsylvania 19104} 
\affiliation{Istituto Nazionale di Fisica Nucleare Pisa, Universities of Pisa, Siena and Scuola Normale Superiore, I-56127 Pisa, Italy} 
\affiliation{University of Pittsburgh, Pittsburgh, Pennsylvania 15260} 
\affiliation{Purdue University, West Lafayette, Indiana 47907} 
\affiliation{University of Rochester, Rochester, New York 14627} 
\affiliation{The Rockefeller University, New York, New York 10021} 
\affiliation{Istituto Nazionale di Fisica Nucleare, Sezione di Roma 1, University of Rome ``La Sapienza," I-00185 Roma, Italy} 
\affiliation{Rutgers University, Piscataway, New Jersey 08855} 
\affiliation{Texas A\&M University, College Station, Texas 77843} 
\affiliation{Istituto Nazionale di Fisica Nucleare, University of Trieste/\ Udine, Italy} 
\affiliation{University of Tsukuba, Tsukuba, Ibaraki 305, Japan} 
\affiliation{Tufts University, Medford, Massachusetts 02155} 
\affiliation{Waseda University, Tokyo 169, Japan} 
\affiliation{Wayne State University, Detroit, Michigan  48201} 
\affiliation{University of Wisconsin, Madison, Wisconsin 53706} 
\affiliation{Yale University, New Haven, Connecticut 06520} 
\author{A.~Abulencia}
\affiliation{University of Illinois, Urbana, Illinois 61801}
\author{D.~Acosta}
\affiliation{University of Florida, Gainesville, Florida  32611}
\author{J.~Adelman}
\affiliation{Enrico Fermi Institute, University of Chicago, Chicago, Illinois 60637}
\author{T.~Affolder}
\affiliation{University of California, Santa Barbara, Santa Barbara, California 93106}
\author{T.~Akimoto}
\affiliation{University of Tsukuba, Tsukuba, Ibaraki 305, Japan}
\author{M.G.~Albrow}
\affiliation{Fermi National Accelerator Laboratory, Batavia, Illinois 60510}
\author{D.~Ambrose}
\affiliation{Fermi National Accelerator Laboratory, Batavia, Illinois 60510}
\author{S.~Amerio}
\affiliation{University of Padova, Istituto Nazionale di Fisica Nucleare, Sezione di Padova-Trento, I-35131 Padova, Italy}
\author{D.~Amidei}
\affiliation{University of Michigan, Ann Arbor, Michigan 48109}
\author{A.~Anastassov}
\affiliation{Rutgers University, Piscataway, New Jersey 08855}
% ...
\collaboration{CDF Collaboration}
\noaffiliation

\date{\today}


\begin{abstract}
We report the first measurements of inclusive $\W$ and $\Z$ boson
cross sections times the corresponding leptonic branching ratios for
$\ppbar$ collisions at $\sqrt{s} =$~1.96~$\TeV$ based on the decays 
of the $\W$ and $\Z$ bosons into electrons and muons.  The data 
were recorded with the CDF~II detector at the Fermilab Tevatron and
correspond to an integrated luminosity of 72.0~$\pm$~4.3~$\pbinv$.  
We test $e$-$\mu$ lepton universality in $\W$ decays by measuring 
the ratio of the $\wmnu$ to $\wenu$ cross sections and determine a 
value of 0.991~$\pm~0.004(stat.)~\pm~0.011(syst.)$ for the ratio 
of $\W-\ell-\nu$ couplings ($g_{\mu}/g_e$).  Since there is no sign 
of non-universality, we combine our cross section measurements in 
the different lepton decay modes and obtain $\sigmabrppwln
=$$~2.749~\pm~0.010(stat.)~\pm~0.053(syst.)~\pm~0.165(lum.)$~$\nb$
and $\sigmabrppzgll
=$$~254.9~\pm~3.3(stat.)~\pm~4.6(syst.)~\pm~15.2(lum.)$~$\pb$ 
for dilepton pairs in the mass range between 66~$\GeVCSq$ and  
116~$\GeVCSq$.  We compute the ratio $R$ of the $\wlnu$ to 
$\zll$ cross sections taking all correlations among channels into 
account and obtain $R=$$~10.84~\pm~0.15(stat.)~\pm~0.14(syst.)$ 
including a correction for the virtual photon exchange component 
in our measured $\zgll$ cross section.  Based on the measured value 
of $R$, we extract values for the $\W$ leptonic branching ratio, 
$Br(\wlnu) =$~0.1082~$\pm$~0.0022; the total width of the $\W$ boson, 
$\Gamma(W) =$~2092~$\pm$~42~$\MeV$; and the ratio of $\W$ and $\Z$ 
boson total widths, $\Gamma(W)/\Gamma(Z) =$~0.838~$\pm$~0.017.  
In addition, we use our extracted value of $\Gamma(W)$ whose value 
depends on various electroweak parameters and certain CKM matrix 
elements to constrain the $\vcs$ CKM matrix element, 
$|\vcs| =$~0.976$\pm$~0.030.
\end{abstract}

\maketitle

\newpage
\section{Introduction}
\label{sec:intro}
Measurements of the production cross sections for both $\W$ and $\Z$ 
bosons in high energy $\ppbar$ collisions are important tests of the
Standard Model (SM) of particle physics. At hadron colliders the $\W$
and $\Z$ bosons can most easily be detected through their leptonic
decay modes. This paper presents measurements of $\sigmabrwln$,
$\sigmabrzll$, and their ratio,
\begin{equation}
R = \frac{\sigmabrwln}{\sigmabrzll}
\label{eq:rdef}
\end{equation}
for $\ell$ = $e$ and $\mu$ based on 72.0 $\pbinv$ of $\ppbar$
collision data collected in 2002-2003 by the upgraded Collider
Detector at Fermilab (CDF) at a center-of-mass energy of 1.96 $\TeV$.
These measurements are also described in ~\cite{int:ourprl}. These
measurements provide a test of SM predictions for the $\W$ and $\Z$
boson production cross sections, $\sigmaw$ and $\sigmaz$, as well as a
precise indirect measurement of the total decay width of the $\W$
boson, $\gam$, within the framework of the SM. This analysis is
sensitive to deviations in $\gam$ from the SM predictions at the level
of about 2~$\!\%$.  We also use our results to extract the leptonic
branching fraction, $Br(\wlnu)$, and the Cabibbo-Kobayashi-Maskawa
(CKM) matrix element, $\vcs$.  Finally, we test the lepton
universality hypothesis for the couplings of the $\W$ boson to $e$ and
$\mu$ leptons.

\subsection{$\W$/$\Z$ production and decay}
\label{sec:I.A}

\begin{figure}[t]
\includegraphics*[width=8.5cm]{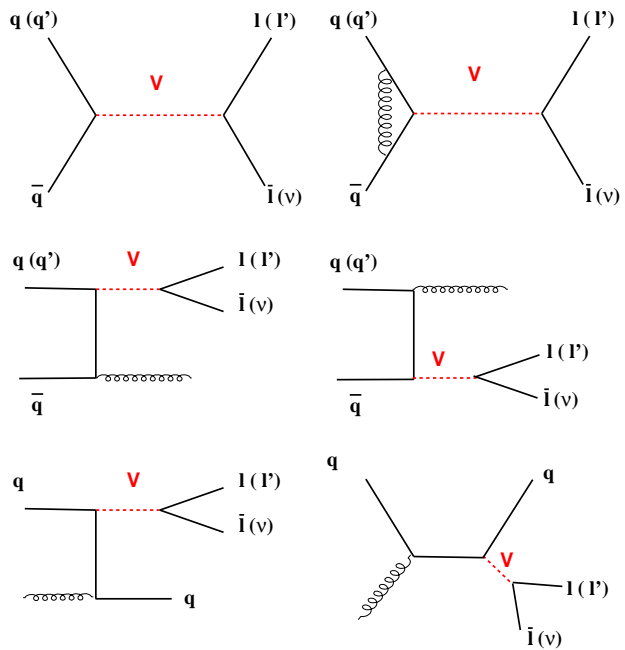}
\caption{Diagrams for production and leptonic decay of 
     a vector boson $V$ = $\W$, $\Z$ at leading (upper left) 
     and next-to-leading order (others).}
\label{fig:vprod}
\end{figure}

The $\W$ and $\Z$ bosons, together with the massless photon
($\gamma$), compose the bosonic fields of the unified electroweak
theory proposed by Weinberg~\cite{int:weinberg},
Salam~\cite{int:salam}, and Glashow~\cite{int:glashow}.  The $\W$ and
$\Z$ bosons were discovered in 1983 using the UA1 and UA2
detectors~\cite{int:wdisc,int:wdisc2,int:zdisc,int:zdisc2} which were designed
and built for this very purpose.  The transverse momentum (\pt)
distribution of the reconstructed leptons in $\wlnu$ events was used
to determine the $\W$ mass, while the $\Z$ mass was determined by
directly reconstructing the invariant mass of dilepton pairs in $\zll$
events.

Present experimental measurements of electroweak parameters including
vector boson masses and decay widths are precise enough to provide
tests of Quantum Chromodynamics (QCD) and the electroweak part of the
Standard Model beyond leading order.  These precise measurements not
only test the electroweak theory but also provide possible windows to
sectors of the theory at mass scales higher than those directly
observable at current accelerator energies. These sectors enter into
the electroweak observables through radiative corrections.  While the
parameters of the $\Z$ boson have been well studied
\cite{res:lepcomb}, the properties of the charged current carriers,
the $\W$ bosons, are known with less precision. In hadron-antihadron
collisions the $\W$ and $\Z$ are predominantly produced via the
processes illustrated in Fig.~\ref{fig:vprod}.  The production of
$\ppbar \rightarrow \zg$ where a quark in one hadron annihilates with
an antiquark in the other hadron to produce the resulting vector boson
is often referred to as the Drell-Yan~\cite{int:drellyan} production
process.

Calculations of the total production cross sections for $\W$ and $\Z$
bosons incorporate parton cross sections, parton distribution
functions, higher-order QCD effects, and factors for the couplings of
the different quarks and antiquarks to the $\W$ and $\Z$
bosons. Beyond the leading order Born processes, a vector boson $V$
can also be produced by $q(\bar{q})g$ interactions, so the Parton
Distribution Functions (PDFs) of the proton and antiproton play an
important role at higher orders.  Theoretical calculations of the $\W$
and $\Z$ production cross sections have been carried out in
next-to-leading order (NLO)~\cite{int:nnlo00,int:nnlo0} and
next-to-next-to-leading order
(NNLO)~\cite{int:nnlo4,int:nnlo1,int:nnlo2,int:nnlo3,acc:slac}.  The NLO and
NNLO computations used in this article are in the modified
minimal-subtraction ($\overline{MS}$)~\cite{msbar,msbar2}
renormalization prescription framework. The full order
$\alpha_{\mathrm{s}}^{2}$ calculation has been made and includes final
states containing the vector boson $V$ and up to two additional
partons.  The two-loop splitting function is used and the running of
$\alpha_{\mathrm{s}}$ includes thresholds for heavy flavors.  The NLO
cross section is $\sim$ 25$\!\%$ larger than the Born-level cross
section, and the NNLO cross section is an additional $\sim$ 3$\!\%$
higher. The main contribution to the calculated cross section is from
$q\overline{q}$ interactions.  The contribution of $q(\bar{q})g$
interactions to the calculated cross section is negative at the
Tevatron collision energy.

The decay modes of the $\W$ boson are $\wlnu$ ($\ell =$ $e$, $\mu$,
and $\tau$) and $q\bar{q}^{\prime}$, where the main modes $u\bar{d}$,
$u\bar{s}$, $c\bar{s}$ and $c\bar{d}$ have branching ratios
proportional to their corresponding CKM matrix elements. The measured
value for the branching fraction of the three combined leptonic modes
is 32.0 $\pm$ 0.4~$\!\%$~\cite{int:pdg}, where the remaining fraction
is assigned to the hadronic decay modes.  The partial width into
fermion pairs is calculated at lowest order to be~\cite{int:pdg}
\begin{equation}
\Gamma_0(W \rightarrow {\rm{f}\bar{f}^{\prime}}) =
|V_{{\rm{ff^{\prime}}}}|^2 N_{\mathrm{C}} G_{\mathrm{F}} M_{\W}^{3} 
/ (6 \sqrt{2} \pi),
\label{eq:vckm}
\end{equation}
\noindent where $V_{{\rm{ff^{\prime}}}}$ is the corresponding 
CKM matrix element for quark pairs or one for leptons.  
$M_{\W}$ is the $\W$ boson mass and $G_{\mathrm{F}}$ is 
the Fermi coupling constant. $N_{\mathrm{C}}$ is the 
corresponding color factor which is three for quarks 
and one for leptons.

The expression for the partial decay widths into quark pairs also 
has an additional QCD correction due to vertex graphs involving 
gluon exchange and electroweak corrections due to next-to-leading 
order graphs which alter the effective coupling at the 
$\W$-fermion vertex for all fermions. Within the context of the 
Standard Model, there are also vertex and bremsstrahlung
corrections~\cite{int:wdecays} that depend on the top quark and 
Higgs boson masses.  The corrections can be summarized in the 
equation 
\begin{eqnarray}
\Gamma(W \rightarrow {\rm{f \bar{f}^{\prime}}})_{\mathrm{SM}} = && 
\Gamma_0(W \rightarrow {\rm{f \bar{f}^{\prime}}}) \nonumber\\
&&\times [1+\delta_{\mathrm{V}} + \delta_{\mathrm{W(0)}} +\delta_\mu],
\end{eqnarray}
where $\delta_{\mathrm{W(0)}}$ is the correction to the width 
from loops at the $\W$-fermion vertex involving the $\Z$ boson 
or a SM Higgs boson, $\delta_{\mathrm{V}}$ arises from the boson 
self-energies, and $\delta_\mu$ is a correction required when the 
couplings are parametrized using the $\W$ mass and the value of 
$G_{\mathrm{F}}$ from muon decay measurements~\cite{int:corrmu,int:corrmu2}.  
Since all of these corrections are small ($\sim$~0.35~$\!\%$), the 
measurement of $\gam$ is not very sensitive to these higher order effects.
Higher order QCD corrections originating from quark mass effects are 
also small.
 
\subsection{Measurement of $\gam$ from the $\W$ and $\Z$ cross sections}
\label{sec:I.B}

The width of the $\W$ boson can be extracted from the measurement
of the ratio $R$, which is defined in Eq.~\ref{eq:rdef}.  This 
method was first proposed by Cabibbo in 1983 as a method to determine 
the number of light neutrino species~\cite{int:cabibbo} and has been 
adopted as a method to indirectly measure the branching ratio 
for the $\wlnu$ decay mode.  The ratio $R$ can be expressed as
\begin{equation}
R = 
\frac{\sigmaw}{\sigmaz}
\frac{\Gamma(\wlnu)}{\Gamma(\zll)}
\frac{\gamz}{\gam}.
\label{eq:ratio}
\end{equation}
On the right hand side of Eq.~\ref{eq:ratio}, the ratio of the $\W$
and $\Z$ production cross sections can be calculated from the boson
couplings and knowledge of the proton structure.  The $\Z$ boson total
width, $\gamz$, and leptonic partial width, $\Gamma(\zll)$, have been
measured very precisely by the LEP experiments~\cite{res:lepcomb}.
With the measured value of $R$ the branching ratio $Br(\wlnu) =
\Gamma(\wlnu)/\gam$ can be extracted directly from Eq.~\ref{eq:ratio}.
The total width of the $\W$ boson, $\gam$, can also be determined
indirectly using the SM prediction for the partial width,
$\Gamma(\wlnu)$.  As shown in Eq.~\ref{eq:vckm}, $\gam$ depends on
electroweak parameters and certain CKM matrix elements.  We also use
our measurement of the total $\W$ width to constrain the associated
sum over CKM matrix elements in the formula for $\gam$ and derive an
indirect value for $\vcs$ which is the least experimentally
constrained element in the sum.  Finally, the ratios of the muon and
electron $\wlnu$ cross section measurements are used to determine the
ratios of the coupling constants of the $\W$ boson to the different
lepton species, providing a test of the lepton universality
hypothesis.  For reference, Table~\ref{tab:history} provides a summary
of previous experimental results for $\sigmabrwln$ and
$\sigmabrzll$ along with the measured values for $R$ and the extracted
values of $\gam$.  The most recent direct measurement of $\gam$
obtained by LEP is 2.150 $\pm$ 0.091
\GeV~\cite{res:lepcomb}.

\begin{table*}[t]
\caption{Previous measurements of the $\W$ and $\Z$ production cross sections times 
branching ratios along with the measured values of $R$ and the extracted values of $\gam$.}
\centering{
\begin{tabular}{l c c c c c r}
\hline
\hline
Experiment         & $\sqrt{s}$       & Mode       & $\sigmabrwln$       & $\sigmabrzll$       & $R$                 & $\gam$            \\ 
                   & ($\TeV$)         &            & ($\nb$)             & ($\pb$)             &                     & ($\GeV$)        \\ \hline
CDF(Run I)~\cite{int:cdf_sigmas,int:cdf_ratio1,int:cdf_ratio2,int:cdf_zmm,acc:cdfzy}  
                   & 1.80             & $e$        & 2.49 $\pm$ 0.12     & 231 $\pm$ 12        & 10.90 $\pm$ 0.43    & 2.064 $\pm$ 0.084 \\
D\O(Run IA)~\cite{int:d0_A}  
                   & 1.80             & $e$        & 2.36 $\pm$ 0.15     & 218 $\pm$ 16        &                     &                   \\
D\O(Run IA)~\cite{int:d0_A}  
                   & 1.80             & $\mu$      & 2.09 $\pm$ 0.25     & 178 $\pm$ 31        &                     &                   \\
D\O(Run IA)~\cite{int:d0_A,int:d0_A2}  
                   & 1.80             & $e+\mu$    &                     &                     & 10.90 $\pm$ 0.49     & 2.044 $\pm$ 0.093 \\
D\O(Run IB)~\cite{int:d0_B}  
                   & 1.80             & $e$        & 2.31 $\pm$ 0.11     & 221 $\pm$ 11        & 10.43 $\pm$ 0.27    & 2.17 $\pm$ 0.07   \\
\hline
\hline
\end{tabular}
}
\label{tab:history}
\end{table*}

\subsection{Overview of this measurement}\label{sec:I.C}

The signature of high transverse momentum leptons from $\W$ and $\Z$
decay is very distinctive in the environment of hadron collisions. As
such, the decay of $\W$ and $\Z$ bosons into leptons provides a clean
experimental measurement of their production rate.  Experimentally,
the cross sections times branching ratios are calculated from
\label{sigmas_formulae}
\begin{equation}
\sigmabrwln 
=\frac{N_W^{\mathrm{obs}}-N_W^{\mathrm{bck}}}{A_W \cdot \epsilon_W \cdot \int \lum dt}
\label{eq:wsigma}
\end{equation} 
\begin{equation}
\sigmabrzll
=\frac{N_Z^{\mathrm{obs}}-N_Z^{\mathrm{bck}}}{A_Z \cdot \epsilon_Z \cdot \int \lum dt},
\label{eq:zsigma}
\end{equation} 
where $N_W^{\mathrm{obs}}$ and $N_Z^{\mathrm{obs}}$ are the numbers of 
$\wlnu$ and $\zll$ candidates observed in the data;  $N_W^{\mathrm{bck}}$ 
and $N_Z^{\mathrm{bck}}$ are the numbers of expected background events 
in the $\W$ and $\Z$ boson candidate samples; $A_W$ and $A_Z$ are the 
acceptances of the $\W$ and $\Z$ decays, defined as the fraction of these 
decays satisfying the geometric constraints of our detector and the 
kinematic constraints of our selection criteria; $\epsilon_W$ and 
$\epsilon_Z$ are the combined efficiencies for identifying $\W$ and $\Z$ 
decays falling within our acceptances; and $\int \lum dt$ is the integrated 
luminosity of our data samples.

In measuring the ratio of the cross sections some of the inputs 
and their experimental uncertainties cancel.  The strategy of 
this measurement is to select $\W$ and $\Z$ boson decays with 
one or both leptons ($e$ or $\mu$) falling within the central 
region of the CDF detector.  This region is well instrumented 
and understood and has good detection efficiencies for both 
lepton species. Using common lepton selection criteria 
(contributing to the factors $\epsilon_W$ and $\epsilon_Z$) 
for the $\W$ and $\Z$ channels has the great advantage of 
decreasing the systematic uncertainty in the measurement of 
$R$.  The resulting smaller systematic uncertainty offsets the 
expected increase in statistical uncertainty originating from 
the requirement of a common central lepton. For each lepton 
species, the selection criteria are optimized to obtain the 
least overall experimental uncertainty.

The measurement of the ratio $R$ is sensitive to new physics processes
which change the $\W$ or $\Z$ production cross sections or the $\wlnu$
branching ratio.  The $\wlnu$ branching ratio could be directly
affected by new decay modes of the $\W$ boson, such as supersymmetric
decays that do not similarly couple to the $\Z$ boson.  A new
resonance at a higher mass scale that decays to $\W$ or $\Z$ bosons
may change the production cross sections.  One example of a particle
with a larger mass is the top quark at $m_{\mathrm{t}} =$ 174.3 $\pm$
5.1 $\GeVCSq$, which decays to a $\W$ boson and a bottom
quark~\cite{int:pdg}.  In $\ppbar$ collisions at $\sqrt{s}$ =
1.8~$\TeV$ the production cross section for $\mathrm{t}
\bar{\mathrm{t}}$ pairs is 6.5$^{+1.7}_{-1.4}$ \pb~\cite{int:cdf_top}, 
about 3000 times smaller than direct $\W$ boson
production~\cite{int:cdf_sigmas}.  The decays of $t \bar{t}$ pairs
which result in the production of two $\W$ bosons should change the
measured value of $R$ by about 7 $\times$ 10$^{-4}$, which is well
below our sensitivity.  The total width of the $\W$ boson can also get
contributions from processes beyond the SM. For example, in
supersymmetry, the decay W$^{+}\rightarrow\chi^{+}\chi^{0}$ may be
possible if the charginos and neutralinos are light~\cite{int:susy}
and so a precise measurement of $\gam$ can constrain the properties of
these particles.

\subsection{Outline of the paper}
\label{sec:I.D}

This paper is organized as follows: in Sec.~\ref{sec:exp} the CDF
detector is described, with particular attention given to the
subdetectors essential in the identification of charged leptons and
the inference of neutrinos. Section~\ref{sec:data} describes the data
samples used in this analysis, and the selection of the $\W$ and $\Z$
candidate events is described in Sec.~\ref{sec:evsel}.
Section~\ref{sec:acc} describes the calculation of the geometric and
kinematic acceptances of our candidate samples, and the methods used
to determine the efficiencies for identifying events within our
acceptances are presented in Sec.~\ref{sec:eff}.  The estimation of
the contributions to our candidate samples from background processes
are discussed in Sec.~\ref{sec:backg}, and finally the calculation of
the cross sections along with the resulting value of $R$ and other
extracted quantities are summarized in Sec.~\ref{sec:results}.

\section{The Experimental Apparatus}
\label{sec:exp}
The data used for the measurements reported in this note were
collected with the upgraded Collider Detector (CDF)~\cite{det:tdr} at
the Fermilab Tevatron $\ppbar$ collider.  Detector upgrades were made
to accommodate the higher luminosities and new beam conditions
resulting from concurrent upgrades to the Tevatron accelerator
complex.  In addition to the increases in luminosity, the $\ppbar$
center of mass energy was also increased from $\sqrt{s} =$ 1.80 $\TeV$
to $\sqrt{s} =$ 1.96 $\TeV$. The relatively small change in beam
energies leads to a substantial increase in the production cross
sections for high-mass objects such as $\W$/$\Z$ bosons
($\sim$~9~$\!\%$) and top quark pairs ($\sim$~30~$\!\%$).  We
highlight the upgrades to the Run I detectors and electronics in the
following sections.

\subsection{The CDF II Detector}
\label{subsec:det}

\begin{figure}[t]
\includegraphics*[width=8.5cm]{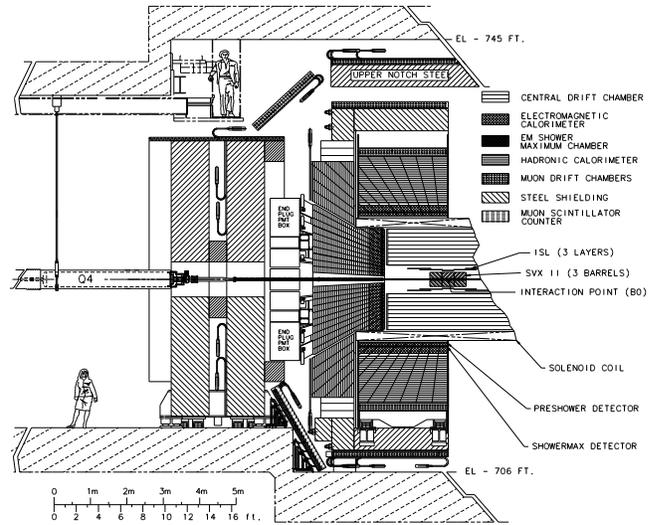}
  \caption{Elevation view of half of the CDF Run II detector.}
\label{fig:det_long}
\end{figure}

CDF is a general-purpose detector~\cite{det:tdr,det:det,det:dettop}
designed to detect particles produced in $\ppbar$ collisions.  As
illustrated in Fig.~\ref{fig:det_long}, the detector has a cylindrical
layout centered on the accelerator beamline.  Tracking detectors are
installed in the region directly around the interaction point to
reconstruct charged-particle trajectories inside a 1.4 $\Tesla$
uniform magnetic field (along the proton beam direction).  The field
is produced by a 5~$\m$ long superconducting solenoid located at the
outer radius of the tracking region (1.5~$\m$).  Calorimeter modules
are arranged in a projective tower geometry around the outside of the
solenoid to provide energy measurements for both charged and neutral
particles.  The outermost part of the detector consists of a series 
of drift chambers used to detect muons which are minimum-ionizing 
particles that typically pass through the calorimeter.

The $z$-axis of the CDF coordinate system is defined to be 
along the direction of the incoming protons.  A particle 
trajectory is then described by $\theta$, the polar angle 
relative to the incoming proton beam; $\phi$, the azimuthal 
angle about this beam axis; and $z_{0}$, the intersection 
point of the particle trajectory with the beam axis.  The 
pseudorapidity of a particle trajectory is defined as 
$\eta = -\ln(\tan(\theta/2))$.  The transverse momentum, 
$\pt$, is the component of the momentum projected on a 
plane perpendicular to the beam axis.  Similarly, the 
transverse energy, $\et$, of a shower or an individual 
calorimeter tower is given by $E \cdot \sin\theta$.  
The total transverse energy in an event is given by a 
sum over all calorimeter towers $\sum_iE_{T}^i\hat{n}_i$ 
where $E_{T}^i$ is the transverse energy measured in 
the $i$th tower and $\hat{n}_i$ is the projection of 
the vector pointing from the event vertex to the i$th$ 
calorimeter tower onto the plane perpendicular to the 
beam axis (unit normalized).  The vector sum of 
transverse energies measured in the calorimeter is 
corrected to account for muons which deposit only a 
fraction of their energy in the calorimeter.  The 
missing transverse energy in an event is the equal 
magnitude vector opposite to this vector sum of 
transverse energies.  Fixed points on the detector are 
described using polar coordinates ($r$,$\phi$,$z$) where 
$r$ is the radial distance from the beam axis, $\phi$ is 
the azimuthal direction about the beam axis, and $z$ is 
the distance from the detector center in the direction 
along the beam axis.  In some cases we also use a detector 
pseudorapidity variable, $\eta_{{\mathrm{det}}}$, to refer 
to fixed locations within the detector.  This variable is 
based on the standard definition of pseudorapidity given 
above where the angle $\theta$ is redefined in the context 
of a fixed location as $\theta = \arctan(r/z)$.  

\subsection{Tracking System}
\label{subsec:track}

All of the detectors in the inner tracking region have been replaced
for Run II.  The new silicon tracking system consists of three
concentric detectors located just outside the beam interaction region.
In combination, these detectors provide high resolution tracking
coverage out to $\absdeteta <$~2.  For the measurements presented
here, silicon tracking information is incorporated solely to aid in
the rejection of cosmic ray events from our muon samples.  The
relevant hit information comes from the Silicon Vertex Detector
(SVX-II) \cite{det:svx} which contains five layers of double-sided
micro-strip detectors at radii of 2.4 to 10.7~$\cm$ from the center of
the detector.  The SVX-II detector consists of three barrels divided
into 12 wedges in $\phi$.  In total, the three barrels cover roughly
45~$\cm$ along the $z$-axis on each side of the detector interaction
point.

The new open-cell drift chamber referred to as the Central Outer
Tracker (COT) \cite{det:cot,det:cot2} sits directly outside of the
silicon tracking detectors in the radial direction.  The measured
momenta and directions of the high $\pt$ lepton candidates in our
event samples are obtained from track reconstruction based solely on
COT hit information.  The chamber consists of eight superlayers of
310~$\cm$ length cells at radii between 40 and 132~$\cm$ from the beam
axis.  Each superlayer contains 12 layers of sense wires strung
between alternating layers of potential wires.  The wires in four of
the superlayers (axial layers) are strung to be parallel to the beam
axis, providing particle track reconstruction in the transverse plane.
In the other four superlayers (stereo layers), the wires are strung at
$\pm$~2 degree angles with respect to the beam axis to allow also for
particle tracking in the $z$-direction.  The two superlayer types are
alternated in the chamber within the eight radial layers starting with
the innermost stereo layer.  The COT chamber has over 30,000 readout
channels, roughly five times the number in the Run I tracking chamber
\cite{det:cotrun1}.  Particles traversing the central region of the
detector ($\absdeteta <$1) are expected to be measured by all eight
superlayers.

\begin{figure}
 \includegraphics*[width=8.5cm]{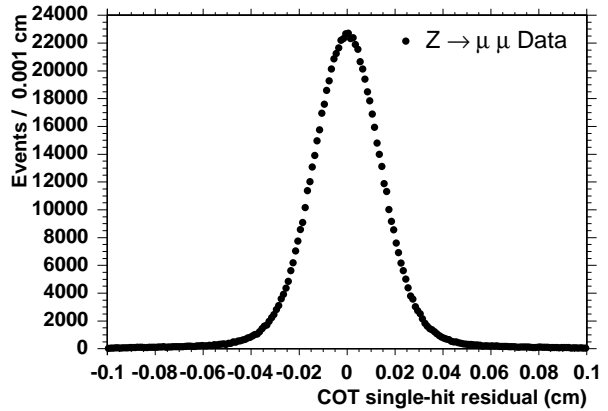}
 \caption{COT single-hit residual distribution obtained from
 $\zmm$ events.}
\label{fig:hitres}
\end{figure}

The COT is filled with a gas mixture of 50$\!\%$ argon and 50$\!\%$
ethane.  This mixture was chosen to ensure a fast drift velocity
($\sim$ 50 $\um/\ns$) compatible with the short interval between beam
bunch crossings and the expected rise in instantaneous luminosity.
The maximum drift distance in the chamber is 0.88~$\cm$ corresponding
to a drift time on the order of 200~$\ns$.  The single-hit resolution
in the chamber has been studied using the high $\pt$ muon tracks in
$\zmm$ candidate events.  The measured offset between the individual
hits associated with these muons and the reconstructed path of the
muon track is shown in Fig.~\ref{fig:hitres}.  Based on this
distribution, we measure a COT single-hit resolution of 180~$\um$.

\begin{figure}
\includegraphics*[width=8.5cm]{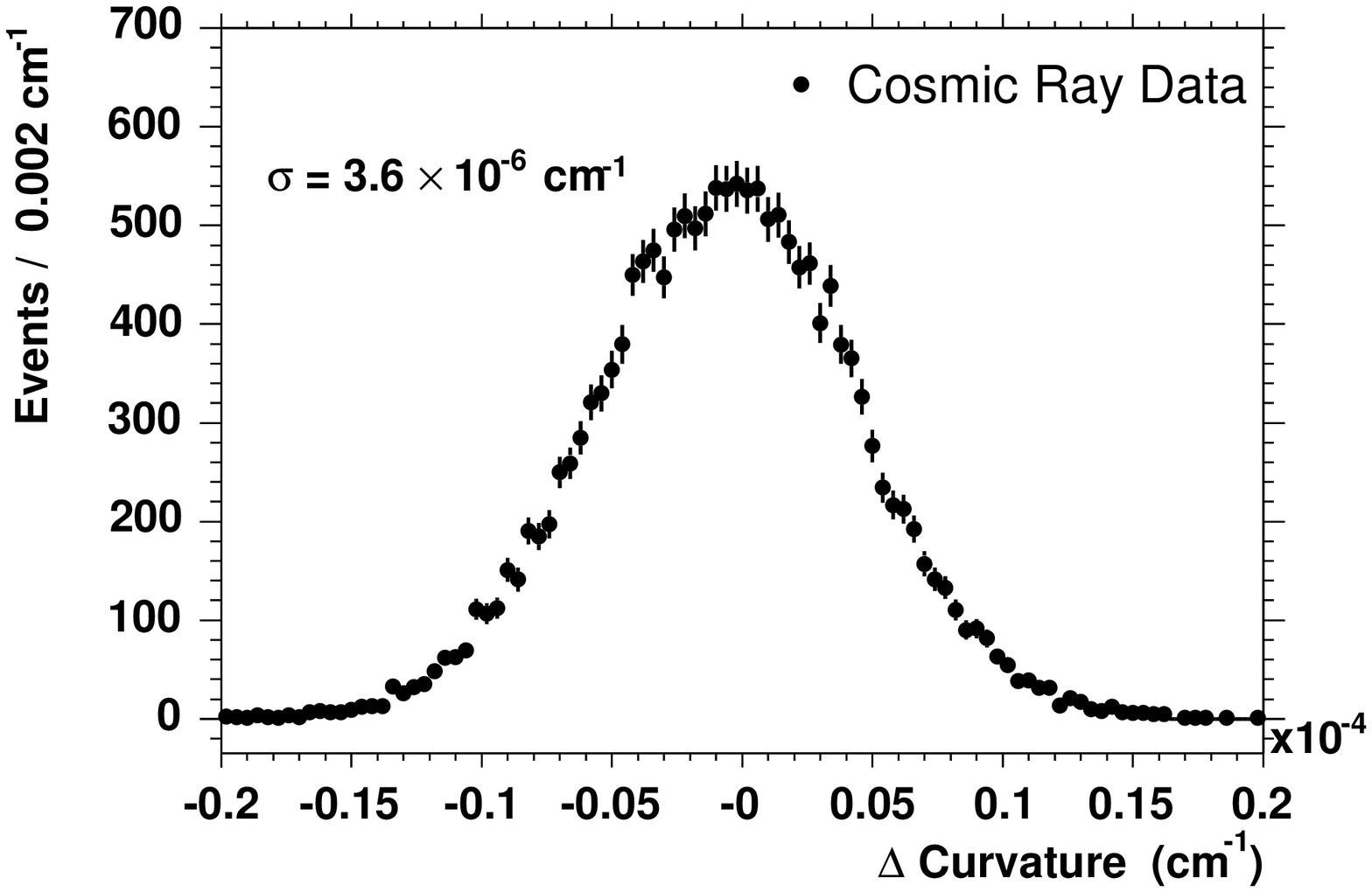} \includegraphics*[width=8.5cm]{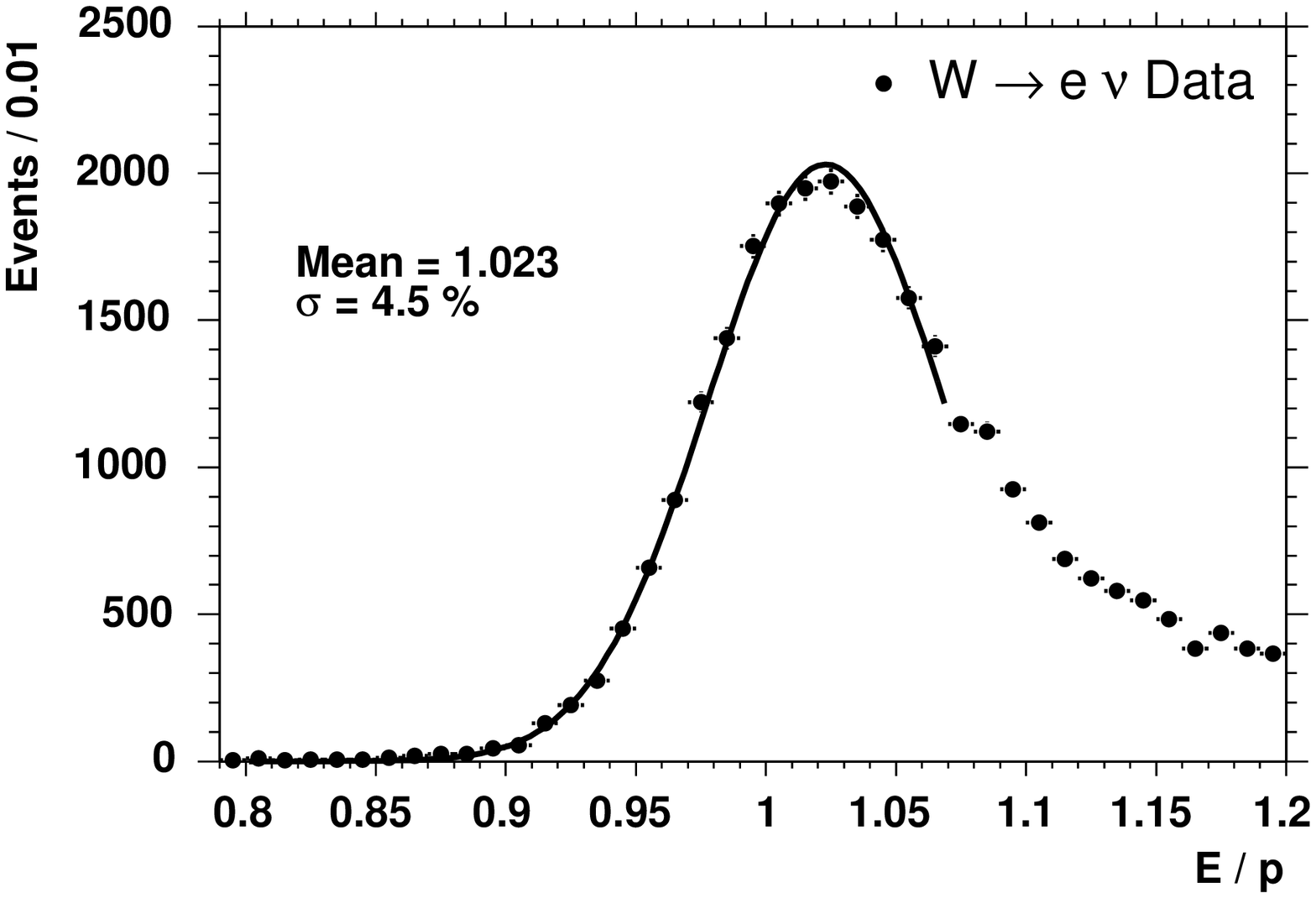}
\caption{Distribution of the difference in 
curvature for the two tracks associated with a cosmic ray event as
reconstructed by the COT, using cosmic ray data (top). Distribution of
the $E/p$ variable defined in Sec.~\ref{sec:evsel} for $\wenu$ events
(bottom).  The mean and $\sigma$ are obtained from the Gaussian fit in the
range $0.8 < E/p < 1.08$.}
\label{fig:curvres}
\end{figure}

The solenoid produces a 1.4~$\Tesla$ magnetic field inside the
tracking volume that is uniform to 0.1~$\!\%$ in the region
$|z|<$~150~$\cm$ and $|r|<$~150~$\cm$.  The transverse momentum of a
reconstructed track, $\pt$ (in $\GeV/c$), is determined from $\pt =
0.3qBr_{c}$ where $B$ (in $\Tesla$) is the magnetic field strength,
the total particle charge is $qe$ ($e$ is the magnitude of the
electron charge), and $r_{c}$ (in $\m$) is the measured radius of
curvature of the track.  The resolution of the COT track momentum
measurement decreases for high $\pt$ tracks which bend less in the
magnetic field.  The curvature resolution has been studied by
comparing the inward and outgoing track legs of reconstructed cosmic
ray events.  The difference in the measured curvature for the two
track legs in these events is shown on the top of
Fig.~\ref{fig:curvres}.  We determine a COT curvature resolution of
3.6~$\times$~10~$^{-6} \cm^{-1}$, estimated from the $\sigma$ of this
distribution divided by $\sqrt{2}$.  This corresponds to a momentum
resolution of $\sigma_{\pt}/\pt^2
\simeq$~1.7~$\times$~10~$^{-3}[\GeVC]^{-1}$.  The COT track momentum
resolution is also studied using the $E/p$ distribution (see
Sec.~\ref{sec:evsel}) of electron candidates in $\wenu$ events.  This
distribution is shown on the bottom of Fig.~\ref{fig:curvres}.  Since
the COT track momentum resolution measurement is less precise at high
$\pt$ than the corresponding calorimeter energy measurement, the
Gaussian width of this distribution for $0.8 < E/p < 1.08$ provides an
additional measure of the curvature resolution.  The resulting value
is in good agreement with that obtained from studying cosmic ray
events.

\subsection{Calorimeters}

Calorimeter modules used to measure the energy of both charged and
neutral particles produced in $\ppbar$ collisions are arranged around
the outer edges of the central tracking volume.  These modules are
sampling scintillator calorimeters with a tower based projective
geometry.  The inner electromagnetic sections of each tower consist of
lead sheets interspersed with scintillator, and the outer hadronic
sections are composed of scintillator sandwiched between sheets of
steel.  The CDF calorimeter consists of two sections: a central barrel
calorimeter ($\absdeteta<$1) and forward end plug calorimeters
(1.1$<\absdeteta<$3.64).  The scintillator planes in the central
barrel lie parallel to the beam line, while those in the forward end
plugs are arranged in the transverse direction.  The central barrel
consists of projective readout towers, each subtending 0.1 in
$\eta_{\mathrm{det}}$ and 15$^{\circ}$ in $\phi$.  Each end plug also
has projective readout towers, the sizes of which vary as a function
of $\eta_{\mathrm{det}}$ (0.1 in $\eta_{\mathrm{det}}$ and
7.5$^{\circ}$ in $\phi$ at $\absdeteta =$~1.1 to 0.5 in
$\eta_{\mathrm{det}}$ and 15$^{\circ}$ in $\phi$ at $\absdeteta
=$~3.64).

The central barrel section of the CDF calorimeter is unchanged from
Run I.  It consists of an inner electromagnetic (CEM) calorimeter and
an outer hadronic (CHA) calorimeter~\cite{det:cem}.  The end-wall
hadronic (WHA) calorimeter completes the coverage of the central
barrel calorimeter in the region 0.6~$< \absdeteta <$~1.0 and provides
additional forward coverage out to $\absdeteta =$~1.3~\cite{det:had}.
As part of the CDF Run II upgrade, the original gas calorimetry of the
end plug region ($\absdeteta >$~1.1) was replaced with scintillator
plate calorimetry using scintillator tiles read out by wavelength
shifting fibers embedded in the
scintillator~\cite{det:plug,det:plug2}.  The new design has an
improved sampling fraction and reduces forward gaps that existed in
the old gas calorimeter system.  The new plug electromagnetic (PEM)
calorimeter provides coverage in the 1.1~$< \absdeteta <$~3.6 region
and the new plug hadronic (PHA) calorimeter provides coverage in the
1.3~$< \absdeteta <$~3.6 region~\cite{det:pha}.  Both the PEM and PHA
incorporate the same polystyrene based scintillator and similar
photomultiplier tubes used in the CEM.

Calorimeter energy resolutions are measured using test-beam data.  The
measured energy resolutions for electrons in the electromagnetic
calorimeters are 14~$\!\%/\sqrt{\et}$ (CEM) and 16~$\!\%/\sqrt{E} ~\oplus$
1~$\!\%$ (PEM) \cite{det:tdr} where the units of $\et$ and $E$ are
$\GeV$.  We also measure the single-particle (pion) energy resolution
in the hadronic calorimeters to be 75~$\!\%$/$\sqrt{E}$ (CHA),
80~$\!\%$/$\sqrt{E}$ (WHA), and 80~$\!\%/\sqrt{E} ~\oplus$ 5~$\!\%$ (PHA)
\cite{det:tdr}.  The energy resolution in the electromagnetic
calorimeters is also determined using $\zee$ candidate events.  The
calorimeter energy scale is set so that the mean of the Gaussian fit
to the dielectron invariant mass peak is 91.1 $\GeV/c^{2}$.  This
procedure results in a CEM energy resolution of 13.5~$\!\%/\sqrt{\et}
~\oplus$ 1.5~$\!\%$, in good agreement with the test-beam
result~\cite{det:calresp}.  Jet energy resolution in the hadronic
calorimeter sections~\cite{det:jetres} is determined using photon-jet
balancing.  In events in which a photon recoils against a jet and no
other activity is observed, the transverse energies associated with
the two objects must be equal and opposite.  The photon energy
measured in the electromagnetic section of the calorimeter provides a
reference point against which the energy deposition associated with
the jet can be compared.  The resolution of the large component of jet
energy deposition in the hadronic calorimeters can be determined based
on this comparison.  The vast majority of hadronic particle showers
are completely contained within the calorimeter.  The combined
longitudinal depth of the central calorimeter module in interaction
lengths is roughly 5.5 $\lambda$ and the equivalent depth in the plug
modules is roughly 8.0 $\lambda$.  However, some small fraction of
hadronic particle showers does leak out from the back end of the
calorimeter, complicating muon identification.

Proportional chambers (CES) are embedded in the electromagnetic
section of the central barrel at a radiation length depth of roughly
6~$X_0$ corresponding to the region of maximum shower intensity for
electrons.  These chambers are used to measure the profile of a shower
and extract the location of the incident particle within a given
tower.  The increased shower position resolution provides additional
selection criteria for electron candidates based on track-shower
matching.  The chambers, two per calorimeter wedge, utilize wires in
the $r$-$\phi$ view and cathode strips in the $z$ view to determine
the three-dimensional position of each shower.  The resolution of the
CES position measurement in $r$-$\phi$ is roughly 0.2~$\cm$. Each
calorimeter module also has a second set of chambers (CPR) situated on
the front of the corresponding electromagnetic section which
presamples each shower to provide additional information useful in
electron identification and pion-photon separation.

The first layer of the plug electromagnetic calorimeter is used as a
preshower detector (PPR).  Its scintillator is polyvinyltoluene-based,
and it is twice as thick as the other sampling layers in the PEM.  It
has the same transverse segmentation as the PEM, but each scintillator
tile in the PPR is read out individually.  The PEM also has a shower
maximum detector (PES) embedded in it at a depth of $\sim$~6~$X_0$
\cite{det:pes}.  The PES consists of two layers of 5~$\mm$ wide
polyvinyltoluene-based scintillator strips, with each layer having a
45$^{\circ}$ crossing angle relative to the other.  The PES provides
coverage in the 1.1 $< \absdeteta <$ 3.5 region.

\subsection{Muon detectors}

In order for a muon to pass through the calorimeter and into 
the most central portion of the CDF muon detector ($\absdeteta \leq$ 
0.6), it must have a minimum $\pt$ on the order of 1.4 $\GeV/c$.  In 
order to reach the outer portion of the central detector or the more 
forward detectors (0.6 $< \absdeteta <$ 1.0), the muon is required 
to pass through an additional layer of steel absorber.  Muons with 
a momentum above 3.0 $\GeV/c$ are essentially 100~$\!\%$ efficient 
for traversing the steel absorber over the entire solid angle of the 
combined muon detector coverage.  The amount of energy deposited in 
the calorimeter by high $\pt$ muons produced in $\W$ and $\Z$ boson 
decays is observed to be Landau distributed with means of 0.3~$\GeV$ 
for deposits in the electromagnetic section and 2.0~$\GeV$ for those
in the hadronic section.  Reconstructed particle tracks in the COT 
matched to minimum ionizing-like energy deposits in the calorimeter 
are treated as ``stubless'' muon candidates even in cases where the 
tracks are not matched with any hits in the muon detectors.  The 
muon offline reconstruction forms stubs based on hit information 
in the muon detector and matches found stubs with the reconstructed 
COT tracks to determine our highest quality muon candidates.  This 
final set of muon candidates includes only a small percentage of 
non-muon fakes originating from other hadronic particles that are 
not fully contained within the calorimeter (hadronic punchthrough).  
Despite the fact that a non-negligible number of hadrons (on the 
order of 1 in 220) pass through the entire calorimeter, the majority 
of those that enter the muon detector are absorbed in the filtering 
steel and do not produce associated hits in the outer sections of 
the detectors.  Conversely, ``stubless'' muon candidates include a 
substantially larger fraction of non-muon fakes, and the presence 
of additional physics objects (such as a second higher quality muon) 
associated with these candidates is typically required to increase 
the purity of the sample. 
 
The CDF muon detector is made up of four independent detector systems
outside the calorimeter modules.  The Central Muon Detector (CMU)
\cite{det:muon} is mounted directly around the outer edge of the
central calorimeter module.  The CMU is an original Run I detector
component containing 2,304 single-wire drift chambers arranged in four
concentric radial layers.  The drift chamber wires are strung parallel
to the direction of the incoming beams, and wire pairs on layers 1 and
3 and layers 2 and 4 project radially back to the nominal beam
interaction point, allowing for a coarse $\pt$ measurement based on
the difference in signal arrival times on the two wires within a pair.
The CMU system provides symmetrical coverage in $\phi$ in the central
part of the detector ($\absdeteta \leq$~0.6).  The drift chambers have
been upgraded to operate in proportional mode (in Run~I these chambers 
were run in streamer mode).  Operating in this mode reduces the high 
voltage settings for the chambers and helps to prevent voltage sagging 
which is an issue due to the higher hit rates at Run II luminosities.  
Precision position measurements in the $\phi$ direction are made by 
converting signal arrival times into drift distances in the plane 
orthogonal to the wire direction.  The wires of cells in neighboring 
stacks are connected via resistive wires at the non-readout end of cells 
to also provide a coarse measurement of each hit position along the 
direction of the wire ($z$ coordinate).  The measurement is made by 
comparing time-over-threshold for the signals observed at the readout 
end of the two neighboring stacks.  The maximum drift time within a CMU 
cell is 800~$\ns$ which is longer than the 396~$\ns$ spacing between bunch 
crossings in the accelerator.  The ambiguity as to which beam crossing a 
particular CMU hit originates from is resolved in both the trigger and the 
offline reconstruction using timing information associated with a matched 
COT track and/or matching energy in the calorimeter.

The Central Muon Upgrade Detector (CMP) and Central Muon Extension
Detector (CMX) were also part of the CDF Run I configuration
\cite{det:newmuon}.  The individual wire drift chambers of these
detectors are identical except for their lengths along the direction
of the wire which is larger for CMP chambers.  These drift cells are 
roughly a factor of two wider than those in the CMU detector resulting 
in a longer maximum drift time of 1.8~$\us$.  Matching scintillator 
detectors (CSP,CSX) installed on the outer edges of these systems 
can in principle provide timing information to resolve the three 
beam-crossing ambiguity arising from the long drift time.  In practice, 
however, occupancies in these chambers are small enough at current 
luminosities to uniquely determine the appropriate beam-crossing from 
COT track matching.  CSX timing information is used in the trigger to 
eliminate out-of-time hits from the beam halo associated with particle 
losses in the accelerator tunnel, but information from the scintillator 
systems is not currently utilized in muon reconstruction in this analysis.  
The CMP/CMX drift chambers are also run in proportional mode.  The CMP
chambers are arranged in a box-like structure around the outside of
the CMU detector and an additional 3$\lambda$ of steel absorber which
is sandwiched between the two detectors.  The additional steel greatly
reduces hadronic punchthrough into the CMP chambers and allows for
cleaner muon identification.  A total of 1,076 drift cells arranged in
four staggered layers form the four-sided CMP structure which provides
additional coverage for the central part of the detector ($\absdeteta
\leq$~0.6) with variable coverage in $\phi$.  Drift cells in the CMX
detector are arranged in conical arrays of eight staggered layers to
extend muon coverage up to $\absdeteta \leq$~1.0.  The partial overlap
between drift tubes in the CMX conical arrangement allows for a rough
hit position measurement in the $z$ coordinate utilizing the different
stereo angles of each cell with respect to the beam axis.  The Run~I
configuration consisted of 1,536 drift cells arranged in four
120$^{\circ}$ sections providing coverage between $-45^{\circ}$ to
75$^{\circ}$ and 105$^{\circ}$ to 225$^{\circ}$ in $\phi$ on both ends
of the detector.  An additional 60$^{\circ}$ of CMX coverage on the
bottom of the detector at both ends has been added for Run~II, but
these new components were still being commissioned in early running
and are not utilized in the measurements reported here.  The Barrel
Muon Upgrade Detector (BMU) is another new addition for Run~II which
provides additional muon coverage in the regions
1.0~$<\absdeteta<$~1.5.  This new detector system was also being
commissioned in the initial part of Run~II and is not used in these
measurements.

\subsection{Cherenkov Luminosity Counters}

The small-angle Cherenkov Luminosity Counters (CLC) detector is used
to measure the instantaneous and integrated luminosity of our data
samples.  This detector system is an additional Run~II
upgrade~\cite{det:CLC_NIM} that allows for high-precision luminosity
measurements up to the highest expected instantaneous luminosities.

The CLC consists of two modules installed around the beampipe at each
end of the detector, which provide coverage in the regions 3.6~$<
\absdeteta <$~4.6.  Each module consists of 48 long, conical gas
Cherenkov counters pointing to the collision region. The counters are
arranged in three concentric layers of 16 counters each, around the
beam-pipe.  The counters in the two outer layers are about 1.8~$\m$
and those in the inner layer are 1.1~$\m$ long.  Each counter is made
of highly reflective aluminized Mylar with a light collector that
gathers the Cherenkov light into fast, radiation hard photomultiplier
tubes with good ultraviolet quantum efficiency.  The modules are
filled with isobutane gas at about 22 $\mathrm{psi}$ which is an
excellent radiator while having good ultraviolet transparency.

The Cherenkov light cone half-angle, $\theta_{\mathrm{c}}$, is 
$3.1^{\circ}$ corresponding to a momentum threshold for light 
emission of 9.3~$\MeVC$ for electrons and 2.6~$\GeVC$ for pions.  
The expected number of photoelectrons, $N_{\mathrm{pe}}$, for 
a single counter is given by $N_{\mathrm{pe}} = N_{\mathrm{o}} 
\cdot L \cdot sin^{2}\theta_{\mathrm{c}}$ where $L$ is the 
distance traversed by the particle in the medium and $N_{\mathrm{o}} 
= 370~\cm^{-1} \eV^{-1} \int \epsilon_{\mathrm{col}} (E) 
\epsilon_{\mathrm{det}} (E) dE$.  The $\epsilon_{\mathrm{det}}$ 
and $\epsilon_{\mathrm{col}}$ terms are defined as the light 
detection and collection efficiencies, respectively, and are 
functions of the energy $E$ of the Cherenkov photon (in $\eV$).  
Our design results in $N_{\mathrm{o}} \sim 200~\cm^{-1}$ 
corresponding to $N_{\mathrm{pe}} \sim 0.6/\cm$~\cite{det:CLC_NIM1}.

The details of the luminosity measurement are described in
Sec.~\ref{sec:data}.

\subsection{Trigger systems}
\label{subsec:trig}
 
The CDF trigger system~\cite{det:trig,det:l3} was redesigned for Run
II because of the changes in accelerator operating conditions. The
upgraded trigger system reduces the raw event rate in the detector
(the nominal 2.5~$\MHz$ beam crossing rate) to 75~$\Hz$, the typical
rate at which events can be recorded.

The corresponding event rejection factor of roughly $3 \times 10^{4}$ 
is obtained using a three-level system where each level is designed 
to provide sufficient rejection to allow for processing with minimal 
deadtime at the subsequent level.  The first level of the trigger 
system (Level~1) utilizes custom hardware to select events based on 
information in the calorimeters, tracking chambers, and muon detectors.  
Three parallel, synchronous hardware processing streams are used to 
create the trigger primitive data required to make the Level~1 decision.  
All detector data are fed into 6~$\us$ pipelines to allow for processing 
time required at Level~1.  The global Level~1 decision must be made and 
returned to the front-end detector hardware before the corresponding 
collision data reach the end of the pipeline.  Trigger decisions are 
made at the 2.5~$\MHz$ crossing rate, providing dead-time free operation.
    
One set of Level~1 hardware is used to find calorimeter objects
(electrons and jets) and calculate the missing transverse energy and
total transverse energy seen by the calorimeter in each event.  At
Level~1, electron and jet candidates are defined as single-tower
energy deposits above some threshold in the electromagnetic or
electromagnetic plus hadronic sections of trigger towers,
respectively.  Calorimeter energy quantities are calculated by summing
the transverse components of all single tower deposits assuming a
collision vertex of $z =$~0.  A second set of hardware is utilized to
select muon candidates from observed hits in the muon detector wire
chamber and scintillator systems.  A loose $\pt$ threshold is applied
based on differences in signal arrival times on pairs of projective
wires in the CMU and CMX chambers.  CMP primitives obtained from a
simple pattern finding algorithm using observed hits on the four drift
cell layers are matched to high $\pt$ CMU candidates, and CSX hits
within a certain time window consistent with collision-produced
particles are matched to CMX candidates.

An important element of the Run II CDF trigger upgrade is the third
set of hardware which identifies COT track candidates within the tight
Level~1 timing constraints.  The eXtremely Fast Tracker
(XFT)~\cite{det:xft} hardware examines hits on each axial superlayer
of the COT and combines them into track segments.  The found segments
on the different layers are then linked to form tracks.  The triggers
used to collect the data samples utilized in these measurements are
based on XFT tracks with reconstructed segments on all four COT axial
superlayers.  As discussed in more detail in Sec.~\ref{sec:evsel},
this requirement has a small effect on the geometrical acceptance for
lepton track candidates in our samples.  The hit requirement for XFT
track segments was changed from hits on 10/12 layers to hits on 11/12
layers during the data collection period for the samples used in these
measurements.  This change led to a few percent drop in the trigger
efficiency for high $\pt$ tracks but provided a substantial increase
in overall Level~1 event rejection.  The XFT hardware reports tracks
in 1.25$^{\circ}$ bins in $\phi$.  If more than one track is
reconstructed within a given $\phi$ bin, the track with the highest
$\pt$ is used.  The XFT feeds its lists of found tracks to another
piece of hardware known as the track extrapolation unit (XTRP).  The
XTRP determines the number of tracks above certain $\pt$ thresholds
and makes this information available for the global Level~1 decision.
It also extrapolates each track based on its reconstructed $\pt$ into
the calorimeter and muon detectors to determine into which $\phi$
slices of each system the track points based on the potential effects
of multiple scattering.  This information is passed to the calorimeter
and muon parts of the Level~1 trigger hardware in two sets of
2.5$^{\circ}$ $\phi$ bins corresponding to groups of tracks above two
programmable $\pt$ thresholds.  Using this information, tracks are
then matched to electron and muon primitives identified in those
pieces of the Level~1 hardware to produce the final lists of electron
and muon objects.
 
The final Level~1 trigger decision is made based on the number of
physics objects (electrons, muons, jets, and tracks) found by the
hardware and the calculated global calorimeter energy quantities.  The
maximum Level~1 event accept rate is roughly 20~$\KHz$ corresponding
to an available Level~2 processing time of 50 $\us$ per event.  Events
accepted at Level~1 are stored in one of four buffers in the front-end
readout hardware.  Multiple event buffers allow for additional Level~1
triggers to be accepted during the Level~2 processing of a previously
accepted event.  The Level~2 trigger system utilizes a combination of
dedicated hardware and modified commercial processors to select
events.  There are two main pieces of dedicated Level~2 hardware.  The
first is the cluster finder hardware which merges the observed energy
deposits in neighboring calorimeter towers to form clusters, and the
second is the silicon vertex tracking hardware (SVT) \cite{det:svt}
which uses silicon detector hit information to search for tracks with
displaced vertices.  These systems are asynchronous in that processing
time is dependent on the amount of input data associated with a given
event.  The output of these systems is passed to the global Level~2
processor along with the input data utilized in the Level~1 decision
and additional hit information from the CES to aid in low $\et$
electron selection.  The data are fed into the Level~2 processor board
and simple selection algorithms, optimized for speed, are run to
determine which events are passed to Level~3.  The processor board has
been designed to simultaneously read in one event while processing
another which streamlines operation and helps to keep data acquisition
deadtime at a minimum ~\cite{det:l3}.

Events selected by the Level~2 trigger hardware are read out of the
front-end detector buffers into the Level~3 processor farm.  The
current maximum Level~2 accept rate for events into Level~3 is roughly
300 $\Hz$.  Level~3 processors run a speed-optimized version of the
offline reconstruction code and impose loose sets of selection cuts on
the reconstructed objects to select the final 75 $\Hz$ of events which
are recorded for further processing.  The Level~3 processor
farm is made up of roughly 300 commercial dual processor computers
running Linux to allow for one second of processing time for each
event.  The software algorithms run at Level~3 take advantage of the
full detector information and improved resolution unavailable at the
lower trigger levels.  The Level~3 algorithms are based on full
three-dimensional track reconstruction (including silicon hit
information) which allows for tighter track matching with
electromagnetic calorimeter clusters and reconstructed stubs in the
muon detector for improved lepton identification.

\section{Data Samples and Luminosity}
\label{sec:data}
The $\wlnu$ and $\zll$ candidate event samples used to make the 
measurements reported here are selected from datasets collected 
using high $\et$ lepton trigger requirements.  Additional data 
samples used in the evaluation of efficiencies and backgrounds 
are discussed in further detail in the corresponding subsequent 
sections.  Here, we present the trigger requirements for events 
contained within the datasets from which our candidate samples 
are selected.  We also briefly describe data processing, the 
event quality criteria applied to our data samples, and the 
measurement of the integrated luminosities corresponding to our 
datasets. 

\subsection{Trigger requirements} 

The datasets used to select our candidate events are composed
of events collected with well-defined trigger requirements at 
each of the three levels within the CDF trigger architecture 
(see Sec.~\ref{sec:exp}).  The specific trigger requirements 
associated with the datasets used to make our measurements are
summarized here.  The measured efficiencies of these trigger 
requirements are presented in Sec.~\ref{sec:eff}.

\subsubsection{Central electron trigger}

The trigger requirements for the dataset used to select $\wenu$ 
and $\zee$ candidate events are described here.  Both candidate
samples are selected from central, high $\et$ electron triggered 
events, corresponding to the region $\absdeteta <$~1.0.

At Level~1, energies in physical calorimeter towers of
0.1$\times$15$^{\circ}$ in $\eta_{\mathrm{det}}$-$\phi$ space are
first summed into 0.2$\times$15$^{\circ}$ trigger towers. At least one
trigger tower is required to have $\et >$~8~$\GeV$ and the ratio of
the hadronic to electromagnetic energies in that tower,
$E_{\mathrm{had}}/E_{\mathrm{em}}$, must be less than 0.125 (for
measured $\et <$~14~$\GeV$).  In addition, at least one COT track with
$\pt >$~8~$\GeVC$ pointing in the direction of the tower must be found
by the XFT hardware.

A clustering algorithm is run at Level~2 to combine associated energy
deposits in neighboring calorimeter towers.  Adjacent ``shoulder''
towers with $\et >$~7.5~$\GeV$ are added to the central ``seed'' tower
found at Level~1.  The total $\et$ of the cluster is required to be
above 16~$\GeV$ and the $E_{\mathrm{had}}/E_{\mathrm{em}}$ ratio of
the cluster is required to be less than 0.125. The presence of an XFT
track with $\pt >$~8~$\GeVC$~ matched to the seed tower of the central
cluster is also reconfirmed.  Finally, in Level~3 an electromagnetic
cluster with $\et >$~18~$\GeV$ and $E_{\mathrm{had}}/E_{\mathrm{em}}
<$~0.125 must be found by the offline reconstruction algorithm.  A
track pointing at the cluster with $\pt >$~9~$\GeVC$ must also be
found by the full three-dimensional COT track reconstruction algorithm
run in the Level~3 processors.

At each level of the trigger, the rate of accepted events 
is significantly reduced.  At typical luminosities 
($\sim$~2.5~$\times$~10$^{31} \cm^{-2} \s^{-1}$), the accepted 
rate of events for the above trigger requirements are 25~$\Hz$, 
3~$\Hz$, and 1~$\Hz$ for Levels~1, 2, and~3, respectively.

\subsubsection{Central muon triggers}

The dataset used to select our $\wmnu$ and $\zmm$ candidate 
samples is made of events collected using two analogous 
sets of trigger requirements.  In the most central region of 
the detector ($\absdeteta <$~0.6), trigger requirements are
designed to select high $\pt$ muon candidates which deposit 
hits in both the CMU and CMP wire chambers.  An independent
but similar set of requirements is used to collect high $\pt$ 
candidates in the extended central region (0.6~$<\absdeteta <$~1.0) 
which produce hits in CMX wire chambers. 

The specific trigger requirements for the central region at Level~1
are matched hits in one or more CMU projective wire pairs with arrival
times within 124~$\ns$ of each other, a pattern of CMP hits on three
of four layers consistent in $\phi$ with the observed CMU hits, and a
matching COT track found by the XFT with $\pt >$~4~$\GeVC$.  For the
early part of the run period corresponding to our datasets we make no
additional requirements at Level~2, but for the later portion we
require at least one COT track with $\pt >$~8~$\GeVC$ in the list of
Level~1 XFT tracks passed to the Level~2 processor boards.  Because no
muon trigger information was available at Level~2 during this run
period, the higher $\pt$ track was not required to match the CMU or
CMP hits associated with the Level~1 trigger.  Finally for Level~3, a
reconstructed three-dimensional COT track with $\pt >$~18~$\GeVC$
matched to reconstructed stubs in both the CMU and CMP chambers is
required based on the offline reconstruction algorithms for muons.

The analogous trigger requirements for the extended central region at
Level~1 are matched hits in one or more CMX projective wire pairs with
arrival times within 124~$\ns$ of each other and a matching COT track
found by the XFT with $\pt >$~8~$\GeVC$.  For the latter part of our
data collection period, a matching hit in the CSX scintillator
counters consistent in time with a beam-produced particle is also
required to help reduce the trigger rate from non-collision
backgrounds.  No additional requirements are made at Level~2.  In
Level~3, a reconstructed three-dimensional COT track with $\pt
>$~18~$\GeVC$ matched to a reconstructed stub in the CMX chambers is
required based on the offline reconstruction algorithms for muons.

At typical luminosities ($\sim$~2.5~$\times$~10$^{31} \cm^{-2} 
\s^{-1}$), the accepted rate of events for the central trigger 
requirements are 30~$\Hz$, 4~$\Hz$, and 0.15~$\Hz$ for Levels~1, 2,
and~3, respectively.  For the extended central muon trigger
requirements, the corresponding rates are 2~$\Hz$, 2~$\Hz$, and
0.1~$\Hz$.

\subsection{Luminosity Measurement}
\label{subsec:lummeas}

The total integrated luminosity ($L$) is derived from the rate 
of the inelastic $\ppbar$ events measured with CLC, $R_{\ppbar}$, 
the CLC acceptance, $\epsilon_{\mathrm{CLC}}$, and the inelastic 
$\ppbar$ cross section at 1.96~$\TeV$, $\sigma_{\mathrm{in}}$, 
according to the expression
\begin{eqnarray}
L = {R_{\ppbar} \over \epsilon_{\mathrm{CLC}} \cdot \sigma_{\mathrm{in}}}.
\end{eqnarray}
The CLC acceptance is measured from tuned simulation and compared
against the value obtained from a second method that relies on 
both data and simulation through the formula
\begin{equation}
\epsilon_{\mathrm{CLC}} = \frac{N_{\mathrm{EW}}}{N_{\mathrm{CLC+Plug}}} 
\cdot \frac{N_{\mathrm{CLC+Plug}}}{N_{\mathrm{inelastic}}},
\end{equation}
where $N_{\mathrm{CLC+Plug}}$ is the number of inelastic events 
tagged by the CLC and plug calorimeter, $N_{\mathrm{EW}}$ is a 
subset of those which contain an east-west hit coincidence and 
pass the online selection criteria, and $N_{\mathrm{inelastic}}$ 
is the total number of inelastic collisions.  The fraction 
$N_{\mathrm{CLC+Plug}}/N_{\mathrm{inelastic}}$ is extracted from 
simulation while the ratio $N_{\mathrm{EW}}/N_{\mathrm{CLC+Plug}}$
is measured from data.  The acceptance calculated using this 
procedure is $\epsilon_{\mathrm{CLC}} =$~60.2~$\pm$~2.6~$\!\%$
which is in good agreement with the value obtained directly from 
simulation.

The value $\sigma_{\mathrm{in}} =$~60.7~$\pm$~2.4~$\mb$ is
obtained by extrapolating the combined result for the inelastic 
$\ppbar$ cross section at $\sqrt{s} =$~1.8~$\TeV$ based on CDF
and E811 measurements (59.3~$\pm$~2.3~$\mb$)~\cite{data:lumi}
to 1.96~$\TeV$.  Using these numbers, and restricting ourselves 
to runs with a good detector status, the total luminosity of 
our datasets is estimated to be 72.0~$\pm$~4.3~$\pbinv$.  The 
6~$\!\%$ quoted uncertainty is dominated by the uncertainty in 
the absolute normalization of the CLC acceptance for a single 
$\ppbar$ inelastic collision~\cite{data:lumi}.  The complete 
list of systematic uncertainties, including uncertainties from 
the inelastic cross section and luminosity detector, is given 
in Table~\ref{tab:lum}.

\begin{table}[t]
\caption{ Systematic uncertainties in the luminosity 
calculation based on the CLC measurement and the combined value of the
CDF and E811 inelastic cross section measurements at $\sqrt{s}
=$~1.80~$\TeV$ extrapolated to $\sqrt{s} =$~1.96~$\TeV$.  The total
uncertainty in the CLC measurement is dominated by the uncertainty in the 
CLC acceptance. The detector instability and calibration uncertainties are
components of the overall CLC measurement uncertainty and therefore
not included in the calculation of the total uncertainty.}
\centering{
\begin{tabular}{l r}
\hline
\hline
Effect                                          & Uncertainty Estimate  \\
\hline
Inelastic Cross Section                         & 4.0~$\!\%$            \\
CLC Measurement                                 & 4.4~$\!\%$            \\
Detector Instability                            & $<$ 2.0~$\!\%$        \\
Detector Calibration                            & $<$ 1.5~$\!\%$        \\
\hline 
Total Uncertainty                               & $\sim$~6.0~$\!\%$     \\
\hline
\hline
\end{tabular}
}
\label{tab:lum}
\end{table}

\section{Event selection}
\label{sec:evsel}
We search for $\W$ bosons decaying into a highly energetic 
charged lepton ($\ell = e, \mu$) and a neutrino, which is
identified via large $\met$ in the detector.  The $\zll$ 
($\ell = e, \mu$) events are selected based on the two 
energetic, isolated leptons originating from a single 
event vertex.  The two leptons produced in the decays 
are oppositely charged, and the charge information for 
leptons is included as part of the selection criteria 
when available.  The reconstructed dilepton invariant 
mass is also required to lie within a mass window 
consistent with the measured $\Z$ boson mass.

The complete set of selection criteria used to identify 
$\wlnu$ and $\zll$ events are described here.  As the
selection of $\W$ and $\Z$ bosons proceeds from lepton 
identification, we also describe in some detail the 
variables used to select good lepton candidates. 

\subsection{Track Selection}
\label{sec:tracksel}
The track quality requirements are common to electron and muon 
selection.  As the silicon tracking information is not vital 
to our measurements, we remove all silicon hits from the tracks
and refit them, including the position of the beamline in the 
transverse direction as an additional constraint in the fit.
The beamline position is measured independently for each run 
period contributing to our datasets using the reconstructed 
COT track data contained within events from that period.  The 
removal of silicon hits from tracks makes our measurements 
insensitive to the time-dependent efficiencies of the individual
pieces of the silicon detector and allows us to include data from 
run periods when the silicon detector was not operational.  The 
resulting beam-constrained COT tracks are used in the 
subsequent analysis work presented here.  All of the kinematic 
track parameters used in these analyses with the one exception 
of the $r-\phi$ track impact parameter variable, $d_0$, used in 
muon selection, are based on these beam-constrained 
COT tracks.

The reconstructed tracks obtained using the method described 
above have small residual curvature biases primarily due 
to COT misalignments that are not currently corrected for 
in our offline tracking algorithms.  We correct our track 
$\pt$ measurement for misalignment effects based on the 
observed $\phi$-dependence of the electron candidate $E/p$ 
distribution (see Sec.~\ref{sec:elesel}).  The form of the 
correction is
\begin{equation}
\frac{Q}{\pt^{\prime}} =
\frac{Q}{\pt} - 0.00037 - 0.00110 \cdot \sin(\phi+0.28),
\end{equation}
where $\pt^{\prime}$ and $\pt$ are the transverse momenta 
in $\GeVC$ of the corrected and uncorrected track,
respectively, $Q$ is the charge of the track, and $\phi$ 
is given in radians.

We apply additional selection criteria on our reconstructed 
tracks to ensure that only high-quality tracks are assigned 
to lepton candidates.  Each track is required to pass a 
set of minimum hit criteria.  The reconstructed tracks are 
required to have a minimum of seven out of twelve possible 
hits on at least three of four axial and stereo superlayers
within the COT.  The minimum hit criteria for reconstructed 
tracks is less restrictive than that used to select Level~1
trigger track candidates (see Sec.~\ref{subsec:trig}) to 
ensure high selection efficiencies for both triggerable 
and non-triggerable track candidates in our events.  In 
addition, to restrict ourselves to a region of high track 
reconstruction efficiency, we require the $z$ coordinate 
of the lepton track intersection with the beam axis in the 
$r-z$ plane, $z_0$, be within 60 $\cm$ of the center of 
the detector.  

To help reduce real muon backgrounds from cosmic rays and $\pi/K$
decays, we impose additional quality requirements on muon track
candidates.  For muon track candidates only, we incorporate silicon
hit information in track reconstruction when available to calculate a
more precise value for the $r-\phi$ impact parameter of the track,
$d_0$.  Cosmic ray muons and muons produced in $\pi/K$ decays are less
likely to point back to the event vertex and therefore will typically
have larger measured impact parameters.  We apply different cuts on
the $d_0$ of muon track candidates depending on whether or not the
tracks contain any silicon hit information; $|d_{0}| <$ 0.2 $\cm$ for
tracks with no silicon hits and $|d_{0}| <$ 0.02 $\cm$ for tracks with
silicon hits.  We also make a requirement on the quality of the final
COT beam-constrained track fit for muon candidates.  The track fit for
muon backgrounds not originating from the event vertex will typically
be worse when the additional constraint of the beamline position is
included.  For muon track candidates we require that
$\chi^2/n_{\mathrm{df}} <$ 2.0 where $n_{\mathrm{df}}$ is the number
of degrees of freedom in the track fit (the number of hits on the
fitted track minus the five free parameters of the fit).

\begin{figure}
\includegraphics*[width=8.5cm]{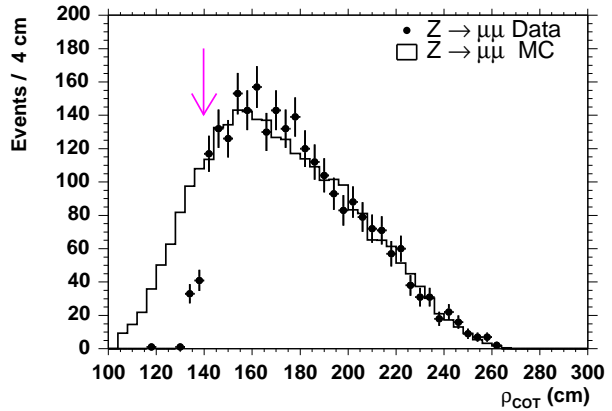}
\caption{The COT exit radius for CMX muons in $\zmm$ 
candidate events.  The points are the data and the histogram is
simulation.  The selected CMX muons from data events are required to
satisfy the high $\pt$ muon trigger criteria, but no trigger
requirement is made on the muons selected from simulation.  The two
histograms are normalized to have the same number of events over the
region 150~$\cm < \rho_{\mathrm{COT}} <$~280~$\cm$.  The arrow
indicates the location of the muon track selection cut made on the
$\rho_{\mathrm{COT}}$ variable.}
\label{fig:cotexitradius}
\end{figure}

We additionally restrict muon track candidates in $\theta$ 
to ensure that the tracks lie in a fiducial region of high 
trigger and reconstruction efficiency well-modeled by our 
detector simulation.  We require that each muon track passes 
through all eight COT superlayers by making a minimum 
requirement on the exit radius of the track at the endplates 
of the COT tracking chamber.  The exit radius is defined as
\begin{equation}
\rho_{\mathrm{COT}} = (z_{\mathrm{COT}} -z_0) \cdot \tan\theta,
\end{equation}
where $z_{\mathrm{COT}}$ is the distance of the COT endplates from 
the center of the detector ($155$~cm for tracks with $\eta >$~0 
and $-155$~cm for those with $\eta <$~0).  Here, $\eta$ and $\theta$ 
are the previously defined pseudorapidity and polar angle of the 
track with respect to the directions of the colliding beams.  A
comparison of the $\rho_{\mathrm{COT}}$ distribution for CMX muons
from $\zmm$ candidate events in data and Monte Carlo (MC) simulation
(see Sec.~\ref{sec:acc}) is shown in Fig.~\ref{fig:cotexitradius}.
The distributions do not match in the region $\rho_{\mathrm{COT}}
<$~140~$\cm$ due to a loss of data events in this region originating
from the XFT track trigger requirements, at least ten (or eleven) hits
out of a possible twelve for each of the four axial COT superlayers,
which is not accounted for in the simulation.  Based on this
comparison, we require $\rho_{\mathrm{COT}} >$~140~$\cm$ for muon
track candidates.

\begin{figure*}
 \includegraphics*[width=8.cm]{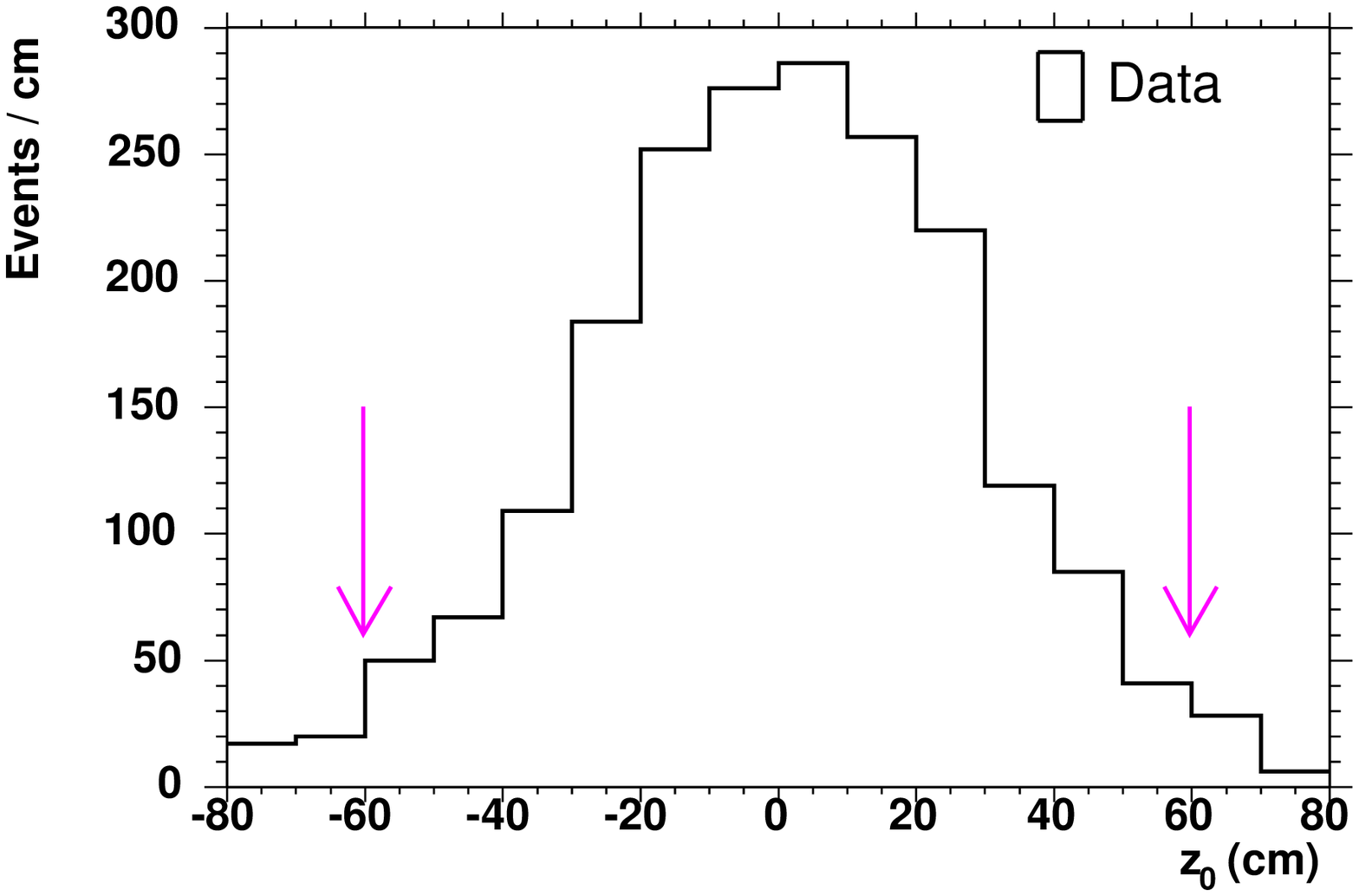} \includegraphics*[width=8.cm]{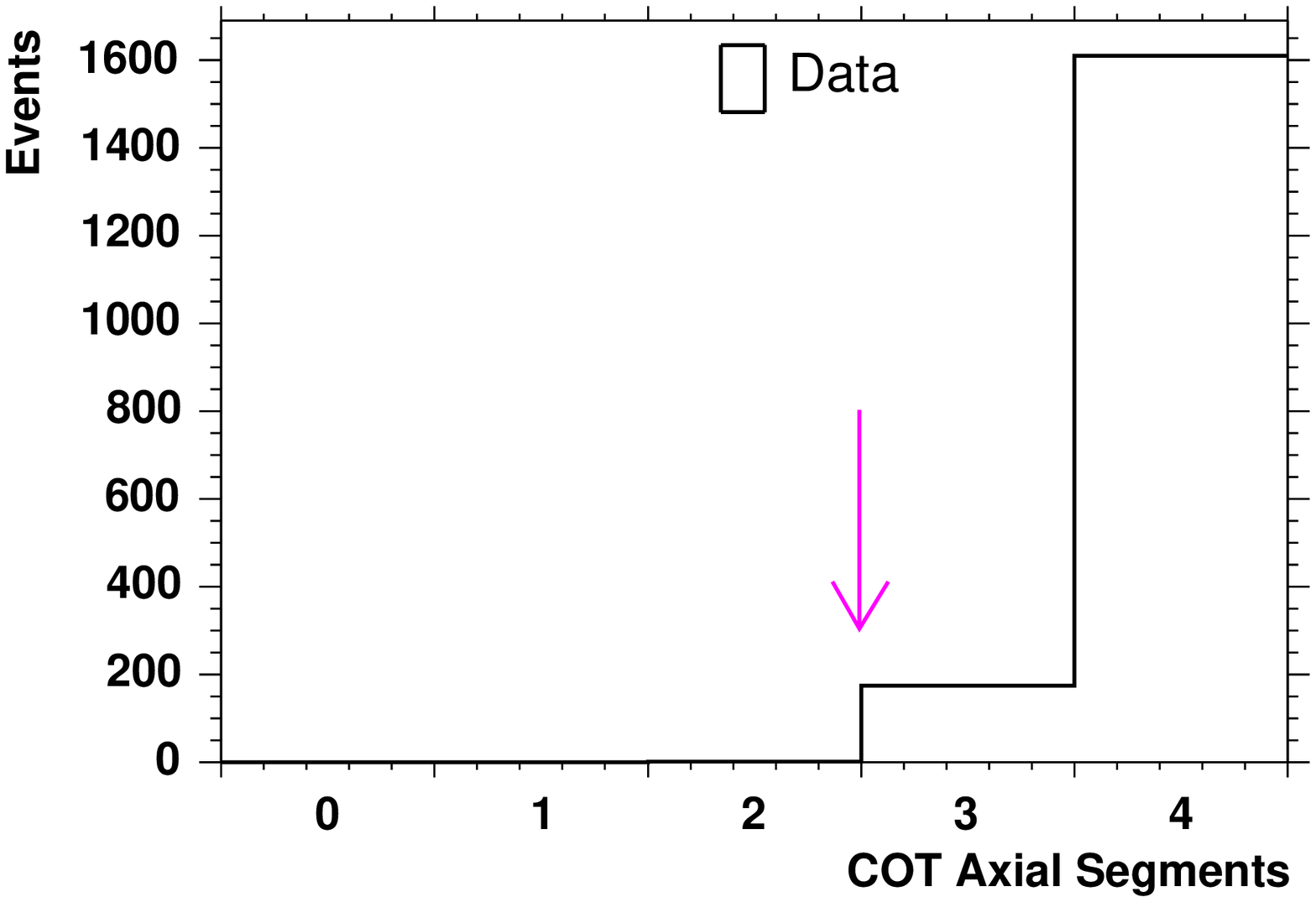} \includegraphics*[width=8.cm]{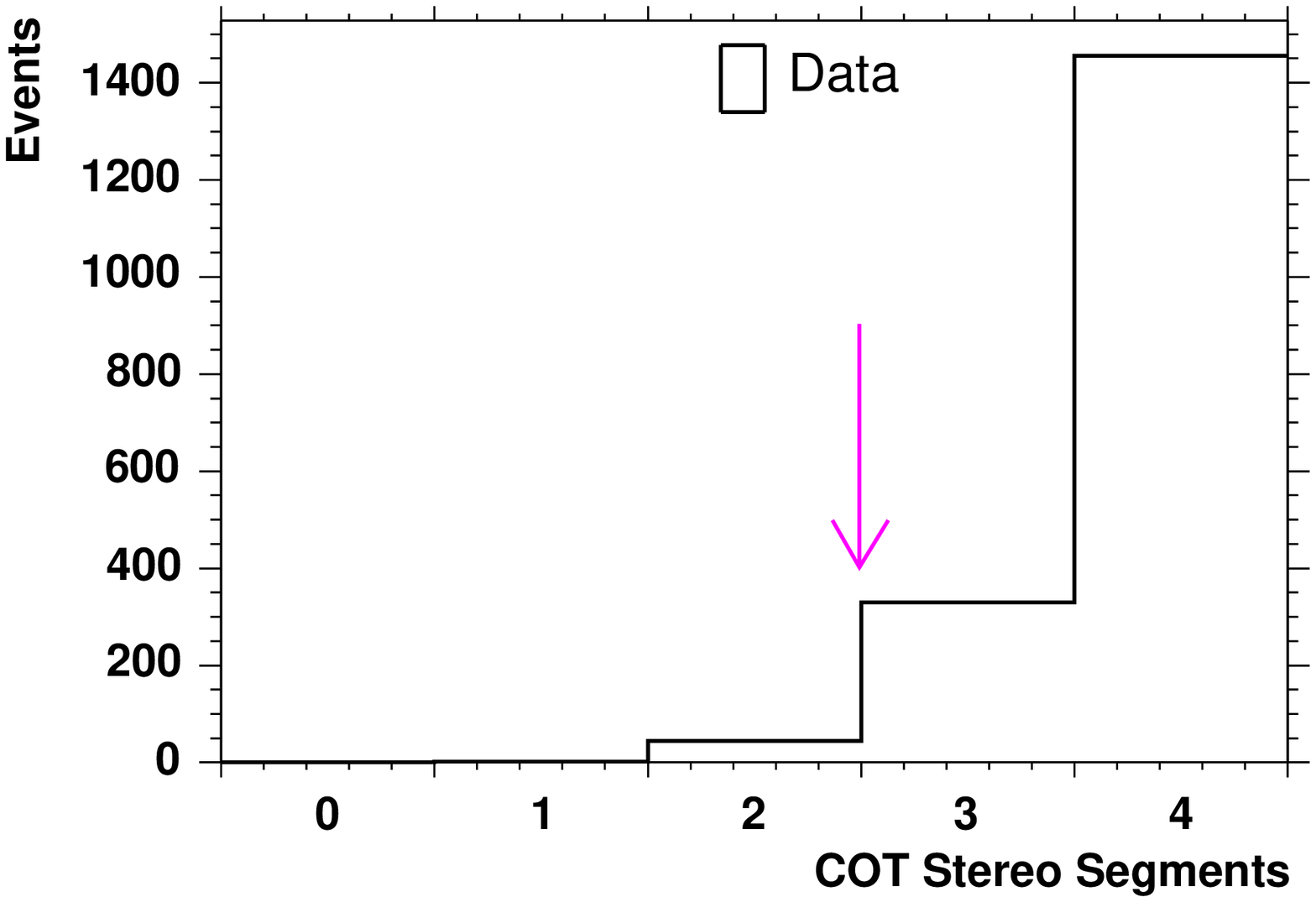}
\caption{Distributions of the $z_0$ and number of 
axial and stereo COT superlayers contributing seven 
or more hits.  These track quality variables are from unbiased, 
second lepton legs of $\zll$ candidate events in 
data.  The arrows indicate the locations of selection 
cuts applied on these variables.}
\label{fig:trkqual}
\end{figure*}

\begin{figure*}
 \includegraphics*[width=8.cm]{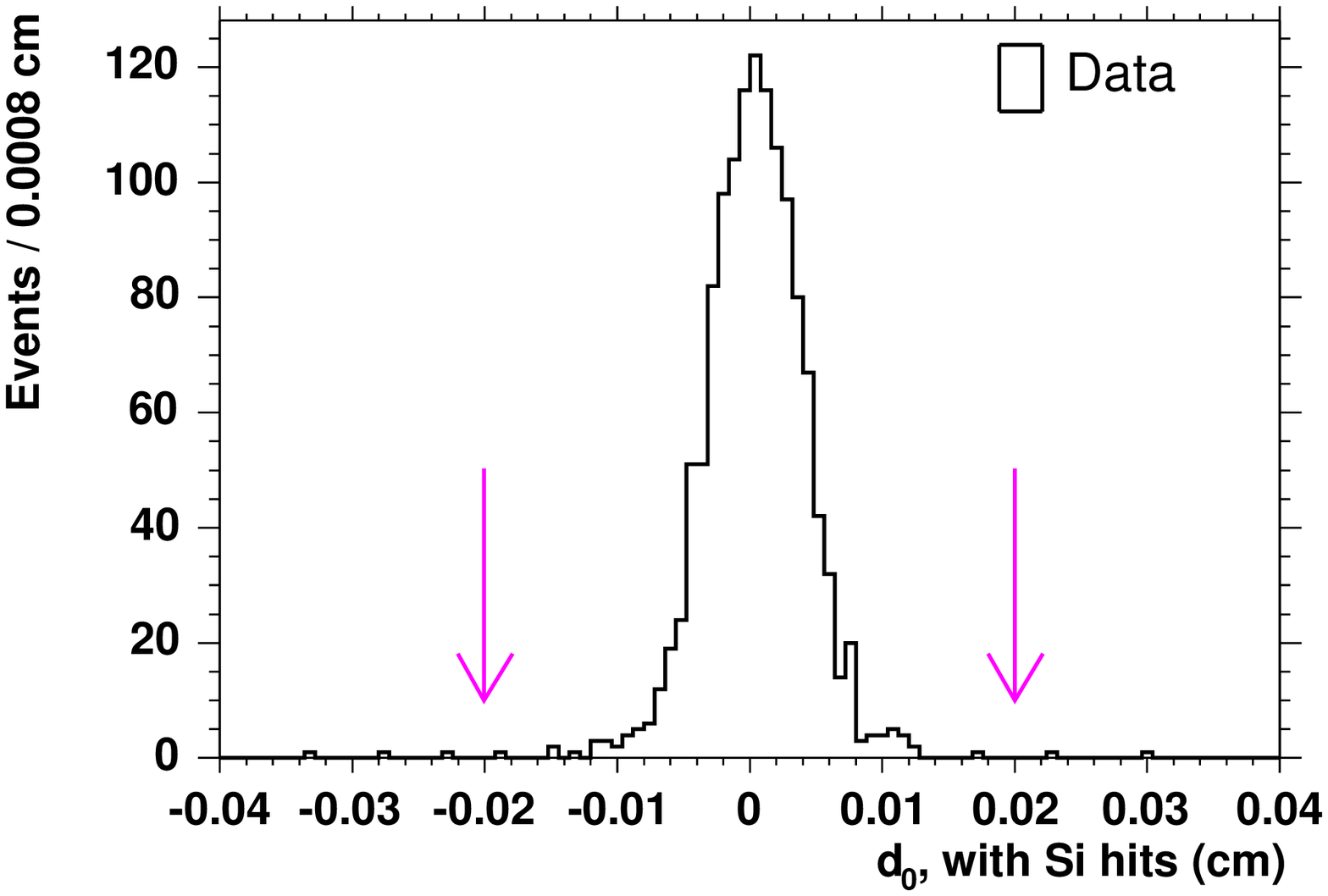} \includegraphics*[width=8.cm]{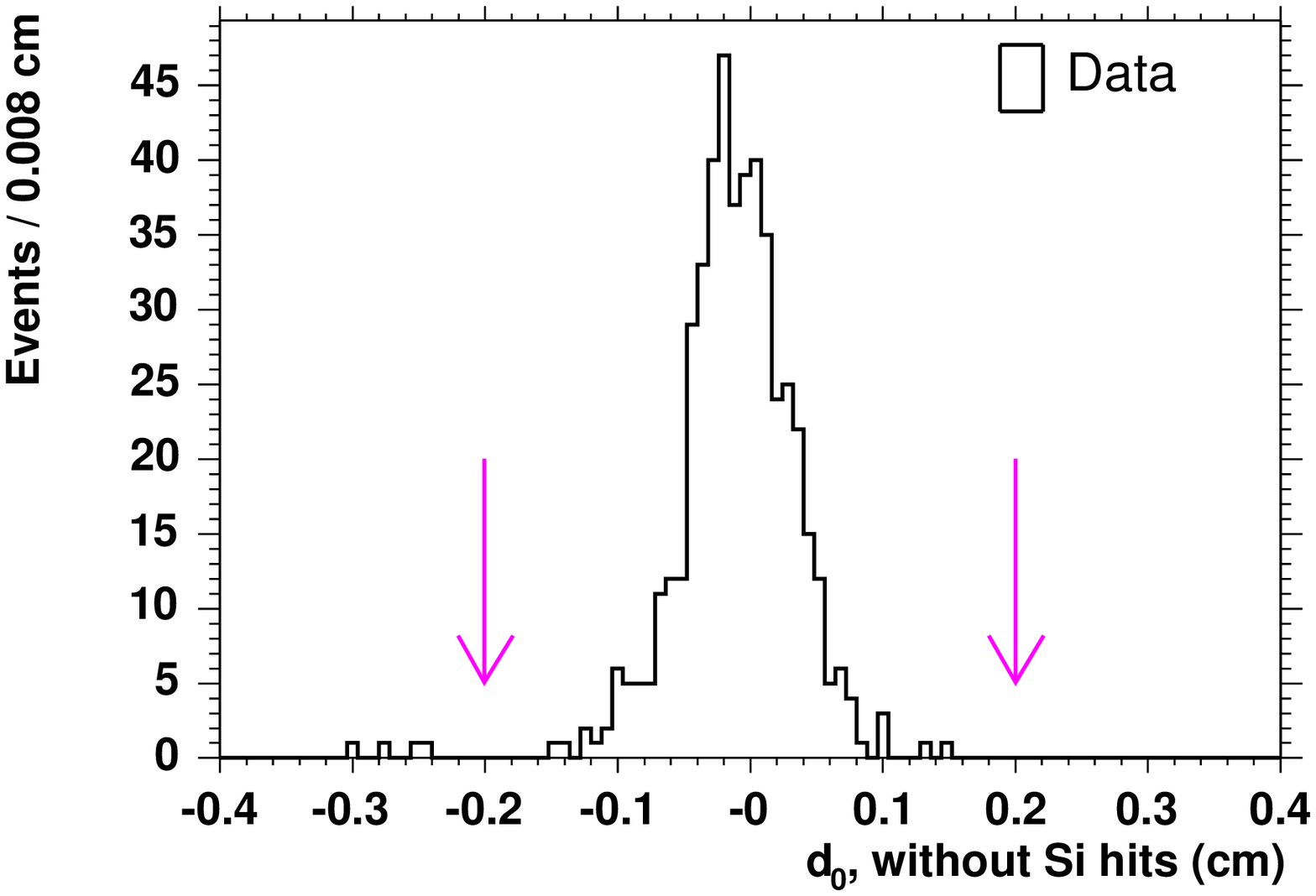} \includegraphics*[width=8.cm]{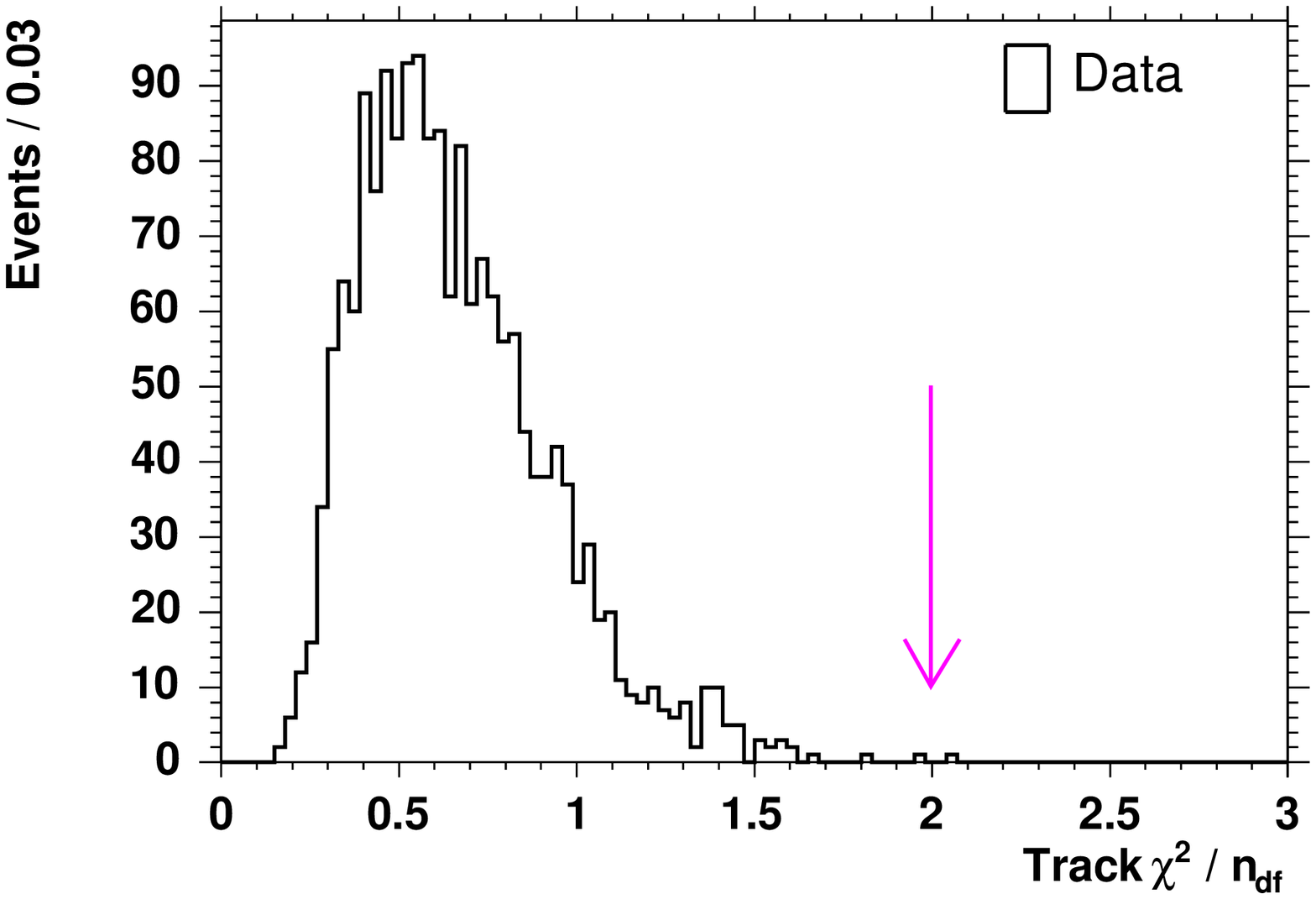} \includegraphics*[width=8.cm]{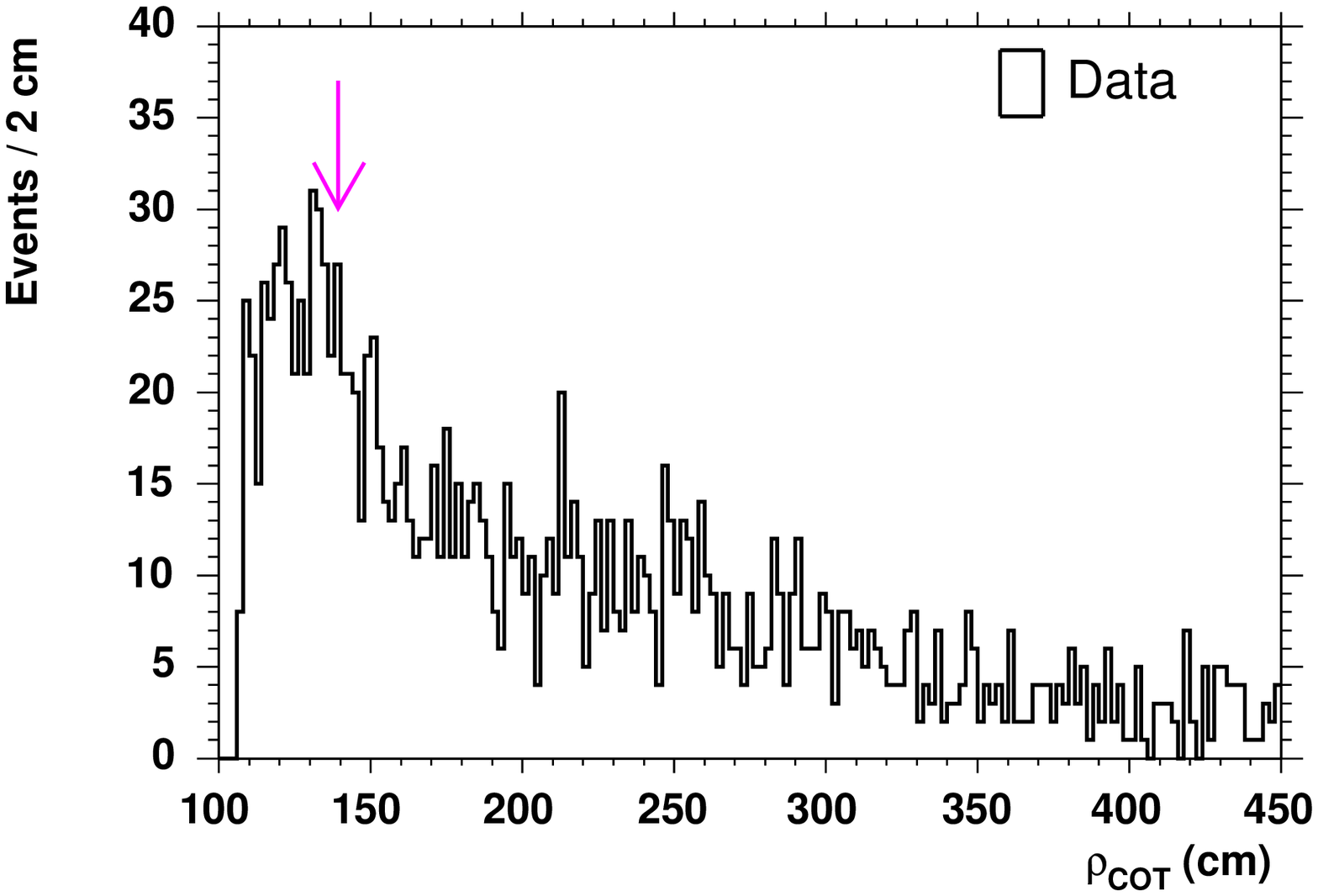}
\caption{Distributions of the $d_0$ (with and without 
attached silicon hits), $\chi^{2}/n_{\mathrm{df}}$, and 
$\rho_{\mathrm{COT}}$.  These track quality variables are for muons 
from unbiased, second muon legs of $\zmm$ candidate
events in data.  The arrows indicate the locations of 
selection cuts applied on these variables.}   
\label{fig:trkqualmu}
\end{figure*}

Track selection requirements are summarized in
Table~\ref{tab:trackqual}.  Distributions of the track 
quality variables used in the selection of all lepton 
tracks are shown in Fig.~\ref{fig:trkqual}, and those 
used solely in the selection of muon tracks are shown 
in Fig~\ref{fig:trkqualmu}.  The distributions are 
constructed from second, unbiased lepton legs in $\zll$ 
candidate data events.  Based on these distributions, 
we expect the measured inefficiencies of our track 
selection criteria (see Sec.~\ref{sec:eff}) to be on 
the order of a few percent.    

\begin{table}
\caption{Summary of track selection requirements.}
\begin{tabular}{l r}
\hline
\hline
Variable                     & Cut                             \\  
\hline
All Tracks:                  &                                 \\
$\#$ Axial COT Superlayers   & $\ge$~3 with $\ge$~7 hits       \\
$\#$ Stereo COT Superlayers  & $\ge$~3 with $\ge$~7 hits       \\
$|z_0|$                      & $<$ 60~$\cm$                    \\
\hline
Muon Tracks:                 &                                 \\ 
$|d_{0}|$                    & $<$ 0.2~$\cm$ (no silicon hits) \\
$|d_{0}|$                    & $<$ 0.02~$\cm$ (silicon hits)   \\
$\chi^{2}/n_{\mathrm{df}}$   & $<$ 2.0                         \\
$\rho_{\mathrm{COT}}$        & $>$ 140~$\cm$                   \\
\hline
\hline
\end{tabular}
\label{tab:trackqual}
\end{table}

\subsection{Electron Selection}
\label{sec:elesel}
Electron candidates are reconstructed in either the 
central barrel or forward plug calorimeters.  The 
clustering algorithms and selection criteria used to 
identify electrons in the two sections are different, 
as we do not make use of tracking information in the 
forward detector region ($\absdeteta >$~1) where 
standalone track reconstruction is less reliable due 
to the smaller number of available tracking layers. 
Here, we discuss the specific identification criteria 
for both central and plug electrons.   

\subsubsection{Central Electron Identification}

Electron objects are formed from energy clusters in neighboring
towers of the calorimeter.  An electron cluster is made from 
an electromagnetic seed tower and at most one additional tower 
that is adjacent to the seed tower in $\eta_{\mathrm{det}}$ 
and within the same $\phi$ wedge.  The seed tower must have 
$\et >$~2~$\GeV$ and a reconstructed COT track which extrapolates 
to that tower.  The hadronic energy in the corresponding towers 
is required to be less than 0.125 times the electromagnetic 
energy of the cluster. 

Electron candidates for these measurements must lie within the
well-instrumented regions of the calorimeter.  The cluster position
within the calorimeter is determined by the location of the associated
CES shower.  The CES shower must lie within 21~$\cm$ of the tower
center in the $r-\phi$ view for the shower to be fully contained
within the active region.  We also exclude electrons reconstructed in
the region where the two halves of the central calorimeter meet ($|z|
<$ 9~$\cm$) and the outer half of the most forward CEM towers ($|z|
>$~230~$\cm$) where there is substantial electron shower leakage into
the hadronic part of the calorimeter.  Finally, we exclude events in
which the electron is reconstructed near the uninstrumented region
surrounding the cryogenic connections to the solenoidal magnet
(0.77~$<
\eta_{\mathrm{det}} <$~1.0, 75$^{\circ}<\phi<$~90$^{\circ}$, 
and $|z| >$~193~$\cm$).

The selection requirements listed in Table~\ref{tab:eidcuts} are
applied to electron candidates in the well-instrumented regions 
of the central calorimeter.  We cut on the ratio of the hadronic
to electromagnetic energies, $E_{\mathrm{had}}/E_{\mathrm{em}}$, 
for the candidate clusters.  Electron showers are typically 
contained within the electromagnetic calorimeter, while hadron 
showers spread across both the hadronic and electromagnetic 
sections of the calorimeter.  We require $E_{\mathrm{had}}
/E_{\mathrm{em}} <$ 0.055 $+$ 0.00045 $\cdot E$ where $E$ 
is the total energy of the cluster in $\GeV$.  The linear term in 
our selection criteria accounts for the increased shower leakage 
of higher-energy electrons into the hadronic calorimeter sections.

We also cut on the ratio of the electromagnetic cluster transverse
energy to the COT track transverse momentum, $E/p$.  This ratio is
nominally expected to be unity, but in cases where the electron 
radiates a photon in the material of the inner tracking volume, the 
measured momentum of the COT track can be less than the measured 
energy of the corresponding cluster in the calorimeter.  In cases
where the electron is highly energetic, the photon and electron
will be nearly collinear and are likely to end up in the same 
calorimeter tower.  The measured COT track momentum will, however, 
correspond to the momentum of the electron after emitting the photon 
and thus be smaller than the original electron momentum.  We require 
$E/p <$ 2.0 which is efficient for the majority of electrons which 
emit a bremsstrahlung photon.  Since this cut becomes unreliable for 
very large values of track $\pt$, we do not apply it to electron 
clusters with $\et >$ 100~$\GeV$.

The lateral shower profile variable, $L_{\mathrm{shr}}$
\cite{det:dettop}, is used to compare the distribution of adjacent CEM
tower energies in the cluster as a function of seed tower energy to
shapes derived from electron test-beam data.  We also perform a
$\chi^{2}$ comparison of the CES lateral shower profile in the $r-z$ view to
the profile extracted from the electron test-beam data.  For central
electrons, we require $L_{\mathrm{shr}} <$ 0.2 and
$\chi^{2}_{\mathrm{strips}} <$ 10.0.

Since central electron candidates include a COT track, we can 
further reduce electron misidentification by cutting on track-shower 
matching variables. We define $Q \cdot \Delta x$ as the distance in 
the $r-\phi$ plane between the extrapolated beam-constrained COT 
track and the CES cluster multiplied by the charge of the track to 
account for asymmetric tails originating from bremsstrahlung radiation.  
The variable $\Delta z$ is the corresponding distance in the $r-z$ 
plane.  We require $-$~3.0~$\cm$ $< Q \cdot \Delta x <$ 1.5~$\cm$ 
and $|\Delta z| <$ 3.0~$\cm$.

\begin{figure*}[t]
\includegraphics*[width=14.cm]{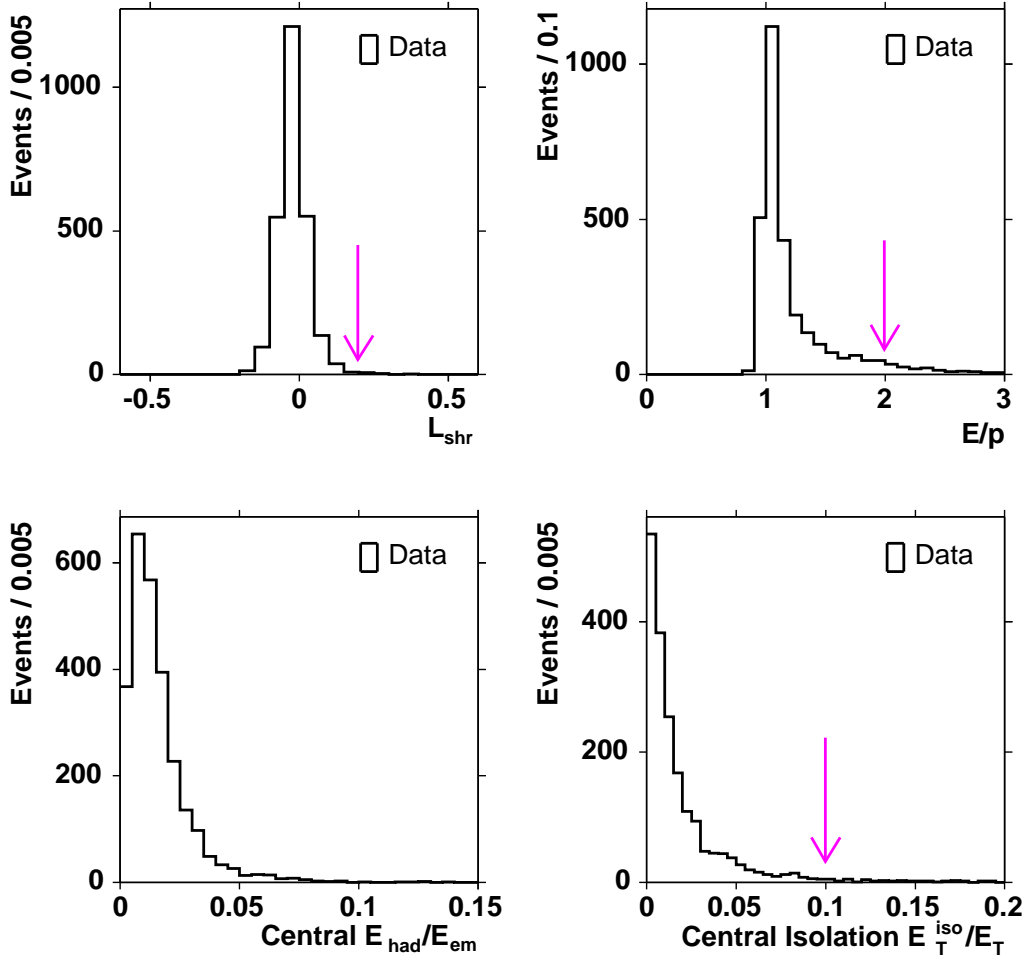}
\caption{Distributions of $L_{\mathrm{shr}}$, 
$E/p$, $E_{\mathrm{had}}/E_{\mathrm{em}}$, and
$\et^{\mathrm{iso}}/\et$ (see Sec.~\ref{sec:wsel}) central calorimeter
electron selection variables from unbiased, second electron legs of
$\zee$ candidate events in data.  The arrows indicate the locations of
selection cuts applied on these variables.  No arrow is shown on the
$E_{\mathrm{had}}/E_{\mathrm{em}}$ distribution since the cut on this
variable is dependent on the electron energy.}
\label{fig:idcalvar}
\end{figure*}

\begin{figure*}[t]
\includegraphics*[width=14.cm]{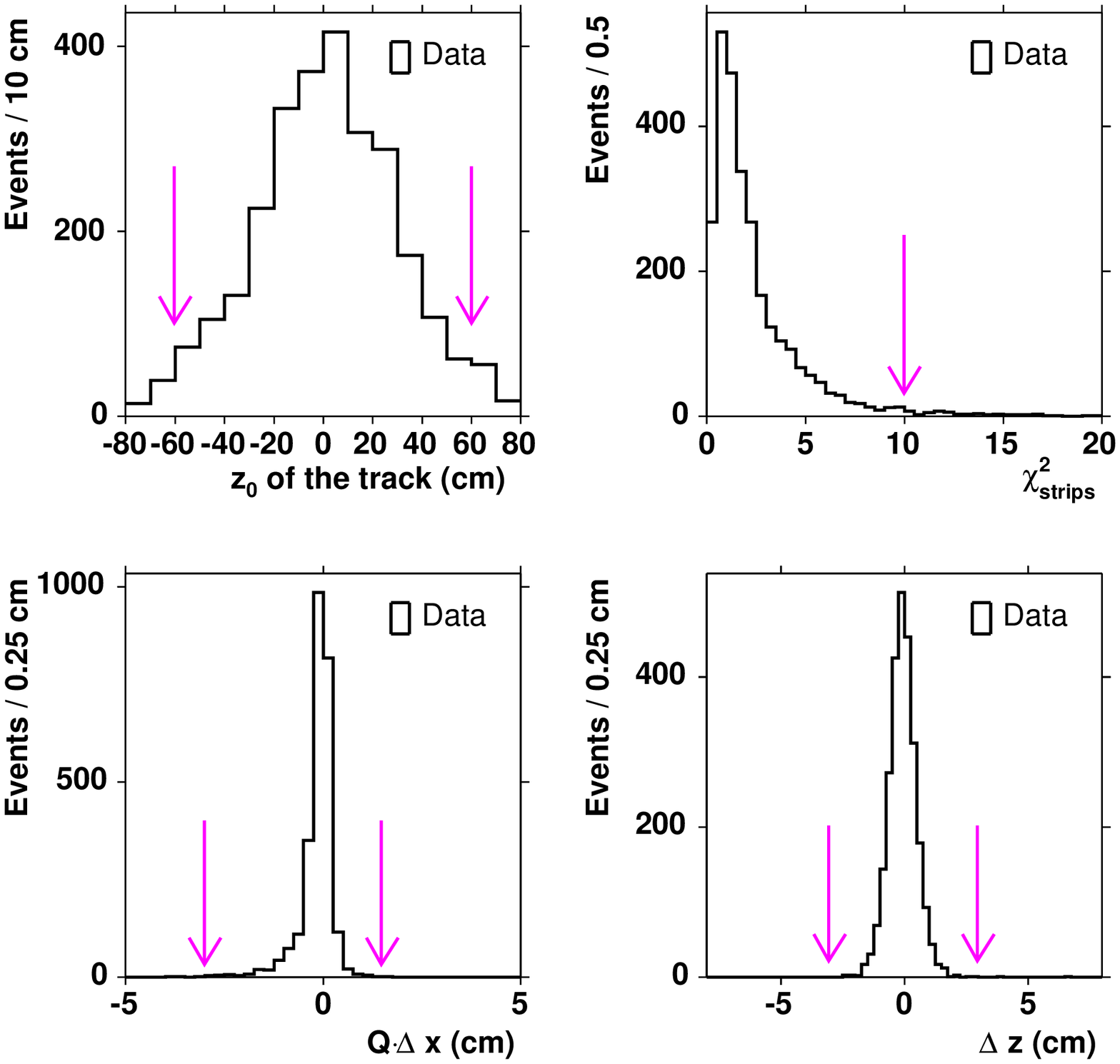}
\caption{Distributions of the $z_0$, $\chi^{2}_{\mathrm{strips}}$,
$Q \cdot \Delta x$, and $\Delta z$ central calorimeter electron 
selection variables from unbiased, second electron legs of 
$\zee$ candidate events in data.  The arrows indicate the 
locations of selection cuts applied on these variables.}
\label{fig:idcaltrkvar}
\end{figure*}

Distributions of central electron identification variables are shown
in Figs.~\ref{fig:idcalvar} and~\ref{fig:idcaltrkvar}.  The plotted 
electron candidates are the unbiased, second electron legs in $\zee$
events in which both electrons are reconstructed within the central 
calorimeter and the first electron is found to satisfy the full set 
of identification criteria.  Based on these distributions, we expect 
a high efficiency for our central electron selection criteria
(see Sec.~\ref{sec:eff}).

\begin{table}[t]
\caption{Calorimeter variables and electron
identification requirements.}
\begin{center}
\begin{tabular}{ l r }  
\hline 
\hline
Variable                           & Cut                             \\ 
\hline
Central                            & $\absdeteta <$ 1.0              \\ 
\hline
$E_{\mathrm{had}}/E_{\mathrm{em}}$ & $<$ 0.055 $+$ 0.00045 $\cdot E [$\GeV$]$ \\
$E/p$ (for $\et <$ 100 $\GeV$)     & $<$ 2.0                         \\
$L_{\mathrm{shr}}$                 & $<$ 0.2                         \\
$Q \cdot \Delta x$                 & $>$ -3.0~$\cm$, $<$ 1.5~$\cm$   \\
$|\Delta z|$                       & $<$ 3.0~$\cm$                   \\
$\chi^{2}_{\mathrm{strips}}$        & $<$ 10.0                        \\ 
\hline
Plug                               & 1.2 $< \absdeteta <$ 2.8        \\ 
\hline
$E_{\mathrm{had}}/E_{\mathrm{em}}$ & $<$ 0.05                        \\
$\chi^2_{\mathrm{PEM}}$            & $<$ 10.0                        \\ 
\hline 
\hline 
\end{tabular}
\label{tab:eidcuts}
\end{center}
\end{table}

\subsubsection{Plug Electron Identification}

Electron candidate clusters in the plug calorimeter are made  
from a seed tower and neighboring towers within two towers 
in $\eta_{\mathrm{det}}$ and $\phi$ from the seed tower.  As 
for central electrons, the hadronic energy of the cluster 
is required to be less than 0.125 times the electromagnetic 
energy.  We also require plug electrons to be reconstructed 
in a well-instrumented region of the detector, defined as 
1.2~$< \absdeteta <$~2.8.  

\begin{figure*}[t]
\includegraphics*[width=17.cm]{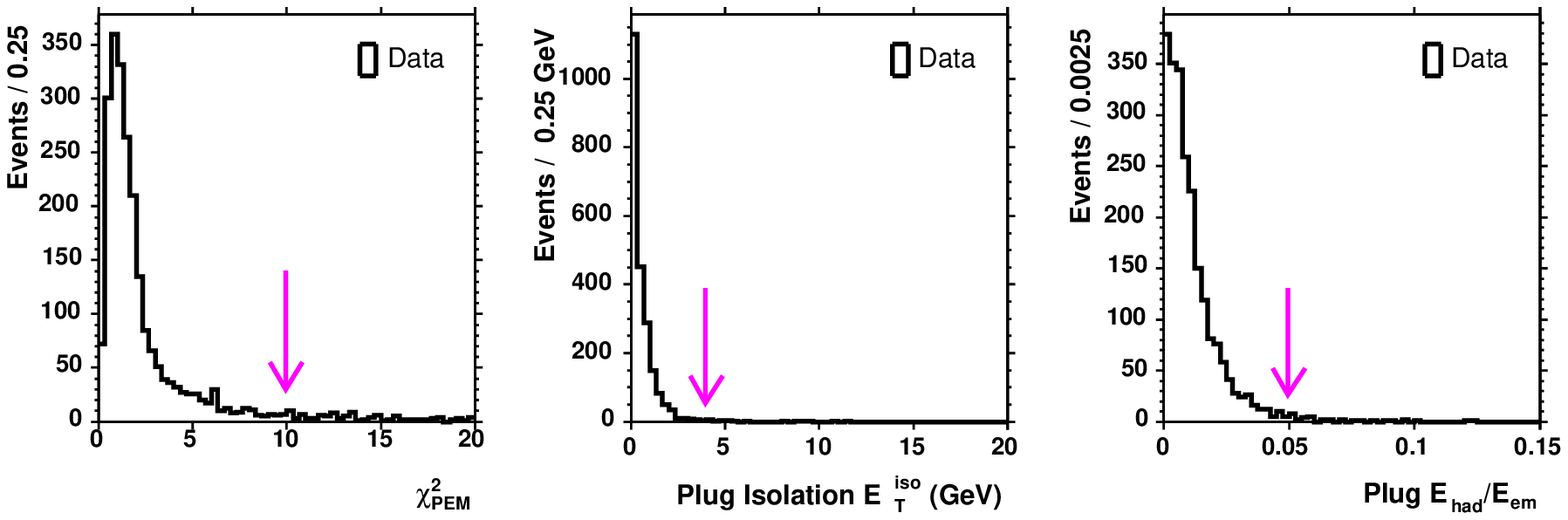}
\caption{Distributions of the $\chi^2_{\mathrm{PEM}}$, 
$\et^{\mathrm{iso}}$ (see Sec.~\ref{sec:wsel}), and 
$E_{\mathrm{had}}/E_{\mathrm{em}}$ plug calorimeter 
electron selection variables from unbiased, second 
electron legs of $\zee$ candidate events in data.  The 
arrows indicate the locations of selection cuts applied 
on these variables.}
\label{fig:plugidvar}
\end{figure*}

The additional selection criteria applied to plug electron 
candidates are summarized in Table~\ref{tab:eidcuts}.  Fewer 
variables are available for selecting plug electrons due to 
the lack of matching track information for candidates in the 
forward region of the detector.  As in the case of central 
electrons, we cut on the ratio of hadronic to electromagnetic 
energies in the cluster, $E_{\mathrm{had}}/E_{\mathrm{em}}$, 
which is required to be less than 0.05.  We also compare the 
distribution of tower energies in a 3$\times$3 array around 
the seed tower to distributions from electron test-beam 
data, forming the variable $\chi_{\mathrm{PEM}}^{2}$ which 
we require to be less than 10.0.  Distributions of the plug 
electron selection variables are shown in Fig.~\ref{fig:plugidvar}.  
The plotted electron candidates are the unbiased, second plug 
electron legs in $\zee$ events in which the first electron 
is reconstructed within the central calorimeter and found
to satisfy a set of more restrictive cuts on the previously
described central electron identification variables.

\subsection{Muon Selection}
\label{sec:muonsel}
Muon candidates used in these measurements must have 
reconstructed stubs in both the CMU and CMP chambers 
(CMUP muons) or a reconstructed stub in the CMX 
chambers.  CMX chambers were offline for the first 
16.5~$\pbinv$ of integrated luminosity corresponding 
to our datasets, and the reduced muon detector 
coverage during this period is taken into account 
in our measured acceptances for events in the muon
candidate samples (see Sec.~\ref{sec:sigaccpythia}).
The muon candidate tracks are required to extrapolate to 
regions of the muon chambers with high single wire hit 
efficiencies to ensure that chamber-edge effects do not 
contribute to inefficiencies in muon stub-reconstruction 
and stub-track matching (see Sec.~\ref{sec:eff}).  We 
measure the location of an extrapolated muon track 
candidate with respect to the drift direction (local $x$) 
and wire axis (local $z$) of a given chamber.  The 
extrapolation assumes that no multiple scattering takes 
place, and in some cases muons that leave hits in the 
muon detectors extrapolate to locations outside of the 
chambers.  In the CMP and CMX chambers, we require that 
the extrapolation is within the chamber volume in local 
$x$, and at least 3~$\cm$ away from the edges of the chamber 
volume in local $z$.  Studies of unbiased muons in $\zmm$
events show that these regions of chambers are maximally 
efficient for hit-finding.  No such requirement is needed 
for the CMU chambers.  Some sections of the upgraded muon 
detectors were not yet fully commissioned for the period 
of data-taking corresponding to our datasets, and we 
exclude all muon candidates with stubs in those sections.

\begin{figure*}[t]
 \includegraphics[width=8.cm]{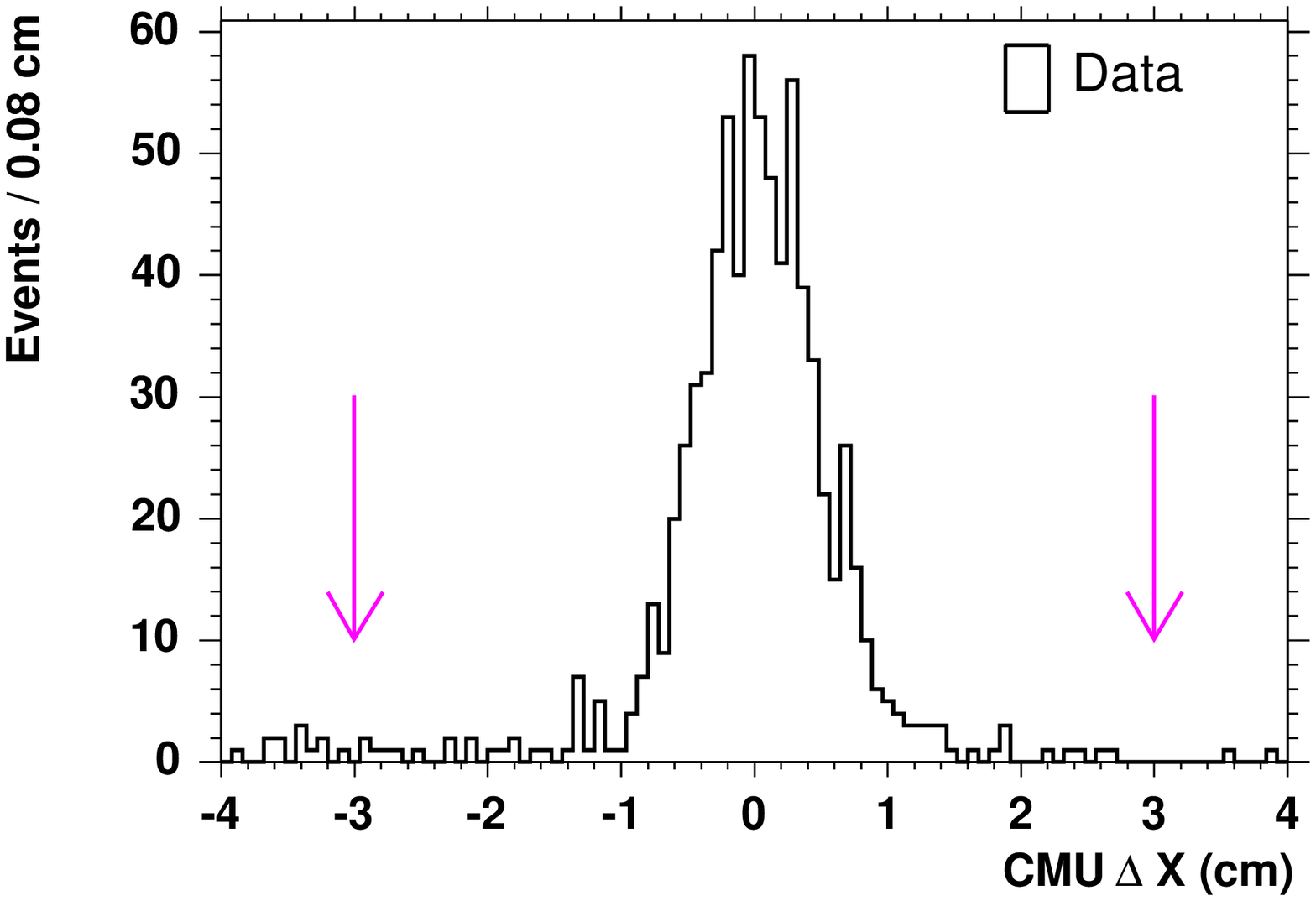} \includegraphics[width=8.cm]{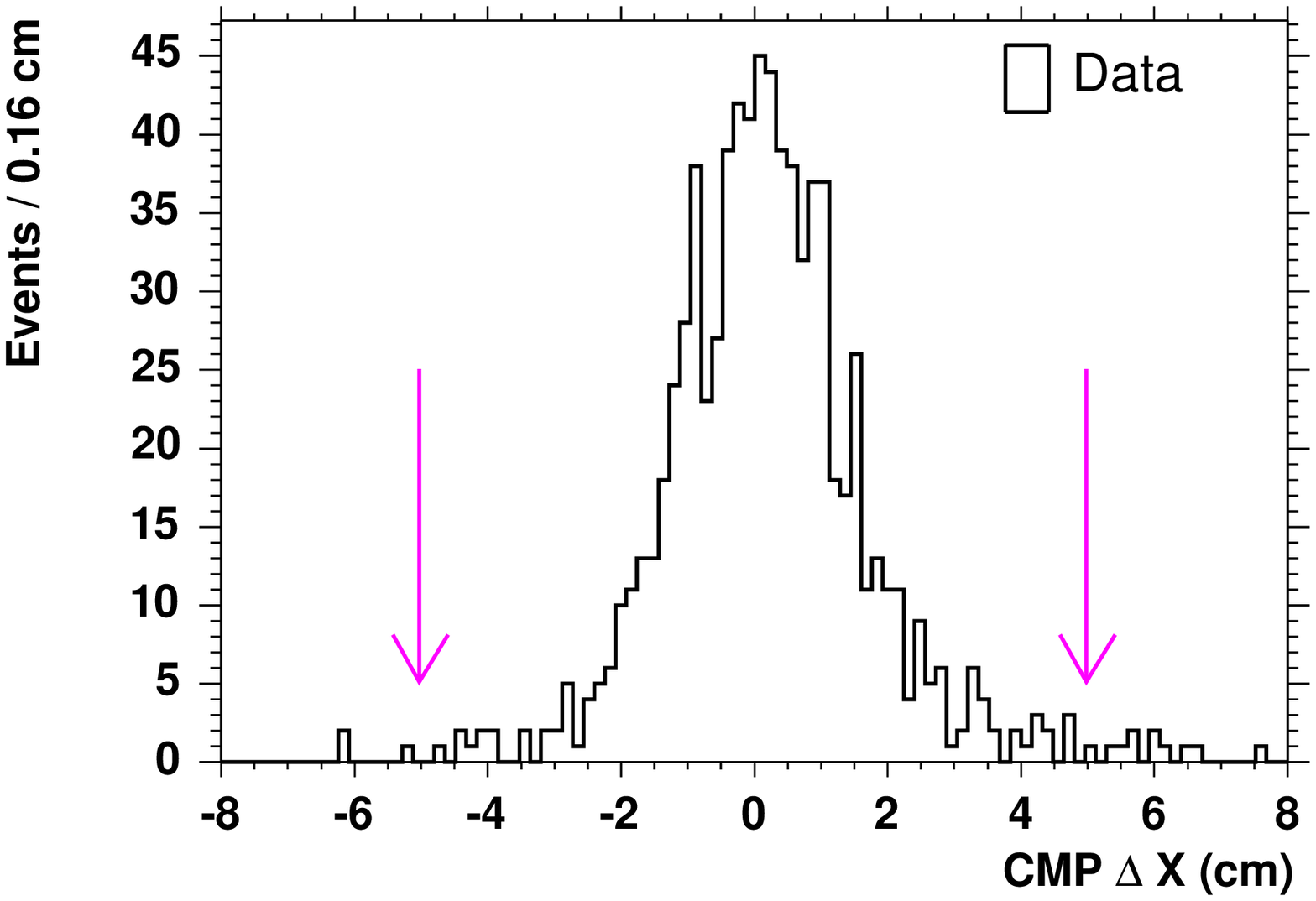} \includegraphics[width=8.cm]{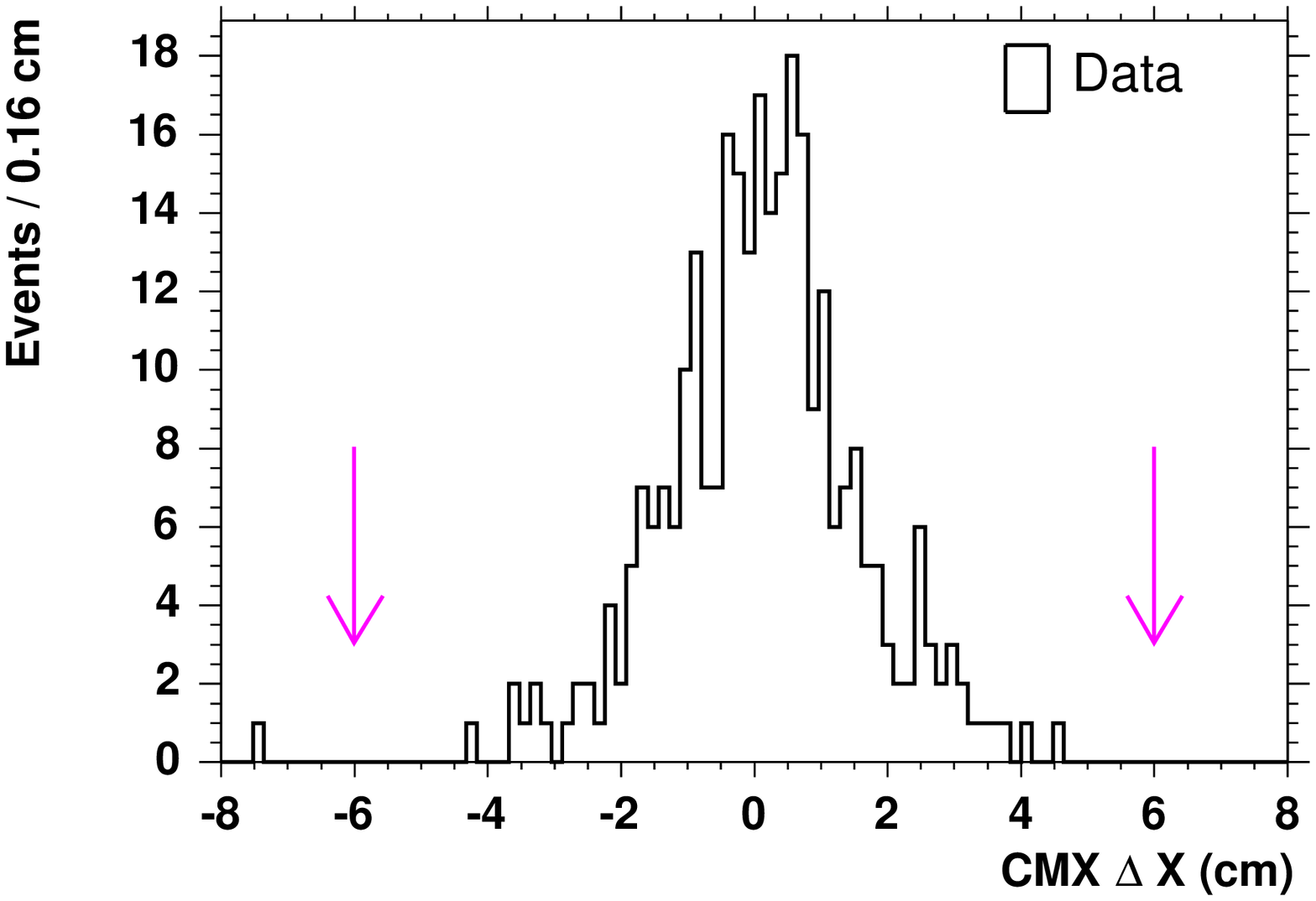}
\caption{Distributions of the CMU, CMP, and CMX $\Delta X$ 
muon selection variables from unbiased, second muon 
legs of $\zmm$ candidate events in data.  The arrows 
indicate the locations of selection cuts applied on these 
variables.}
\label{fig:dxmuon}
\end{figure*}

The selection criteria applied to muon candidates are 
summarized in Table~\ref{tab:muoncuts}.  We require that 
the measured energy depositions in the electromagnetic 
and hadronic sections of the calorimeters along 
the muon candidate trajectory, $E_{\mathrm{em}}$ and 
$E_{\mathrm{had}}$, are consistent with those expected 
from a minimum-ionizing particle.  The positions of the 
reconstructed chamber stubs are required to be near the 
locations of the extrapolated tracks.  The track-stub 
matching variable $|\Delta X|$ is the distance in the 
$r-\phi$ plane between the extrapolated COT track and 
the CMU, CMP, or CMX stub.  Fig.~\ref{fig:dxmuon} 
shows the $\Delta X$ distributions for unbiased, CMU, 
CMP and CMX second muons in $\zmm$ events.  

Energetic cosmic ray muons traverse the detector at a 
significant rate, depositing hits in both muon chambers
and the COT, and can in a small fraction of cases satisfy 
the requirements of the high $\pt$ muon trigger paths
and the offline selection criteria.  We remove cosmic ray
events from our sample using the previously discussed 
track quality cuts for muon candidates and a cosmic ray 
tagging algorithm (see Sec.~\ref{sec:cosmicwbkg}) based 
on COT hit timing information.

\begin{table}[t]
\caption{Calorimeter and muon chamber variables used in 
muon identification.  The fiducial distance variables 
are defined as the extrapolated position of the muon 
track candidate with respect to the edges of a given 
muon chamber.  The fiducial distance is negative if this 
position lies within the chamber and positive otherwise.  
CMUP muon candidates are those with reconstructed stubs 
in both the CMU and CMP detectors.  CMX muon candidates 
have reconstructed stubs in the CMX detector.}
\begin{tabular}{ l r }  
\hline 
\hline
Variable                                  & Cut                                     \\ 
\hline
Minimum Ionizing Cuts:                    &  ($\GeV$)                                        \\
\hline 
$E_{\mathrm{em}}$ ($p \leq$ 100~$\GeVC$)     & $<$ 2                            \\
$E_{\mathrm{em}}$ ($p >$ 100~$\GeVC$)     & $<$ 2 + ($p$$-$100) $\cdot$ 0.0115 \\
$E_{\mathrm{had}}$ ($p \leq$ 100~$\GeVC$)    & $<$ 6                            \\
$E_{\mathrm{had}}$ ($p >$ 100~$\GeVC$)    & $<$ 6 + ($p$$-$100) $\cdot$ 0.0280 \\
\hline
Muon Stub Cuts:                           & ($\cm$)                                         \\
\hline 
$|\Delta X_{\mathrm{CMU}}|$ (CMUP)        & $<$ 3.0                            \\
$|\Delta X_{\mathrm{CMP}}|$ (CMUP)        & $<$ 5.0                            \\
$|\Delta X_{\mathrm{CMX}}|$ (CMX)         & $<$ 6.0                            \\
CMP $x$-fiducial distance (CMUP)          & $<$ 0.0                            \\
CMP $z$-fiducial distance (CMUP)          & $<$ $-3.0$                           \\
CMX $x$-fiducial distance (CMX)           & $<$ 0.0                            \\
CMX $z$-fiducial distance (CMX)           & $<$ $-3.0$                           \\ 
\hline
\hline
\end{tabular}
\label{tab:muoncuts}
\end{table}

\subsection{$\wlnu$ Selection}
\label{sec:wsel}
$\wlnu$ events are selected by first requiring a high-$\pt$ charged
lepton in the central detectors, as described above.  Electrons 
must have electromagnetic-cluster $\et$ greater than 25 $\GeV$ and 
COT track $\pt$ greater than 10 $\GeVC$.  Muons must have COT track 
$\pt$ greater than 20 $\GeVC$.  The leptons from decays of $\W$ and 
$\Z$ bosons are often isolated from hadronic jets, in contrast 
to leptons originating from decays of heavy-flavor hadrons.  We 
therefore require that the calorimeter energy in a cone of radius 
$\Delta{R} = \sqrt{\Delta{\eta}^{2} + \Delta{\phi}^{2}} \leq$~0.4 
around the lepton excluding the energy associated with the 
lepton, $\et^{\mathrm{iso}}$, be less than 10~$\!\%$ the energy 
of the lepton ($\et$ for electrons and $\pt$ for muons).  
Fig.~\ref{fig:idcalvar} shows the isolation distribution for 
unbiased central electrons from $\zee$ decays.

We also require evidence for a neutrino in $\W$ candidate events in
the form of an imbalance of the measured momentum of the event since
neutrinos do not interact with our detector.  The initial state of the
colliding partons has $\pt \simeq 0$, but unknown $p_z$ due to the
unknown value of initial parton momentum.  Therefore, we identify the
missing transverse energy, $\met$, in the event with the $\pt$ of the
neutrino.  In muon events, we correct the $\met$ measured in the
calorimeter to account for the energy carried away by the muon, a
minimum-ionizing particle.  The muon momentum is used in place of the
calorimeter energy deposits observed along the path of the muon.  For
$\wmnu$ candidate events we require $\met >$~20~$\GeV$ and tighten the
requirement for $\wenu$ events, $\met >$~25~$\GeV$, to further reduce
backgrounds from hadron jets.

A background to $\wlnu$ is the $\zll$ channel, when one of the 
leptons falls into an uninstrumented region of the detector, 
creating false $\met$.  This is a bigger problem in the muon 
channel since the coverage of the muon detectors is in general
less uniform than that of the calorimeter.  Therefore, in the 
muon channel we reject events with additional minimum-ionizing 
tracks in the event with $\pt >$~10~$\GeVC$, 
$E_{\mathrm{em}} <$~3~$\GeV$ ($E_{\mathrm{em}} <$~3 + 
0.0140~$\cdot$~($p$$-$100)~$\GeV$ if $p >$~100~$\GeVC$) and 
$E_{\mathrm{had}} <$~6~$\GeV$ ($E_{\mathrm{had}} <$~6 +
0.0420~$\cdot$~($p$$-$100)~$\GeV$ if $p >$~100~$\GeVC$).  
Studies of simulated $\wmnu$ and $\zmm$ event samples (see 
Sec.~\ref{sec:acc}) show that this additional rejection 
criteria removes 54.7~$\!\%$ of the $\zmm$ background while 
retaining 99.6~$\!\%$ of the $\wmnu$ signal.  Further 
discussion of backgrounds to the $\wlnu$ channels is found 
in Section~\ref{sec:backg}.

\subsection{$\zll$ Selection}
\label{sec:zsel}
We select events which contain an electron or muon that passes 
the same identification requirements as the lepton in $\wlnu$ 
candidate events.  As described in Sec.~\ref{sec:intro}, 
systematic uncertainties are reduced by using a common lepton 
selection in the $\W$ and $\Z$ analyses.  We identify a second 
lepton in these events using less restrictive (``loose'') 
selection criteria to increase our $\zll$ detection efficiency.  
The invariant mass of the two leptons is required to be between 
66 and 116~$\GeVCSq$.

After selecting the first electron, $\zee$ events are identified by
the presence of another isolated electron in the central calorimeter
with $\et >$~25~$\GeV$ passing less restrictive selection criteria or 
an isolated electron in the plug calorimeter with $\et >$~20~$\GeV$.  
The selection criteria for each type of electron are summarized in 
Table~\ref{tab:zeecuts}.  In the calculation of $\et$ for the plug 
electron, the $z$-vertex is taken from the COT track of the central 
electron in the event.  In the case of two central electrons, we also 
require they be oppositely charged, with both electron tracks passing 
the track quality criteria listed in Table~\ref{tab:trackqual}.

After selecting the first muon, $\zmm$ events are identified by the 
presence of another oppositely charged, isolated muon track with 
$\pt >$~20~$\GeV$ originating from a common vertex. The muon-stub 
criteria are dropped for the second leg to gain signal acceptance 
with very little increase in background; this second muon is merely 
a minimum-ionizing track.  Table~\ref{tab:zmmcuts} shows the 
complete set of selection criteria used to identify $\zmm$ events.  
Again, we require both tracks to pass the track requirements of 
Table~\ref{tab:trackqual}.

\begin{table*}[t]
\caption{$\zee$ selection criteria.}
\begin{tabular}{ l c c r }  
\hline 
\hline
Variable                           & ``Tight'' Central $e$                  & ``Loose'' Central $e$                  & Plug $e$      \\ 
\hline 
$\et$                              & $>$ 25~$\GeV$                          & $>$ 25~$\GeV$                          & $>$ 20~$\GeV$ \\ 
$\pt$                              & $>$ 10~$\GeVC$                         & $>$ 10~$\GeVC$                         &               \\ 
$\et^{\mathrm{iso}}$               & $<$ 0.1 $\cdot \et^{\mathrm{cluster}}$ & $<$ 0.1 $\cdot \et^{\mathrm{cluster}}$ & $<$ 4~$\GeV$  \\ 
$E_{\mathrm{had}}/E_{\mathrm{em}}$ & $<$ 0.055 + 0.00045 $\cdot$ E          & $<$ 0.055 + 0.00045 $\cdot$ E          & $<$ 0.05      \\ 
$E/p$                              & $<$ 2.0 (or $\pt >$ 50~$\GeVC$)        &                                        &               \\ 
$L_{\mathrm{shr}}$                 & $<$ 0.2                                &                                        &               \\ 
$Q \cdot \Delta x$                 & $>$ $-3.0$ $\cm$, $<$ 1.5 $\cm$          &                                        &               \\ 
$|\Delta z|$                       & $<$ 3.0 $\cm$                          &                                        &               \\ 
$\chi^{2}_{\mathrm{strips}}$        & $<$ 10.0                                 &                                        &               \\  
$\chi^2_{\mathrm{PEM}}$            &                                        &                                        & $<$ 10.0      \\ 
\hline 
\hline 
\end{tabular}
\label{tab:zeecuts}
\end{table*}

\begin{table}
\caption{$\zmm$ selection criteria.}
\begin{tabular}{ l r }  
\hline 
\hline
Variable                                 & Cut                                     \\ 
\hline
Fiducial and Kinematic:            &                                         \\
$|\eta_{\mu}^{(1)}|$                     & $<$ 1.0 (CMUP+CMX)                      \\  
$|\eta_{\mu}^{(2)}|$                     & $<$ 1.2 (Track)                         \\  
$\pt^{\mu(1)}$                           & $>$ 20~$\GeVC$                          \\ 
$\pt^{\mu(2)}$                           & $>$ 20~$\GeVC$                          \\ 
\hline
Both Muon Legs:                          &                                         \\ 
$E_{\mathrm{em}}$ ($p \leq$ 100~$\GeVC$)    & $<$ 2~$\GeV$                            \\
$E_{\mathrm{em}}$ ($p >$ 100~$\GeVC$)    & $<$ 2 + ($p$$-$100) $\cdot$ 0.0115~$\GeV$ \\
$E_{\mathrm{had}}$ ($p \leq$ 100~$\GeVC$)   & $<$ 6~$\GeV$                            \\
$E_{\mathrm{had}}$ ($p >$ 100~$\GeVC$)   & $<$ 6 + ($p$$-$100) $\cdot$ 0.0280~$\GeV$ \\
$\et^{\mathrm{iso}}$                     & $<$ 0.1 $\cdot \pt$                     \\ 
\hline
First Muon Leg:                          &                                         \\ 
$|\Delta X_{\mathrm{CMU}}|$                    & $<$ 3.0 $\cm$ (CMUP)                    \\
$|\Delta X_{\mathrm{CMP}}|$                    & $<$ 5.0 $\cm$ (CMUP)                    \\
$|\Delta X_{\mathrm{CMX}}|$                    & $<$ 6.0 $\cm$ (CMX)                     \\
CMP $x$-fiducial distance                & $<$ 0.0 $\cm$ (CMUP)                    \\
CMP $z$-fiducial distance                & $<$ $-3.0$ $\cm$ (CMUP)                   \\
CMX $x$-fiducial distance                & $<$ 0.0 $\cm$ (CMX)                     \\
CMX $z$-fiducial distance                & $<$ $-3.0$ $\cm$ (CMX)                    \\ 
\hline 
\hline
\end{tabular}
\label{tab:zmmcuts}
\end{table}

\subsection{Event Selection Summary}
\label{sec:evselsummary}
Using the selection criteria described here, we find a total of 
37,584 $\wenu$ candidate events.  In the muon channel, we find 
21,983 $\W$ boson candidates with CMUP muons and 9,739 with 
CMX muons for a grand total of 31,722 $\wmnu$ candidates.  In 
the $\Z$ boson decay channel, we find 1,730 events with two 
reconstructed electrons in the central calorimeter and 2,512 
events in which the second electron is reconstructed in the 
plug calorimeter giving a total of 4,242 $\zee$ candidates.  
From our high $\pt$ muon dataset, we find 1,371 CMUP + track 
and 677 CMX + track $\zmm$ candidates. There is an overlap of 
263 events between these two samples in which one candidate 
track is matched to stubs in the CMU and CMP muon chambers 
and the other is matched to a stub in the CMX chamber.  Taking 
this overlap into account, we obtain a total of 1,785 $\zmm$ 
candidate events.

\section{Signal Acceptance}
\label{sec:acc}
\subsection{Introduction}
\label{sec:accintro}

The acceptance terms in Eqs.~\ref{eq:wsigma} and~\ref{eq:zsigma} 
are defined as the fraction of $\wlnu$ or $\zll$ events produced 
in $\ppbar$ collisions at $\sqrt{s} =$ 1.96 $\TeV$ that satisfy 
the geometrical and kinematic requirements of our samples.  
Lepton reconstruction in our detector is limited by the finite 
fiducial coverage of the tracking, calorimeter, and muon systems.  
Several kinematic requirements are also made on candidate events 
to help reduce backgrounds from non-$\W/\Z$ processes.  The 
reconstructed leptons in these events are required to pass minimum 
calorimeter cluster $\et$ and/or track $\pt$ criteria.  In addition, 
a minimum requirement on the total measured missing $\et$ is made 
on events in our $\wlnu$ candidate samples, and the invariant mass 
of $\zll$ candidate events is restricted to a finite range around 
the measured $\Z$ boson mass.

The fraction of signal events that satisfy the geometrical and 
kinematic criteria outlined above for each of our samples is 
determined using simulation.  One geometrical cut on candidate 
events for which we measure the acceptance directly from data
is the requirement that the primary event vertex for each event 
lies within 60.0~$\cm$ of the detector origin along the $z$-axis 
(parallel to the direction of the beams).  Our simulation does 
include a realistic model of the beam interaction region, but we 
obtain a more accurate estimation of the selection efficiency for 
the event vertex requirement from studies of minimum bias events 
in the data as described in Sec.~\ref{sec:eff}.  Since the 
geometrical and kinematic acceptance for candidate events with 
a primary vertex outside our allowed region is significantly 
smaller, we remove the subset of simulated events with vertices
outside this region from our acceptance calculations to avoid 
double-counting correlated inefficiencies.  

There is one additional complication involved in determining the
kinematic and geometrical acceptances for our $\zll$ candidate
samples.  Because we make our $\zgll$ production cross section 
measurements in a specific invariant mass range, 66~$\GeVCSq$ to
116~$\GeVCSq$, we need to account for events outside this mass 
range that are reconstructed in the detector to sit within this 
range due to the effects of detector resolution and final state
radiations.  To include events of this type in our acceptance 
calculations, we use simulated $\zgll$ event samples generated
over a wider invariant mass range ($M_{\ell\ell} >$~30~$\GeVCSq$).  
In order for an event to contribute to the denominator of our 
acceptance calculations, we require that the invariant mass of 
the lepton pair at the generator level prior to application 
of final state radiative effects lies within the range of our 
measurement (66~$\GeVCSq < M_{\ell\ell} <$~116~$\GeVCSq$).  The
generator-level invariant mass requirement is not made on events
contributing to the numerator of our acceptance calculations,
however, so that $\zgll$ events generated outside the invariant 
mass range of our measurement which have reconstructed masses 
within this range are properly accounted for in the acceptance 
calculations.

\subsection{Event and Detector Simulation}
\label{sec:eventsim}

The simulated events used to estimate the acceptance of our samples
were generated with {\sc pythia} 6.203 ~\cite{acc:pythia}.  The
default set of Parton Distribution Functions (PDFs) used in the
generation of these samples is CTEQ5L~\cite{acc:cteq5}.  {\sc
pythia} generates the hard, leading-order (LO) QCD interaction,
$q+\bar{q} \rightarrow \gamma^{*}/Z$ (or $q+
\bar{q^\prime} \rightarrow W$), simulates initial-state QCD 
radiation via its parton-shower algorithms, and generates the decay,
$\zgll$ (or $\wlnu$).  No restrictions were placed at the generator
level on the $\pt$ of the final-state leptons or on their
pseudorapidity.  Both initial- and final-state radiation were turned
on in the event simulation.  In order to model the data accurately,
the beam energy was set to 980~$\GeV$, and the beam parameters were
adjusted to provide the best possible match with data.  The profile of
the beam interaction region in $z$ was matched to data by setting the
mean of the vertex distribution to 3.0~$\cm$ in the direction along
the beams and the corresponding Gaussian spread to 25.0~$\cm$.  The
offset of the beam from the nominal center of the detector in the
$r-\phi$ plane is also taken into account.  In the simulation, the
position of the beams at $z=0$ is offset by -0.064~$\cm$ in $x$ and
+0.310~$\cm$ in $y$ to provide a rough match with the measured offsets
in data.  The location of a given vertex within the $r-\phi$ plane is
also observed to depend on its location along the $z$-axis due to the
non-zero angle between the beams and the central axis of the detector.
Slopes of -0.00021 and 0.00031 are assigned in the simulation to the
direction of the beams relative to the $y-z$ and $x-z$ detector planes
to model this effect.

\begin{figure}
\includegraphics[width=3.5in]{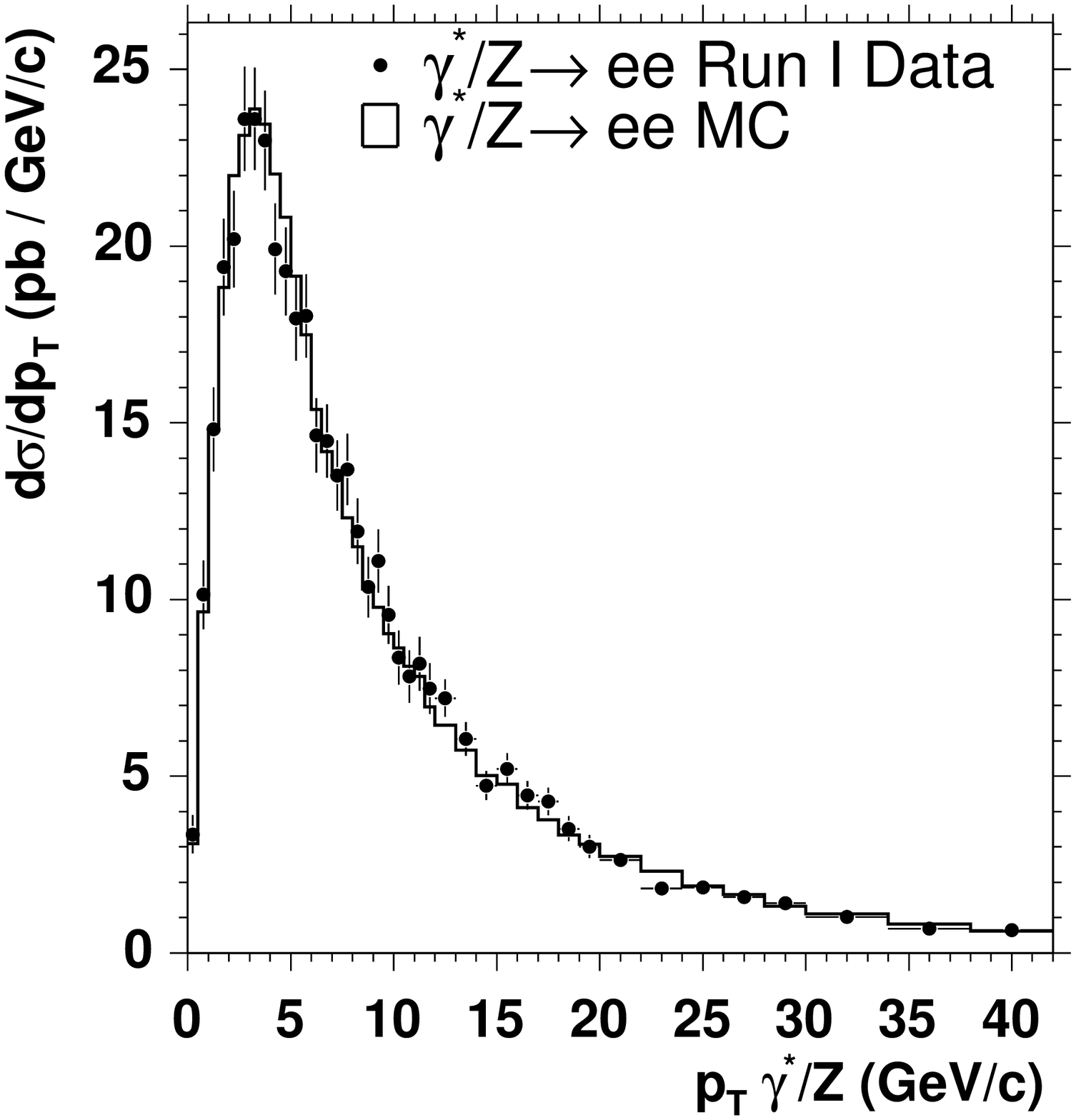}
\caption{\label{fig:bosonpt} Tuned {\sc pythia} 6.21 $d \sigma 
/ d \pt$ in $\pb$ per $\GeVC$ (on average) of $\zgee$ pairs in the mass region
66~$\GeVCSq < M_{ee} <$~116$\GeVCSq$ (histogram) versus the
measurement made by CDF in Run I (points).}
\end{figure}

The intermediate vector boson $\pt$ distribution in the simulation 
is tuned to match the CDF Run~I measurement of the $d \sigma / d \pt$
spectrum for electron pairs in the invariant mass range between
66~$\GeVCSq$ and 116~$\GeVCSq$~\cite{acc:cdfzpt}.  The tuning is
done using {\sc pythia}'s nonperturbative ``$K_{T}$ smearing''
parameters, {\sc parp(91)} and {\sc parp(93)}, and shower
evolution $Q^2$ parameters, {\sc parp(62)} and {\sc parp(64)}.
The {\sc parp(91)} parameter affects the location of the peak in
the $d \sigma / d \pt$ distribution in the vicinity of 3~$\GeVC$, 
and the {\sc parp(62)} and {\sc parp(64)} parameters affect 
the shape of the distribution in the region between 7~$\GeVC$ and
25~$\GeVC$.  A comparison between the ``tuned'' $\gamma^* / Z$ 
$\pt$ distribution from simulation and the measured Run~I spectrum 
is shown in Fig.~\ref{fig:bosonpt}.  We assume that the optimized 
{\sc pythia} tuning parameters obtained from data collected at 
the Run~I center of mass energy ($\sqrt{s} =$ 1.80 $\TeV$) remain 
valid at the increased Run~II center of mass energy ($\sqrt{s} =$ 
1.96 $\TeV$).  The underlying event model in {\sc pythia} is 
also tuned based on observed particle distributions in minimum 
bias events \cite{acc:ricktunea}.

\begin{figure}
\includegraphics[width=3.5in]{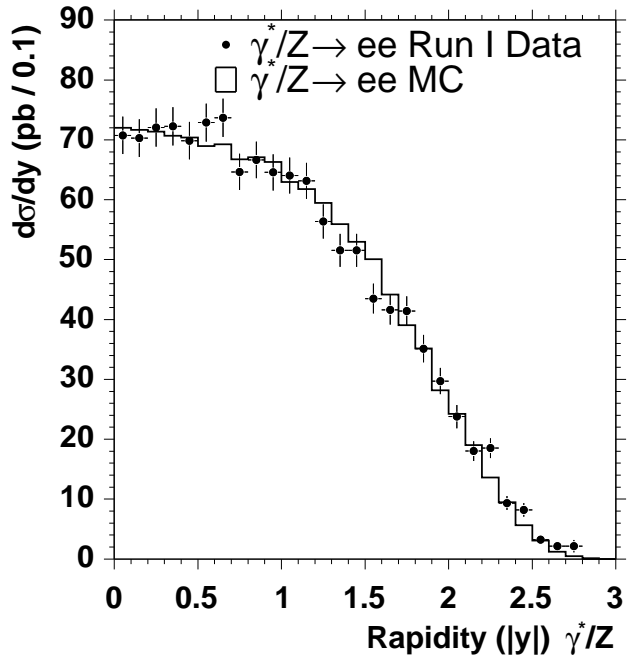}
\caption{\label{fig:bosony} Tuned {\sc pythia} 6.21 $d \sigma / dy$ 
in $\pb$ per 0.1 of $\zgee$ pairs in the mass region 66~$\GeVCSq < M_{ee} 
<$~116~$\GeVCSq$ (histogram) versus the measurement made by 
CDF in Run~I (points).}
\end{figure}

The shape of the boson rapidity distribution is strongly dependent on 
the choice of PDFs. The shape of the $d \sigma /dy$ distribution for 
$\zgee$ pairs in the mass region, 66~$\GeVCSq < M_{ee} <$~116~$\GeVCSq$, 
was measured by CDF in Run~I\cite{acc:cdfzy}.  The good agreement 
observed between the measured shape of $d \sigma / dy$ with that 
obtained from simulation using CTEQ5L PDFs motivates the selection of 
this PDF set for our event generation.  A comparison between the shape 
of the Run~I measured $d \sigma / dy$ distribution and the shape of 
the same distribution from {\sc pythia} 6.21 simulated event samples 
generated with CTEQ5L is shown in Fig.~\ref{fig:bosony}.

A detector simulation based on {\sc geant3}~\cite{acc:geant,acc:geant2}
is used to model the behavior of the CDF detector.  The
{\sc gflash}~\cite{acc:gflash} package is used to decrease the simulation
time of particle showers within the calorimeter.  

\subsection{Signal Acceptances from {\sc pythia}}
\label{sec:sigaccpythia}

Additional tuning is performed after detector simulation to improve
modeling of the data further.  A detailed description of the
techniques used to determine the post-simulation tunings described
here and the associated acceptance uncertainties is provided in
sections~\ref{subsec:remod} and~\ref{subsec:epres}.

The tuned, simulated event samples are used to determine the
acceptances of each $\W$ and $\Z$ event sample.  As discussed in
Sec.~\ref{sec:accintro}, events with a primary event vertex outside
our allowed region ($|z_{\mathrm{vtx}}| <$ 60~$\cm$) are removed from
both the numerator and denominator of our acceptances.  The $\wlnu$
acceptance calculations are outlined in Tables~\ref{tab:welacc}
and~\ref{tab:wmuacc} for the electron and muon candidate samples.  The
geometric and kinematic requirements listed in each table define the
acceptances for the corresponding samples.  The number of simulated
events which satisfy each of the successive, cumulative criteria is
shown in the tables along with the resulting net acceptances based on
the total number of events with primary vertices inside our allowed
region.  The $\wmnu$ events which contain reconstructed muons with
stubs in the CMX region of the muon detector are assigned a weight 
of 55.5/72.0 in the numerator of the acceptance calculation to 
account for the fact that the CMX detector was offline during the 
first 16.5~$\pbinv$ of integrated luminosity that define our samples.
The largest uncertainties attached to the individual luminosity 
measurements (see Sec.~\ref{subsec:lummeas}) cancel in our weighting 
ratio for CMX events and the residual uncertainty on this ratio has
a negligible effect on the overall acceptance uncertainty.

\begin{table*}
\caption{$\wenu$ selection acceptance from {\sc pythia} Monte Carlo simulation.  Statistical uncertainties are shown.}
\begin{tabular}{l c c}
\hline
\hline 
Selection Criteria & Number of Events & Net Acceptance \\
\hline
Total Events                                   & 1933957 & -                    \\
$|z_{\mathrm{vtx}}| <$ 60~$\cm$                & 1870156 & -                   \\
Central EM Cluster                             & 927231  & 0.4958 $\pm$ 0.0004 \\
Calorimeter Fiducial Cuts                      & 731049  & 0.3909 $\pm$ 0.0004 \\
Electron Track $\pt >$ 10~$\GeVC$              & 647691  & 0.3463 $\pm$ 0.0003 \\
EM Cluster $\et >$ 25~$\GeV$                   & 488532  & 0.2612 $\pm$ 0.0003 \\
Event $\met >$ 25~$\GeV$                       & 447836  & 0.2395 $\pm$ 0.0003 \\
\hline
\hline 
\end{tabular}
\label{tab:welacc}
\end{table*}

\begin{table*}
\caption{$\wmnu$ selection acceptance from {\sc pythia} Monte Carlo simulation. Statistical uncertainties are shown.}
\begin{tabular}{l c c}
\hline 
\hline
Selection Criteria & Number of Events & Net Acceptance \\
\hline
Total Events                                   & 2017347 & -                   \\
$|z_{\mathrm{vtx}}| <$ 60~$\cm$                & 1951450 & -                  \\
CMUP or CMX Muon                               & 545221  & 0.2794 $\pm$ 0.0003 \\
Muon Chamber Fiducial Cuts                     & 523566  & 0.2683 $\pm$ 0.0003 \\
Muon Track $\pt >$ 20~$\GeVC$                  & 435373  & 0.2231 $\pm$ 0.0003 \\
Muon Track Fiducial Cuts                       & 411390  & 0.2108 $\pm$ 0.0003 \\
Event $\met >$ 20~$\GeV$                       & 383787  & 0.1967 $\pm$ 0.0003 \\
\hline
\hline 
\end{tabular}
\label{tab:wmuacc}
\end{table*}

The $\zll$ acceptance calculations are outlined in
Tables~\ref{tab:zelacc} and~\ref{tab:zmuacc} for the corresponding
electron and muon candidate samples.  As previously stated in
Sec.~\ref{sec:accintro}, the acceptances that we define for these
samples are for $\zgll$ in the invariant mass range 66~$\GeVCSq <
M_{\ell\ell} <$~116~$\GeVCSq$.  The simulated event samples used to
estimate the $\Z$ boson acceptances are generated with a looser
invariant mass requirement, $M_{\ell\ell} >$ 30~$\GeVCSq$.  Generated
events with an invariant mass outside our allowed range do not
contribute to the denominator of these acceptance calculations but can
contribute to the numerator if the final reconstructed invariant mass
turns out to lie within our allowed region due to radiative and/or
detector resolution effects.  In order for an event to contribute to
the denominator of the $\Z$ boson acceptance calculations, we require
that the invariant mass of the dilepton pair at the generator level
before application of any final state radiative effects lies within
the correct invariant mass range, 66~$\GeVCSq <
M_{\ell\ell}$(Gen)~$<$~116~$\GeVCSq$.  As in the case of the $\wmnu$
acceptance calculation, events in the numerator of the $\zmm$
acceptance calculation must be weighted to account for the fact that
the CMX portion of the muon detector was offline for the first subset
of integrated luminosity that defines our samples.  In order to
account for this effect, a weight of (55.5/72.0) is applied to events
contributing to the numerator of the $\zmm$ acceptance calculation
which contain a CMX muon candidate satisfying the three muon geometric
and kinematic requirements listed in Table~\ref{tab:zmuacc} and no
CMUP muon candidates satisfying these same three requirements.

\begin{table*}[t]
\caption{$\zee$ selection acceptance from {\sc pythia} Monte Carlo simulation.  Statistical uncertainties are shown.}
\begin{tabular}{l c c}
\hline 
\hline
Selection Criteria & Number of Events & Net Acceptance \\
\hline
Total Events                                                    & 507500 & -                   \\
$|z_{\mathrm{vtx}}| <$ 60~$\cm$                                 & 490756 & -                   \\
66~$\GeVCSq < M_{ee}$(Gen) $<$ 116~$\GeVCSq$                    & 376523 & -                   \\
\hline
Central EM Cluster                                              & 363994 & 0.9667 $\pm$ 0.0003 \\
Calorimeter Fiducial Cuts                                       & 299530 & 0.7955 $\pm$ 0.0007 \\
Electron Track $\pt >$ 10~$\GeVC$                               & 252881 & 0.6716 $\pm$ 0.0008 \\
EM Cluster $\et >$ 25~$\GeV$                                    & 186318 & 0.4948 $\pm$ 0.0008 \\
Second EM cluster (Central or Plug)                             & 176417 & 0.4685 $\pm$ 0.0008 \\
Second Cluster Calorimeter Fiducial Cuts                        & 146150 & 0.3882 $\pm$ 0.0008 \\
Second Electron Track $\pt >$ 10~$\GeVC$ (Central)              & 138830 & 0.3687 $\pm$ 0.0008 \\
Second EM Cluster $\et >$ 25~$\GeV$ (Central), 20~$\GeV$ (Plug) & 125074 & 0.3322 $\pm$ 0.0008 \\
Second EM Cluster $E_{\mathrm{had}}/E_{\mathrm{em}}$ $<$ 0.125 (Plug)                       & 124881 & 0.3317 $\pm$ 0.0008 \\
66~$\GeVCSq < M_{ee}$(Rec) $<$ 116~$\GeVCSq$                    & 120575 & 0.3202 $\pm$ 0.0008 \\
Opposite Charge (Central-Central)                               & 119925 & 0.3185 $\pm$ 0.0008 \\
\hline
\hline 
\end{tabular}
\label{tab:zelacc}
\end{table*}

\begin{table*}[t]
\caption{$\zmm$ selection acceptance from {\sc pythia} Monte Carlo simulation.  Statistical uncertainties are shown.}
\begin{tabular}{l c c}
\hline 
\hline
Selection Criteria & Number of Events & Net Acceptance \\
\hline
Total Events                                     & 507500 & -                   \\
$|z_{\mathrm{vtx}}| <$ 60~$\cm$                  & 490755 & -                   \\
66~$\GeVCSq < M_{\mu\mu}$(Gen) $<$ 116~$\GeVCSq$ & 375981 & -                   \\ 
\hline
CMUP or CMX Muon                                 & 217041 & 0.5773 $\pm$ 0.0008 \\
Muon Chamber Fiducial Cuts                       & 209693 & 0.5577 $\pm$ 0.0008 \\
Muon Track Fiducial Cuts                         & 199940 & 0.5318 $\pm$ 0.0008 \\
Muon Track $\pt >$ 20~$\GeVC$                    & 157244 & 0.4182 $\pm$ 0.0008 \\
Second Track with $\pt >$ 10~$\GeVC$             &  91048 & 0.2422 $\pm$ 0.0007 \\
Second Track Fiducial Cuts                       &  62663 & 0.1667 $\pm$ 0.0006 \\
Second Track $\pt >$ 20~$\GeVC$                  &  56459 & 0.1502 $\pm$ 0.0006 \\
66~$\GeVCSq < M_{\mu\mu}$(Rec) $<$ 116~$\GeVCSq$ &  52160 & 0.1387 $\pm$ 0.0006 \\
\hline
\hline 
\end{tabular}
\label{tab:zmuacc}
\end{table*}

\subsection{Improved Acceptance Calculations}

The tuned {\sc pythia} simulated event samples are designed to
provide the best possible model for our $\W$ and $\Z$ boson candidate
samples.  However, the actual boson production cross section
calculation made by {\sc pythia} is done only at leading order
(LO); see Fig.~\ref{fig:vprod}.  The complex topologies of
higher-order contributions are modeled using a backward shower
evolution algorithm which includes initial-state radiative effects and
a separate, post-generation algorithm for including final-state
radiation.  A better description of boson production can be obtained
from recently developed NNLO theoretical calculations of the
double-differential production cross sections, $d^{2}\sigma/dydM$, as
a function of boson rapidity ($y$) and mass ($M$), for both $W^{\pm}$
and Drell-Yan production~\cite{acc:slac}.  The calculations are based
on the MRST 2001 NNLL PDF set~\cite{int:mrst1} and electroweak
parameters taken from~\cite{int:pdg}.  The single differential cross
sections, $d\sigma/dy$, are obtained by integrating over the mass
range, 66~$\GeVCSq < M_{\ell\ell} <$ 116~$\GeVCSq$ for Drell-Yan
production and 40~$\GeVCSq < M_{\ell\nu} <$ 240~$\GeVCSq$ for
$W^{\pm}$ production.

\begin{figure}
\includegraphics[width=3.3in]{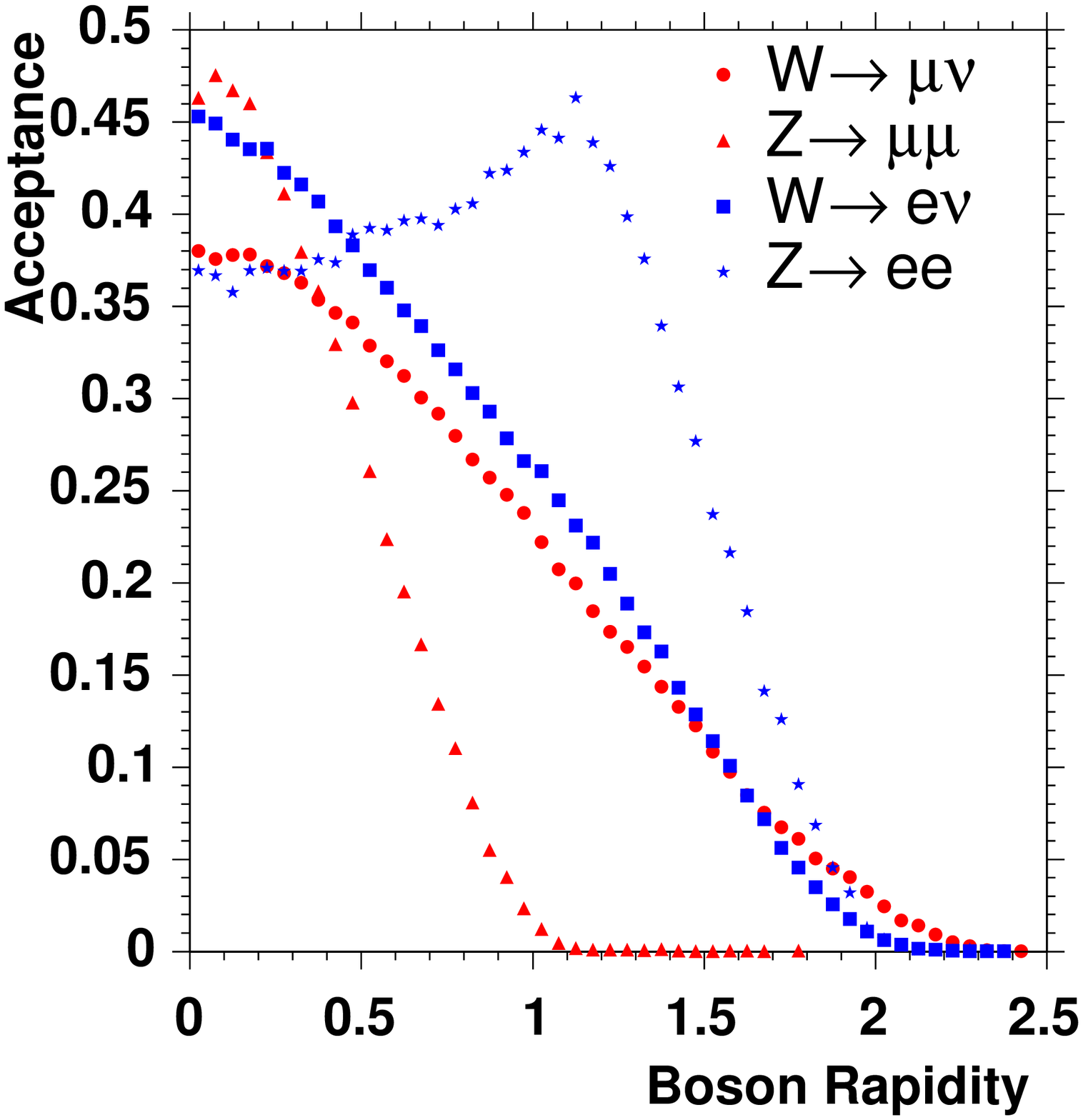}
\caption{\label{fig:ay} Acceptance as a function of boson rapidity, $A(y)$, 
for our four candidate samples: $\wenu$ (squares), $\wmnu$ (points), 
$\zee$ (stars), and $\zmm$ (triangles).}
\end{figure}

We use these NNLO theoretical calculations of $d\sigma/dy$ for
Drell-Yan and $W^{\pm}$ production to obtain improved acceptance
estimates for our candidate samples.  First, the tuned {\sc pythia}
event simulation is used to create acceptance functions for each
candidate sample as a function of boson rapidity, $A(y)$.  These
functions provide the acceptance in each boson rapidity bin based on
our modeling of the CDF detector contained in the event simulation.
Fig.~\ref{fig:ay} shows the $A(y)$ acceptance functions for each of
our four candidate samples.  The $\zee$ sample has larger acceptance
at higher boson rapidity due to the plug calorimeter modules which
provide additional coverage in the forward part of the detector for
the second electron in these events.  Based on these distributions,
the acceptance of each sample, $\overline{A}$, is then calculated as

\begin{eqnarray}
\overline{A} = \frac{\int \frac{d\sigma}{dy} \cdot A(y) \cdot dy}{\int \frac{d\sigma}{dy} \cdot dy}.
\label{eq:accclc}
\end{eqnarray}

The acceptance values obtained with this approach are shown in
Table~\ref{tab:accres} and compared with values obtained directly from
the {\sc pythia} simulated event samples.  The results all agree
within 0.4~$\!\%$ indicating that the shapes of the NNLO $d\sigma/dy$
distributions are very similar to those computed with the {\sc pythia}
simulation.  The acceptance values obtained using the NNLO theoretical
differential cross section calculations are used for our measurements.

\begin{table}
\caption{Central acceptance values for our candidate samples based on 
$d\sigma/dy$ distributions obtained from both NNLO and {\sc pythia}
simulation.}
\begin{tabular}{l c c r}
\hline 
\hline
Acceptance  & NNLO Calc. & {\sc pythia} & Difference ($\!\%$) \\
\hline
$A_{\wmnu}$             & 0.1970                 & 0.1967       & +0.15       \\
$A_{\wenu}$             & 0.2397                 & 0.2395       & +0.08       \\
$A_{\zmm}$              & 0.1392                 & 0.1387       & +0.36       \\
$A_{\zee}$              & 0.3182                 & 0.3185       & -0.09      \\
$A_{\zmm}/A_{\wmnu}$    & 0.7066                 & 0.7054       & +0.17       \\
$A_{\zee}/A_{\wenu}$    & 1.3272                 & 1.3299       & -0.20       \\
\hline
\hline 
\end{tabular}
\label{tab:accres}
\end{table}

\subsection{Uncertainties in NNLO Calculation}

Uncertainties in the NNLO calculations of the differential 
boson production cross sections lead to uncertainty on our
calculated acceptance values.  The theoretical calculations 
require a large number of input parameters taken from 
world average experimental results that have their own 
associated uncertainties.  The renormalization scale used 
in the calculations is another source of uncertainty.  The 
default renormalization scales used in the calculations are 
$M_{Z}$ for Drell-Yan production and $M_{W}$ for $W^{\pm}$ 
production.  To study the effect of this scale on our 
central acceptance values, we recalculate the $d\sigma/dy$ 
production cross sections using renormalization scales twice 
and one-half of the default values.  For both cases, we find 
the net change in our calculated acceptances to be less than 
0.1~$\!\%$ which has a negligible effect on our overall 
acceptance uncertainty.

We perform a computational consistency check on the NLO component of
the NNLO $d\sigma/dy$ calculation~\cite{acc:slac} with a different
$\overline{MS}$ NLO computation of
$d\sigma/dy$~\cite{int:nnlo00,int:nnlo0,int:nnlo4,int:nnlo2}.  We find that the
resulting acceptance values differ by no more than 0.1~$\!\%$.  Based
on this agreement between the two calculations, we assign no
additional uncertainty to our acceptance values based on the
calculation itself.  However, our default calculation is still
susceptible to uncertainties from higher order effects beyond NNLO.
To place a conservative limit on the magnitude of higher-order
uncertainties, we compare acceptance values based on the NLO and NNLO
versions of our default $d\sigma/dy$ production cross section
calculations and assign an uncertainty based on the differences.  The
results are shown in Table~\ref{tab:therr}.  The largest difference is
seen in the acceptance for the $\zmm$ candidate sample, which has the
narrowest acceptance window in boson rapidity.

\begin{table}[t]
\caption{Comparison of acceptances for our candidate samples 
based on $d\sigma/dy$ distributions from the NNLO and NLO 
versions of our default theoretical calculation.  The 
difference is taken as an uncertainty on higher-order 
contributions.}
\begin{tabular}{l c c r}
\hline 
\hline
Acceptance  & NNLO & NLO & Difference ($\!\%$) \\
\hline 
$A_{\wmnu}$            & 0.1970                 & 0.1975                & 0.25        \\
$A_{\wenu}$            & 0.2397                 & 0.2404                & 0.29        \\
$A_{\zmm}$             & 0.1392                 & 0.1402                & 0.72        \\
$A_{\zee}$             & 0.3182                 & 0.3184                & 0.06        \\
$A_{\zmm}/A_{\wmnu}$   & 0.7066                 & 0.7101                & 0.50        \\
$A_{\zee}/A_{\wenu}$   & 1.3272                 & 1.3246                & 0.20        \\
\hline
\hline 
\end{tabular}
\label{tab:therr}
\end{table}

\subsection{Uncertainties from PDF Model}

The largest uncertainties on our acceptance values arise from
uncertainties on the momentum distributions of quarks and gluons
inside the proton modeled in the PDF sets used as inputs to our
theoretical calculations.  The choice of PDF input has a significant
effect on the shape of the $d\sigma/dy$ distributions, and
consequently a significant effect on the calculated acceptances for
our candidate samples.  As noted earlier, our theoretical calculations
use the best-fit MRST 2001 NNLL PDF set \cite{int:mrst1}.  The input
PDF sets are created by fitting relevant experimental results to
constrain the parameters which describe the quark/gluon momentum
distributions in the proton.  Currently, the NNLL PDF set provided by
the MRST group is the only one available to us.  NLL PDF sets are
available from both groups (MRST01E~\cite{int:mrst1,int:mrst2} and
CTEQ6.1~\cite{int:cteq}), however.  To investigate differences between
the CTEQ and MRST PDF sets, we calculate $d\sigma/dy$ at NLO using
each group's NLL PDF set and check for differences in the acceptance
values for our candidate samples based on each calculation.  The
results are shown in Table~\ref{tab:pdferr1}.  The differences are
significant, especially for the $\zmm$ candidate sample.

\begin{table}[t]
\caption{Comparison of acceptances for our candidate samples 
based on $d\sigma/dy$ distributions from NLO theoretical   
calculations using NLL MRST and CTEQ PDF sets.}
\begin{tabular}{l c c r}
\hline 
\hline
Acceptance & MRST & CTEQ & Difference ($\!\%$) \\
\hline 
$A_{\wmnu}$            & 0.1976                 & 0.1960                 & 0.82        \\
$A_{\wenu}$            & 0.2405                 & 0.2385                 & 0.84        \\
$A_{\zmm}$             & 0.1401                 & 0.1376                 & 1.82        \\
$A_{\zee}$             & 0.3183                 & 0.3164                 & 0.60        \\
$A_{\zmm}/A_{\wmnu}$   & 0.7088                 & 0.7021                 & 0.95        \\
$A_{\zee}/A_{\wenu}$   & 1.3235                 & 1.3264                 & 0.22        \\
\hline
\hline 
\end{tabular}  
\label{tab:pdferr1}
\end{table}

Another recent development from the CTEQ and MRST groups is the
release of ``error'' PDF sets at NLL which map out the space of
potential PDF parameter values based on the uncertainties of the
experimental results used to constrain them. The CTEQ PDF
parameterization is based on 20 parameters, $P_{\mathrm{i}}$, which
are tuned to their most likely values based on a minimization of the
$\chi^2$ of a global fit to the experimental data.  The equivalent
MRST parameterization uses only 15 parameters.  As the covariance
matrix of the $P_{\mathrm{i}}$ is not diagonal at the minimum, it is
difficult to propagate fit errors on the $P_{\mathrm{i}}$ directly
into uncertainties on experimentally measured quantities such as
acceptances.  However, both groups construct different sets of
eigenvectors, $Q_{\mathrm{i}}$, which do diagonalize the covariance
matrix of the fit in the vicinity of the minimum.  The
$Q_{\mathrm{i}}$ are linearly independent by design, which allows
experimental uncertainties based on deviations in each parameter to be
added in quadrature.  The MRST and CTEQ groups transform individual
$\pm$~1~$\sigma$ variations of each $Q_{\mathrm{i}}$ back into the
$P_{\mathrm{i}}$ parameter space and generate sets of ``up'' and
``down'' error PDFs.  This procedure outputs two PDF sets per
parameter for a total of 40 CTEQ (30 MRST) error PDF sets.  The
$\pm$~1~$\sigma$ variations of each eigenvector for the MRST01E and
CTEQ6.1 error PDFs are different.  These variations are based on the
following values for the global fit $\chi^{2}$ from its minimum:
$\Delta \chi^{2} =$ 50 for MRST01E and $\Delta
\chi^{2} =$ 100 for CTEQ6.1. 

To determine the uncertainty on the acceptance values for
our candidate samples based on the CTEQ and MRST error
PDF sets, we perform the NLO $d\sigma/dy$ production 
cross section calculations for each error PDF set 
and check how much the acceptance values based on each 
calculation deviate from the values obtained using the
best-fit NLL PDF set.  The uncertainty associated with 
each $Q_{\mathrm{i}}$ is determined from the changes 
in acceptance between the best-fit PDF set and both 
the ``up'' and ``down'' error PDF sets associated with 
the given parameter, $\Delta A^{\mathrm{i}}_{\uparrow}$ 
and $\Delta A^{\mathrm{i}}_{\downarrow}$.  In most 
cases the two acceptance differences lie in opposite 
directions and can be treated independently, but in a 
small number of cases both differences lie in the 
same direction and a different procedure needs to be 
followed.  Table~\ref{tab:errpro} defines both the 
positive and negative uncertainties assigned to the acceptance 
uncertainty for each $Q_{\mathrm{i}}$ based on the 
relative signs of $\Delta A^{\mathrm{i}}_{\uparrow}$ 
and $\Delta A^{\mathrm{i}}_{\downarrow}$.

\begin{table*}[t]
\caption{Contributions to the positive and negative acceptance uncertainties 
based on acceptance differences between the ``up'' and ``down'' error 
PDF sets associated with a given $Q_{\mathrm{i}}$ and the best-fit PDF 
set.}  
\begin{tabular}{l c c }
\hline 
\hline
Direction of Acceptance Shifts           & $+$ Uncertainty & $-$ Uncertainty \\
\hline
\vspace{-0.4cm}
& \\
\vspace{0.1cm}
$\Delta A^{\mathrm{i}}_{\uparrow} >$~0 and $\Delta A^{\mathrm{i}}_{\downarrow} >$~0 & $\sqrt{({\Delta A^{\mathrm{i}}_{\uparrow}}^{2}+{\Delta A^{\mathrm{i}}_{\downarrow}}^{2})/2}$ & 0 \\
\vspace{0.1cm}
$\Delta A^{\mathrm{i}}_{\uparrow} >$~0 and $\Delta A^{\mathrm{i}}_{\downarrow} <$~0 & $\Delta A^{\mathrm{i}}_{\uparrow}$  & $\Delta A^{\mathrm{i}}_{\downarrow}$ \\
\vspace{0.1cm}
$\Delta A^{\mathrm{i}}_{\uparrow} <$~0 and $\Delta A^{\mathrm{i}}_{\downarrow} >$~0 & $\Delta A^{\mathrm{i}}_{\downarrow}$  & $\Delta A^{\mathrm{i}}_{\uparrow}$ \\
\vspace{0.1cm}
$\Delta A^{\mathrm{i}}_{\uparrow} <$~0 and $\Delta A^{\mathrm{i}}_{\downarrow} <$~0 & 0 & $\sqrt{({\Delta A^{\mathrm{i}}_{\uparrow}}^{2}+{\Delta A^{\mathrm{i}}_{\downarrow}}^{2})/2}$ \\
\hline
\hline 
\end{tabular}
\label{tab:errpro}
\end{table*}

The positive and negative uncertainties associated with each of 
the individual $Q_{\mathrm{i}}$ (20 CTEQ and 15 MRST) are 
summed in quadrature to determine the overall PDF model 
uncertainty on our acceptance values.  The results of 
these calculations using both the CTEQ and MRST error
PDF sets are shown in Table~\ref{tab:pdferr2}. We note 
that the MRST uncertainties are a factor of 2-3 lower than the 
CTEQ uncertainties which is most likely related to different 
choices for the $\Delta \chi^{2}$ values used by the two 
groups to choose the $\pm$~1~$\sigma$ points associated 
with each of the $Q_{\mathrm{i}}$.  We choose to use the 
larger CTEQ uncertainties based on the fact that the 
magnitude of those uncertainties is more consistent with the 
differences observed between the acceptance values for 
our samples calculated with the best-fit NLL CTEQ and MRST 
PDF sets (see Table~\ref{tab:pdferr1}).  Based on the 
technique outlined above, we also determine the PDF model 
uncertainties associated with three additional quantities useful 
in the calculation of $\gam$ and $g_{\mu}/g_{e}$ detailed 
in Sec.~\ref{sec:results}.  These values are given in 
Table~\ref{tab:pdferr3}.   
 
\begin{table*}[t]
\caption{PDF model acceptance uncertainties based on the CTEQ and 
MRST error PDF sets.}  
\begin{tabular}{l c c c c}
\hline
\hline                                                       
                       & CTEQ    & CTEQ    & MRST    & MRST    \\
Acceptance             & $+$ Uncertainty & $-$ Uncertainty & $+$ Uncertainty & $-$ Uncertainty \\
                       & ($\!\%$) & ($\!\%$) & ($\!\%$) & ($\!\%$) \\ 
\hline 
$A_{\wmnu}$            & 1.13      & 1.47      & 0.46      & 0.57      \\
$A_{\wenu}$            & 1.16      & 1.50      & 0.48      & 0.58      \\
$A_{\zmm}$             & 1.72      & 2.26      & 0.67      & 0.87      \\
$A_{\zee}$             & 0.69      & 0.84      & 0.27      & 0.33      \\
$A_{\zmm}/A_{\wmnu}$   & 0.67      & 0.86      & 0.26      & 0.31      \\
$A_{\zee}/A_{\wenu}$   & 0.74      & 0.56      & 0.29      & 0.23      \\
\hline
\hline 
\end{tabular}
\label{tab:pdferr2}
\end{table*}

\begin{table*}[t]
\caption{Additional PDF model acceptance uncertainties based on 
the CTEQ and MRST error PDF sets.} 
\begin{tabular}{l c c c c }
\hline
\hline
                                                         & CTEQ    & CTEQ    & MRST    & MRST    \\
Acceptance                                               & $+$ Uncertainty & $-$ Uncertainty & $+$ Uncertainty & $-$ Uncertainty \\
                                                         & ($\!\%$) & ($\!\%$) & ($\!\%$) & ($\!\%$) \\ 
\hline
${\sigmawen \cdot A_{\wmnu}}/{\sigmazee \cdot A_{\zmm}}$ & 1.03       & 1.06       & 0.52       & 0.42       \\
${\sigmawen \cdot A_{\wenu}}/{\sigmazee \cdot A_{\zee}}$ & 0.70       & 1.06       & 0.42       & 0.62       \\
$A_{\wenu}/A_{\wmnu}$                                    & 0.03       & 0.04       & 0.01       & 0.01      \\
\hline
\hline 
\end{tabular}
\label{tab:pdferr3}
\end{table*}

\subsection{Uncertainties from Boson $\pt$ Model}

As discussed in Sec.~\ref{sec:eventsim}, the boson $\pt$ distributions
in our {\sc pythia} simulated event samples are tuned based on the
CDF Run I measurement of the $d \sigma / d
\pt$ spectrum of electron pairs in the mass region between 
66~$\GeVCSq$ and 116~$\GeVCSq$ (see Sec.~\ref{sec:eventsim}).  The
simulated $\gamma^{*}/Z$ $\pt$ distribution at $\sqrt{s} =$ 1.8~$\TeV$
after tuning is shown in Fig.~\ref{fig:bosonpt} along with the
measured distribution from Run~I.  The values for the four parameters
we use in {\sc pythia} for this tuning are chosen using a
$\chi^{2}$ comparison of the $\Z$ boson $\pt$ spectrum measured in
Run~I and the {\sc pythia} generated spectra obtained while varying
the values of our tuning parameters.

The acceptance uncertainties related to our boson $\pt$ model come
primarily from the Run~I measurement uncertainties.  We quantify the
effect of these uncertainties on our measured acceptances using the
uncertainties returned from the $\chi^{2}$ fits used to obtain the four
{\sc pythia} tuning parameters.  We choose to use conservative
$\pm$~3~$\sigma$ fit errors since the fit values for each of the
tuning parameters, {\sc parp(64)} in particular, are somewhat
inconsistent with expectation.  We study the effects of changes in the
boson $\pt$ distributions on our measured acceptances by re-weighting
events in the default simulated event samples based on differences
between the default boson $\pt$ distribution and those obtained from
$\pm$~3~$\sigma$ changes in our individual tuning
parameters. Table~\ref{tab:parp} summarizes the best fit values and
$\pm$~3~$\sigma$ variations obtained for each tuning parameter and the
corresponding acceptance uncertainties for each candidate sample.
Changes to the {\sc parp(93)} tuning parameter were found to have a
negligible effect on the boson $\pt$ spectrum and the measured
acceptances.  Uncertainties associated with the other three tuning
parameters are taken in quadrature to determine an overall acceptance
uncertainty associated with the boson $\pt$ model.

\begin{table*}[t]
\caption{Fit results for {\sc pythia} boson $\pt$ tuning 
parameters and corresponding uncertainties on the measured 
acceptances of our candidate samples.}
\begin{tabular}{l c c c c c c } 
\hline
\hline  
          &          & $\pm$~3~$\sigma$ &                    & & & \\               
Parameter & Best Fit & Variation        & $\Delta A_{\wmnu}$ &
$\Delta A_{\wenu}$ & $\Delta A_{\zmm}$ & $\Delta A_{\zee}$ \\ 
          &          &                  & ($\!\%$)           & 
($\!\%$)           & ($\!\%$)          & ($\!\%$)          \\
\hline  
{\sc parp(62)} & 1.26 & 0.30 & 0.01  & 0.00  & 0.01  & 0.01  \\
{\sc parp(64)} & 0.2  & 0.03 & 0.03  & 0.04  & 0.08  & 0.06  \\ 
{\sc parp(91)} & 2.0  & 0.3  & 0.02  & 0.02  & 0.00  & 0.02  \\ 
{\sc parp(93)} & 14   & 3    & -         & -         & -         & -         \\ 
Combined          &      &      & 0.04  & 0.04  & 0.08  & 0.06  \\
\hline
\hline
\end{tabular}  
\label{tab:parp}
\end{table*}

\subsection{Uncertainties from Recoil Energy Model}
\label{subsec:remod}

An accurate model of the event recoil energy in the 
simulation is important for estimating the acceptance 
of the event $\met$ criteria applied to our $\wlnu$ 
candidate events.  Simulated recoil energy distributions 
are dependent on the models for hadronic showering, the 
boson recoil energy, and the underlying event energy.  
In addition, the simulation used in these measurements 
does not model other mechanisms that also contribute to 
the residual recoil energy in data events such as multiple 
interactions and accelerator backgrounds.  To account for 
the effects of these differences, the simulated 
recoil energy distributions are tuned to match those in 
the data.

As discussed in Sec.~\ref{sec:sigaccpythia}, the event recoil energy
is defined as the total energy observed in the calorimeter after
removing the energy deposits associated with the high $\pt$ leptons in
our $\wlnu$ and $\zll$ candidates.  To tune the simulated recoil
energy distributions, we separate the observed recoil $\et$ in each
event into components that are parallel and perpendicular to the
transverse direction of the highest $\pt$ lepton in the event.  The
two components, $U^{\mathrm{recl}}_{\parallel}$ and
$U^{\mathrm{recl}}_{\perp}$, are each assigned energy shift~($C$) and
scale~($K$) corrections in the form
\begin{eqnarray}
(U^{\mathrm{recl}}_{\parallel})^{\prime} & 
= & ( K_{\parallel} \times U^{\mathrm{recl}}_{\parallel}) + C_{\parallel},
\label{eq:fixpar}
\\
(U^{\mathrm{recl}}_{\perp})^{\prime} & 
= & ( K_{\perp} \times U^{\mathrm{recl}}_{\perp}) + C_{\perp}.
\label{eq:fixperp}
\end{eqnarray}
The scale corrections are used to account for 
 problems in the calorimeter response model 
and the effects of multiple interactions, the 
underlying event model, and accelerator backgrounds 
which are in principle independent of the lepton 
direction.  The shift corrections are 
designed to account for simulation deficiencies  
that have a lepton-direction dependence such as 
the $\W$ boson recoil model and the model for lepton 
energy deposition in the calorimeter.  Based on the 
nature of these effects, we expect that the scaling 
corrections in both directions, $K_{\parallel}$ and 
$K_{\perp}$, should be equivalent and that the shift 
correction in the perpendicular direction, $C_{\perp}$, 
should be zero.  We check these assumptions, however,
by keeping each parameter independent in the fitting 
procedure used to determine the best values for tuning
the recoil energy in simulated events.  

\begin{figure}
\includegraphics[width=3.5in]{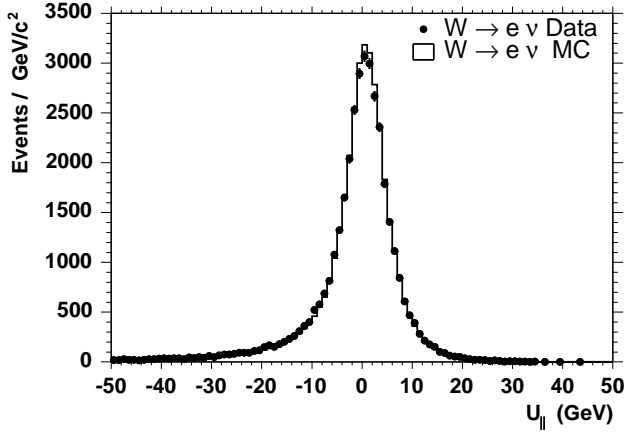}
\caption{Comparison of $U^{\mathrm{recl}}_{\parallel}$ recoil 
energy distributions for $\wenu$ candidate events in 
tuned simulation and data.}
\label{fig:upar_el}
\end{figure}

\begin{figure}
\includegraphics[width=3.5in]{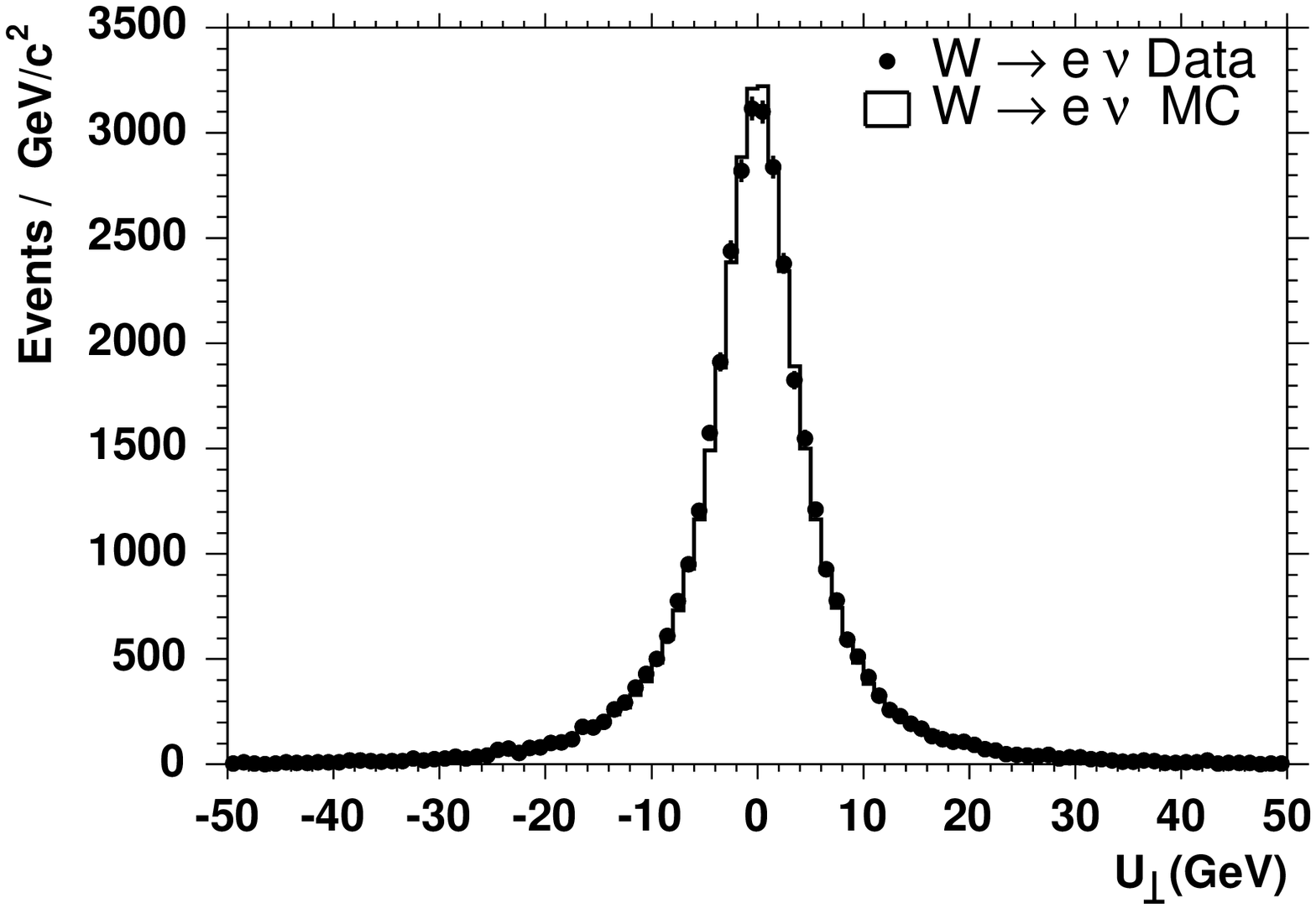}
\caption{Comparison of $U^{\mathrm{recl}}_{\perp}$ recoil 
energy distributions for $\wenu$ candidate events 
in tuned simulation and data.}
\label{fig:uper_el}
\end{figure}

To determine the best values for these scaling and shifting constants,
we perform $\chi^{2}$ fits between the data recoil energy
distributions and corrected distributions from the simulation based on
a range of scaling and shifting constants.  An iterative process is
used in which we first determine the best possible shifting constants
and then fit for scaling constants based on those values.  We repeat
this process until the $\chi^{2}$ fits for both the scaling and
shifting constants stabilize at set values.  No effects are expected
which can give rise to shifts in the energy perpendicular to the
lepton momentum and the fitted shifts of these distributions are
consistent with zero.  We set $C_{\perp}$ to zero.  We also find that
the fitted scale factors for both recoil energy components agree well
with each other in both the electron and muon candidate samples.
Based on this agreement, we also make a combined fit to both
components for a single correction scale factor.  We use this single
scaling factor to correct both recoil energy components.  A comparison
of the $U^{\mathrm{recl}}_{\parallel}$ and $U^{\mathrm{recl}}_{\perp}$
distributions for $\wenu$ candidate events in tuned simulation and
data are shown in Figs.~\ref{fig:upar_el} and~\ref{fig:uper_el}.

The uncertainties on our measured acceptances related 
to the recoil energy model in the simulation are  
estimated using the $\pm$~3~$\sigma$ values of the scale 
and shift correction factors returned from our fit 
procedure.  As in the case of boson $\pt$ model
uncertainties, we choose to use the $\pm$~3~$\sigma$ 
values rather than the $\pm$~1~$\sigma$ values as we 
are using these parameters to cover a wide range of 
effects that are potentially incorrectly modeled in 
our simulated event samples.  Since the tuning 
parameters are not directly related to the underlying 
mechanisms that affect the recoil energy distributions,
we choose to be conservative in how we estimate the 
associated acceptance uncertainties via this procedure.  
We recalculate the acceptance of our candidate samples 
with each of the individual tuning parameters changed 
to its $\pm$~3~$\sigma$ values and assign an uncertainty 
based on the differences between these results and our
default acceptance values.  The changes in acceptance 
found from modifying the overall scale correction 
$K$ and the shift corrections for both directions,
$C_{\parallel}$ and $C_{\perp}$, are added in 
quadrature to estimate the total uncertainties on 
our measured acceptances due to the recoil energy model
in the simulation.  To be conservative we choose to 
include an uncertainty based on fit results for $C_{\perp}$ 
even though this parameter is set to zero for tuning the 
simulated recoil energy distributions.

Table~\ref{tab:recerr} summarizes the best fit values 
and $\pm$~3~$\sigma$ variations with respect to the 
best fit values obtained for each of the scaling and 
shifting parameters used to tune the recoil energy 
model in simulation and the corresponding acceptance 
uncertainties for the $\wlnu$ candidate samples. 

\begin{table*}[t]
\caption{Summary of simulation recoil energy tuning 
parameter values and uncertainties obtained from our 
fit procedure and the corresponding uncertainties on 
our measured acceptance values.}
\begin{tabular}{l c c c c c c } 
\hline
\hline  
Tuning    & $\wenu$      & $\wenu$                    & $\wmnu$   & $\wmnu$ & & \\ 
Parameter & Fit Value    & $\pm$~3~$\sigma$ variation & Fit Value & $\pm$~3~$\sigma$ variation
                         & $\Delta A_{\wenu}$ & $\Delta A_{\wmnu}$              \\ 
          &              &                            &           &                                     & ($\!\%$)           & ($\!\%$)                         \\
\hline 
$K_{\parallel}$ & 1.06               & 0.02                             & 1.06            & 0.03 & -         & -         \\ 
$K_{\perp}$     & 1.04               & 0.02                             & 1.05            & 0.02 & -         & -         \\
$K$             & 1.05               & 0.02                             & 1.05            & 0.02 & 0.17  & 0.20  \\
$C_{\parallel}$ & -0.4               & 0.1                              & -0.1            & 0.1  & 0.18  & 0.29  \\ 
$C_{\perp}$     &  0.0               & 0.1                              &  0.0            & 0.1  & 0.00  & 0.00  \\ 
Combined        &                    &                                  &                 &      & 0.25  & 0.35  \\ 
\hline
\hline
\end{tabular}
\label{tab:recerr}
\end{table*}

\subsection{Uncertainties from Energy and Momentum Scale/Resolution}
\label{subsec:epres}

The modeling of COT track $\pt$ scale and resolution in the simulation
affects our acceptance estimates for the minimum track $\pt$
requirements made on muon and electron candidates in our samples.
Similarly, the model of cluster $\et$ scale and resolution for the
electromagnetic sections of the calorimeter can change the acceptance
estimates for the minimum cluster $\et$ requirements on electrons.
Lepton energy and momentum measurements can also alter the event
$\met$ calculation, and incorrect modeling of these quantities can
therefore also affect our acceptance estimates for the minimum $\met$
criteria applied to our $\wlnu$ samples.

\begin{figure}
\includegraphics[width=3.5in]{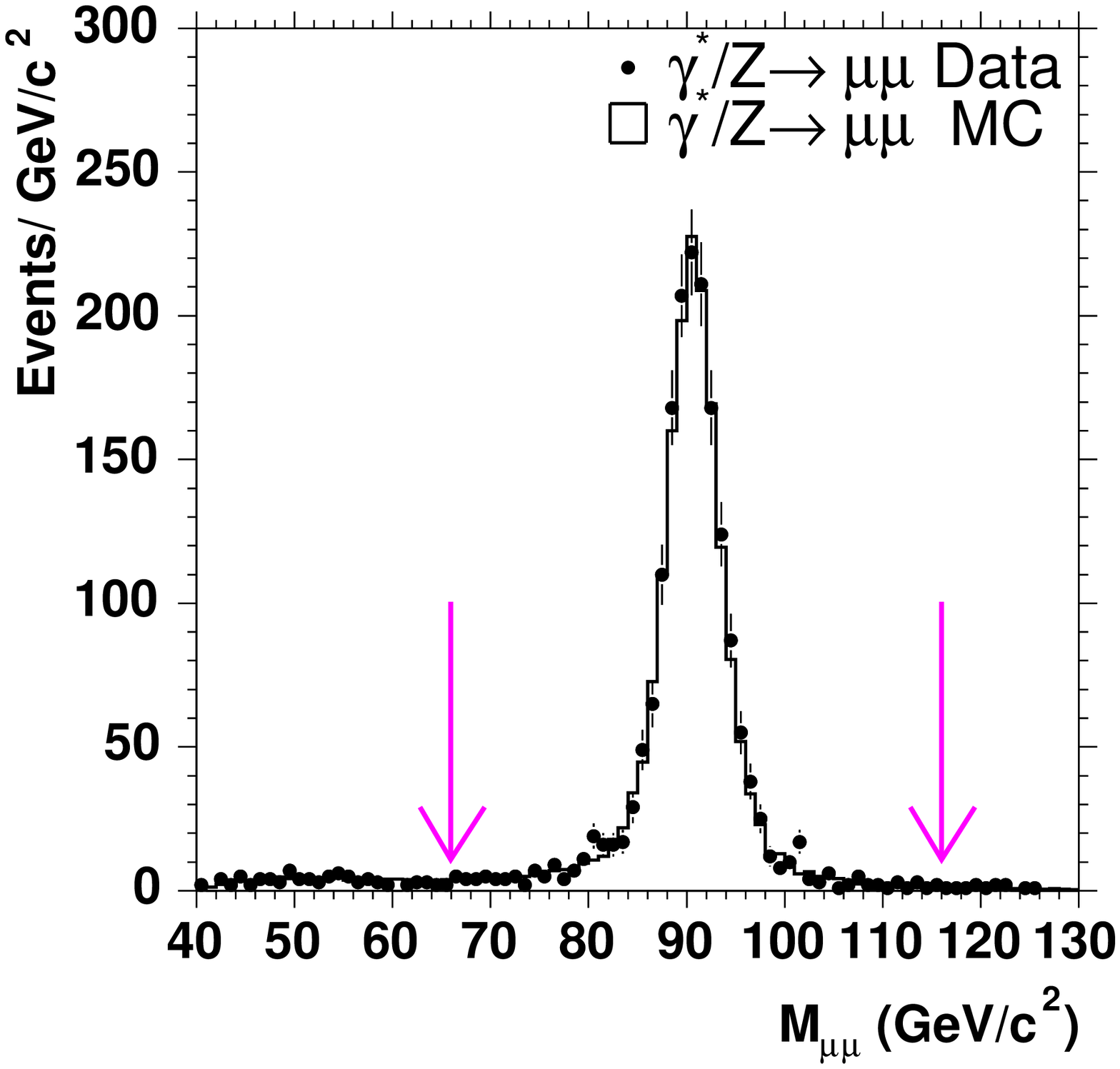}
\caption{$\zgmm$ invariant mass distribution in data and 
tuned simulation normalized to the data.  The arrows indicate the
invariant mass range of our $\gamma^{*}/Z$ cross section measurement.}
\label{fig:ptfit} 
\end{figure} 

We check the scale and resolution of the track 
$\pt$ and cluster $\et$ measurements in the 
simulation using the invariant mass distributions
of $\zgmm$ and $\zgee$ candidate events.  A direct 
comparison of these distributions in data and 
simulation is possible due to the small level of 
background contamination in these samples.  We 
first define scale factors for COT track $\pt$ 
($K_{\pt}$) and cluster $\et$ ($K_{\et}$) in the 
simulation via the expressions 
\begin{eqnarray}
p^{\prime}_{T} & = & K_{\pt} \times \pt,
\label{eq:fixpt}
\\
E^{\prime}_{T} & = & K_{\et} \times \et.
\label{eq:fixet}
\end{eqnarray}
The best values for these scale factors are 
determined by making a series of $\chi^{2}$ fits 
between the $\zgll$ invariant mass distributions 
in data and tuned simulation based on a range 
of values for the scale factors.  The 
best $\chi^{2}$ fit for the track $\pt$ scale
factor is $K_{\pt} =$ 0.997.  Since the mean of the 
$\zgmm$ invariant mass peak in the simulation is 
centered on the measured $\Z$ boson mass, the best 
fit value for $K_{\pt}$ is indicative of the fact 
that the current $\pt$ scale for reconstructed 
tracks in the data is low.  This result is 
consistent with track $\pt$ scaling factors for 
simulation obtained from similar fits to the $J/\Psi 
\rightarrow \mu \mu$ and $\Upsilon \rightarrow 
\mu \mu$ invariant mass peaks indicating that 
the resulting scale factor is not $\pt$ dependent.  
The best fits for the cluster $\et$ scale factors 
in the central and plug calorimeter modules are 
$K_{\et}$(central)~$=$~1.000 and 
$K_{\et}$(plug)~$=$~1.025 indicative of a model 
that underestimates energy deposition in the plug 
modules but is accurate for the central module.

\begin{figure}
\includegraphics[width=3.5in]{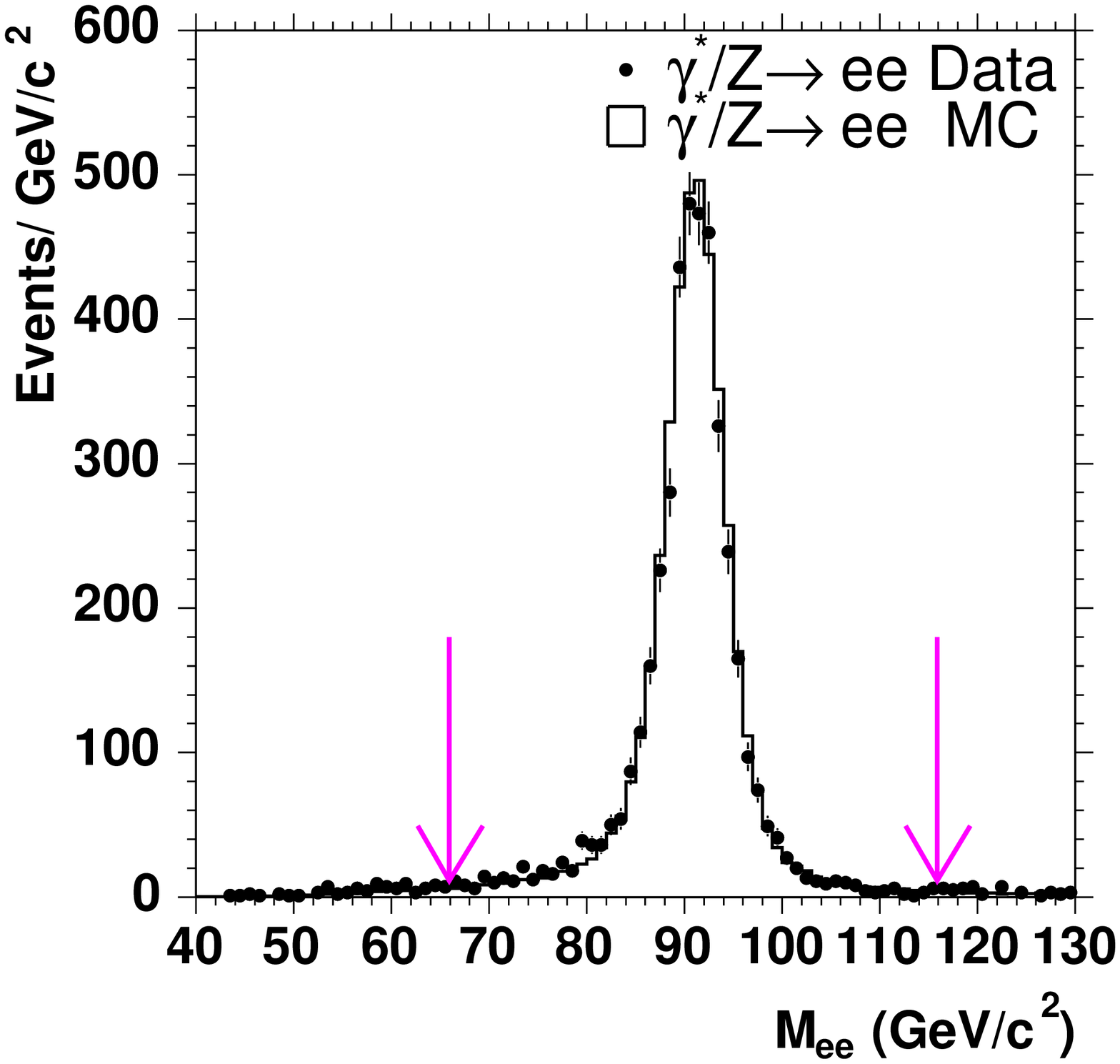}
\caption{$\zgee$ invariant mass distribution in data and 
tuned simulation normalized to the data. The arrows indicate the
invariant mass range of our $\gamma^{*}/Z$ cross section measurement.}
\label{fig:etfit} 
\end{figure}

Comparisons of the $\zgll$ invariant mass distributions in the data
and the simulation are used to tune the track $\pt$ and cluster $\et$
resolution in the simulation.  We smear these values in simulated
events by generating a random number from a Gaussian distribution with
mean equal to one and width equal to a chosen $\sigma$ for each lepton
candidate in our samples.  The resolution smearing is obtained by
multiplying the track $\pt$ and/or cluster $\et$ by the different
random numbers obtained from our distribution.  Setting $\sigma$ equal
to zero adds no smearing since each generated random number equals one
by definition.  The best values for $\sigma$ are obtained from
$\chi^{2}$ fits between the $\zgll$ invariant mass distributions in
data and tuned simulation corresponding to a range of values for
$\sigma$.  The best $\chi^{2}$ fits for track $\pt$ and central
calorimeter $\et$ resolution are found to be for the case of $\sigma$
equal to zero indicating that these resolutions are well-modeled in
the simulation.  The best $\chi^{2}$ fit for plug calorimeter $\et$
resolution is for a value of $\sigma$ above zero indicating that the
simulation model for $\et$ resolution in the plug modules needs to be
degraded to match the data better.  Fig.~\ref{fig:ptfit} and 
Fig.~\ref{fig:etfit} show comparisons between the $\zgmm$ and $\zgee$
invariant mass distributions in data and tuned simulation.

\begin{table*}[t]
\caption{Summary of simulation track $\pt$ scale 
and resolution tuning parameters and corresponding  
uncertainties on our measured acceptance values.} 
\begin{tabular}{l c c c c c c c c} 
\hline
\hline  
Tuning                   & $\zmm$ & $\zmm$ & $\zee$ & $\zee$ & & & & \\ 
Parameter                & Fit Value & $\pm$~3~$\sigma$ variation & Fit Value & $\pm$~3~$\sigma$ variation
                         & $\Delta A_{\wenu}$ & $\Delta A_{\wmnu}$
                         & $\Delta A_{\zee}$  & $\Delta A_{\zmm}$  \\ 
                         &           &                            &           &                         & ($\!\%$)           & ($\!\%$)                                                & ($\!\%$)           & ($\!\%$)           \\
\hline 
$K_{\pt}$                & 0.997  & 0.003 & - & - & 0.03  & 0.21  & 0.04  & 0.05  \\ 
$\sigma_{\pt}$           & 1.000  & 0.003 & - & - & 0.00  & 0.00  & 0.00  & 0.00  \\
$K_{\et}$ (Central)      & - & - & 1.000  & 0.003 & 0.34  & 0.00  & 0.23  & 0.00  \\
$\sigma_{\et}$ (Central) & - & - & 1.000  & 0.015 & 0.03  & 0.00  & 0.05  & 0.00  \\
$K_{\et}$ (Plug)         & - & - & 1.025  & 0.006 & 0.00  & 0.00  & 0.11  & 0.00  \\
$\sigma_{\et}$ (Plug)    & - & - & 1.027  & 0.011 & 0.00  & 0.00  & 0.05  & 0.00  \\ 
\hline
\hline
\end{tabular}
\label{tab:perr}
\end{table*}

The effects of uncertainties in the simulation model for 
the scale and resolution of track $\pt$ and cluster 
$\et$ on our measured acceptances are estimated based 
on the $\pm$~3~$\sigma$ values of the corresponding tuning 
parameters obtained from our fit procedure.  Our choice 
of using the $\pm$~3~$\sigma$ values to estimate 
acceptance uncertainties is conservatively based on the 
idea that these tuning parameters are not directly related 
to the underlying mechanisms that set the scale and 
resolution of track $\pt$ and cluster $\et$ in the detector.      
The acceptance uncertainties are estimated by observing 
the changes in measured acceptance for each candidate 
sample that occur when each individual tuning parameter
is changed between its default and $\pm$~3~$\sigma$ values. 
A summary of the fitted values and uncertainties of the 
scale and resolution tuning parameters for track $\pt$
and cluster $\et$ is given in Table~\ref{tab:perr}
along with the estimated uncertainties on the measured 
acceptances of our candidate samples associated with 
each parameter.

\subsection{Uncertainties from Detector Material Model}

The acceptances of the kinematic selection criteria applied to
electron candidates are dependent on the amount of material in the
detector tracking volume since electrons can lose a significant
fraction of their energy prior to entering the calorimeter via
bremsstrahlung radiation originating from interactions with detector
material.  

The electron $E/p$ distribution, because of its sensitivity to
radiation, is used to compare the material description in the detector
simulation in the central region with that of the real detector as
observed in data.  One measure of the amount of material that
electrons pass through in the tracking region is the ratio of the
number of events in the peak of the $E/p$ distribution (0.9~$< E/p
<$~1.1) to the number of events in the tail of the distribution
(1.5~$< E/p <$~2.0).  We study the uncertainty in the amount of
material in the simulation by varying the thickness of a cylindrical
layer of material in the detector simulation geometry description in
the region between the silicon and COT tracking volumes.  We choose to
use copper as the material for this cylindrical layer as it best
describes the silicon tracker copper readout cables and is also
supported by independent studies of muon energy loss in the
calorimeter.  Based on electron candidates produced in decays of both
$\W$ and $\Z$ bosons, we determine that the matching of the $E/p$
distribution between data and simulation has an uncertainty
corresponding to $\pm$~1.5~$\!\%$ of a radiation length ($X_0$) of
copper.  This variation in the thickness of the cylindrical layer is
used to model the acceptance uncertainties originating from the model
of the detector material in the simulation.

This result is cross-checked by counting the fraction of electrons in
$\wenu$ candidate events which form ``tridents'' (see
Sec.~\ref{sec:backg}).  The probability of finding a trident, created
when an electron radiates a photon which immediately converts into an
electron-positron pair, is strongly dependent on the amount of
material traversed by the electron inside the tracking volume.  We
also compare the resolution of the $\zee$ invariant mass peak in data
and simulation which is sensitive to the rate of radiative
interactions within the tracking volume.  The results of these studies
are consistent with the $E/p$ results.  

\begin{figure}
\includegraphics[width=3.5in]{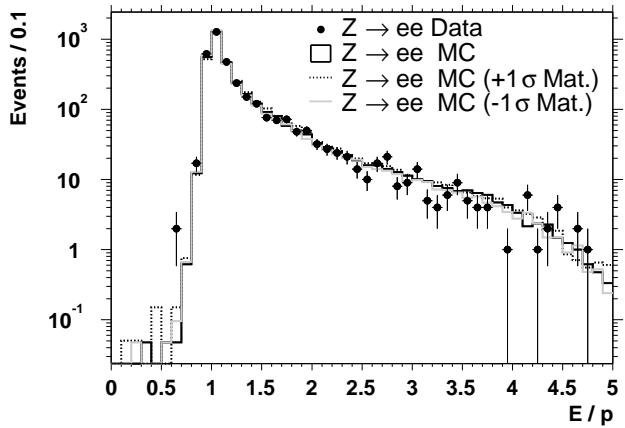}
\caption{\label{fig:epmatch} Comparison of $E/p$ distribution for
electron candidates in $\zee$ events in data and simulation.  The
$\pm$~1~$\sigma$ samples are simulated with $\pm$~1.5~$\!\%$ of a
radiation length of copper in the tracking volume.}
\end{figure}

Fig.~\ref{fig:epmatch} shows the $E/p$ distributions for electron
candidates in our $\zee$ data and simulated event samples.  The
$\pm$~1~$\sigma$ material samples are simulated using $\pm$~1.5~$\!\%$
of a radiation length of copper.  We observe good agreement between
data and our default simulation in the region below $E/p =$ 2.5.  In
the high $E/p$ tail above this value, the comparison is biased by
dijet background events in the data.

The tracks associated with electron candidates in the calorimeter plug
modules have a low reconstruction efficiency due to the limited number
of tracking layers in the forward region.  Therefore, the plug
preradiator detector is used to study the detector material in the
simulation for plug electron candidates.  The amount of energy
deposited in the plug preradiator depends on the shower evolution of
the electron in front of the calorimeter which is itself dependent on
the amount of material the electron passes through before entering the
calorimeter.  On average, electrons passing through more material
inside the tracking volume will have more evolved showers at the inner
edges of the calorimeter and therefore deposit more energy in the plug
preradiator.

\begin{figure*}[t]
\includegraphics[width=6.in]{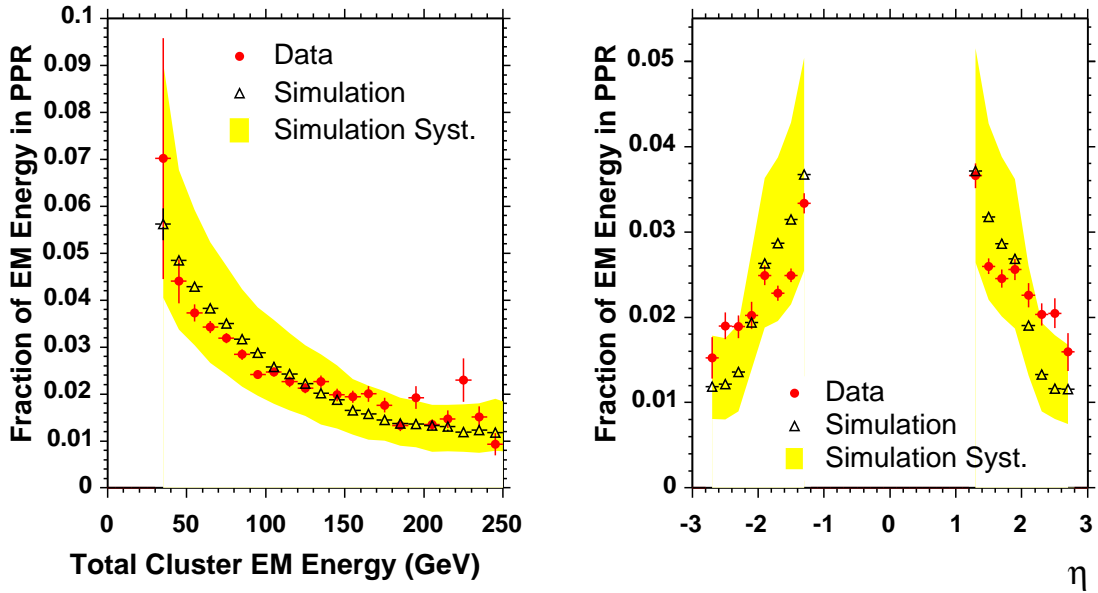}
\caption{\label{fig:pprmatch} Comparison of observed ratio of 
energies in plug preradiator and plug electromagnetic calorimeter 
for electron candidates as a function of the combined energy (left) 
and pseudorapidity (right) of candidates.  Data distributions are 
denoted by the filled circles.  The open triangles and associated 
shaded band show the distribution and uncertainty range obtained 
from simulation when $\pm$~16.5~$\!\%$ of a  radiation length thick 
iron disk is used in the detector material description.}
\end{figure*}

To study the detector simulation material description in the forward
part of the tracking volume, we compare the ratio of energies observed
in the plug preradiator and remaining plug calorimeter sections for
forward electron candidates in data and simulation.  As in the central
region, we study the material in the simulation by varying the
thickness of an iron disk in the volume between the tracking chamber
endplate and the inner edge of the plug calorimeter.  These studies
indicate that our model for detector material in the forward region
has an uncertainty corresponding to $\pm$~16.5~$\!\%$ of a radiation
length ($X_0$) in the iron disk.  Fig.~\ref{fig:pprmatch} shows the
ratio of energies observed in the plug preradiator (PPR) and the plug
electromagnetic calorimeter (PEM) for electron candidates in data and
simulation as a function of both the combined energy (PPR + PEM) and
pseudorapidity of the candidates.

Acceptance uncertainties coming from the simulation material model are
determined by generating simulated event samples with the thicknesses
of the extra material layers set one at a time in the simulation to
the lower and upper limits of their uncertainty ranges.  The changes
in measured acceptance for the $\wenu$ and $\zgee$ samples relative to
the default simulation for the modified detector material models are
summarized in Table~\ref{tab:matsyst}.

\begin{table}
\caption{Summary of acceptance uncertainties due to
detector tracking volume material model in simulation.} 
\begin{tabular}{l c r} 
\hline
\hline  
Material Model       & $\Delta A_{\wenu}$        & $\Delta A_{\zee}$   \\ 
\hline 
Central              & 0.73~$\!\%$               & 0.94~$\!\%$         \\  
Plug                 & -                         & 0.21~$\!\%$         \\ 
\hline
\hline
\end{tabular}
\label{tab:matsyst}
\end{table}

\subsection{Acceptance Uncertainty Summary}

The acceptance uncertainties on our event 
samples are summarized in Table~\ref{tab:accerr}. 

\begin{table*}[t]
\caption{Summary of estimated uncertainties on the 
measured acceptances for our four candidate samples.}
\begin{tabular}{l c c c c } 
\hline
\hline 
Uncertainty Category             & $\Delta A_{\wenu}$        & $\Delta A_{\wmnu}$
                                 & $\Delta A_{\zee}$         & $\Delta A_{\zmm}$               \\ 
                                 & ($\!\%$)                  & ($\!\%$)                                         & ($\!\%$)                  & ($\!\%$)                        \\
\hline
NNLO $d\sigma/dy$ Calculation    & 0.29  & 0.25  & 0.06  & 0.72        \\
PDF Model (positive)             & 1.16  & 1.13  & 0.69  & 1.72        \\
PDF Model (negative)             & 1.50  & 1.47  & 0.84  & 2.26        \\
Boson $\pt$ Model                & 0.04  & 0.04  & 0.06  & 0.08        \\ 
Recoil Energy Model              & 0.25  & 0.35  & 0.00  & 0.00        \\
Track $\pt$ Scale/Resolution     & 0.03  & 0.21  & 0.04  & 0.05        \\ 
Cluster $\et$ Scale/Resolution   & 0.34  & 0.00  & 0.26  & 0.00        \\
Detector Material Model          & 0.73  & 0.00  & 0.96  & 0.00        \\
Simulated Event Statistics       & 0.13  & 0.14  & 0.24  & 0.41        \\
Total (positive)                 & 1.46  & 1.22  & 1.23  & 1.94        \\
Total (negative)                 & 1.75  & 1.57  & 1.26  & 2.44        \\ 
\hline
\hline
\end{tabular}
\label{tab:accerr}
\end{table*}

\section{Efficiency}
\label{sec:eff}
\subsection{Introduction}

The acceptance values estimated from our simulated samples are
corrected for additional inefficiencies from event selection criteria
that are either not modeled in the simulation or are better measured
directly from data.  We determine a combined efficiency,
$\epsilon_{\mathrm{tot}}$, for each candidate sample based on measured
efficiencies for the individual selection criteria.  We account for
correlations between different selection criteria by having a specific
order in which individual efficiency measurements are made.  The
efficiency measurement for a given selection criterion is made using a
subset of candidates that passes the full set of selection criteria
ordered prior to the one being measured.  In addition, since the
efficiency is applied as a correction to the acceptance, candidates
used to measure efficiencies are also required to meet the geometrical
and kinematic requirements used to define these acceptances.  The
ordering and definitions of the individual selection criteria
efficiencies are presented in this introductory section.  The
following two sections describe how these individual efficiencies are
combined to obtain the total event efficiencies for our $\wlnu$ and
$\zll$ candidate samples.  The remaining sections describe how each of
the individual efficiency terms is measured.

The first efficiency term is $\epsilon_{\mathrm{vtx}}$, 
the fraction of $\ppbar$ collisions that occur within 
$\pm$~60~$\cm$ of the center of the detector along the 
$z$-axis.  We impose this requirement as a fiducial cut 
to ensure that $\ppbar$ interactions are well-contained 
within the geometrical acceptance of the detector.  The 
$z$-coordinate of the event vertex for a given event 
is taken from the closest intersection point of the 
reconstructed high $\pt$ lepton track(s) with the $z$-axis.  
Since event selection criteria can bias our samples 
against events originating in the outer interaction region, 
the efficiency of our vertex position requirement, 
$\epsilon_{\mathrm{vtx}}$, is measured directly from the 
observed vertex distribution in minimum-bias events.

We define $\epsilon_{\mathrm{trk}}$ as the efficiency 
for reconstructing the track of the high $\pt$ lepton in 
the COT and $\epsilon_{\mathrm{rec}}$ as the efficiency 
for matching the found track to either a reconstructed 
electromagnetic cluster in the calorimeter (electrons) 
or a reconstructed stub in the muon chambers (muons).  
The $\epsilon_{\mathrm{rec}}$ term incorporates both the 
reconstruction efficiency for the cluster or stub and 
the matching efficiency for connecting the reconstructed 
cluster or stub with its associated COT track.

For reconstructed leptons (tracks matched to clusters or 
stubs), $\epsilon_{\mathrm{id}}$ is the efficiency of the 
lepton identification criteria used to increase the purity 
of our lepton samples.  To increase the number of events 
in our $\zll$ candidate samples, we use a looser set of 
identification criteria on the second lepton leg in these 
events.  The loose lepton selection criteria are a subset 
of the set of cuts applied to the single lepton in 
$\wlnu$ events and the first lepton leg in $\zll$ events.  
The combined efficiency for the loose subset of cuts is 
referred to as $\epsilon_{\mathrm{lid}}$, and we define 
$\epsilon_{\mathrm{tid}}$ as the efficiency for the set 
of remaining identification cuts not included in the loose 
subset.  The efficiency of our lepton isolation requirement, 
which helps to reduce non-$\W/\Z$ backgrounds in our samples, 
is defined independently as $\epsilon_{\mathrm{iso}}$.  It 
is important to avoid double-counting correlated efficiency 
losses when measuring the efficiencies for our two sets of 
identification cuts and the isolation requirement.  We 
eliminate this problem by defining a specific ordering of 
these terms ($\epsilon_{\mathrm{lid}}$, $\epsilon_{\mathrm{iso}}$, 
$\epsilon_{\mathrm{tid}}$) and measuring each efficiency term 
using the subset of lepton candidates that meets the requirements 
associated with all of the efficiency terms ordered prior to that 
being measured.  A natural consequence of using this procedure 
is that the total lepton identification efficiency, 
$\epsilon_{\mathrm{id}}$, is necessarily equal to the product of 
$\epsilon_{\mathrm{lid}}$ and $\epsilon_{\mathrm{tid}}$.  

As discussed previously, the high $\pt$ electron and muon data
samples used to make the production cross section measurements 
are collected with lepton-only triggers.  We define 
$\epsilon_{\mathrm{trg}}$ as the efficiency for an isolated, 
high quality reconstructed lepton to have satisfied all of the 
requirements of the corresponding lepton-only trigger path.  
CDF has a three-level trigger system, and the value of 
$\epsilon_{\mathrm{trg}}$ is determined from the product of the 
efficiencies measured for each of the levels.  The measured 
efficiency for a specific level of the trigger is based on the 
subset of reconstructed track candidates that satisfy the trigger 
requirements of the levels beneath it.  This additional
requirement is made to avoid double-counting correlated losses 
in efficiency observed in the different trigger levels.

Finally, there are two efficiencies that are applied only in 
measurements made in the muon decay channels.  We define 
$\epsilon_{\mathrm{cos}}$ as the efficiency for signal 
events not to be tagged as cosmic ray candidates via our 
tagging algorithm.  The cosmic ray tagging algorithm is not 
based on the properties of a single muon, but rather on the 
full set of tracking data available from the COT in each event.  
As a result, $\epsilon_{\mathrm{cos}}$ is determined as an 
overall event efficiency rather than an additional lepton 
efficiency.  Due to topological differences between $\wmnu$ 
and $\zmm$ events, the fraction of signal events tagged by the 
algorithm as cosmic rays is different for the two candidate 
samples.  We refer to the efficiency term for the $\wmnu$ sample 
as $\epsilon^{W}_{\mathrm{cos}}$ and that for the $\zmm$ sample 
as $\epsilon^{Z}_{\mathrm{cos}}$.  One additional event selection 
made only in the case of our $\wmnu$ candidate sample is the 
$\Z$-rejection criteria.  Due to the non-uniform coverage of 
the muon chambers, we find cases in which only one of the two 
high $\pt$ muon tracks originating from a $\Z$-boson decay has 
a matching stub in the muon detector.  The additional selection 
criteria made to eliminate these events from our $\wmnu$ 
candidate sample has a corresponding efficiency defined as 
$\epsilon_{\mathrm{z-rej}}$.

\subsection{$\wlnu$ Efficiency Calculation}

The efficiency of detecting a $\wlnu$ decay that satisfies the 
kinematic and geometrical criteria of our samples is obtained 
from the formula shown in Eq.~\ref{eq:weffcalc}.

\begin{eqnarray}
\epsilon_{\mathrm{tot}} = && \epsilon_{\mathrm{vtx}} \times \epsilon_{\mathrm{trk}} \times 
\epsilon_{\mathrm{rec}} \times \epsilon_{\mathrm{id}} \nonumber\\
&& \times \epsilon_{\mathrm{iso}} \times 
\epsilon_{\mathrm{trg}} \times \epsilon_{\mathrm{z-rej}} \times \epsilon^{W}_{\mathrm{cos}}
\label{eq:weffcalc}
\end{eqnarray}
 
As described in detail above, the ordering of the cuts, as shown
by their left to right order in the formula, is important.  Each 
efficiency term is an efficiency for the subset of $\wlnu$ events 
that satisfies the kinematic and geometric criteria of our samples 
as well as the requirements associated with each of the efficiency 
terms to the left of the term under consideration.  For example, 
the trigger efficiency term in the formula, $\epsilon_{\mathrm{trg}}$, 
is an efficiency for reconstructed leptons that satisfy the 
geometrical, kinematic, identification, and isolation criteria 
used to select the high $\pt$ lepton in our $\wlnu$ candidate 
events.  As noted previously, the $\epsilon_{\mathrm{z-rej}}$ and 
$\epsilon^{W}_{\mathrm{cos}}$ terms in the formula apply to the 
$\wmnu$ candidate sample only.  Table~\ref{tb:ceneff} summarizes 
the measurements of the individual efficiency terms (described in 
detail below) and the resulting combined efficiencies for our 
$\wlnu$ candidate samples.  The electron efficiencies shown 
in Table~\ref{tb:ceneff} are for central calorimeter electrons 
only since our $\wenu$ cross section measurement is also 
restricted to candidates in this part of the detector.     

\begin{table*}[t]
\caption{Summary of the individual efficiency terms for $\wlnu$.}
\begin{tabular}{l c c r} 
\hline
\hline 
Selection Criteria       & Label                 & $\wenu$           & $\wmnu$             \\ 
\hline
Fiducial Vertex          & $\epsilon_{\mathrm{vtx}}$      & 0.950 $\pm$ 0.004 & 0.950 $\pm$ 0.004   \\
Track Reconstruction     & $\epsilon_{\mathrm{trk}}$      & 1.000 $\pm$ 0.004 & 1.000 $\pm$ 0.004   \\
Lepton Reconstruction    & $\epsilon_{\mathrm{rec}}$      & 0.998 $\pm$ 0.004 & 0.954 $\pm$ 0.007   \\
Lepton ID                & $\epsilon_{\mathrm{id}}$       & 0.840 $\pm$ 0.007 & 0.893 $\pm$ 0.008   \\
Lepton Isolation         & $\epsilon_{\mathrm{iso}}$      & 0.973 $\pm$ 0.003 & 0.982 $\pm$ 0.004   \\
Trigger                  & $\epsilon_{\mathrm{trg}}$      & 0.966 $\pm$ 0.001 & 0.925 $\pm$ 0.011   \\
$\Z$-Rejection Cut       & $\epsilon_{\mathrm{z-rej}}$    & -                 & 0.996 $\pm$ 0.002   \\ 
Cosmic Ray Tagging       & $\epsilon^{\W}_{\mathrm{cos}}$ & -                 & 0.9999 $\pm$ 0.0001 \\
\hline
Total                    & $\epsilon_{\mathrm{tot}}$      & 0.749 $\pm$ 0.009 & 0.732 $\pm$ 0.013   \\ 
\hline
\hline
\end{tabular}
\label{tb:ceneff}
\end{table*}

\subsection{$\zll$ Efficiency Calculation}

For both electrons and muons, we define a loose set of lepton 
selection criteria for the second leg of $\zll$ events to 
increase the size of our candidate samples.  The efficiency 
calculation for these samples is complicated by the fact that 
in many events both leptons from the $\Z$ boson decay can 
satisfy the tight lepton selection criteria which are required 
for only one of the two legs.  

In the electron channel, we allow for two different types of loose 
lepton legs.  The second leg can be either a central calorimeter   
electron candidate passing a looser set of selection criteria or 
an electron reconstructed in the forward part of the calorimeter 
(plug modules).  For $\zmm$ candidates, a loose track leg is 
not required to have a matching reconstructed stub in the muon 
detectors.  For this sample, the second muon leg is simply 
required to be a high $\pt$, isolated track satisfying the subset 
of muon identification cuts corresponding to the track itself.  
The breakdown of lepton identification cut efficiencies between 
the loose and tight criteria is shown in Table~\ref{tb:looseeff} 
for both muons and central electrons.  There is no reconstruction 
inefficiency associated with loose muon legs since track candidates 
are not required to have a matching muon detector stub. 
  
\begin{table}[t]
\caption{Breakdown of loose and tight lepton identification efficiencies.}
\begin{tabular}{l c c c} 
\hline
\hline 
Selection Criteria       & Label                     & Central Electron       & Muon              \\ 
\hline
Loose Lepton ID      & $\epsilon_{\mathrm{lid}}$ & 0.960 $\pm$ 0.004      & 0.933 $\pm$ 0.006 \\
Tight Lepton ID      & $\epsilon_{\mathrm{tid}}$ & 0.876 $\pm$ 0.007      & 0.957 $\pm$ 0.005 \\
All Lepton ID        & $\epsilon_{\mathrm{id}}$  & 0.840 $\pm$ 0.007      & 0.893 $\pm$ 0.008 \\ 
\hline
\hline
\end{tabular}
\label{tb:looseeff}
\end{table}

Efficiencies for loose plug electrons are given in 
Table~\ref{tb:plugeff}.  There is no track reconstruction 
component in the plug electron selection efficiency since 
a matched track is not required for candidates in the plug
region of the calorimeter.  Also, since no matching between
tracks and clusters is done in this region, the plug lepton 
reconstruction efficiency is 100~$\!\%$.  There are no dead
calorimeter towers in the data-taking period used in these 
measurements.  We also find that kinematic distributions for 
tight central electron legs in our central-plug $\zee$ event 
sample are somewhat different from those in the central-central 
sample.  These kinematic differences have a small effect on 
the electron identification efficiencies for the central legs 
in central-plug $\zee$ events.  In order to correct for 
this effect, we measure a central leg scale factor, 
$S^{\mathrm{plug}}_{\mathrm{cl}}$, which is the ratio of 
central leg efficiencies in central-plug $\zee$ events to 
those in central-central events.  The value of this scale 
factor given in Table~\ref{tb:plugeff} is determined from 
simulation and is applied as an extra term in the overall 
selection efficiency for plug electrons.    
  
\begin{table}
\caption{Plug electron efficiencies.}
\begin{tabular}{l c c} 
\hline
\hline 
Selection Criteria       & Label                                     & Plug Electron       \\ 
\hline
Lepton Reconstruction    & $\epsilon^{\mathrm{plug}}_{\mathrm{rec}}$ & 1.000               \\ 
Lepton ID                & $\epsilon^{\mathrm{plug}}_{\mathrm{id}}$  & 0.876 $\pm$ 0.015   \\ 
Lepton Isolation         & $\epsilon^{\mathrm{plug}}_{\mathrm{iso}}$ & 0.993 $\pm$ 0.003   \\
Central Leg Scale Factor & $S^{\mathrm{plug}}_{\mathrm{cl}}$         & 1.014 $\pm$ 0.002   \\
Total                    & $\epsilon^{\mathrm{plug}}_{\mathrm{tot}}$ & 0.883 $\pm$ 0.015   \\ 
\hline
\hline
\end{tabular}
\label{tb:plugeff}
\end{table}

To determine a total event selection efficiency for 
$\zee$ events, we first calculate efficiencies for the 
central-central and central-plug samples which are 
independent of one another by definition.  The total
efficiency is a weighted sum of the efficiencies for the 
two samples.  The weighting factors are determined from 
the relative numbers of central-central and central-plug
events in our simulated sample.  The fraction of 
central-plug events, $f_{\mathrm{cp}}$, is determined to 
be 0.655~$\pm$~0.001.  Eq.~\ref{eq:cceff} shows the 
efficiency calculation for central-central $\zee$ events:
\begin{eqnarray}
\epsilon_{\mathrm{tot}}^{\mathrm{cc}} = && \epsilon_{\mathrm{vtx}} \times \epsilon^{2}_{\mathrm{trk}} \times 
\epsilon^{2}_{\mathrm{rec}} \times \epsilon^{2}_{\mathrm{lid}} \times \epsilon^{2}_{\mathrm{iso}} \nonumber\\
&& \times
[\epsilon_{\mathrm{tid}} \times (2 - \epsilon_{\mathrm{tid}})] \nonumber\\
&& \times [\epsilon_{\mathrm{trg}} \times
(2 - \epsilon_{\mathrm{trg}})].  
\label{eq:cceff}
\end{eqnarray}
The squared terms in the formula apply to efficiency terms 
that are applied twice (we require two reconstructed central
electrons passing loose identification and isolation criteria).
In order for this treatment to be correct, the efficiencies 
of the two electron legs in the $\zee$ candidates are required 
to be uncorrelated.  Using our sample of simulated $\zee$
events, we look for correlations between the efficiencies for 
the two electron legs and find them to be negligible.  The 
tight identification and trigger criteria can be satisfied by 
either of the two electrons.  The combined efficiency for one 
of two objects to satisfy a particular requirement can be 
written as $\epsilon^{2} + 2 \times \epsilon \times (1 - \epsilon) 
= \epsilon \times (2 - \epsilon)$.  The efficiency calculation 
for central-plug $\zee$ events is given in Eq.~\ref{eq:cpeff}.
\begin{eqnarray}
\epsilon_{\mathrm{tot}}^{\mathrm{cp}} = && \epsilon_{\mathrm{vtx}} \times \epsilon_{\mathrm{trk}} 
\times \epsilon_{\mathrm{rec}} \times \epsilon_{\mathrm{lid}} \times \epsilon_{\mathrm{tid}} \times 
\epsilon_{\mathrm{iso}} \times \epsilon_{\mathrm{trg}} \nonumber\\
&& \times \epsilon^{\mathrm{plug}}_{\mathrm{rec}} 
\times \epsilon^{\mathrm{plug}}_{\mathrm{id}} \times S^{\mathrm{plug}}_{\mathrm{cl}} \times 
\epsilon^{\mathrm{plug}}_{\mathrm{iso}}  
\label{eq:cpeff}
\end{eqnarray}    
In these events only the central electron leg can satisfy the 
tight identification and trigger criteria so these efficiencies
are only applied to the one central leg.  Similarly, the plug 
efficiencies are applied only to the plug electron leg.  Based 
on Eqs.~\ref{eq:cceff} and~\ref{eq:cpeff} the event efficiency 
for our combined $\zee$ sample takes the form:
\begin{eqnarray}
\epsilon^{\zee}_{\mathrm{tot}} = && \epsilon_{\mathrm{vtx}} \times \epsilon_{\mathrm{trk}} 
\times \epsilon_{\mathrm{rec}} \times \epsilon_{\mathrm{lid}} \times \epsilon_{\mathrm{tid}} 
\times \epsilon_{\mathrm{iso}} \times \epsilon_{\mathrm{trg}}\nonumber\\
&& \times 
[(1 - f_{\mathrm{cp}}) \times \epsilon_{\mathrm{trk}} \times \epsilon_{\mathrm{rec}} 
\times \epsilon_{\mathrm{lid}} \times \epsilon_{\mathrm{iso}} \nonumber\\
&& \times (2 - \epsilon_{\mathrm{tid}}) 
\times (2 - \epsilon_{\mathrm{trg}}) \nonumber\\
&& +  f_{\mathrm{cp}} \times \epsilon^{\mathrm{plug}}_{\mathrm{rec}} \times \epsilon^{\mathrm{plug}}_{\mathrm{id}} 
\times S^{\mathrm{plug}}_{\mathrm{cl}} \times \epsilon^{\mathrm{plug}}_{\mathrm{iso}}].    
\label{eq:zeeeff}
\end{eqnarray}

The calculation of the total selection criteria efficiency for 
$\zmm$ candidate events is similar to that for events in the 
electron channel but involves some additional complications.  
As discussed above we increase our acceptance for $\zmm$ events 
by releasing the muon detector stub requirements for one of the 
two candidate track legs.  The second muon leg in our candidate 
events can be any COT track passing the track quality, isolation, 
and minimum ionizing calorimeter energy deposition criteria 
used in this analysis for selecting muon track candidates.  
Since the track selection criteria are applied to both muon legs 
in our candidate events, the corresponding terms in the overall 
efficiency formula are squared.  Only one of the two muon 
track candidates is required to have a matching stub in the 
muon detectors that satisfies our stub selection criteria.  For 
roughly 40~$\!\%$ of our candidate events, both of the muon track 
legs point to active regions of the muon detectors.  In these 
cases, either of the two legs can have a matching stub in the muon 
detectors and satisfy the tight leg criteria.  In other cases, one 
of the two legs will not point to an active detector region, and 
the stub-matching criteria must be satisfied by the one leg that 
is pointed at the muon detectors.  In order to determine the total 
efficiency for $\zmm$ candidate events, we first determine the 
total selection efficiencies for both of these event classes.  
The event selection efficiency for the combined sample is then 
extracted as the weighted sum of the efficiencies for the two 
different event types.

The efficiency calculation for the subset of $\zmm$ events in 
which only one of the two muon tracks points to an active region 
of the muon detector is shown in Eq.~\ref{eq:zeffmu1}.  The 
efficiencies corresponding to selection criteria applied to both 
muon legs (track reconstruction, loose identification, and isolation) 
enter into the formula as squared terms.  The track leg pointing at  
the inactive regions of the muon detector can not have an associated
reconstructed stub so the other track leg in the event must have a 
matching stub for the event to satisfy the $\zmm$ selection criteria.  
This leg must also satisfy the tight muon identification and event 
trigger requirements since an associated reconstructed muon detector 
stub is a necessary pre-condition for a muon leg to meet these criteria.  
Since the muon stub reconstruction, tight identification, and trigger 
selection criteria can only be satisfied by one of the two muon 
legs in these events, the corresponding efficiency terms enter into 
Eq.~\ref{eq:zeffmu1} linearly.  As previously mentioned, the efficiency 
for $\zmm$ events not to be tagged as cosmics, $\epsilon^{Z}_{\mathrm{cos}}$, 
is independent of the measured value for $\wmnu$ events.  As described 
subsequently, we measure this efficiency in $\zmm$ events to be 
$\epsilon^{Z}_{\mathrm{cos}} =$~0.9994~$\pm$~0.0006.

\begin{eqnarray}
\epsilon^{\mu \mathrm{trk}}_{\mathrm{tot}} = && \epsilon_{\mathrm{vtx}} \times \epsilon^{Z}_{\mathrm{cos}} 
\times \epsilon^{2}_{\mathrm{trk}} \times \epsilon^{2}_{\mathrm{lid}} \times \epsilon^{2}_{\mathrm{iso}} \nonumber\\
&& \times \epsilon_{\mathrm{rec}} \times \epsilon_{\mathrm{tid}} \times \epsilon_{\mathrm{trg}} 
\label{eq:zeffmu1}
\end{eqnarray}

This situation is more complicated for the class of $\zmm$ events 
where both muon legs point to active regions of the muon detector.  
For these events both legs can individually satisfy the stub 
reconstruction, tight identification, and trigger criteria of 
the sample.  In order to simplify the efficiency calculation, 
we require that at least one of the two muon legs in each 
candidate event satisfies the requirements associated with 
all three of the above criteria.  With this additional restriction, 
the overall event selection efficiency in the subset of $\zmm$ 
candidates where both muon legs point at active regions of the 
muon detector can be written as shown in Eq.~\ref{eq:zeffmu2}.  
The combined efficiency for a muon leg to satisfy the stub 
reconstruction, tight identification, and trigger criteria 
($\epsilon^{*} = \epsilon_{\mathrm{rec}} \times \epsilon_{\mathrm{tid}} 
\times \epsilon_{\mathrm{trg}}$) enters into Eq.~\ref{eq:zeffmu2} in 
the form $\epsilon^{*} \times (2 - \epsilon^{*})$ which, as described 
above, is the resulting efficiency for a set of criteria required for 
one of two identical objects within an event.      

\begin{eqnarray}
\epsilon^{\mu \mu}_{\mathrm{tot}} = && \epsilon_{\mathrm{vtx}} \times \epsilon^{Z}_{\mathrm{cos}} 
\times \epsilon^{2}_{\mathrm{trk}} \times \epsilon^{2}_{\mathrm{lid}} \times \epsilon^{2}_{\mathrm{iso}} \nonumber\\
&& \times [\epsilon_{\mathrm{rec}} \times \epsilon_{\mathrm{tid}} \times \epsilon_{\mathrm{trg}} \nonumber\\
&& \times (2 - \epsilon_{\mathrm{rec}} \times \epsilon_{\mathrm{tid}} \times \epsilon_{\mathrm{trg}})] 
\label{eq:zeffmu2}
\end{eqnarray}

In order to combine Eqs.~\ref{eq:zeffmu1} and~\ref{eq:zeffmu2} 
into a formula for the total event efficiency of our combined 
sample, we need to introduce an additional parameter, 
$f_{\mathrm{dd}}$, which is defined as the fraction of $\zmm$ 
events within our geometric and kinematic acceptance in which 
both muon legs are found to point at active regions of the 
muon detector.  This quantity is determined from the 
simulated event sample.  For our candidate sample, we obtain 
$f_{\mathrm{dd}} = 0.3889 \pm 0.0021$, which is a luminosity 
weighted average of the values for the different run periods      
in which the CMX was either offline or online.  Using this 
additional factor, we determine a formula for the total event 
efficiency of our candidate sample by adding the expressions in 
Eqs.~\ref{eq:zeffmu1} and~\ref{eq:zeffmu2} weighted by factors 
of $1 - f_{\mathrm{dd}}$ and $f_{\mathrm{dd}}$ respectively.  
Finally, we obtain the expression shown in Eq.~\ref{eq:zeffmu3} 
for the total selection efficiency for events in our $\zmm$ 
candidate sample.

\begin{eqnarray}
\epsilon^{\zmm}_{\mathrm{tot}} = && \epsilon_{\mathrm{vtx}} \times \epsilon^{Z}_{\mathrm{cos}} 
\times \epsilon^{2}_{\mathrm{trk}} \times \epsilon^{2}_{\mathrm{lid}} \times \epsilon^{2}_{\mathrm{iso}} \nonumber\\
&& \times \epsilon_{\mathrm{rec}} \times \epsilon_{\mathrm{tid}} \times \epsilon_{\mathrm{trg}} \times [1 + f_{\mathrm{dd}} \nonumber\\
&& \times (1 - \epsilon_{\mathrm{rec}} \times \epsilon_{\mathrm{tid}} 
\times \epsilon_{\mathrm{trg}})] 
\label{eq:zeffmu3}
\end{eqnarray}

Based on the expressions in Eqs.~\ref{eq:zeeeff} and~\ref{eq:zeffmu3},
we can substitute our measured values for the individual efficiency 
terms and determine the combined event selection efficiencies for 
our $\zll$ candidate samples.  The resulting values are shown in 
Table~\ref{tb:zefffin}.  

\begin{table}[h]
\caption{Results of $\zll$ combined event efficiency calculations.}
\begin{center}
\begin{tabular}{l r} 
\hline
\hline 
Candidate Sample       & $\epsilon_{\mathrm{tot}}$ \\ 
\hline
$\zee$                 & 0.713 $\pm$ 0.012         \\ 
$\zmm$                 & 0.713 $\pm$ 0.015         \\ 
\hline
\hline
\end{tabular}
\label{tb:zefffin}
\end{center}
\end{table}

\subsection{Vertex Finding Efficiency}

Our requirement that the $z$-position of the primary event 
vertex be within 60~$\cm$ of the center of the CDF detector 
($|Z_{\mathrm{vtx}}| \leq$~60~$\cm$) limits the event 
acceptance to a fraction of the full luminous region for 
$\ppbar$ collisions.  However, the luminosity estimate 
used in our cross section measurements is based on the 
full luminous range of the beam interaction region.  We 
use minimum-bias data to measure the longitudinal profile 
of the $\ppbar$ luminous region, and this profile is 
subsequently used to estimate the fraction of interactions 
within our fiducial range in $z$.

\begin{figure}
\includegraphics[width=3.5in]{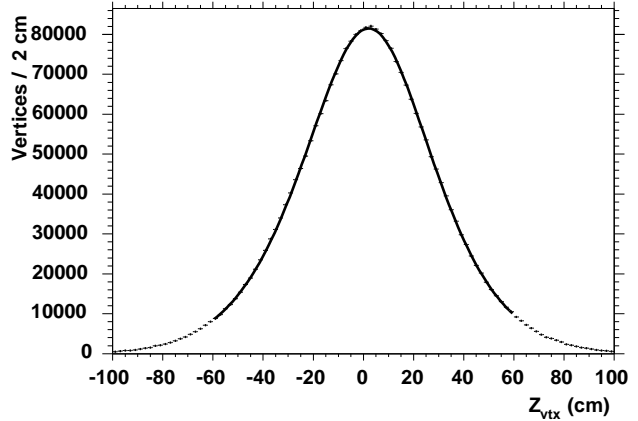}
\caption{The measured $Z_{\mathrm{vtx}}$ distribution. The 
units on the horizontal axis are $\cm$ and there are a total of 100
bins from $-$100~$\cm$ to $+$100~$\cm$. The curve is the fit to the
luminosity function (Eq.~\ref{eq:zvtx}) for $|z| <$~60~$\cm$, and the
resulting fit with 55 degrees of freedom has a $\chi^2$ of 119.}
\label{zvSetA}
\end{figure}

Fig.~\ref{zvSetA} shows the distribution of measured positions
along the $z$-axis (parallel to beams) for reconstructed primary 
vertices in minimum-bias events.  The minimum-bias events are 
taken from the same set of runs from which our candidate samples
are constructed.  In addition, the minimum-bias data is weighted 
to ensure that it has the same run-by-run integrated luminosity 
as the cross section event samples.  We fit the distribution in 
Fig.~\ref{zvSetA} using the following form of the Tevatron beam 
luminosity function: 
\begin{equation}
\frac{d {\cal L}(z)}{d z} = N_0 \:\:
      \frac{\exp{(-z^{2}/2 \sigma_{z}^{2})}}
           {\sqrt{ [1 + (\frac{ z - z_{01} }{\beta^*})^2 ]
                   [1 + (\frac{ z - z_{02} }{\beta^*})^2 ]
           }} 
\label{eq:zvtx}
\end{equation}
The five free parameters of the fit are $N_0$, $\sigma_{z}$, 
$z_{01}$, $z_{02}$, and $\beta^{*}$.  The $Z_{\mathrm{vtx}}$ 
distribution has some biases at large values of $z$ due 
to increased contamination from non-$\ppbar$ interactions 
such as those originating from beam-gas collisions and due 
to the decrease of COT tracking acceptance far away from the 
center of the detector.  We avoid these biases by only fitting 
the measured $Z_{\mathrm{vtx}}$ distribution to our function 
for $d {\cal L}(z) / d z$ in the region where $|z| <$~60~$\cm$.  
Within this finite range in $z$, the fraction of events not 
from $\ppbar$ collisions is negligible and the COT tracking 
acceptance is high and uniform. 

The acceptance of our requirement on the $z$-position of the
primary event vertex ($|Z_{\mathrm{vtx}}| <$~60~$\cm$) is 
calculated as 
\begin{equation}
   \epsilon_{\mathrm{vtx}}(|z| < 60) =
           \frac{ \int_{- 60}^{+ 60}
                  \: [ d {\cal L}(z) / dz ] \: dz }
                { \int_{- \infty}^{+ \infty}
		  \;\; [ d {\cal L}(z) / dz ] \: dz } \: . 
\end{equation}
We perform the fit to the data and evaluate the acceptance 
for both the full sample and several sub-samples of our 
minimum-bias data set.  We observe slight differences in 
the various sub-samples indicating small changes over time 
in the $z$-profile of $\ppbar$ collisions in the interaction 
region of our detector.  The maximum shift seen in the 
measured acceptance among the various sub-samples is
0.6~$\!\%$, and we assign half of this value as a systematic 
uncertainty on the efficiency measurement.  The statistical 
uncertainty on the measurement is assigned based on fit 
errors obtained from the $z_{\mathrm{vtx}}$ fit for the 
full minimum-bias sample.  Using the techniques described 
above, we measure the signal acceptance of our cut on the 
$z$-position of the primary event vertex to be  
\begin{displaymath}
  \epsilon_{\mathrm{vtx}} = 0.950 \pm 0.002 \: (stat.)
                             \pm 0.003 \: (syst.) \: .
\end{displaymath}

\subsection{Tracking Efficiency}

We define tracking efficiency as the fraction of high $\pt$ 
leptons contained within our geometrical acceptance for which 
our offline tracking algorithm is able to reconstruct the 
lepton track from hits observed in the COT.  We measure this 
quantity using a sample of clean, unbiased $\wenu$ candidate 
events based on a tight set of calorimeter-only selection 
criteria.  The events for this sample were collected using 
a trigger path based on calorimeter $\met$ requirements to 
ensure that the sample is unbiased with respect to XFT 
tracking requirements in the hardware portion of the trigger 
and track reconstruction in the software portion.  Events are 
required to have an electromagnetic calorimeter cluster with 
$\et >$~20~$\GeV$ and overall event $\met >$~25~$\GeV$.  
Since we can not use a track matching requirement to help 
reduce non-electron backgrounds, we apply a tighter than 
normal set of electron identification criteria on the 
electromagnetic cluster itself.  We also remove candidate 
events containing additional reconstructed jets in the 
calorimeter with $\et >$~3~$\GeV$ and require that the $\pt$ 
of the reconstructed $\W$ boson is above 10~$\GeVC$.  These 
cuts are designed to remove background events in our sample 
originating from QCD dijet processes.  
  
Our tracking efficiency measurement is obtained from the 
fraction of events in this candidate sample which have a COT 
track pointing to the electromagnetic cluster.  Matching, 
reconstructed tracks in the COT are required to point within 
5~$\cm$ of the calorimeter electromagnetic cluster seed tower.  
In order to be absolutely sure that we are not including 
track-less background events in our efficiency calculation, 
we also require that our candidate events have a 
reconstructed track based entirely on hits in the silicon 
tracking detector (independent of the central outer tracking 
chamber) pointing at the electromagnetic cluster.  A total 
of 1368 candidate events in our 72.0~$\pbinv$ sample have a 
matching silicon track.  Of these, 1363 events also contain 
a matching reconstructed track based solely on hits in the 
central outer tracker yielding a COT tracking efficiency of 
$\epsilon_{\mathrm{trk}}(Data) =$~0.9963~$^{+0.0035}_{-0.0040}$.  
The uncertainty on the measurement is primarily systematic 
and is based on studies of both silicon-only track fake rates 
and correlated failures in COT and silicon based tracking 
algorithms.

We compare the tracking efficiency measured in the data with
an equivalent measurement based on our $\wenu$ simulated event 
sample.  Using the same technique, we obtain a simulation 
tracking efficiency of $\epsilon_{\mathrm{trk}}(MC) =$ 
0.9966~$^{+0.0015}_{-0.0024}$, consistent with our measured 
value from data.  A study of failing simulated events reveals 
that the small tracking inefficiency we measure is mainly due 
to bremsstrahlung radiation where the silicon-only track points 
in the direction of the hard photon and the COT track follows 
the path of the soft electron (pointing away from the high 
$\et$ electromagnetic cluster).  Since the loss of events due 
to this process is already accounted for in our acceptance 
calculation, we avoid double-counting by taking the ratio of 
the tracking efficiency measured in data to that measured in 
simulation as our net tracking efficiency.  Based on this 
approach, our final value for the COT tracking efficiency 
is $\epsilon_{\mathrm{trk}} =$~1.000~$\pm$~0.004 where the 
uncertainty is the combined statistical and systematic 
uncertainty of our measurement technique.  

\subsection{Reconstruction Efficiency}

The lepton reconstruction efficiency is defined as the fraction 
of real leptons that is within our geometrical acceptance and 
has matching, reconstructed COT tracks which are subsequently 
reconstructed as leptons by our offline algorithms.  In the 
case of electrons, this efficiency corresponds to the combined 
probability for forming the electromagnetic cluster and 
matching it to the associated COT track.  For muons, it is the 
probability for reconstructing a stub in the muon detectors 
and matching the stub to the corresponding COT track.  

The reconstruction efficiency is measured using the unbiased, 
second legs of $\zll$ decays.  The leptons from $\Z$ boson 
decays have a similar momentum spectrum to those originating 
from $\W$ boson decays and are embedded in a similar event 
environment.  Events are required to have at least one fully 
reconstructed lepton leg that satisfies the complete set of 
lepton identification criteria used in the selection of our 
candidate samples.  The same lepton leg must also satisfy the 
requirements of the corresponding high $\pt$ lepton trigger 
path to ensure that the second leg is unbiased with respect to
the trigger.  A lepton leg satisfying these requirements is 
then paired with a second opposite-sign, high $\pt$ track in 
the event.  If the invariant mass of a lepton-track pair lies 
within the $\Z$ boson mass window, 80~$\GeVCSq$~$< M_{\ell\ell} 
<$~100~$\GeVCSq$, the second track leg is utilized as a 
candidate for testing the lepton reconstruction efficiency.  
In the case of $\zmm$ candidate events only, the second track 
leg is also required to have associated calorimeter energy 
deposition consistent with a minimum ionizing particle which 
reduces backgrounds from fake muons without biasing the 
measurement.  In the subset of $\zll$ candidate events where 
each track leg is a reconstructed lepton passing the full set 
of identification and trigger criteria, both legs are unbiased 
lepton candidates and included in the efficiency measurement. 

\begin{figure}
\includegraphics[width=8.5cm]{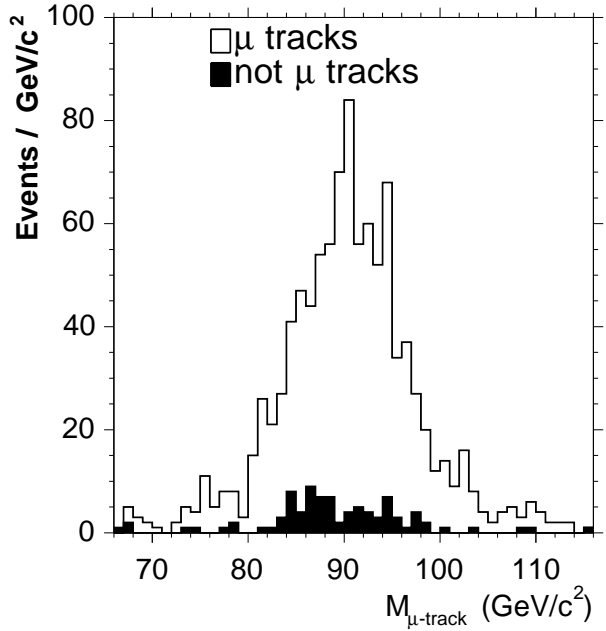}
\caption{\label{fig:recoeff_cmup}Invariant mass of muon-track 
pairs for the muon reconstruction-efficiency measurement.  We show the
distribution for pairs in which the track is a reconstructed as a muon
track (open histogram) and for pairs in which the track is not reconstructed
as a muon track (solid histogram).  Only the region between 80~$\GeVCSq$ and
100~$\GeVCSq$ is used for the efficiency calculation.}
\end{figure}
 
Each candidate track leg is extrapolated to determine if it 
points at an active area of the calorimeter or muon detectors
as appropriate.  If the track does point at an active detector
region, it is expected to be reconstructed as a lepton.  The 
fraction of this subset of candidate tracks which are in 
fact reconstructed as leptons provides our measurement of 
the reconstruction efficiency.  Fig.~\ref{fig:recoeff_cmup} 
shows the invariant mass distributions for muon-track pairs 
in cases where the second track is and is not reconstructed 
as a muon.  The small peak seen near the $\Z$ boson mass in 
the latter case indicates that we do observe a non-negligible 
muon reconstruction inefficiency in the data.  However, some 
of the measured inefficiency is attributable to the effects 
of multiple scattering.  A particle associated with a track 
that points at an active detector region can in some cases 
pass outside of this region due to the cumulative effects 
of interactions with material in the detector.  This effect 
is modeled using the simulated event samples.  All real 
reconstruction inefficiencies observable in the simulation 
are accounted for in the acceptance calculation and must not 
be double-counted in the lepton reconstruction efficiency 
measurement.  Therefore, we determine our net lepton 
reconstruction efficiency by dividing the value measured in 
data by the value obtained from an equivalent measurement 
using simulation.  The lepton reconstruction efficiency 
measurements for electrons and muons are summarized in 
Table~\ref{tb:leptrec}.  Plug electron candidates are not 
required to have a matching reconstructed track and therefore 
by our definition have a fixed reconstruction efficiency of 
100~$\!\%$.  We make additional checks to confirm that the 
leptons in our test samples are a good match for the leptons 
in our candidate samples and based on this agreement take 
the statistical uncertainty of our measurements as the total 
uncertainty on the reconstruction efficiencies. 

\begin{table*}[t]
\caption{Summary of lepton reconstruction efficiency measurements.
         Because plug electron candidates are not required to 
         have a matching reconstructed track, the corresponding 
         reconstruction efficiency is one by definition.} 
\begin{tabular}{l c c c} 
\hline
\hline 
Lepton           & Data Efficiency       & Simulation Efficiency       & Net Efficiency       \\ 
\hline 
Central Electons & 0.990 $\pm$ 0.004     & 0.992 $\pm$ 0.001           & 0.998 $\pm$ 0.004    \\
Plug Electrons   & 1.000                 & 1.000                       & 1.000                \\
Muons            & 0.935 $\pm$ 0.007     & 0.980 $\pm$ 0.001           & 0.954 $\pm$ 0.007    \\ 
\hline
\hline 
\end{tabular}
\label{tb:leptrec}
\end{table*}

\subsection{Lepton Identification and Isolation Cut Efficiencies}

The efficiencies of our lepton identification and isolation
cuts are also determined directly from the data using $\zll$ 
events.  We use slightly different techniques for measuring 
these efficiencies for electrons and muons.  The motivation 
for using separate methods is the non-negligible fraction 
of background events in the $\zee$ candidate sample in which 
at least one of the reconstructed electrons is either a fake
or the direct semileptonic decay product of a hadron.  In 
order to accurately measure the selection efficiencies for 
electrons originating from $\W$ and $\Z$ boson decays, it is 
important to correct for the contribution of these background 
events to our efficiency calculation.  Since these types of 
backgrounds are negligible in our $\zmm$ candidate sample, 
we are able to use a more aggressive approach which maximizes 
the statistical size of the muon candidates used to determine 
these efficiencies.

As previously mentioned, the identification and isolation 
efficiencies for leptons are determined in a specific order 
to avoid double-counting correlated inefficiencies between 
different groups of selection criteria.  The order we 
employ in making these measurements is the following: 
efficiencies from loose identification cuts, isolation cut 
efficiencies, and efficiencies from tight identification 
cuts.  This ordering is chosen to simplify the extraction 
of combined selection efficiencies for $\zll$ events from 
our individual, measured efficiency terms.  To protect 
this ordering, we require that lepton candidates used to
measure each group of selection efficiencies satisfy the 
selection criteria associated with all groups defined to 
be earlier within our assigned order.

To minimize backgrounds in the $\zee$ event sample used to 
make the efficiency measurements, we require that at least 
one of the two reconstructed electrons passes the full set 
of identification and isolation criteria used in the $\wenu$ 
analysis.  The second electron leg in each event, referred 
to here as the probe electron, is simply required to satisfy 
the geometric and kinematic cuts that define the acceptance 
of our candidate samples.  In addition, the invariant mass 
of the electron pair is required to be within a tight window 
centered on the measured $\Z$ boson mass (75~$\GeVCSq 
< M_{ee} <$~105~$\GeVCSq$), which further reduces non-$\Z$ 
backgrounds in the sample.  By definition, the electron 
passing the complete set of identification and isolation 
criteria is a central electron.  Central-central $\zee$ 
events satisfying the criteria listed above are used to 
measure central electron efficiencies, and central-plug 
events are used to measure plug electron efficiencies.

We define the number of central-central $\zee$ candidates 
passing our criteria as $N_{\mathrm{tc}}$.  As mentioned 
above, each event has at least one electron passing the 
full set of identification and isolation criteria.  
Electrons of this type are referred to as tight.  In some 
number of events in our candidate sample, $N_{\mathrm{tt}}$, 
both electrons are found to satisfy the tight criteria.  In 
the remaining events, the probe electron necessarily fails 
at least one part of our selection criteria.  However, some 
number of these remaining events will satisfy a particular 
subset of the identification and isolation requirements 
corresponding to an efficiency term that we want to measure.  
The total number of events where the probe leg is found 
to satisfy a given subset of cuts is referred to as 
$N_{\mathrm{t}i}$.  In this case, the corresponding 
efficiency for the subset of cuts being studied is determined 
from the expression given in Eq.~\ref{eq:eid1}.  The variable 
$i$ in this expression refers to the three sets of selection 
cut efficiencies to be measured (1 = loose identification, 
2 = isolation, and 3 = tight identification).  In the second 
two cases, we limit our sample of probe electrons to those 
that satisfy the criteria associated with the lower numbered 
efficiency terms to avoid the double-counting problem 
discussed above.  The net result is that for the second 
two cases $N_{\mathrm{tc}} = N_{\mathrm{t}(i-1)}$ and 
$N_{\mathrm{t}i}$ is re-defined as the number of events where 
the probe leg is found to pass the requirements associated 
with the efficiency term being measured and those numbered 
below it.  This new definition implies that for the final 
case $N_{\mathrm{t}i}$ is simply equal to $N_{\mathrm{tt}}$.     

\begin{eqnarray}
\epsilon_{i} &=& {{N_{\mathrm{t}i} + N_{\mathrm{tt}}} \over {N_{\mathrm{tc}} + N_{\mathrm{tt}}}},
\label{eq:eid1}
\end{eqnarray}

One additional complication is that we must subtract the 
contribution of background to each of the input event 
totals in Eq.~\ref{eq:eid1} to accurately measure the 
efficiencies for electrons produced in $\W$ and $\Z$ 
boson decays.  For central-central $\zee$ events, the 
background in each event subset is determined from the 
number of equivalent same-sign events observed in the 
data sample.  A correction for tridents (real $\zee$ 
events where the charge of one electron is measured 
incorrectly due to the radiation of a hard bremsstrahlung 
photon) in the same-sign event totals is made based on 
the relative numbers of opposite-sign and same-sign events 
in our $\zee$ simulated event sample.  The event counts 
and background corrections for each of the input 
parameters used in the efficiency calculations are given 
in Table~\ref{tb:ceid}. 

\begin{table*}[t]
\caption{$\zee$ event counts used as inputs to the calculation of electron identification and isolation efficiencies.}
\begin{tabular}{l c c c c c c c c c} 
\hline
\hline 
Efficiency Measurement       & Symbol            & $i$               & $N_{\mathrm{tc}}$ & $N_{\mathrm{t}i}$ & $N_{\mathrm{tt}}$ 
                             & $N_{\mathrm{tc}}$ & $N_{\mathrm{t}i}$ & $N_{\mathrm{tt}}$ & Efficiency        \\ 
                             &                   &                   &                   &                   &                
                             & Background        & Background        & Background        &                   \\ 
\hline
Loose Identification Cuts    & $\epsilon_{\mathrm{lid}}$ & 1 & 1901 & 1751 & 1296 & 28.3 & 6.1  & 0.6 & 0.960 $\pm$ 0.004 \\
Isolation Cut                & $\epsilon_{\mathrm{iso}}$ & 2 & 1751 & 1663 & 1296 & 6.1  & -0.4 & 0.6 & 0.973 $\pm$ 0.003 \\ 
Tight Identification Cuts    & $\epsilon_{\mathrm{tid}}$ & 3 & 1663 & 1296 & 1296 & -0.4 & 0.6  & 0.6 & 0.876 $\pm$ 0.007 \\ 
\hline
\hline
\end{tabular}
\label{tb:ceid}
\end{table*}

The fraction of background events in the central-plug 
$\zee$ candidate sample used to measure plug electron 
efficiencies is much larger than that in the 
central-central sample.  In order to eliminate some 
of this additional background, we make an even 
tighter set of requirements on the isolation and 
electron quality variables associated with the central 
electron to pick the candidate events used to measure 
these efficiencies.  As the probe leg in these 
candidates is the only plug electron of interest in 
the event, efficiencies are measured simply as the 
fraction of probe legs that satisfy the associated 
set of selection criteria.  In the analyses reported 
here, plug electrons are utilized only as loose 
second legs for selecting $\zee$ candidate events.  
There is therefore no corresponding tight 
identification cut efficiency to measure for plug 
electrons.  However, the ordering of the loose 
identification and isolation cuts for plug electrons 
is identical to that used for electrons in the central 
region.  We account for this ordering by requiring that 
the probe electrons used to measure the efficiency of 
the isolation cut satisfy the full set of loose plug 
electron identification cuts.  We correct the number 
of probe legs in both the numerator and denominator of 
our efficiency calculations for the residual backgrounds 
remaining in our candidate sample.  These backgrounds 
are estimated using electron fake rate calculations 
outlined in Sec.~\ref{sec:backg}.  Based on this method, 
we obtain independent estimates for the background 
contributions from both QCD dijet and $\wenu$ plus jet 
processes and sum them to obtain our final background 
estimates.  The inputs to our plug electron efficiency 
measurements and the resulting efficiency values are 
summarized in Table~\ref{tb:peid}.     

\begin{table*}[t]
\caption{Input parameters to plug electron identification and isolation efficiency measurements using central-plug $\zee$ candidates.} 
\begin{tabular}{l c c c c c c} 
\hline
\hline 
Efficiency                    & Symbol          & Number of         & Number passing 
                              & Probe Electron  & Passing Electron  & Efficiency          \\ 
Measurement                   &                 & Probe Electrons   & Selection Criteria 
                              & Background      & Background        &                     \\ 
\hline
Identification Cuts           & $\epsilon^{\mathrm{plug}}_{\mathrm{id}}$  & 2517 & 2126 & 108.4 & 15.0 & 0.876 $\pm$ 0.015 \\
Isolation Cut                 & $\epsilon^{\mathrm{plug}}_{\mathrm{iso}}$ & 2126 & 2111 & 15.0  & 14.1 & 0.993 $\pm$ 0.003 \\ 
\hline
\hline
\end{tabular}
\label{tb:peid}
\end{table*}

The calculation of muon identification and isolation 
efficiencies is simplified by the lack of significant 
backgrounds in our $\zmm$ candidate samples.  To obtain 
the largest possible sample of probe muons for measuring 
these efficiencies, we make only a minimal set of 
requirements on the first muon leg in these events.  In 
order to avoid selection biases, we simply require that 
at least one muon leg in each event satisfies both the 
trigger requirements and loose cuts used to select events 
into our high $\pt$ muon sample from which the candidate 
events are chosen.  The second muon leg in each of these 
events is then utilized as an unbiased probe leg for 
measuring our selection efficiencies.  In the subset of 
candidate events where both muon legs satisfy the trigger 
and loose selection requirements of our sample, both 
muons are unbiased and included in our sample of probe 
muons.  To ensure that we are selecting probe muons from 
a clean (low background) sample of $\zmm$ candidate 
events, we do require that the invariant mass of each muon 
pair lies within a tight window around the measured $\Z$ 
boson mass (80~$\GeVCSq < M_{\mu\mu} <$~100~$\GeVCSq$) 
and remove any events identified by our tagging algorithm 
as cosmic ray candidates.  After applying these criteria, 
we find that only 3 of over 1,500 probe muons come from 
same-sign candidate events confirming the negligible 
background fraction in the event sample used for these 
measurements.

As in the case of electrons, the full set of muon 
identification cuts is divided into loose and tight 
subsets to simplify the calculation of the combined  
event selection efficiency for $\zmm$ candidate events.  
The second muon track leg in these events is not 
required to have a matching stub in the muon detector.  
Therefore, the identification cuts for muons which we 
refer to as loose are those that are applied to the 
track itself.  The remaining tight selection cuts are 
those applied only to muon track legs with matching 
muon detector stubs.  In some sense, the reconstruction 
of a matching stub in the muon detector is therefore 
also a tight selection criteria although we choose to 
treat the efficiency for this requirement separately.  
We use the same ordering of selection criteria (loose 
identification, isolation, and tight identification) 
as that used for electrons to avoid the double-counting 
of correlated muon inefficiencies.  Muon probe legs 
used to measure the efficiency for each set of 
selection criteria are required to satisfy all 
selection cuts corresponding to previously ordered 
efficiency terms.  Table~\ref{tb:mid} summarizes the 
inputs to the muon efficiency calculations and the 
resulting efficiency values.     

\begin{table*}[t]
\caption{Input parameters to muon identification and isolation efficiency measurements using $\zmm$ candidates.} 
\begin{tabular}{l c c c c c c} 
\hline
\hline 
Efficiency Measurement        & Symbol       & Number of         & Number passing           & Efficiency        \\ 
                              &              & Probe Muons       & Selection Criteria       &                   \\ 
\hline
Loose Identification Cuts     & $\epsilon_{\mathrm{lid}}$ & 1574 & 1469                     & 0.933 $\pm$ 0.006 \\
Isolation Cut                 & $\epsilon_{\mathrm{iso}}$ & 1469 & 1443                     & 0.982 $\pm$ 0.003 \\
Tight Identification Cuts     & $\epsilon_{\mathrm{tid}}$ & 1443 & 1381                     & 0.957 $\pm$ 0.005 \\ 
\hline
\hline
\end{tabular}
\label{tb:mid}
\end{table*}

\subsection{Trigger Efficiency}

As described in Sec.~\ref{sec:data}, the data samples used 
to select our candidate events are collected via high $\pt$ 
lepton-only trigger paths.  The three-level trigger system 
utilized by the upgraded CDF data acquisition system reduces 
the 2.5~$\MHz$ beam-interaction rate into a final event 
collection rate on the order of 75~$\Hz$.  The first two 
levels utilize dedicated hardware to select events for 
readout from the detector, and the third level is a processor 
farm that runs a fast version of the full event reconstruction 
to pick out the final set of events to be written to tape.  
Level~1 lepton triggers are constructed from high $\pt$ COT 
tracks identified in the fast tracking hardware matched with 
single tower electromagnetic energy deposits in the calorimeter 
(electrons) or groups of hits in the outer wire chambers (muons).  
Level~2 hardware is used to perform a more sophisticated 
calorimeter energy clustering algorithm on electron candidates 
to obtain improved $\et$ resolution.  The improved $\et$ 
variable is utilized at Level~2 to make tighter kinematic cuts 
on the electron candidates.  No additional requirements are 
made on muon candidates at Level~2.  Events selected at Level~2 
are read out of the detector and passed to the Level~3 processor
farm.  A fast version of the offline lepton reconstruction
algorithms are run on each event, and the identified leptons 
are subjected to both kinematic and loose quality selection 
cuts.  

The measurement of trigger efficiencies for electrons is 
simplified by the availability of secondary trigger paths 
that feed into our $\wenu$ candidate sample.  A trigger 
path based solely on calorimeter quantities is used to 
measure the efficiency of tracking requirements at each 
of the three trigger levels.  This path utilizes identical 
calorimeter cluster requirements to those in the default 
electron path but does not require matching tracks to be 
found at any level.  Instead, events are selected based 
on the presence of large $\met$ in the calorimeter 
(15~$\GeV$ at Level~1/Level~2 and 25~$\GeV$ at Level~3) 
associated with the high energy neutrino in the $\W$ 
boson decays.  For $\wmnu$ candidate events, the muon 
deposits only a small fraction of its energy into 
the calorimeter and hence the residual $\met$ in the 
calorimeter is too small to allow for an equivalent 
trigger path for muon candidates.  To measure the 
efficiencies of the electron trigger path track 
requirements, we select events from the secondary trigger 
path that pass the complete set of $\wenu$ selection 
criteria.  The fraction of events in this unbiased sample 
that satisfy the track requirements of our lepton-only 
trigger path at each of the three levels gives the 
corresponding efficiency for  those requirements.  The 
double-counting of correlated inefficiencies between the 
different trigger levels is avoided by requiring that 
events used to measure higher level trigger efficiencies 
pass all of the tracking requirements associated with 
levels below that being measured. 

Due to slight changes in the track trigger requirements  
over time, the corresponding efficiencies are measured 
in three run ranges.  A final efficiency is determined
by taking the luminosity weighted average of the results 
obtained for each run range.  The event samples used to 
make these measurements were studied to look for possible 
trigger efficiency dependencies on other event variables
such as electron isolation, number of additional jets in
the events, total event energy, and electron charge.  No
dependencies were found for these variables, within the 
statistical uncertainties of our sample.  We did observe 
a small trigger efficiency dependence as a function of the 
measured pseudorapidity of the electron track.  We observe 
a small inefficiency for tracks near $\eta_{\mathrm{det}} 
\sim$~0 due to wire spacers in the tracking chamber and 
reduced overall charge collection due to the shorter track 
path length through the chamber.  However, the effect of 
this dependence on our final efficiency results was found 
to be negligible within our measurement uncertainties.  
The final efficiency results for the electron trigger path 
tracking requirements at each trigger level are shown in
Table~\ref{tb:etrgeff1}.

\begin{table*}[t]
\caption{Efficiencies for tracking requirements in high $\et$ electron trigger path.}
\begin{tabular}{l c r} 
\hline
\hline 
Trigger Level       & Track Requirement                       & Measured Efficiency \\ 
\hline 
Level~1             & Fast Tracker ($\pt >$ 8~$\GeVC$)        & 0.974 $\pm$ 0.002   \\
Level~2             & Fast Tracker ($\pt >$ 8~$\GeVC$)        & 1.000 $\pm$ 0.000   \\	
Level~3             & Full Reconstruction ($\pt >$ 9~$\GeVC$) & 0.992 $\pm$ 0.001   \\
Combined            & Level~1 $\rightarrow$ Level~3           & 0.966 $\pm$ 0.002   \\ 
\hline
\hline 
\end{tabular}
\label{tb:etrgeff1}
\end{table*}

In order to measure the total efficiency of our 
electron trigger path, we additionally need to measure 
the efficiencies of the calorimeter cluster requirements 
at each level of the trigger.  The requirement of an 
electromagnetic cluster with $\et >$ 8~$\GeV$ at Level~1 
is studied using reconstructed electromagnetic objects 
found in muon-triggered events.  We determine the highest 
energy trigger tower associated with each object and check 
to see if the Level~1 trigger bit corresponding to this 
tower is turned on in the data.  We measure a turn-on 
efficiency of 99.5~$\!\%$ for trigger towers with a measured 
electromagnetic energy between 8~$\GeV$ and 14~$\GeV$ and 
100~$\!\%$ for those measured above 14~$\GeV$.  The small 
inefficiency observed for towers with measured energies 
below 14~$\GeV$ is due to an additional Level~1 requirement 
placed on the ratio of hadronic and electromagnetic energies 
($E_{\mathrm{had}}/E_{\mathrm{em}} <$~0.05) in towers with 
energies below this cut-off value.  The effect of this 
inefficiency on the fully reconstructed electrons in our 
$\wenu$ candidate events is determined by checking how often 
the associated trigger tower with the highest electromagnetic 
$\et$ has a measured energy below 14~$\GeV$.  We find that 
less than 1~$\!\%$ of the reconstructed electrons in our 
candidate sample ($\et >$ 25~$\GeV$) do not have at least 
one associated trigger tower with $\et >$ 14~$\GeV$.  Based 
on these numbers, we estimate the overall trigger efficiency 
for the Level~1 electromagnetic cluster requirement to be 
100~$\!\%$ for the events in our candidate samples.  

Additional secondary trigger paths are used to measure 
the efficiencies of the Level~2 and Level~3 cluster 
requirements in our default electron trigger path.  
The efficiency of the Level~2 cluster requirement is 
obtained using events collected with two additional 
secondary trigger paths that have no Level~2 selection
requirements other than simple prescales.  The Level~1
and Level~3 trigger requirements in these paths are 
equivalent to in one case those of the default path and 
in another those of the path used to collect events for
measuring the efficiencies of track requirements.  The 
subset of these events that pass our full set of $\wenu$ 
selection criteria are also found to satisfy the Level~2 
cluster trigger criteria.  Based on these samples we 
conclude that the efficiency of the Level~2 electron 
cluster requirement is 100~$\!\%$ for reconstructed 
electrons also satisfying our selection criteria for 
tight central electrons.  Since the electron clustering 
algorithm run in the Level~3 processor farms is nearly 
identical to that used in offline reconstruction, we 
expect candidate events with high $\et$ electrons to 
also satisfy the Level~3 cluster requirements of our 
trigger path.  However, due to slight differences in 
the calorimeter energy corrections applied at Level~3 
and offline, it is possible that we could observe 
trigger inefficiencies close to the $\et$ threshold 
utilized for Level~3 clusters.  To check for this 
inefficiency, we collect events on an additional 
secondary trigger path which is based on the Level~1 
and Level~2 requirements of our default electron 
trigger path but no requirements at Level~3 other 
than a simple prescale.  We find that all of the 
events collected on this path which satisfy our 
event selection criteria also satisfy the Level~3 
cluster criteria of our default trigger path.  Based 
on this study,  the efficiency of the Level~3 cluster 
requirement for events in our candidate samples is 
also 100~$\!\%$.  Since we do not measure inefficiencies 
for the cluster requirements of our trigger path at 
any of the three levels, we conclude that the overall 
efficiency of our default trigger path for electrons 
is completely determined by the measured efficiencies 
of the track criteria given in Table~\ref{tb:etrgeff1}.

As mentioned above we do not have the benefit of
an equivalent set of secondary trigger paths for
collecting $\wmnu$ candidate events to measure the 
efficiencies of our muon trigger path requirements.  
Instead, we use $\zmm$ candidate events in which 
both muons satisfy the full set of isolation and 
identification cuts used to define our samples.  
To avoid background events we require that the 
invariant mass of the dimuon pair lies in a tight 
window around the $\Z$ boson mass 
(76~$\GeVCSq < M_{\mu\mu} <$~106~$\GeVCSq$) and 
that the event has not been identified as a cosmic 
ray by our tagging algorithm.  In this sample we 
know that at least one of the two muons in the event 
satisfied the muon trigger path requirements and can 
make a measurement of the muon trigger efficiency 
based on the fraction of events in which both muons 
meet the criteria of our trigger path.  If we define 
$\epsilon_{\mathrm{trg}}$ as the single muon trigger 
efficiency we want to measure, then 
$(\epsilon_{\mathrm{trg}})^{2}$ is the fraction of 
events containing two triggered muons, and 
$2(\epsilon_{\mathrm{trg}})(1-\epsilon_{\mathrm{trg}})$ 
is the fraction of events with only one triggered muon.  
There is also a remaining fraction of events 
$(1-\epsilon_{\mathrm{trg}})^{2}$ which contain no 
triggered muons, but these events do not make it into 
our $\zmm$ candidate sample.  Based on these definitions, 
the number of candidate events in our sample in which 
both muons meet the trigger criteria, $N_{2\mathrm{trg}}$ 
divided by the total number of events in the sample, 
$N_{\mathrm{tot}}$, can be expressed with 
Eq.~\ref{eq:mutrg1}.  From this expression we obtain the 
formula shown in Eq.~\ref{eq:mutrg2} which gives the muon 
trigger efficiency as a function of this fraction $F$.

\begin{equation}
F = \frac{N_{2\mathrm{trg}}}{N_{\mathrm{tot}}} = \frac{(\epsilon_{\mathrm{trg}})^{2}}{(\epsilon_{\mathrm{trg}})^{2} 
+ 2(\epsilon_{\mathrm{trg}})(1-\epsilon_{\mathrm{trg}})}    
\label{eq:mutrg1}
\end{equation}

\begin{equation}
\epsilon_{\mathrm{trg}} = \frac{2 \cdot F}{1 + F}
\label{eq:mutrg2}
\end{equation}

To check whether an individual muon in our candidate 
sample satisfies the requirements of our muon trigger
path, we first look at the hits on the reconstructed 
muon stub to determine the position of the muon with 
respect to the 144 Level~1 muon trigger towers (2.5 
degrees each in $\phi$) defined in the hardware.  We 
then check to see if the trigger bits corresponding 
to each individual requirement of our trigger path 
are set for the matched trigger tower.  The Level~1
requirements of our trigger path include both a high 
$\pt$ COT track identified in the fast tracking 
hardware and a sufficient set of matching hits in the 
muon detector wire chamber(s) along the path of the 
reconstructed muon.  Matching CSX scintillator hits 
are additionally required in the region of the muon 
detector between 0.6 and 1.0 in $\eta_{\mathrm{det}}$ 
(CMX region).  No significant additional trigger 
requirements are made at Level~2 for muon candidates.  
In order to measure the efficiency of the muon 
reconstruction algorithms at Level~3, we use the 
subset of events in the $\zmm$ candidate sample 
described above in which both muons are found to 
satisfy the Level~1 trigger criteria.  This 
restriction is made to ensure that we do not 
double-count correlated inefficiencies between 
the different trigger levels.  In addition, we 
require that one of the two muons is found in the 
region of the muon detector between 0.0 and 0.6 in 
$\eta_{\mathrm{det}}$ (CMUP region) while the other 
is found in the region between 0.6 and 1.0 in 
$\eta_{\mathrm{det}}$ (CMX region).  Since different 
Level~3 muon reconstruction algorithms are run in 
these two regions, it is simple to check if both or 
only one of the muons in these events satisfy the 
Level~3 requirements of our muon trigger path.  The 
input parameters to our muon trigger path efficiency 
calculations are shown in Table~\ref{tb:mutrgeff} 
along with the final results of these calculations.    

\begin{table*}[t]
\caption{Efficiencies for high $\pt$ muon trigger path.}
\begin{tabular}{l c c c} 
\hline
\hline 
Trigger Level       & Number of $\zmm$       & Number of Events           & Efficiency        \\ 
                    & Candidate Events       & with 2 Muon Triggers       &                   \\ 
\hline
Level~1             & 338                    & 293                        & 0.929 $\pm$ 0.011 \\
Level~3             & 138                    & 137                        & 0.996 $\pm$ 0.004 \\
Combined            & -                      & -                          & 0.925 $\pm$ 0.011 \\ 
\hline
\hline 
\end{tabular}
\label{tb:mutrgeff}
\end{table*}

\subsection{Cosmic Tagger Efficiency}

The tagging algorithm used to remove cosmic ray events in 
our $\wmnu$ and $\zmm$ candidate samples is discussed in 
Sec.~\ref{sec:cosmicwbkg}.  We measure the fraction of real 
events tagged as cosmic rays by this algorithm for both 
candidate samples using the corresponding electron decay 
mode samples.  The tagging algorithm is based solely on 
the hit timing information associated with reconstructed 
tracks in the COT.  Since the kinematics of $\W$ and $\Z$ 
boson decays into electrons and muons are nearly identical, 
we expect that the reconstructed electron tracks in $\wenu$ 
and $\zee$ candidate events are a good model for the muon 
tracks in the corresponding decay channels.  Unlike the 
muon channels, however, the electron decay mode candidate 
samples have a negligible cosmic background.  Therefore, 
we obtain a measurement of the fraction of real $\wmnu$ 
and $\zmm$ signal events tagged as cosmic ray candidates
directly from the observed fraction of events in the 
corresponding electron channels which our algorithm 
identifies as cosmic ray candidates.  In order to make 
the tracks in the electron events match as closely as 
possible with those in the muon events, we first apply 
the muon track impact parameter cut described in 
Sec.~\ref{sec:evsel} to each of the electron candidate 
tracks in these samples.  This additional requirement 
reduces the number of events in the $\wenu$ candidate 
sample to 37,070.  Of the remaining events, only five 
are tagged as cosmic ray candidates by our modified 
version of the cosmic tagging algorithm.  The resulting 
efficiency for a $\W$ boson decay not to be tagged as a 
cosmic by our algorithm is $\epsilon^{\W}_{\mathrm{cos}} 
=$~0.9999~$\pm$~0.0001.  Applying the track impact 
parameter cut to the $\zee$ sample reduces the total 
number of candidate events to 1,680.  Of these events, 
only one is tagged as a cosmic by our modified tagging 
algorithm.  The resulting efficiency for a $\Z$ boson 
decay not to be tagged as a cosmic by our algorithm is 
$\epsilon^{\Z}_{\mathrm{cos}} =$~0.9994~$\pm$~0.0006.
  
\subsection{Over-Efficiency of $\Z$-Rejection Criteria}

The criteria for rejecting $\zmm$ events in our $\wmnu$ 
candidate sample are defined in Sec.~\ref{sec:evsel}.  
A small fraction of real signal events are also removed 
from our candidate sample via this selection criteria.  
We measure the efficiency for signal events to survive 
the $\Z$-rejection cuts directly from simulation.  The
resulting value, 0.9961~$\pm$~0.0001, is determined by 
the number of $\wmnu$ candidate events in our simulated 
sample that exclusively fail the $\Z$-rejection criteria. 

The systematic uncertainty on this efficiency is based on a 
comparison of the shape of the invariant mass spectrum for 
the muon plus track candidate events rejected solely due to 
this criteria to the shape of the same spectrum obtained 
from $\zgmm$ simulated events.  A comparison of the ratio 
of rejected events inside and outside the $\Z$-mass window 
(66~$\GeVCSq < M_{\mu\mu} <$~116~$\GeVCSq$) to that found 
in the $\gamma^{*}/Z$ simulation sample provides a good 
measure whether our rejected events are a relatively pure 
sample of $\gamma^{*}/Z$ decays.  Based on this approach, 
we measure an additional systematic of $\pm$~0.17~$\!\%$ to
apply to the $\Z$-rejection efficiency value obtained from 
simulation.  The final result is $\epsilon_{\mathrm{z-rej}} 
=$~0.9961~$\pm$~0.0017.

\section{Backgrounds}
\label{sec:backg}
Other physics processes can produce events that mimic the signature of
$\wlnu$ and $\zll$ events in our detector. Some processes have similar
final-state event topologies to those of our signal samples and others
can fake similar topologies if a non-lepton object within the event is
misidentified as an electron or muon.  In this section, the sources of
backgrounds to $\W$ and $\Z$ events are discussed.  We separate the
background sources into three main categories: events in which
hadronic jets fake leptons; events from other electroweak processes;
and events from non-collision cosmic ray backgrounds.  The techniques
used to estimate the contribution to our candidate samples from each
background source are given in this section along with the final
estimates.

\subsection{Hadron Jet Background in $\wlnu$}
\label{sec:qcdwbkg}
Extracting the contribution of events to the $\wlnu$ 
candidate samples in which real or fake leptons from hadronic 
jets are reconstructed in the detector is one of the more 
challenging components of our measurements.  Real leptons 
are produced both in the semileptonic decay of hadrons and 
by photon conversions in the detector material.  Some events 
also contain other particles in hadronic jets which are 
misidentified and reconstructed as leptons.  Typically, these 
types of events will not be accepted into our $\W$ candidate 
samples because we require large event $\met$.  In a small 
fraction of these events, however, a significant energy 
mismeasurement does reproduce the $\met$ signature of 
our samples.  Because of the large total cross section 
for hadronic jets in our detector, even this small fraction 
results in a substantial number of background events 
in our $\W$ candidate samples.  These events are particularly 
difficult to model in the simulation since the associated 
energy mismeasurement makes them unrepresentative of typical 
hadronic events.  In order to estimate the background 
contribution of these sources to our samples, we release the 
selection criteria on lepton isolation and event $\met$ and 
use events with low lepton isolation and low $\met$ as a model 
of the background in the low lepton isolation and high $\met$ 
$\W$ signal region.  The contributions in the low and high 
$\met$ regions are normalized to the number of events in those 
regions with high lepton isolation based on the assumption that 
there is no correlation between lepton isolation and $\met$ in 
the hadronic background.  

\begin{figure}[t]
\begin{center}
 \includegraphics[width=3.5in]{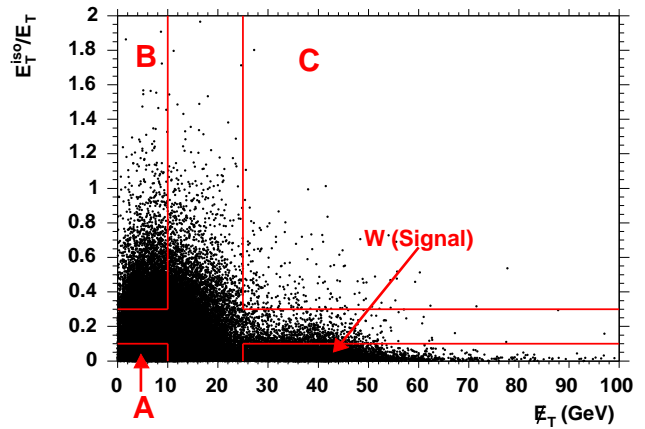}
\end{center}
\caption{$\et^{\mathrm{iso}}/\et$ versus event $\met$ for $\wenu$ 
candidates (no cuts on the lepton isolation fraction variable or 
the event $\met$).  The definitions of regions A, B, and C which 
are used in the calculation of the hadronic background are provided 
in the text.}
\label{fig:IsoVsMet}
\end{figure}

Fig.~\ref{fig:IsoVsMet} shows the lepton isolation fraction 
variable plotted against event $\met$ for $\wenu$ candidates 
(no cuts on lepton isolation fraction or event $\met$).  In 
the lepton isolation fraction versus $\met$ parameter space, 
we define four regions as follows:
\begin{itemize}
\item{Region A: $\et^{\mathrm{iso}}/\et <$ 0.1 and $\met <$ 10 $\GeV$}
\item{Region B: $\et^{\mathrm{iso}}/\et >$ 0.3 and $\met <$ 10 $\GeV$}
\item{Region C: $\et^{\mathrm{iso}}/\et >$ 0.3 and $\met >$ 25 $\GeV$ (20 $\GeV$ for $\wmnu$)}
\item{Region W: $\et^{\mathrm{iso}}/\et <$ 0.1 and $\met >$ 25 $\GeV$ (20 $\GeV$ for $\wmnu$)}
\end{itemize}

Region W is the $\wlnu$ signal region and the others contain 
mostly hadronic background events.  The background 
contribution to the $\W$ signal region, $N_W^{\mathrm{bck}}$,
is estimated using
\begin{eqnarray}
\frac{N_W^{\mathrm{bck}}}{N_{\mathrm{evt}}^{\mathrm{C}}} = \frac{N_{\mathrm{evt}}^{\mathrm{A}}}{N_{\mathrm{evt}}^{\mathrm{B}}},
\label{eqn:back}
\end{eqnarray}

\noindent where $N_{\mathrm{evt}}^{\mathrm{A}}$,
$N_{\mathrm{evt}}^{\mathrm{B}}$, $N_{\mathrm{evt}}^{\mathrm{C}}$ are
the number of events in regions A, B and C, respectively, as defined
above.  This technique has been previously described
in~\cite{det:dettop,int:cdf_ratio2} and more recently in~\cite{back:qcdrun2}.

\begin{figure}[t!]
\includegraphics[width=3.5in]{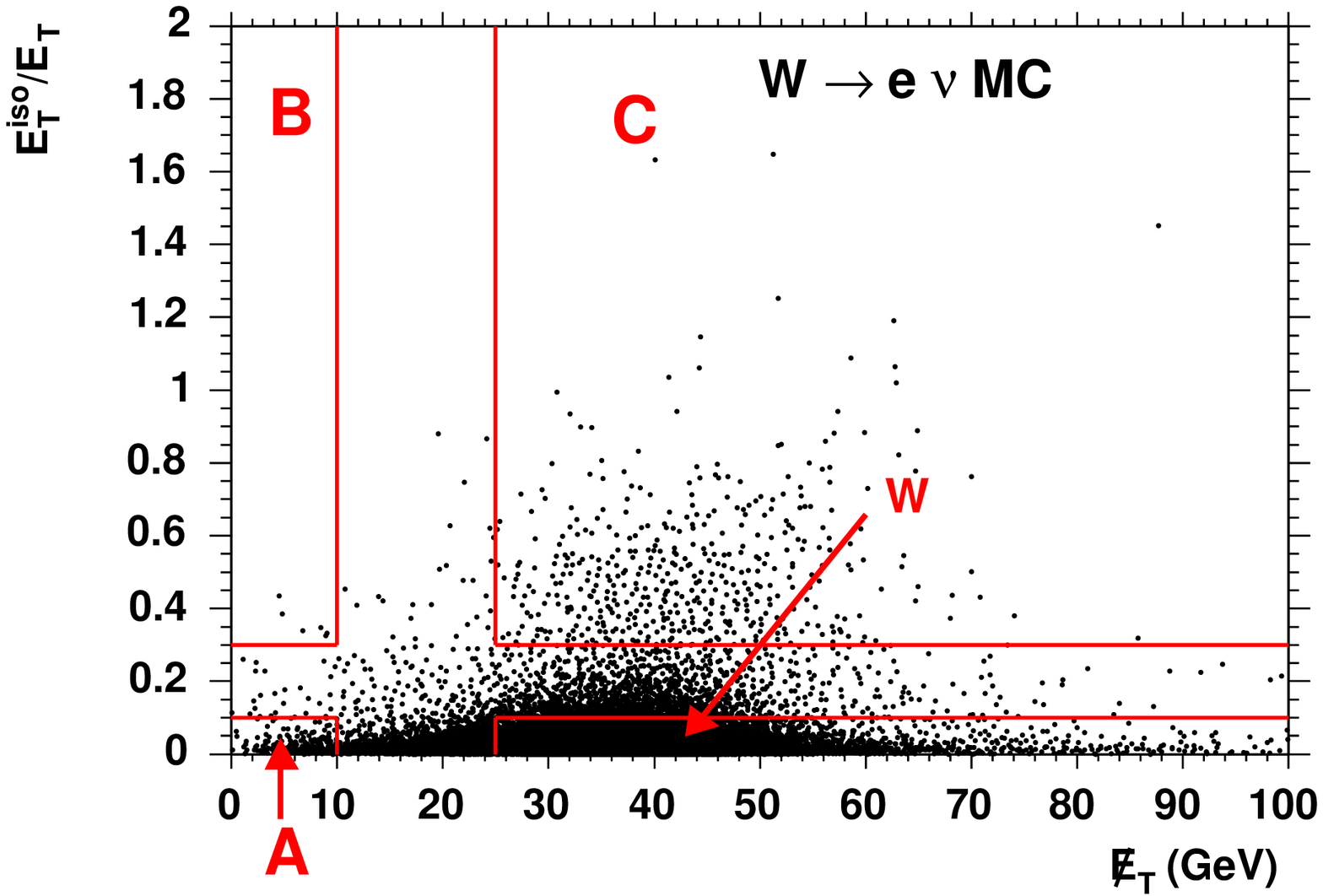} \includegraphics[width=3.5in]{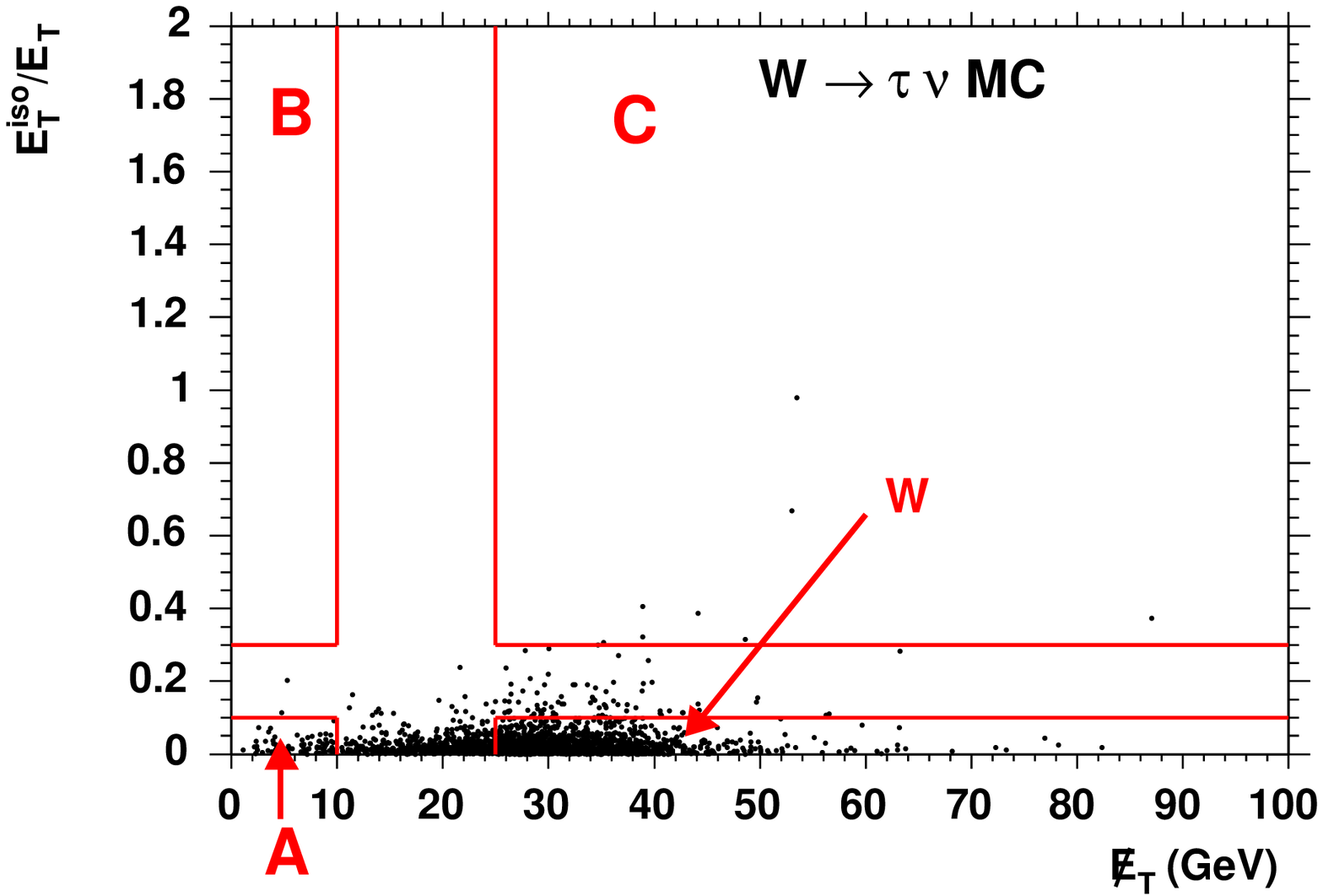} \includegraphics[width=3.5in]{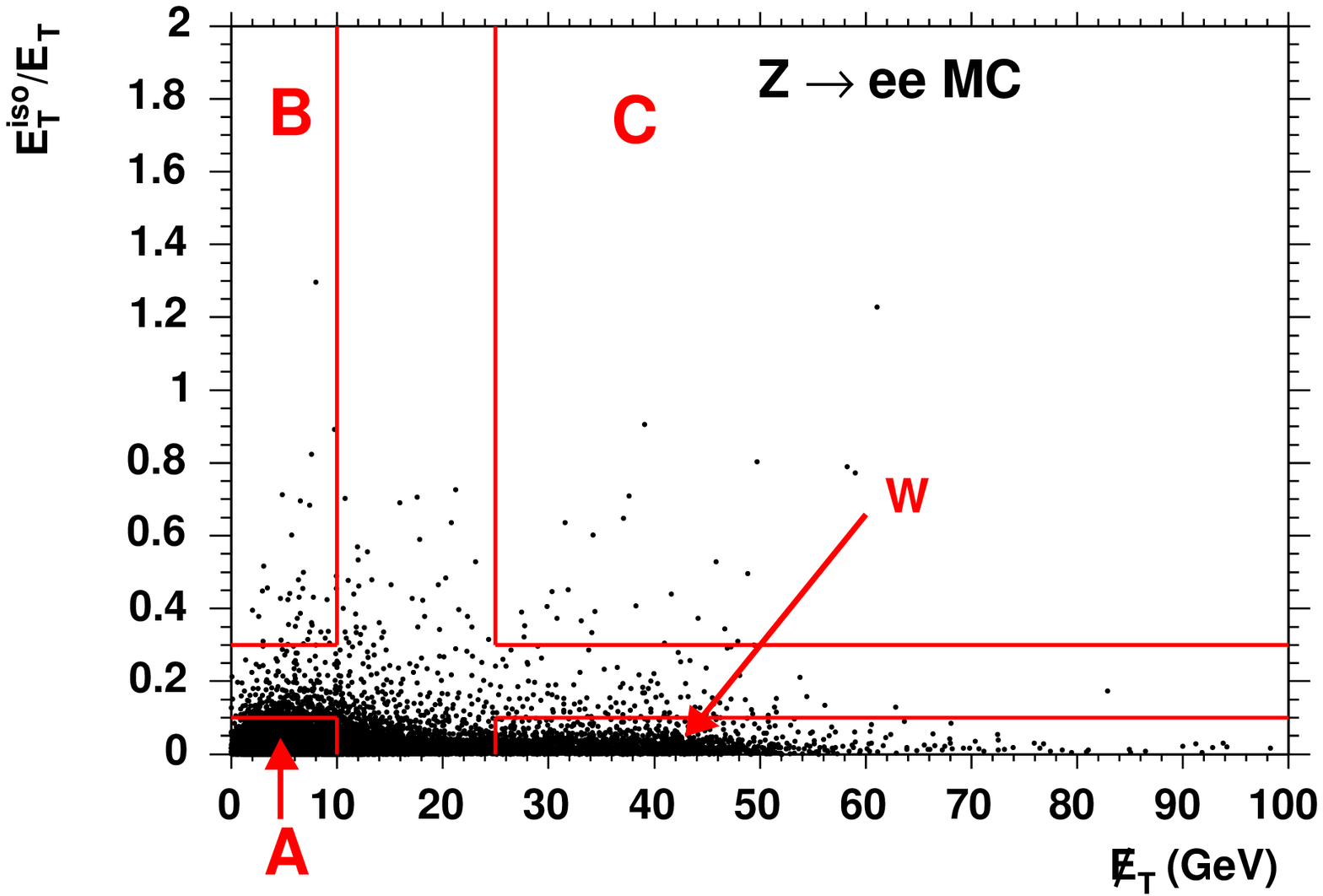}
\caption{$\et^{\mathrm{iso}}/\et$ versus event $\met$ for the 
simulated $\wenu$ signal, $\wtnu$ background, and $\zee$ background
samples.  We correct the observed number of data events in regions 
A, B, C to account for events from these processes when estimating 
the hadronic background in the $\wlnu$ candidate samples.}
\label{fig:isometmc}
\end{figure}

A simple approach would be to assume that all of the events in regions
A, B, and C are hadronic background events.  In that case, the
observed number of data events in each region can be used directly in
Eq.~\ref{eqn:back} to extract the hadronic background contribution to
the $\W$ signal region.  We further improve our estimate, however, by
accounting for the fact that these regions contain small fractions of
signal events and events from other electroweak background processes
such as $\zll$ and $\wtnu$ in addition to hadronic background
events. Fig.~\ref{fig:isometmc} shows distributions of lepton
isolation fraction versus $\met$ for simulated events passing the full
set of selection criteria (no cuts on lepton isolation fraction or
$\met$) for the $\wenu$ signal, $\zee$ background, and $\wtnu$
background samples.  From these distributions, and the equivalent ones
for $\wmnu$ candidates, we obtain modeled event fractions in regions
A, B, and C relative to the signal region for the signal and other
electroweak background processes.  Based on these fractions and our
estimates for the relative contributions of $\wlnu$, $\zll$, and
$\wtnu$ in the signal region (see Sec.~\ref{sec:ewkwbkg}), we correct
the observed numbers of events in regions A, B, and C to remove the
contributions from non-hadronic backgrounds.  A more accurate estimate
of the hadronic background in the $\W$ signal region is then obtained
from Eq.~\ref{eqn:back} using these corrected inputs.
Table~\ref{tab:qcdwback} summarizes both the corrected and uncorrected
hadronic background estimates for the $\W$ signal region obtained from
Eq.~\ref{eqn:back} for the $\wenu$ and $\wmnu$ decay channels.

\begin{table*}[t]
\caption{Summary of hadronic background event contribution estimates 
to the $\wenu$ and $\wmnu$ candidate samples.  The statistical and 
systematic uncertainties are indicated.}
\begin{tabular}{l c c c c} 
\hline 
\hline
                    & Uncorrected & Corrected   & Uncorrected & Corrected     \\ 
                    & $\wenu$     & $\wenu$     & $\wmnu$     & $\wmnu$       \\ 
\hline 
Region A            & 30023             & 26655             & 3926              & 3575                \\ 
Region B            & 5974              & 5972              & 5618              & 5615                \\ 
Region C            & 228               & 131               & 496               & 345                 \\ 
Region W            & 37584             & 37584             & 31722             & 31722               \\ 
\hline
Hadronic Background & 1146              & 587               & 346               & 220                 \\
Statistical Error   & 78                & 52                & 17                & 13                  \\
Systematic Error    & -                 & 294               & -                 & 110                 \\
Background Fraction & 3.0 $\pm$ 0.2$\!\%$ & 1.6 $\pm$ 0.8$\!\%$ & 1.1 $\pm$ 0.1$\!\%$ & 0.7 $\pm$ 0.4$\!\%$ \\ 
\hline 
\hline
\end{tabular}
\label{tab:qcdwback}
\end{table*}

\begin{figure}[t!]
\includegraphics[width=3.5in]{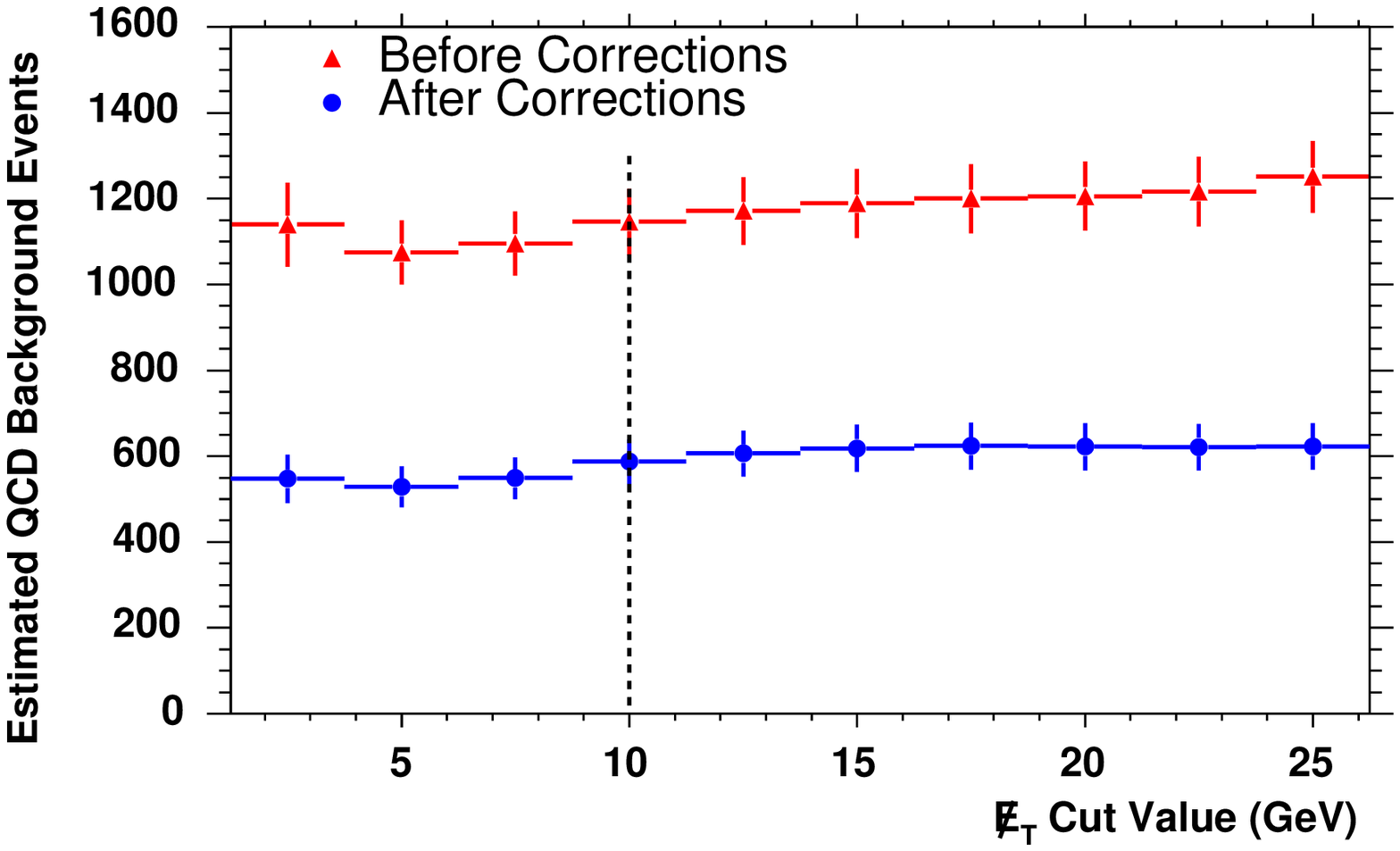} \includegraphics[width=3.5in]{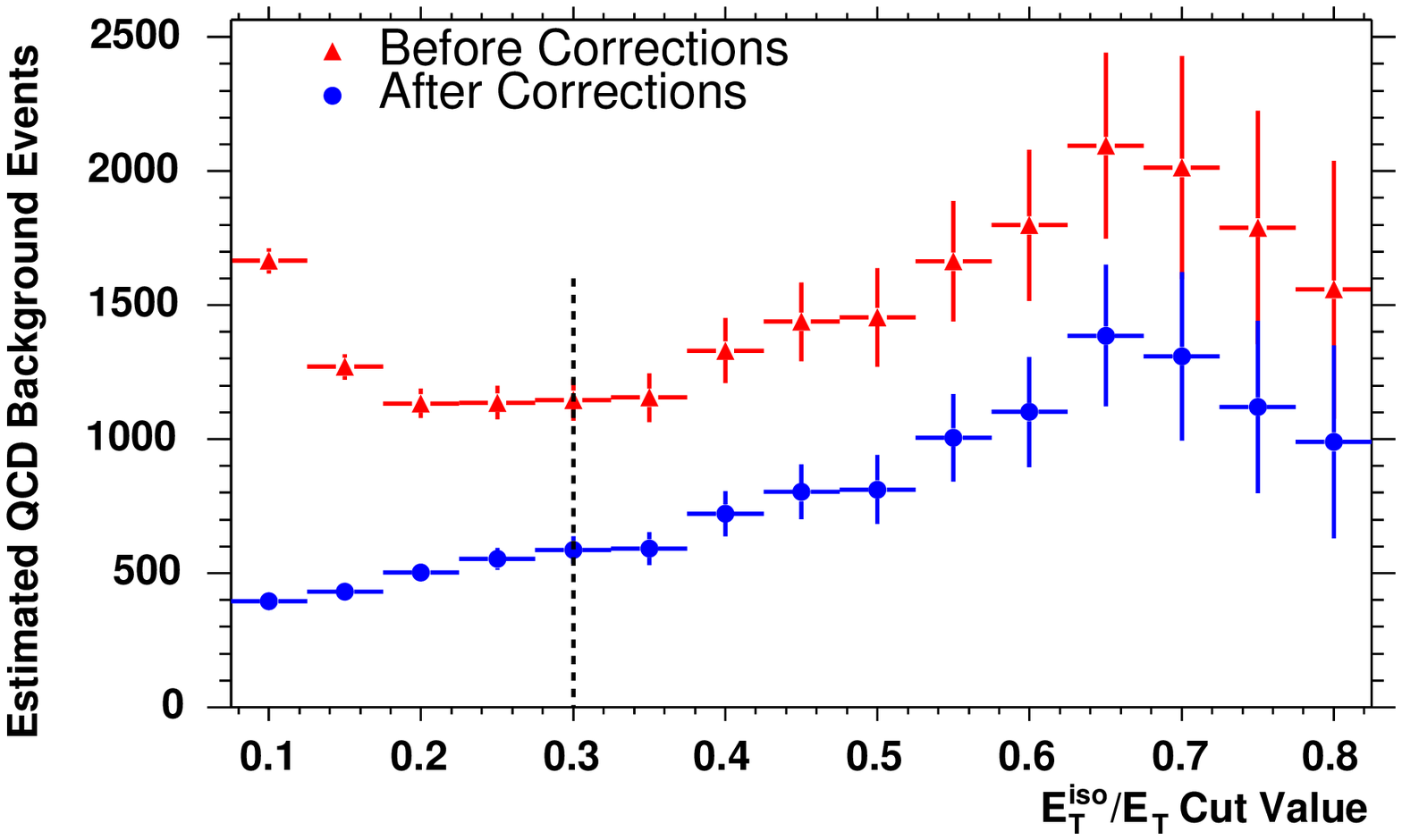}
\caption{Dependence of hadronic background estimate on the 
$\et^{\mathrm{iso}}/\et$ and event $\met$ cut values used 
to define the control regions for $\wenu$.  The results 
both before and after corrections for signal and electroweak 
background contributions to regions A, B, and C are applied 
are shown in triangles and circles, respectively.}
\label{fig:backdep}
\end{figure}

Since the lower limit on lepton isolation fraction and upper limit 
on event $\met$ used to define regions A, B, and C are arbitrary 
choices, we check the robustness of our technique for obtaining 
the hadronic background estimates by raising and lowering the 
cuts used to define these regions.  We take observed changes in 
the estimated hadronic backgrounds as a systematic uncertainty on 
our measurement technique.  Fig.~\ref{fig:backdep} shows the 
dependence of the estimated hadronic background contribution to 
the signal region as a function of the lepton isolation fraction 
and event $\met$ values used to define the non-signal regions both 
before and after correcting the number of observed events in 
regions A, B, C for $\wenu$ signal and other background processes.  
We observe similar dependencies using the $\wmnu$ candidate sample.  
The background estimate is mostly independent of the selection of 
the lower $\met$ border for regions A and B but does depend on the
location of the upper lepton isolation fraction border for regions 
B and C.  Although we observe some evidence from simulated event 
samples that the observed fluctuations are a feature of the 
hadronic background, we choose to use a conservative systematic 
uncertainty that covers the full range of the fluctuations seen 
in Fig.~\ref{fig:backdep}.  We estimate the range of the observed
fluctuations to be within 50~$\!\%$ of our central values 
corresponding to uncertainty estimates of $\pm$~294 events in the 
$\wenu$ candidate event sample and $\pm$~110 events in the $\wmnu$ 
candidate sample (see Table~\ref{tab:qcdwback}).

We make an independent cross-check of the estimated hadronic 
background in $\wenu$ events by applying a measured rate for 
jets faking electrons to a generic hadronic jet sample.  The 
rate for jets faking electrons is measured from events with 
at least two jets with $\et >$ 15~$\GeV$, $\met <$ 15~$\GeV$, 
and no more than one loose electron.  These requirements 
ensure that the input sample has a negligible contribution 
from real $\W$ and $\Z$ events.  From this sample, the jet 
fake rate is defined as the fraction of reconstructed jets 
with $\et >$ 30~$\GeV$ that are also found to pass the 
standard set of tight electron cuts.

\begin{figure}[t]
\begin{center}
 \includegraphics[width=3.5in]{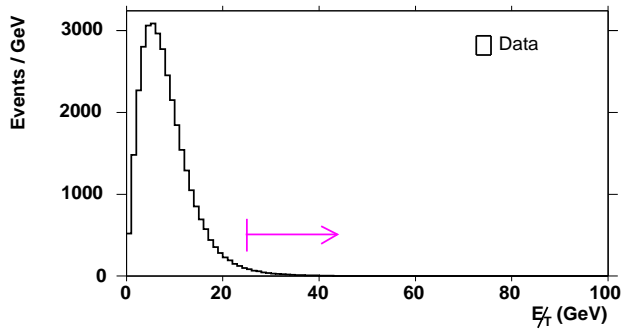}
\end{center}
\caption{$\met$ distribution for inclusive jet sample weighted by 
measured jet fake rate.  The arrow indicates the location of the 
selection cut on $\met$ used to select $\wenu$ candidate events.}
\label{fig::jetfakerate}
\end{figure}

We use the $\et$ dependence of the jet fake rate in the 
background estimate.  Because of differences in the 
clustering algorithms used for electrons and jets, the 
reconstructed energies of electrons originating from hadronic 
jets are smaller than the reconstructed energies of the jets.
Scale factors are applied to convert the measured jet energies 
into corresponding electron cluster energies, and as a consequence 
the lowest $\et$ bins are not included in the fitted constant 
for the jet fake rate. A significant uncertainty on the final
background estimate is assigned, however, based on the results
obtained using different models for fitting the $\et$ dependence of
the jet fake rate.  The measured fake rate is applied to jets in an
inclusive jet data sample to determine how often these types of
hadronic events with fake electrons satisfy the additional selection
criteria of our $\wenu$ candidate sample.  Jets in the inclusive
sample are required to have $\et^{\mathrm{scaled}} >$ 25~$\GeV$ where
$\et^{\mathrm{scaled}}$ is the jet $\et$ scaled down to the $\et$ of 
the fake electron to match the electron selection criteria of our 
sample.  The distribution in Fig.~\ref{fig::jetfakerate} is the
resulting $\met$ distribution for the inclusive jet sample weighted 
by the jet fake rate.  The events above the candidate sample $\met$ 
cut of 25~$\GeV$ are integrated giving 800~$\pm$~300 background
events, consistent with the result obtained using our default
technique.

\subsection{Hadron Jet Background in $\zgll$}
\label{sec:qcdzbkg}
Our $\zll$ candidate samples have smaller overall contributions
from background sources than the $\wlnu$ samples.  One common 
background source is events in which one or both leptons are 
either real or fake leptons from hadronic jets.  We expect 
that the two leptons in these types of events have no charge 
correlation so that the numbers of opposite-sign and same-sign 
lepton pairs from this source are roughly equal.  Based on
this assumption, we use the number of same-sign lepton pair 
candidates to place an upper limit on the number of hadronic 
background events in our opposite-sign dilepton pair candidate
samples.  This approach is only viable for events with two 
central leptons where the lepton charge is taken from the 
reconstructed track.  As discussed later in this section, the 
background contribution to $\zee$ events with one central 
electron and one plug electron is measured using a variation 
of the jet fake rate method described previously.

Since the calorimeter energy associated with muon candidates is 
required to be consistent with a minimum-ionizing particle, the 
probability for a hadronic jet to fake a muon is significantly 
smaller than that for an electron.  Despite the fact that we  
make no opposite-sign charge requirement on the two muon legs in 
our $\zmm$ candidate events, none of the 1,785 events in this 
sample are observed to contain a same-sign muon pair.  Based on 
finding no such events, we estimate a background contribution 
of 0.0~$\!^{+1.1}_{-0.0}$ events from muons produced in hadronic 
jets.

\begin{figure}[t!]
\begin{center}
\includegraphics[width=3.4in,height=3.in]{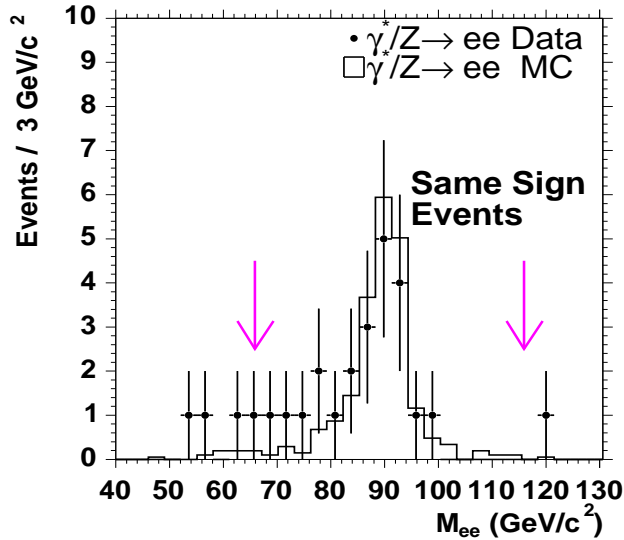}
\caption{Reconstructed invariant mass of two central electrons
in $\zee$ candidate events in data (points) and simulation
(histogram).  This distribution is for events in which the electrons
are reconstructed with the same charge. The distribution for events
with two electrons of opposite sign is shown in Fig.~\ref{fig:etfit}.
The number of events in the simulated distributions are normalized so
that the number of opposite-sign events in the simulated sample is
equal to the number of opposite-sign events in the data. The arrows
indicate the location of the invariant mass cuts used to select our
candidate samples.}
\label{fig:invmass}
\end{center}
\end{figure}

The number of same-sign events observed in the $\zee$ candidate sample
needs to be corrected for a fraction of real $\zee$ events that are
reconstructed as same-sign electron pairs.  We observe a total of 22
events with same-sign electron pairs corresponding to our sample 1730
$\zee$ candidate events with two central electrons.  The invariant
mass distributions for both the opposite-sign and same-sign electron
pairs in our candidate sample are shown in Fig.~\ref{fig:etfit} and
Fig.~\ref{fig:invmass}.  Both distributions show a peak in the $\Z$
boson mass window indicating that at least some fraction of the
same-sign electron pairs are produced in $\Z$ decays.  These events
result from decays in which one of the electrons radiates a high $\et$
photon which subsequently converts in the detector material producing
an electron-positron pair.  We call this type of event a ``trident''
event.  If the track associated with the positron from the photon
conversion is matched to the corresponding electron cluster, both
electrons in the event will be assigned the same charge.  We remove
the contribution of real $\zee$ events from the number of observed
same-sign electron pairs by subtracting the observed number of
opposite-sign events in the data scaled by the fraction of same-sign
to opposite-sign candidates in our simulated samples.  The remaining
number of same-sign electron pair candidate events is then used to
estimate the background contribution from electrons produced in
hadronic jets to the opposite-sign candidate sample.

Using this technique, we estimate 20.4 same-sign events from 
$\Z$ decays in the invariant mass window between 66~$\GeVCSq$
and 116~$\GeVCSq$.  Subtracting this estimate from the total  
number of observed same-sign events (22), we estimate the 
contribution from electrons originating from hadronic jets 
to be 1.6~$\!^{+4.7}_{-1.6}$ where the uncertainty is based 
solely on the statistics of our sample. 

The dominant source of systematic uncertainty on the background 
contribution from events with electrons originating from hadronic
jets comes from the simulation detector material model.  The 
probability for an electron to radiate a bremsstrahlung photon 
prior to entering the calorimeter is strongly dependent on the 
amount of material in the tracking volume.  We study the effect 
of the material model using the two previously described samples
of simulated events generated with $\pm$~1.5~$\!\%$ of a radiation 
length of copper added in a cylinder between the silicon and 
COT tracking detectors.  We estimate the systematic uncertainty 
based on differences in the number of same-sign events observed  
after subtracting the predicted number of real $\zee$ events
based on the default and modified simulations.  The resulting 
systematic uncertainty on our estimate is 5.2 events which when 
added in quadrature with the statistical error results in a 
final background estimate of 1.6~$\!^{+7.0}_{-1.6}$.

\begin{figure}[t]
\begin{center}
\includegraphics*[width=3.5in]{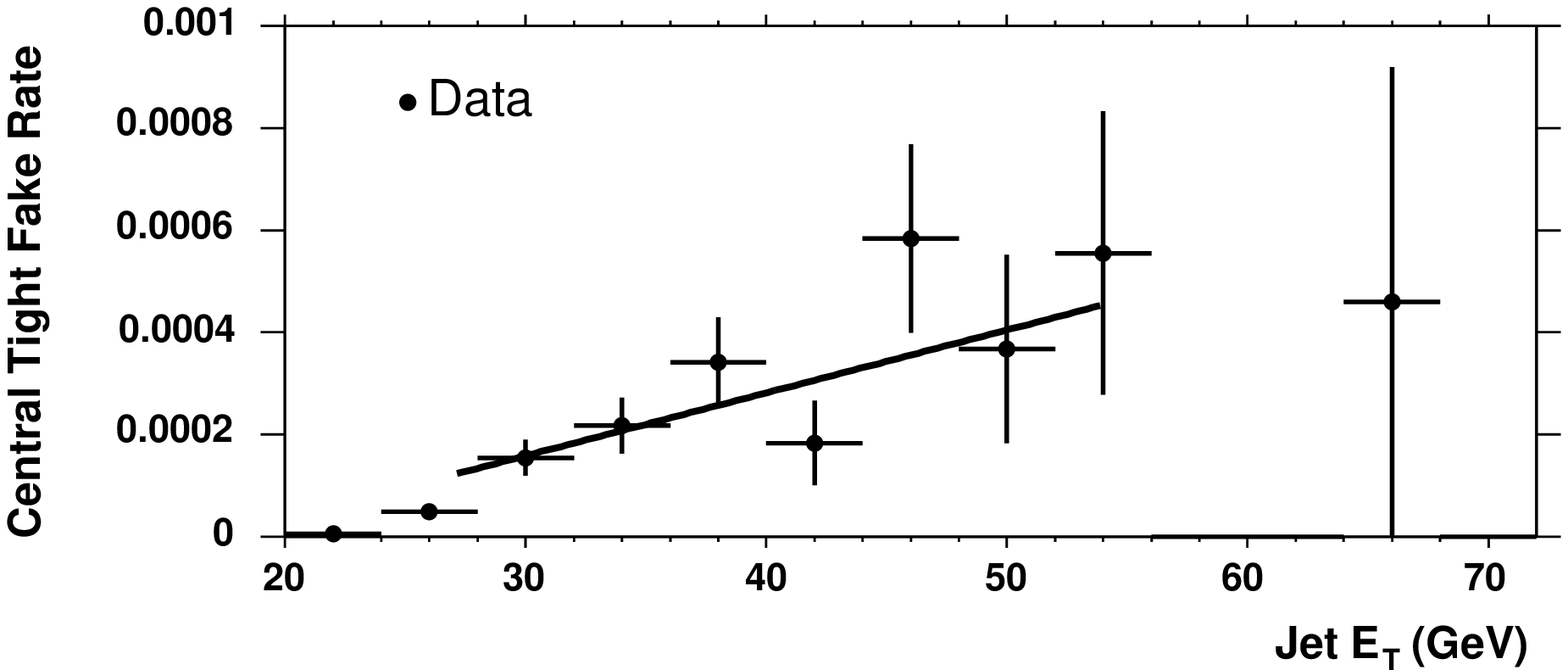} 
\includegraphics*[width=3.5in]{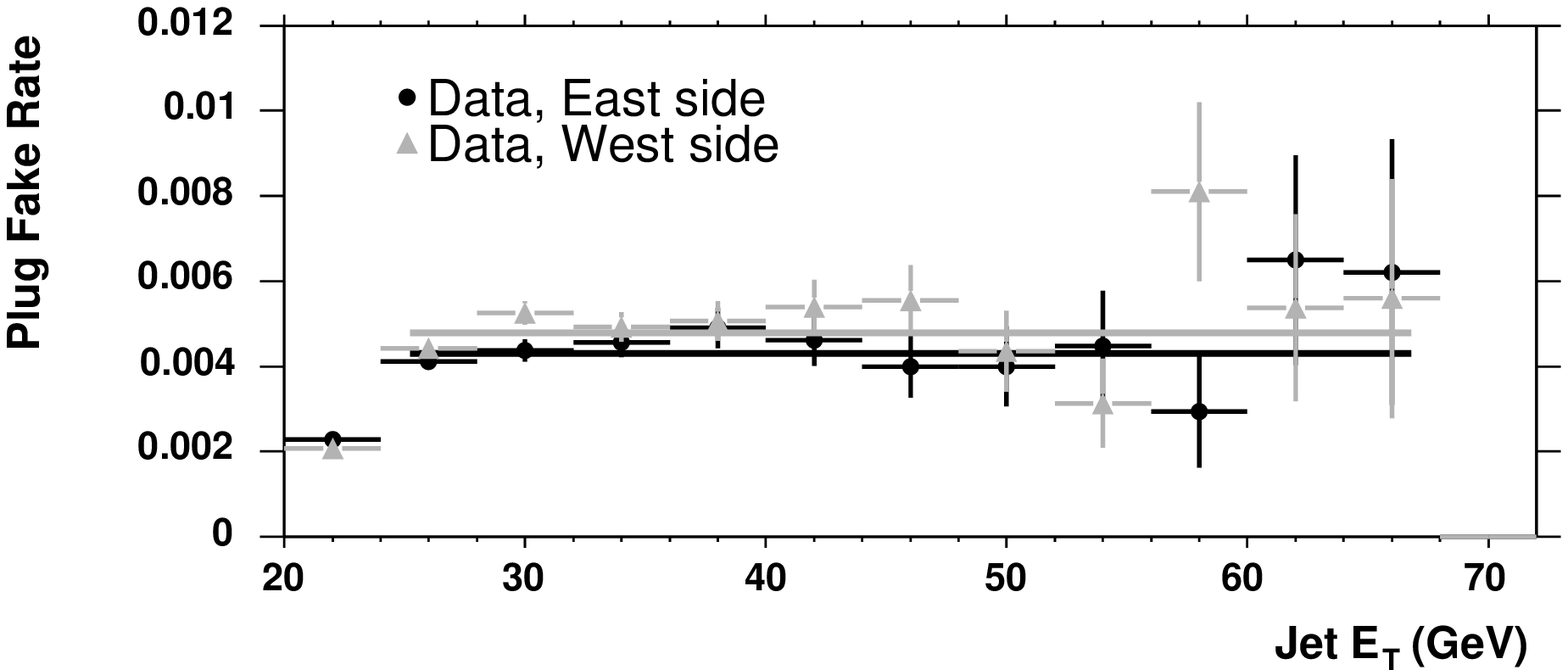}
\end{center}
\caption{Measured tight central and plug 
electron jet fake rates as a function of jet $\et$.}
\label{fig::fakerates}
\end{figure}

This technique outlined above can not be used to estimate the
background contribution from electrons originating from hadronic jets
in $\zee$ candidates with one central and one plug electron owing to
the undetermined charge of the plug candidate.  Instead, we estimate
the background contamination based on a variation of the previously
described method using measured jet fake rates.  In order to measure
the background contribution to the combined $\zee$ sample from
hadronic events producing two fake electrons, we need to measure the
jet fake rates for tight central, loose central, and plug electrons.
We remove $\W$ and $\Z$ boson candidates from the inclusive jet sample
used to make the fake rate measurements by selecting events with no
more than one loose electron and $\met <$ 15~$\GeV$.  Based on this
inclusive sample, the jet fake rates are defined as the fractions of
central jets reconstructed as either tight or loose electrons and plug
jets reconstructed as plug electrons.  The measured jet fake rates for
reconstructed tight central and plug electrons as a function of jet
$\et$ are shown in Fig.~\ref{fig::fakerates}.

\begin{figure}[hbt]
\begin{center}
\includegraphics*[width=3.5in]{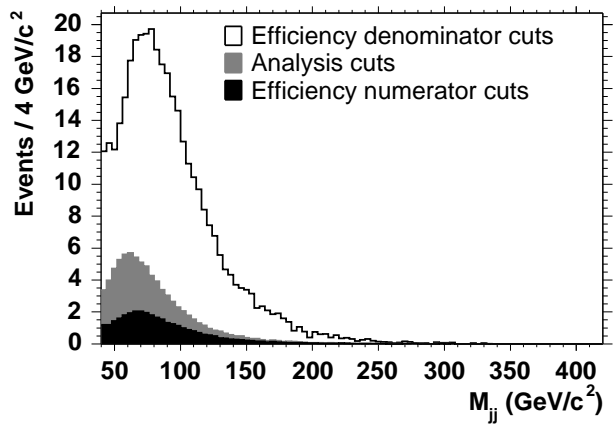}
\caption{Di-jet invariant mass distributions for central-central 
and central-plug events in generic multi-jet data.  Events are 
weighted by the product of the measured jet fake rates for each 
jet.  The scaled energies of both jets must pass the electron 
$\et$ requirements of our $\zee$ candidate sample (25~$\GeV$ for
central electrons and 20~$\GeV$ for plug electrons).} 
\label{fig::mjjE}
\end{center}
\end{figure}

As previously mentioned, the reconstructed energy of the electrons 
produced by hadronic jets is smaller than the reconstructed 
energy of the jets themselves.  To account for these differences, 
we fit the distributions of $\et^{\mathrm{ele}}/\et^{\mathrm{jet}}$ 
to a Gaussian for the jets reconstructed as tight central, loose 
central, and plug electrons.  The means of the fits are used as 
scaling factors to convert raw jet energies into scaled electron 
energies, $\et^{\mathrm{scaled}}$.  To obtain the background 
contribution of events with two electrons originating from hadronic 
jets to the $\zee$ sample, we apply the measured jet fake rates and 
energy scalings to a generic multi-jet data sample.  Events containing 
either two central jets with $\et^{\mathrm{scaled}} >$~25~$\GeV$ 
or one central jet and a plug jet with $\et^{\mathrm{scaled}} 
>$~20~$\GeV$ are used to extract dijet invariant mass distributions 
to model the hadronic background for $\zee$.  The weights assigned 
to each event in these distributions is set equal to the product of 
the jet fake rates for the two jets based on the parameterizations 
shown in Fig.~\ref{fig::fakerates}.  The final weighted dijet 
invariant mass distributions for central-central and central-plug 
events are shown in Fig.~\ref{fig::mjjE}.  The resulting distributions
are integrated over the invariant mass window of our measurements
(66~$\GeVCSq < M_{ee} <$~116~$\GeVCSq$) to obtain an estimate for the 
number of background events in the $\zee$ candidate sample (after
scaling upward by the trigger prescale used to collect events in 
the generic multi-jet sample).

As illustrated in Fig.~\ref{fig::fakerates}, the jet fake rates 
measured as a function of jet $\et$ need to be assigned an 
additional uncertainty based on the assumed shape of the fit.  
We fit the jet fake rate distributions using several different 
functional forms and assign an additional systematic uncertainty 
of 30~$\!\%$ based on the resulting spread in background estimates.
Based on this technique the measured background contribution of 
events with two electrons originating from hadronic jets to 
the $\zee$ candidate sample is 2.4~$\pm$~1.0 central-central 
and 39~$\pm$~17 central-plug events.  The estimated number of 
central-central events is in good agreement with the result 
obtained using the observed number of same-sign events in our 
candidate sample.  Using the central-central background estimate
based on same-sign events and the central-plug estimate based on
the jet fake rate method, we obtain a combined estimate for the 
background contribution of events with two electrons originating 
from hadronic jets of 41~$\pm$~18 events.

\subsection{Electroweak Backgrounds in $\wlnu$}
\label{sec:ewkwbkg}
$\zll$ events mimic the signature of $\wlnu$ events in cases where one
of the two leptons passes through an uninstrumented region of the
detector creating an imbalance in the observed event $\et$.  The
$\wlnu$ signature can also be reproduced by $\wtnu$ events in which
the $\tau$ lepton subsequently decays into an electron or muon.
Background contributions from both diboson and $t \bar{t}$ production
processes are negligibly small.

The contribution of these electroweak background sources to our 
$\wlnu$ candidate samples are obtained from simulation.  The 
$\zgll$ and $\wtnu$ simulated event samples are obtained from the 
equivalent {\sc pythia} event generation and detector simulation 
used to produce the signal samples (see Sec.~\ref{sec:acc}).  
The complete set of $\wlnu$ selection criteria as described in   
Sec.~\ref{sec:evsel} are applied to the simulated events in these 
samples to obtain the fraction of events from each process that 
satisfy the criteria of our candidate samples.  Then, based on   
Standard Model predictions for the relative production rates 
of our signal process and the two background processes, we use 
the estimated acceptances from simulation to obtain the relative 
contributions of each process to our candidate samples.

The Standard Model predicts equivalent production cross sections 
for $\wenu$, $\wmnu$ and $\wtnu$, while the $\zll$ production 
cross sections are related to the corresponding $\wlnu$ cross 
sections via the ratio $R$ defined in Eq.~\ref{eq:rdef}.  In 
order to extract the relative contributions of $\zgll$ events 
to our $\wlnu$ candidate samples, an input value for $R$ is
required.  We choose to use the value $R =$~10.67~$\pm$~0.15 
for $\W$ and $\Z$ boson production at $\sqrt{s} =$~1.96~$\TeV$ 
obtained from the NNLO theoretical calculation
~\cite{int:nnlo4,int:nnlo1,int:nnlo2,int:nnlo3}.  
However, to be conservative we inflate the uncertainty on the 
predicted value for $R$ based on the CDF Run~I measured value 
of $R =$~10.90~$\pm$~0.43~\cite{int:cdf_ratio1,int:cdf_ratio2}.  
The difference in the values of $R$ at $\sqrt{s} =$~1.80~$\TeV$ and
1.96~$\TeV$ is expected to be negligible.  Based on this prediction,
the measured value is in good agreement with the theoretical value.
To account for the current level of experimental uncertainty, we add
an additional 3.9~$\!\%$ systematic uncertainty to the NNLO prediction
resulting in a value of $R =$~10.67~$\pm$~0.45.

\begin{table}[t]
\caption{Estimated $\wlnu$ backgrounds from other electroweak 
production processes.}
\begin{center}
\begin{tabular}{l c c} 
\hline 
\hline
Source       & $\wenu$          & $\wmnu$          \\
             & Background       & Background       \\ 
\hline
$\zll$       & 426~$\pm$~19     & 2229~$\pm$~96    \\ 
$\wtnu$      & 749~$\pm$~17     & 988~$\pm$~24     \\ 
\hline 
\hline
\end{tabular}
\label{tab:ewkwback}
\end{center}
\end{table}

The relative contributions of $\wlnu$, $\zll$, and $\wtnu$ 
in our $\wlnu$ candidate samples are estimated based on the
above value for $R$ and the simulated acceptances for each 
process.  The relative acceptances are normalized to the 
total number of events in each candidate sample after 
subtracting contributions from non-electroweak backgrounds 
(events with reconstructed leptons originating from hadronic 
jets and cosmic rays).  The final background estimates for 
electroweak backgrounds in the $\wlnu$ candidate samples are 
summarized in Table~\ref{tab:ewkwback}.

\subsection{Electroweak Backgrounds in $\zgll$}
\label{sec:ewkzbkg}
Several electroweak processes also contribute background events to 
our $\zll$ candidate samples.  $\ztt$ events mimic the $\zll$ event 
signature when both $\tau$ leptons decay into or are reconstructed 
as an electron or muon pair with a reconstructed invariant mass 
within the mass window of our $\zll$ measurements.  As in the 
previous section, this background is estimated using a simulated 
$\ztt$ event sample obtained from the equivalent {\sc pythia} 
event generation and detector simulation used to produce the $\zll$
signal samples.  The full set of $\zll$ selection criteria is 
applied to the simulated $\ztt$ and $\zll$ samples to determine the 
relative acceptances.  Based on the Standard Model prediction of 
equivalent production cross sections for $\zee$, $\zmm$, and $\ztt$,
the number of $\ztt$ background events in each candidate sample is
extracted using the relative acceptances from the total number of  
events after removing non-electroweak background contributions.

\begin{figure}[t]
\begin{center}
\includegraphics[width=3.5in]{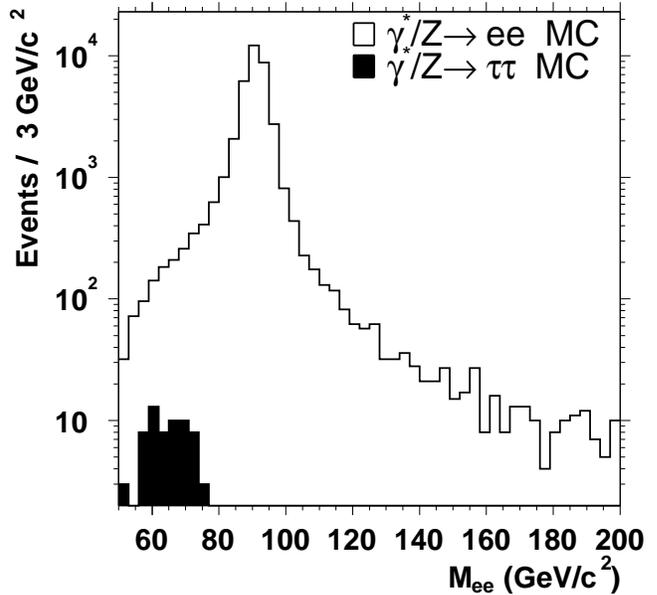}
\caption{Reconstructed invariant mass distribution for simulated 
$\zgee$ (open histogram) and $\zgtt$ (solid histogram) events 
satisfying the $\zee$ candidate sample selection criteria.} 
\label{fig:invmassbackg}
\end{center}
\end{figure}

A comparison of the reconstructed invariant mass distributions for 
simulated $\zgee$ and $\zgtt$ events passing the $\zee$ selection 
criteria is shown in Fig.~\ref{fig:invmassbackg}.  The majority of 
$\zgtt$ events are observed to have a reconstructed invariant mass
below the mass window used in our measurements.  As a result, the 
contribution of this background source to our candidate samples is 
small, 3.7~$\pm$~0.4 events in the $\zee$ sample and 1.5~$\pm$~0.3
events in the $\zmm$ sample.  An identical approach is used to
estimate $\zll$ background contributions from both top quark and 
diboson production.  The estimated background contributions from 
each of these sources is found in all cases to be less than one 
event and therefore considered to be negligible.  

An additional source of background events to the $\zee$ candidate 
sample is $\wenu$ events with an associated hadronic jet that 
results in a second reconstructed electron within the event.  We
use our simulated $\wenu$ sample to estimate the background 
contribution from this source by applying previously determined 
jet fake rates for the hadronic jets in these events with scaled 
$\et$ above the corresponding electron thresholds.  The relative 
acceptance of simulated $\wenu$ events, weighted by the measured 
jet fake rates, and $\zee$ signal events are used to extract the
number of background events from this process based on the value 
for $R$ presented in the previous section.  Once again, the 
relative acceptances are applied to the final candidate sample 
after subtracting the estimated number of background events 
from non-electroweak sources.  The estimated number of $\wenu$ 
background events in the $\zee$ sample is 16.8~$\pm$~2.8 events.

\subsection{Cosmic Ray Backgrounds in $\wmnu$}
\label{sec:cosmicwbkg}
Energetic cosmic ray muons traverse the detector at a significant rate, 
depositing hits in both muon and COT chambers, and in some cases can 
mimic the signatures of our $\wmnu$ and $\zmm$ candidate events.  A 
cosmic ray muon passing through the detector is typically reconstructed 
as a pair of incoming and outgoing legs relative to the beam line of 
the detector.  The reconstructed muon legs tend to be isolated and pass 
our muon selection criteria.  In some cases, one of the two cosmic legs
is not reconstructed due to fiducial and/or timing constraints.  These 
events typically satisfy both the $Z$-rejection and $\met$ criteria of 
our $\wmnu$ candidate sample due to the lack of an additional track and 
the resulting transverse momentum imbalance.    

We remove cosmic ray events from our $\wmnu$ candidate sample using a
tagging algorithm based on the timing information associated with hits 
in the COT.  The algorithm uses a multi-parameter fit over the full 
set of hits left by the incoming and outgoing cosmic legs. The leg 
belonging to the reconstructed muon serves as the seed track for the 
fit.  The other leg is referred to as the opposite-side track.  The 
algorithm performs the following steps to determine if an event is 
consistent with the cosmic ray hypothesis:
\begin{itemize}
\item Hits belonging to the seed track are refitted with the five helix 
parameters and a floating global time shift, $t_{0}$.
\item Based on the best fit values, an incoming or outgoing hypothesis 
is assigned to the seed track.
\item The refitted seed track is used to search for the hits belonging 
to the second cosmic leg on the opposite side of the COT. 
\item If enough hits are found on the other side of the COT, a similar 
fit procedure is performed to identify the opposite-side track.
\item If both legs are found, a simultaneous fit is performed to combine 
all hits from the seed and opposite-side legs into a single helix.
\end{itemize}
The final decision of the cosmic tagger depends on the quality of the
simultaneous fit to the hits on both legs.  If one leg is recognized
as incoming and fits well to an outgoing leg on the other side of the
detector, the event is tagged as a cosmic ray.  As described in
greater detail below, we observe that our tagging algorithm identifies
most of the cosmic background events in our candidate sample.  We also
find that the algorithm tags very few real events as cosmic rays (see
Sec.~\ref{sec:eff}).

After removing tagged events from our $\wmnu$ sample, we need to
estimate the remaining background from cosmic rays.  This estimate is
made by searching for hits in the muon chambers on the opposite side
of the reconstructed muon track in our final candidate events.  These
hits are present for a large fraction of cosmic ray muons even in
cases where the second leg is not identified by our algorithm.  Since
the muon chamber hits are not used in the tagging algorithm, their
presence is unbiased with respect to its decision.  The $\Delta \phi$
distribution for matched hits produced by cosmic ray muons with
respect to the direction of the muon candidate track is sharply peaked
in the region around 180~$\!^{\circ}$.  These events sit on top of a
flat event background in $\Delta \phi$ originating from random
coincidences between the muon track and unrelated matched hits in the
muon chambers.  The contribution of cosmic ray events to the candidate
$\Delta \phi$ distribution is determined by counting the number of
events with matched muon chamber hits in a 10~$\!^{\circ}$ window
centered on $\Delta \phi =$~180~$\!^{\circ}$ and subtracting a fitted
contribution from the flat background.  Using this approach, we would
estimate a cosmic background contribution of 54.7~$\pm$~5.0 events in
our 31,722 event $\wmnu$ candidate sample.

Some of the cosmic ray background events in our candidate sample, 
however, do not have opposite-side muon chamber hits due to gaps 
in the muon detector coverage.  In order to estimate the total 
cosmic ray background in our candidate sample from the observed 
number of events with matched opposite-side hits, we apply an 
acceptance correction based on the fraction of $\wmnu$ candidate 
events in which the reconstructed muon track points at an active 
region of muon chambers when extrapolated to the opposite side 
of the detector.  We extrapolate the 31,722 muon tracks in our 
$\wmnu$ candidate events to the opposite side of the detector 
and find that 58~$\pm$~30~$\!\%$ point at active regions of 
the muon chambers.  Our acceptance correction assumes that the   
spacial distribution of muons originating from cosmic rays is
similar to that of our $\wmnu$ candidate sample.  We assign 
a large systematic uncertainty on the measured acceptance to 
account for the non-uniform spacial distribution (most enter 
from the top side of the detector) of cosmic rays and the 
reconstruction biases associated with their entry locations 
and angles of incidence on the detector.      

To complete the cosmic background measurement for our $\wmnu$
candidate sample, we also need to estimate the contribution of
$\zmm$ events to the observed excess of events in the window 
around $\Delta \phi =$~180~$\!^{\circ}$.  $\zmm$ events that 
contain a second reconstructed track passing a loose set of 
minimum ionizing cuts are rejected from our candidate sample 
via the $Z$-rejection selection criteria.  However, a small 
fraction of muon tracks from $\zmm$ events are embedded in 
jets and fail the loose minimum ionizing cuts or in other 
cases are simply not reconstructed.  Since the muons in 
$\zmm$ decays are typically produced in roughly opposite 
directions to one another, the non-identified tracks in these 
events can also produce muon chamber hits on the opposite side 
of the one reconstructed muon in these events.  This background 
is estimated from our simulated $\zmm$ event sample.  Based 
on this sample, we estimate the number of $\zmm$ background 
events in our $\wmnu$ candidate sample with matched muon 
chamber hits in the 10~$\!^{\circ}$ window centered on $\Delta 
\phi =$~180~$\!^{\circ}$ to be 35.4~$\pm$~9.1.  The uncertainty 
assigned to this background is based on our use of different 
techniques for looking at opposite side muon chamber hits in
data and simulation. 

The final estimate of the cosmic ray background in the $\wmnu$
sample, $N^{\mathrm{cos}}_{\mathrm{bg}}$, is obtained from 
\begin{eqnarray}
N^{\mathrm{cos}}_{\mathrm{bg}} = \frac{N^{\mathrm{MH}}_{\mathrm{evt}} - N^{\mathrm{MH}}_{\zmm}}{A^{\mathrm{opp}}_{\mu}}  
\end{eqnarray}
where $N^{\mathrm{MH}}_{\mathrm{evt}}$ is the number of $\wmnu$ 
candidate events with matched hits in the tight window centered 
on $\Delta \phi =$~180~$\!^{\circ}$, $N^{\mathrm{MH}}_{\zmm}$ is 
the predicted number of $\zmm$ background events with matched 
hits in the same window, and $A^{\mathrm{opp}}_{\mu}$ is muon 
chamber acceptance for muon tracks in $\wmnu$ candidate events 
extrapolated to the opposite side of the detector.  Using the 
input values obtained above, we estimate a total cosmic 
background of 33.1~$\pm$~22.9 events for our $\wmnu$ candidate 
sample.

\subsection{Cosmic Ray Backgrounds in $\zmm$}
\label{sec:cosmiczbkg}
\begin{figure}[t]
\begin{center}
\includegraphics[width=3.5in]{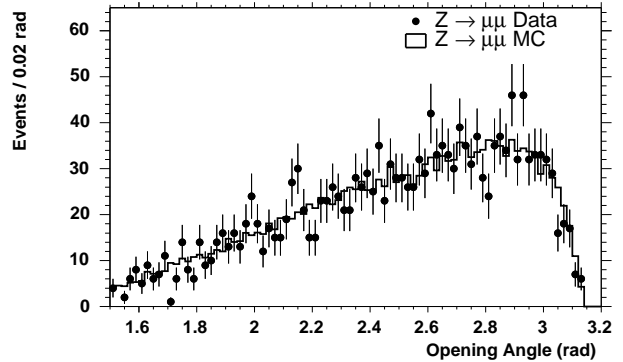}
\caption{Comparison of the three-dimensional opening angle 
distribution for muon tracks in $\zmm$ candidate events with 
the same distribution from simulated events. The simulated 
distribution is scaled to match the data in the region below 
2.8~radians.}
\label{fig:cosmic4}
\end{center}
\end{figure}

Cosmic rays also contribute to the $\zmm$ candidate sample.  The
majority of these events are removed using the cosmic ray tagging
algorithm described in the previous section.  The remaining cosmic ray
background is estimated based on the distribution of the
three-dimensional opening angle between the muon tracks in candidate
events.  The two reconstructed muon legs in the cosmic ray background
events are typically back-to-back with opening angles at or near
~180~$\!^{\circ}$.  The residual background is estimated by fitting
the opening angle distribution for data events in the region below
2.8~radians (assumed to be background free) to the same distribution
for simulated $\zmm$ events.  The output of the fit is a scale factor
for the distribution from simulation which is also applied in the
region above 2.8~radians.  The number of scaled simulation events with
an opening angle greater than 2.8~radians is compared to the number of
data candidate events in the same region.  The observed excess in data
over simulation is taken as our estimate of the cosmic ray background.
Using this technique, we estimate a total of 12~$\pm$~12 cosmic ray
background events in our $\zmm$ candidate sample where the quoted
uncertainties are based on the statistics of our data sample.  A
comparison of the opening angle distribution between data and scaled
simulation is shown in Fig.~\ref{fig:cosmic4}.
   
\subsection{Background Summary}
\label{sec:bkgsummary}
Based on the information presented in the preceding sections, the
estimated background contributions to the $\wenu$, $\wmnu$, $\zee$ 
and $\zmm$ candidate samples are summarized in Table~\ref{tab:backsum}.

\begin{table*}[t]
\caption{Summary of background event estimates for the $\wlnu$ and $\zll$ candidate samples.}
\begin{center}
\begin{tabular}{l c c c c} 
\hline 
\hline
Background source       & $\wenu$        & $\wmnu$        & $\zee$         & $\zmm$        \\ 
\hline
Multi-jet               & 587 $\pm$ 299  & 220 $\pm$ 112  & 41 $\pm$ 18    & $0^{+1}_{-0}$ \\
$\zll$                  & 426 $\pm$ 19   & 2229 $\pm$ 96  & -              & -             \\ 
$\ztt$                  & negl.          & negl.          & 3.7 $\pm$ 0.4  & 1.5 $\pm$ 0.3 \\
$\wtnu$                 & 749 $\pm$ 17   & 988 $\pm$ 24   & negl.          & negl.         \\
$\wlnu$                 & -              & -              & 16.8 $\pm$ 2.8 & negl.         \\ 
Cosmic rays             & negl.          & 33 $\pm$ 23    & negl.          & 12 $\pm$ 12   \\ 
\hline 
Total                   & 1762 $\pm$ 300 & 3469 $\pm$ 151 & 62 $\pm$ 18    & 13 $\pm$ 13   \\ 
\hline 
\hline
\end{tabular}
\label{tab:backsum}
\end{center}
\end{table*}

\section{Results}
\label{sec:results}
Using the measured event counts, kinematic and geometric 
acceptances, event selection efficiencies, background 
estimates, and integrated luminosities for our candidate 
samples, we extract the $\W$ and $\gamma^{*}/Z$ boson 
production cross sections multiplied by the leptonic 
($e$ and $\mu$) branching ratios.  We also determine a 
value for the ratio of $\wlnu$ to $\zll$ cross sections, 
$R$, taking advantage of correlated uncertainties in the 
two cross section measurements which cancel in the ratio.     
To test for lepton universality, we use the measured 
ratio of $\wlnu$ cross sections in the muon and electron 
channels to extract a ratio of the $\wlnu$ coupling 
constants, $g_{\mu}/g_e$.  Then, based on the assumption 
of lepton universality, we increase the precision of 
our results by combining the production cross section 
and cross section ratio measurements obtained from the 
electron and muon candidate samples.  The resulting 
combined value of $R$ is used to extract the total decay 
width of the $\W$ boson, $\gam$, and the $\W$ leptonic 
branching ratio, $Br(\wlnu)$, which are compared with 
Standard Model predictions.  The measurement of $\gam$ 
is also used to constrain the CKM matrix element $\vcs$.

\subsection{$\wlnu$ Cross Section}    
\label{subsec:res_wlnu}

The cross section $\sigmappw$ times the branching ratio $Br(\wlnu)$ 
is calculated using Eq.~\ref{eq:wsigma} given in Sec.~\ref{sec:intro}.  
The measurements of the required input parameters for the electron 
and muon candidate samples are described in the previous sections 
and summarized in Table~\ref{tab:wlnusigma}.  Based on these values, 
we obtain
\begin{eqnarray}
\sigmabrwen = 2.771 &\pm& 0.014({\it stat.}) \nonumber\\
                    &\pm& ^{0.062}_{0.056}({\it syst.})\nonumber\\
                    &\pm& 0.166({\it lum.})~\nb \nonumber\\  
\end{eqnarray}
and
\begin{eqnarray}
\sigmabrwmn = 2.722 &\pm& 0.015({\it stat.}) \nonumber\\
                    &\pm& ^{0.066}_{0.061}({\it syst.}) \nonumber\\
                    &\pm& 0.163({\it lum.})~\nb~{\rm .}\nonumber\\
\end{eqnarray}
We compare our measurements to a recent NNLO total cross section
calculation for $\sqrt{s} =$~1.96~$\TeV$~\cite{int:mrst1} which utilizes the
MRST 2002 NNLL PDF set~\cite{int:mrst1,int:mrst2}.  The resulting
predicted $\wlnu$ cross section is 2.687~$\pm$~0.054~$\nb$, which
agrees well with our measured values in both lepton channels.  The
uncertainty on the predicted cross section is mostly due to PDF model
uncertainties derived from the MRST error PDF sets.  We also perform
an independent calculation of the uncertainty on the total $\wlnu$
cross section originating from uncertainties in the PDF model using
the method described in Sec.~\ref{sec:acc}.  Based on this method, we
obtain a consistent 1.3~$\!\%$ uncertainty based on the MRST error PDF
sets and a 3.9~$\!\%$ uncertainty based on the CTEQ6 error PDF sets.

\begin{table}[t]
\caption{Summary of the input parameters to the $\wlnu$ 
cross section calculations for the electron and muon 
candidate samples.}
\begin{center}
\begin{tabular}{l c c}
\hline
\hline
                                  & $\wenu$                       & $\wmnu$                       \\
\hline
$N_{W}^{\mathrm{obs}}$            & 37584                         & 31722                         \\
$N_{W}^{\mathrm{bck}}$            & 1762 $\pm$ 300                & 3469 $\pm$ 151                \\
$A_W$                             & 0.2397 $^{+0.0035}_{-0.0042}$ & 0.1970 $^{+0.0024}_{-0.0031}$ \\
$\epsilon_W$                      & 0.749 $\pm$ 0.009             & 0.732 $\pm$ 0.013             \\
$\int \mathcal{ L} dt$ ($\pbinv$) & 72.0 $\pm$ 4.3                & 72.0 $\pm$ 4.3                \\
\hline
\hline
\end{tabular}
\label{tab:wlnusigma}
\end{center}
\end{table}

\begin{figure*}[t]
\begin{center}
\includegraphics[width=3.2in]{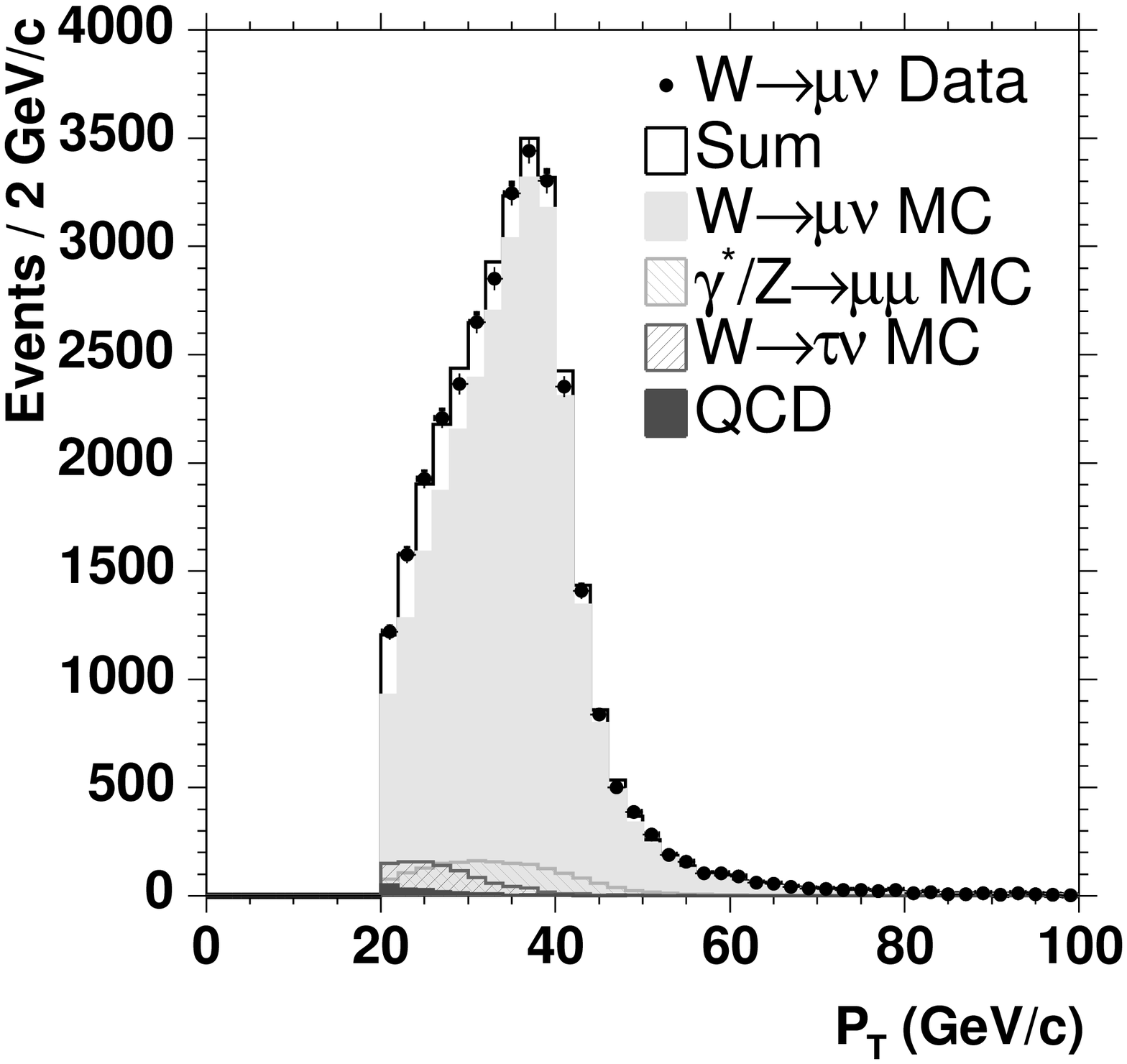} \includegraphics[width=3.2in]{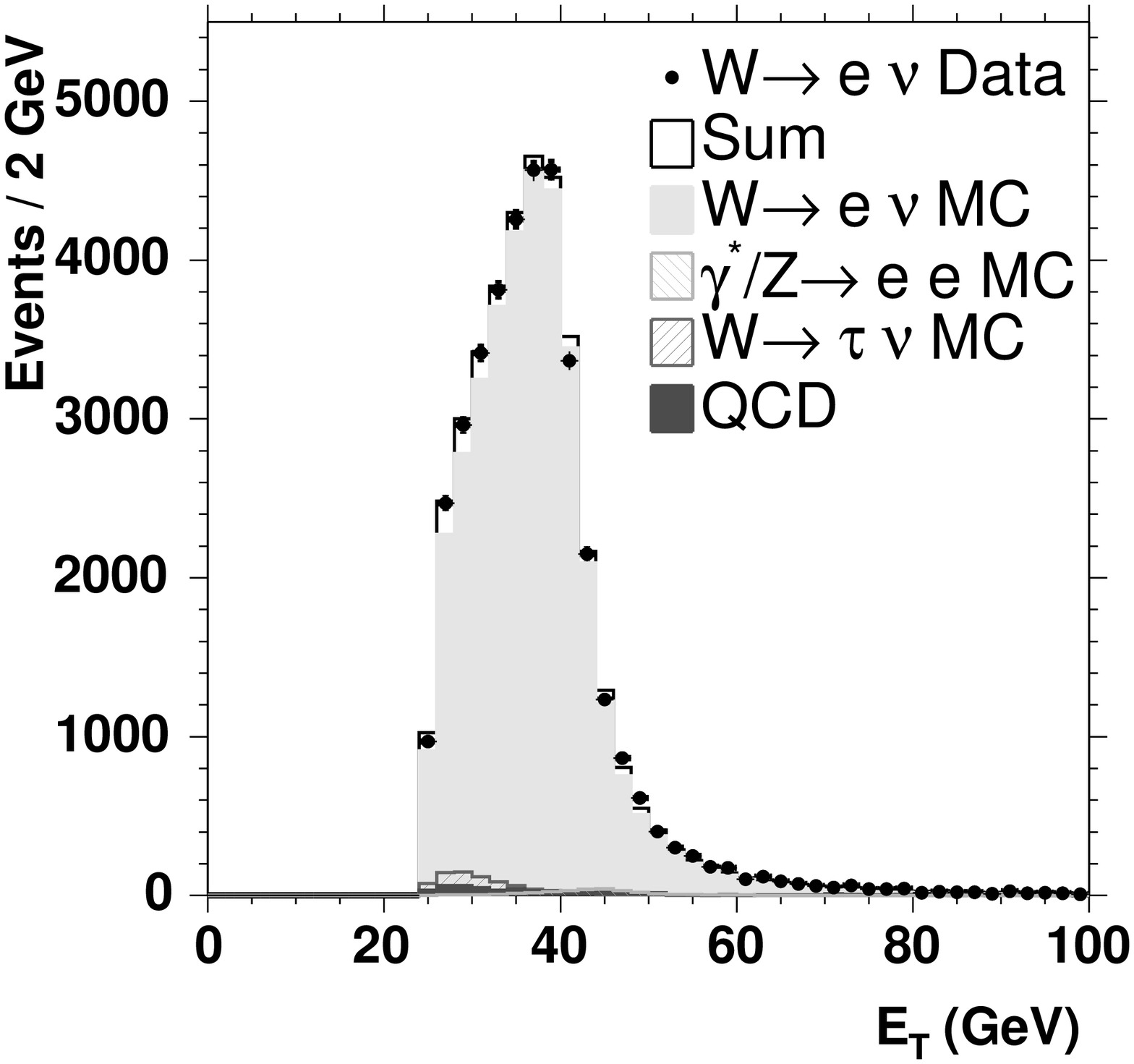}
\caption{Muon $\pt$ (left) and electron $\et$ (right) distributions 
for $\wlnu$ candidate events in data (points).  The solid lines 
are the sum of the predicted shapes originating from the signal and 
background processes weighted by their estimated contributions to our 
candidate samples.  The separate contributions originating from the 
signal and each individual background process are also shown.}
\label{fig:ptmuo_etele}
\end{center}
\end{figure*}

\begin{figure*}[t]
\begin{center}
\includegraphics[width=3.2in]{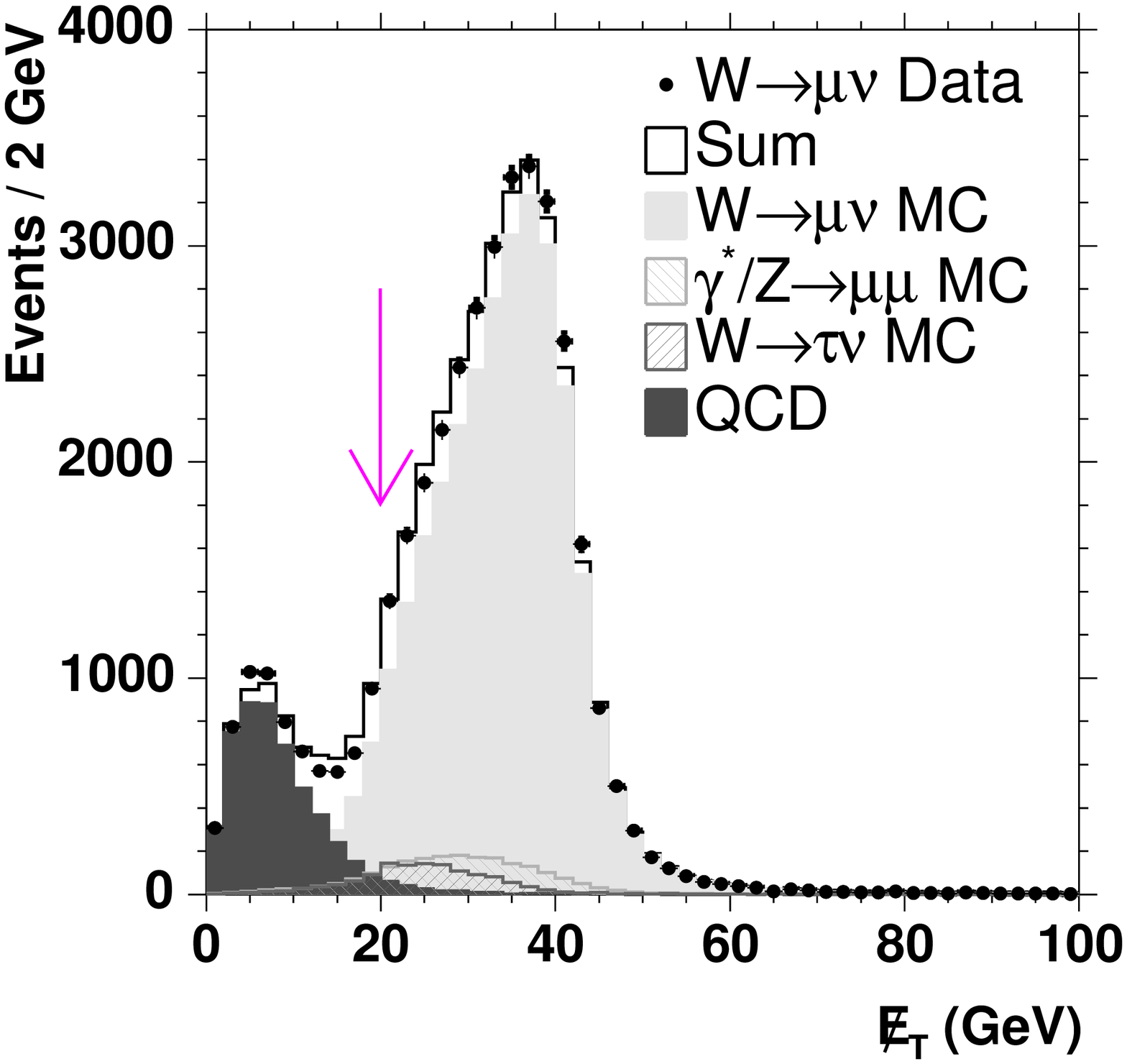} \includegraphics[width=3.2in]{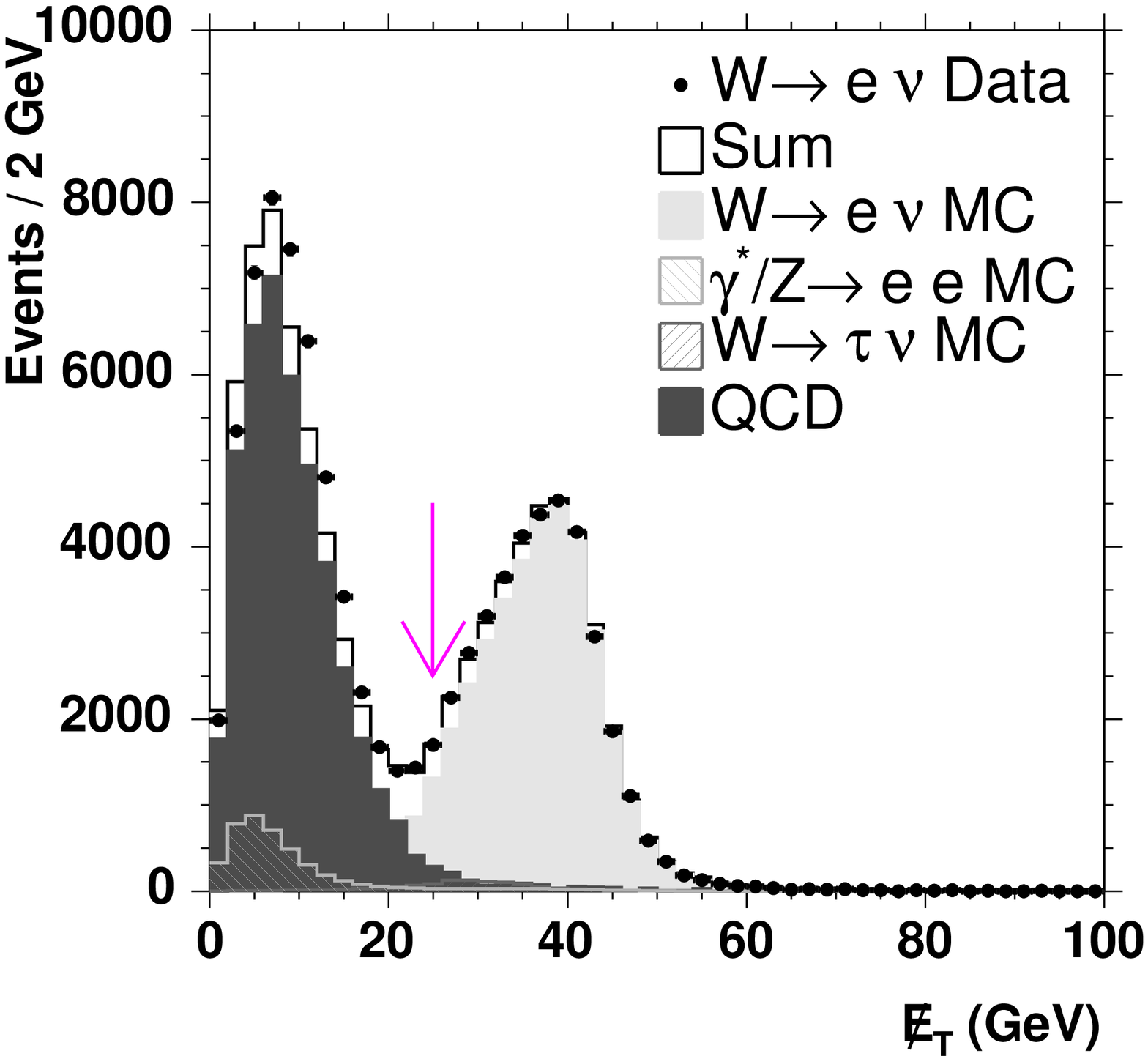}
\caption{Event $\met$ distributions for $\wlnu$ candidate events 
in data (points).  The selection requirement on event $\met$ 
has been removed to include candidate events with low $\met$.  
The solid lines are the sum of the predicted shapes originating 
from the signal and background processes weighted by their estimated 
contributions to our candidate samples.  The separate contributions 
originating from the signal and each individual background process 
are also shown.  The arrows indicate the location of the event 
$\met$ selection criteria used to define our candidate samples.}
\label{fig:wmet}
\end{center}
\end{figure*}

\begin{figure*}[t]
\begin{center}
\includegraphics[width=3.2in]{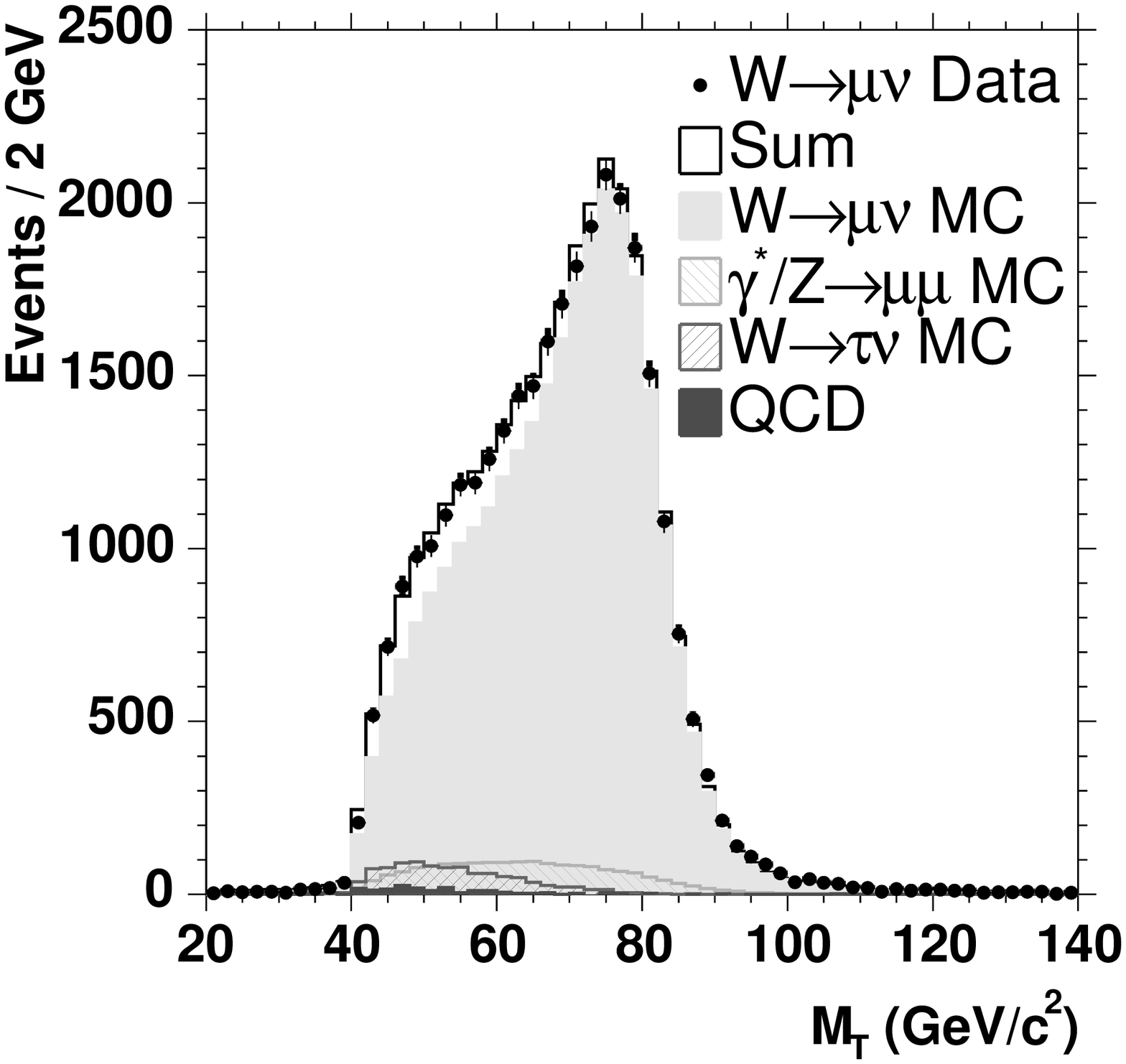} \includegraphics[width=3.2in]{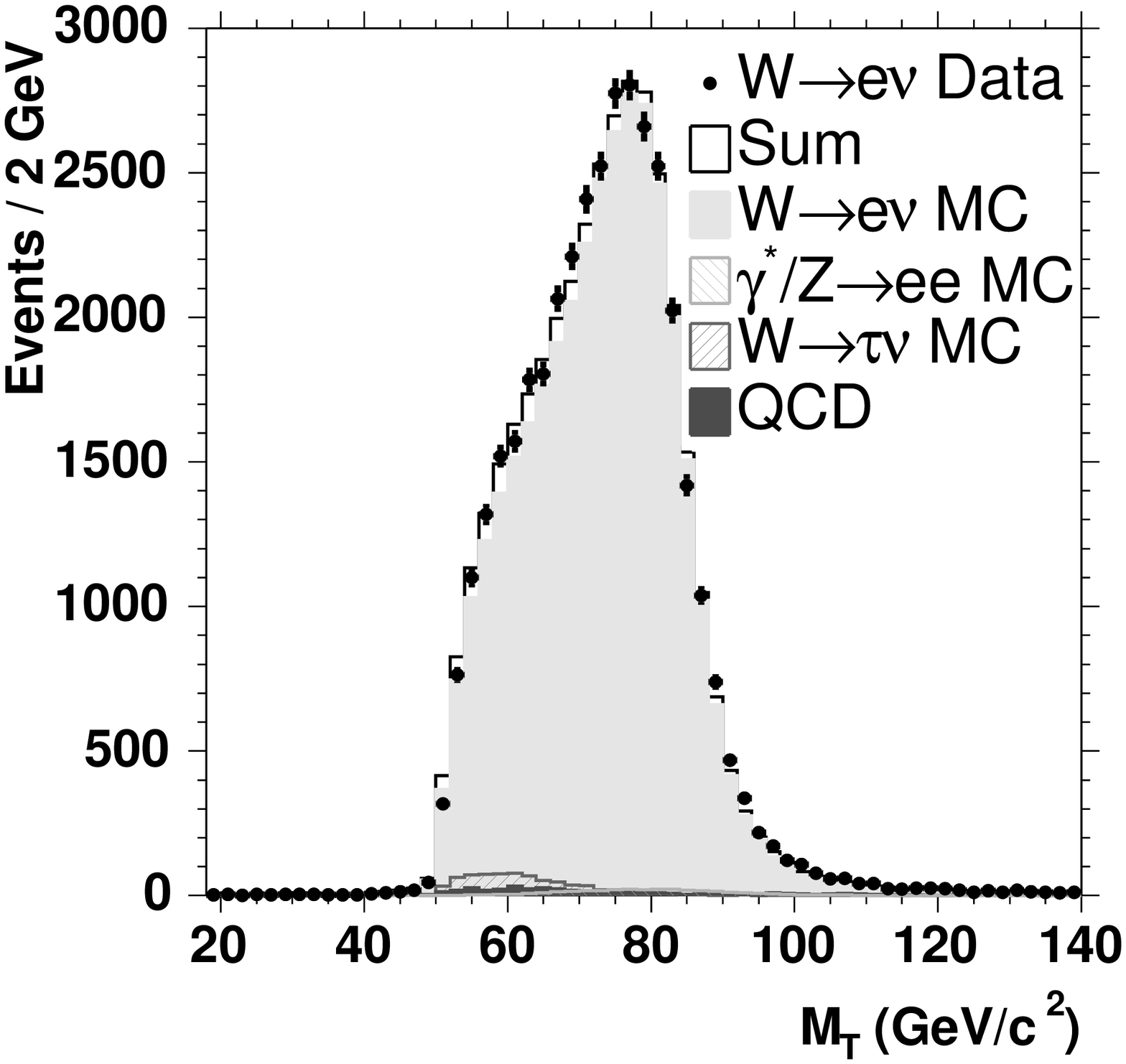}
\caption{Transverse mass ($M_{T}$) distributions for $\wlnu$ 
candidate events in data (points).   The solid lines are 
the sum of the predicted shapes originating from the signal 
and background processes weighted by their estimated 
contributions to our candidate samples.  The separate 
contributions originating from the signal and each individual 
background process are also shown.}
\label{fig:wmt}
\end{center}
\end{figure*}

Distributions of electron $\et$, muon $\pt$, event $\met$, and
$\W$ transverse mass $(M_{T} = \sqrt{2 [\et\met - (\ex \metx + \ey\mety)]})$ for events in our $\wlnu$ candidate 
samples are shown in Figs.~\ref{fig:ptmuo_etele} -~\ref{fig:wmt}.  
The data distributions are compared against a sum of the predicted 
shapes of these distributions for the $\wlnu$ signal and each 
contributing background process ($\zll$, $\wtnu$, and hadronic jets).  
The predicted shapes are obtained from our simulated event samples 
except in the case of the background arising from hadronic jets, 
which is modeled using events in the data containing non-isolated 
leptons that otherwise satisfy the $\wlnu$ selection criteria.  In
the sum, the predicted shape obtained for each process is weighted 
by the estimated number of events in our $\wlnu$ candidate samples  
originating from that process (see Table~\ref{tab:backsum}).  In the 
case of the $\met$ distribution, we remove the selection cut on the 
$\met$ variable to include events with low $\met$ in the comparison
and highlight the significant background contribution from hadronic 
jets in this region.

\subsection{$\zgll$ Cross Section}     
\label{subsec:res_zll}

Similarly, the cross section $\sigmappzg$ times the branching 
ratio $Br(\zgll)$ is calculated using Eq.~\ref{eq:zsigma} given 
in Sec.~\ref{sec:intro}.  The measurements of the required 
input parameters for the electron and muon candidate samples 
are described in the previous sections and summarized in 
Table~\ref{tab:zllsigma}.  Based on these values, we obtain
\begin{eqnarray}
\sigmabrzgee = 255.8 &\pm& 3.9({\it stat.}) \nonumber\\
                     &\pm& ^{5.5}_{5.4}({\it syst.}) \nonumber\\
                     &\pm& 15.3({\it lum.})~\pb \nonumber\\
\end{eqnarray}
and
\begin{eqnarray}
\sigmabrzgmm = 248.0 &\pm& 5.9({\it stat.}) \nonumber\\
                     &\pm& ^{8.0}_{7.2}({\it syst.}) \nonumber\\
                     &\pm& 14.8({\it lum.})~\pb~{\rm .} \nonumber\\
\end{eqnarray}
These measurements are the cross sections for dileptons produced in
the mass range 66~$\GeVCSq < M_{\ell\ell} <$~116~$\GeVCSq$ where both
$\gamma^{*}$ and $\Z$ boson exchange contribute.  A correction factor
of $F =$~1.004~$\pm$~0.001 determined from a NNLO $d\sigma/dy$
calculation, {\sc phozpr}~\cite{int:nnlo1,int:nnlo2,int:nnlo3}, using MRST 2002
NNLL PDFs~\cite{int:mrst2}, is needed to convert these measured cross
sections into those for pure $\Z$ boson exchange over the entire
dilepton mass range; the measured cross sections need to be multiplied
by $F$.  We compare the corrected cross sections for pure $\Z$ boson
exchange to the recent NNLO total cross section calculations for
$\sqrt{s} =$~1.96~$\TeV$~\cite{int:mrst1}.  The $\zll$ production
cross section predicted by these calculations is
251.3~$\pm$~5.0~$\pb$, which is in good agreement with the corrected,
measured values obtained in both lepton channels.  The uncertainty on
the predicted $\Z$ boson production cross section is also primarily
due to uncertainties in the PDF model derived from the MRST error PDF
sets.  Our independent estimates for these uncertainties using the
method described in Sec.~\ref{sec:acc} are a consistent 1.2~$\!\%$
uncertainty based on the MRST error PDF sets and a somewhat larger
3.7~$\!\%$ uncertainty based on the CTEQ6 error PDF sets.

Fig.~\ref{fig:ptfit} and Fig.~\ref{fig:etfit} show the invariant 
mass distributions for events in our $\zll$ candidate samples.  
The data distributions are compared against predicted shapes 
from our simulated $\zll$ event samples.  The predicted shapes 
are normalized to the total number of events in the candidate 
samples.  In making these comparisons, we ignore background 
processes which account for less than 1~$\!\%$ of the events 
in these samples (see Table~\ref{tab:backsum}).

\begin{table}[t]
\caption{Summary of the input parameters to the $\zgll$ 
cross section calculations for the electron and muon 
candidate samples.}
\begin{center}
\begin{tabular}{l c c}
\hline
\hline
                                  & $\zgee$                       & $\zgmm$                       \\
\hline
$N_{Z}^{\mathrm{obs}}$            & $4242$                        & 1785                          \\
$N_{Z}^{\mathrm{bck}}$            & 62 $\pm$ 18                   & 13 $\pm$ 13                   \\
$A_Z$                             & 0.3182 $^{+0.0039}_{-0.0041}$ & 0.1392 $^{+0.0027}_{-0.0033}$ \\
$\epsilon_Z$                      & 0.713 $\pm$ 0.012             & 0.713 $\pm$ 0.015             \\
$\int \mathcal{ L} dt$ ($\pbinv$) & 72.0 $\pm$ 4.3                & 72.0 $\pm$ 4.3                \\
\hline
\hline
\end{tabular}
\label{tab:zllsigma}
\end{center}
\end{table}

\subsection{Ratio of $\wlnu$ to $\zll$}
\label{subsec:res_eratio}

\begin{table}[t]
\caption{Summary of the input parameters to the $R$ calculations 
for the electron and muon candidate samples.}
\begin{center}
\begin{tabular}{ l c c } 
\hline 
\hline
                                & R$_e$               & R$_{\mu}$           \\ 
\hline 
$N_{W}^{\mathrm{obs}}$          & 37584               & 31722               \\
$N_{W}^{\mathrm{bck}}$          & 1762 $\pm$ 300      & 3469 $\pm$ 151      \\
$N_{Z}^{\mathrm{obs}}$          & 4242                & 1785                \\
$N_{Z}^{\mathrm{bck}}$          & 62 $\pm$ 18         & 13 $\pm$ 13         \\
$\frac{A_Z}{A_W}$               & 1.3272 $\pm$ 0.0109 & 0.7066 $\pm$ 0.0068 \\
$\frac{\epsilon_Z}{\epsilon_W}$ & 0.952 $\pm$ 0.011   & 0.974 $\pm$ 0.010   \\
$F$                             & 1.004 $\pm$ 0.001   & 1.004 $\pm$ 0.001   \\
\hline 
\hline 
\end{tabular}
\label{tab:finalr}
\end{center}
\end{table}

Precision measurements of the ratio of $\wlnu$ to $\zll$ 
production cross sections, $R$, are used to test the Standard 
Model.  The Standard Model parameters $\gam$ and $Br(\wlnu)$ 
can be extracted from our measured values of this ratio and 
are sensitive to non-Standard Model processes that result in 
additional decay modes for the $\W$ boson.  A new high-mass 
resonance which decays to either $\W$ or $\Z$ bosons could 
also have a direct effect on the measured value for $R$.  

The ratio of cross sections can be expressed in terms of 
measured quantities: 
\begin{equation}
R = \frac{1}{F} \cdot \frac{N_{W}^{\mathrm{obs}}-N_{W}^{\mathrm{bck}}}
{N_{Z}^{\mathrm{obs}}-N_{Z}^{\mathrm{bck}}} \cdot
\frac{A_Z}{A_W} \cdot
\frac{\epsilon_Z}{\epsilon_W},  
\label{eq::rex}
\end{equation} 
where $F$ is the correction factor for converting the 
measured $\zgll$ cross section into the cross section 
for pure $\Z$ boson exchange and the other parameters are 
as defined for the $\W$ and $\Z$ production cross section 
measurements.  The integrated luminosity terms in the $\W$ 
and $\Z$ cross section calculations along with their 
associated uncertainties cancel completely in the 
$R$ calculation, allowing for a significantly more precise 
measurement of the ratio than is possible for the individual 
cross sections.  In addition, we take advantage of many
correlated uncertainties in the event selection efficiencies
and kinematic and geometric acceptances of our $\W$ and $\Z$ 
candidate samples which cancel in the ratios $A_{Z}/A_{W}$
and $\epsilon_{Z}/\epsilon_{W}$.  For example, uncertainties 
on the acceptances arising from the PDF model are 
significantly smaller for the ratio of the $\zll$ and $\wlnu$ 
acceptances than for either individual acceptance.  The 
calculation of $A_{Z}/A_{W}$ and $\epsilon_{Z}/\epsilon_{W}$ 
for our electron and muon candidate samples and the treatment 
of the correlated uncertainties in these ratios are discussed 
in Secs.~\ref{sec:acc} and~\ref{sec:eff}.  The event counts 
and background estimates for the $\wlnu$ and $\zll$ candidate 
samples are the same as those used in the individual cross 
section calculations.  Table~\ref{tab:finalr} summarizes the 
input parameters used to calculate $R$ using the electron 
and muon candidate samples.  Substituting these values into 
Eq.~\ref{eq::rex}, we obtain
\begin{equation}
R_{e} = 10.79 \pm 0.17({\it stat.}) \pm 0.16({\it syst.})
\end{equation}
and
\begin{equation}
R_{\mu} = 10.93 \pm 0.27({\it stat.}) \pm 0.18({\it syst.})~{\rm .}
\end{equation}

Based on the calculations of the production cross sections for $\wlnu$
and $\zll$ provided by
\cite{int:nnlo0,int:nnlo4,int:nnlo1,int:nnlo2,int:nnlo3}, the expected
value for $R$ at $\sqrt{s} =$~1.96~$\TeV$ is 10.69.  To obtain an
accurate estimate for the uncertainty on this prediction, we need to
account for correlated uncertainties in the individual cross section
predictions.  The error originating from PDF model uncertainties has
the largest contribution to the total uncertainty.  We estimate the
magnitude of this contribution using the previously defined method in
Sec.~\ref{sec:acc} and obtain a 0.45~$\!\%$ uncertainty based on the
MRST error PDF sets and a larger 0.56~$\!\%$ uncertainty based on the
CTEQ6 error PDF sets.  We also need to account for the effect of
additional uncertainties in the values of the electroweak parameters
and CKM matrix elements used in the cross section calculations.  We
estimate these uncertainties using the $\overline{MS}$ NNLO total
cross section calculation, {\sc zwprod}
\cite{int:nnlo1,int:nnlo2}. We have updated the calculation code to
incorporate the CTEQ and MRST PDFs and variations of the electroweak
parameters and CKM matrix elements.  We obtain an uncertainty of
0.15~$\!\%$ for the $\sigmazee$ calculation and 0.40~$\!\%$ for the
$\sigmawen$ calculation.  The larger uncertainty associated with the
$\sigmawen$ calculation is due primarily to experimental uncertainties
on the CKM matrix values.  To be conservative, we add the larger PDF
model uncertainty (0.56~$\!\%$) in quadrature with the individual
cross section calculation uncertainties (0.15~$\!\%$ and 0.40~$\!\%$)
to obtain a combined uncertainty on the prediction for $R$ of
0.70~$\!\%$.  The resulting prediction, 10.69 $\pm$ 0.08, agrees with
the measured values of $R$ in both lepton channels.

\subsection{$\mu$-e Universality in $\W$ Decays}     
\label{subsec:gmuge}

Stringent tests of lepton universality at LEP provide strong 
evidence for lepton universality in $\zll$ production.  We 
make a similar test for lepton universality in $\wlnu$ 
production by extracting the ratio of $\wlnu$ couplings, 
$g_{\mu}/g_e$, from the measured ratio of the $\wmnu$ and 
$\wenu$ cross sections.  The $\wlnu$ couplings are related 
to the measured ratio $U$ of the cross sections, defined as 
\begin{equation}
  U \equiv \frac{\sigmabrwmn}
                {\sigmabrwen}  = 
     \frac{\Gamma(\wmnu)}{\Gamma(\wenu)} =
     \frac{g_{\mu}^2}{g_{e}^2}~{\rm .}
\label{eq:gratio}
\end{equation}

As in the case of the $R$ measurements described in the 
previous section, many of the uncertainties associated 
with the individual cross section measurements cancel 
in the ratio.  Table~\ref{tab:Usys} summarizes the 
uncorrelated uncertainties between the two cross 
section measurements that contribute to the overall 
uncertainty on $g_{\mu}/g_e$.  The uncertainties 
due to the PDF model cancel almost completely in the 
ratio.  The major remaining contributions to the systematic 
uncertainty  come from the uncorrelated event selection 
efficiencies for the electron and muon candidate samples.  
Since these efficiencies are measured directly from $\zll$ 
candidate events in the data, the associated uncertainties
will decrease as additional data are analyzed.  In this 
sense, the remaining uncertainty on $g_{\mu}/g_e$ is 
primarily statistical in nature and can be reduced with 
larger data samples.  Using the input parameters to our 
$\wlnu$ cross section measurements, we obtain 
\begin{equation}
   \frac{g_\mu}{g_e} = 0.991 \pm 0.012~{\rm .}
\end{equation}
Using Eq.~\ref{eq:gratio} and the current world average of 
experimental results for $Br(\wmnu) =$~0.1057~$\pm$~0.0022 
and $Br(\wenu) =$~0.1072~$\pm$~0.0016~\cite{int:pdg}, the 
expected value of $g_{\mu}/g_{e}$ is 0.993~$\pm$~0.013
which is in good agreement with our measured value.

\begin{table}[t]
\caption{Uncertainties on the measured ratio of $\wmnu$ and 
$\wenu$ cross sections, $U$.}
\begin{center}
\begin{tabular}{l r}
\hline 
\hline
Category                   & Uncertainty            \\ 
\hline
Statistical Uncertainty    & 0.0075                 \\
\hline
Acceptance Ratio:          &                        \\   
Simulation Statistics      & 0.0019                 \\
Boson $\pt$ Model          & 0.0001                 \\
PDF Model                  & $^{+0.0003}_{-0.0004}$ \\
$\pt$ Scale and Resolution & 0.0018                 \\
$\et$ Scale and Resolution & 0.0034                 \\
Material Model             & 0.0072                 \\
Recoil Energy Model        & 0.0010                 \\
\hline
Efficiency Ratio:          &                        \\
Uncorrelated               & 0.0199                 \\
\hline
Backgrounds:               &                        \\
Hadronic                   & 0.0043                 \\
Electroweak                & 0.0030                 \\
Cosmic Ray                 & 0.0008                 \\
\hline 
\hline
\end{tabular}
\label{tab:Usys}
\end{center}
\end{table}

\subsection{Combined Results from the Electron and Muon Channels}
\label{subsec:res_emu}

Since our measurement of $g_\mu/g_e$ supports the conclusion 
of lepton universality in $\wlnu$ production, we proceed to 
combine our measurements of the $\wlnu$ and $\zll$ production 
cross section measurements in the electron and muon channels
to increase the overall precision of these results.  We 
also combine our measurements of $R_{e}$ and $R_{\mu}$ to 
determine a precision value for $R$ which is used to test 
the Standard Model.

\subsubsection{Combination of the Cross Sections}
\label{subsec:res_combcs}

\begin{table}
\caption{Uncertainty categories for the inclusive $\W$ 
cross section measurements.  These values are absolute 
contributions to $\sigmawen$ in $\pb$.  The uncertainties
in the electron and muon channels for each category 
are treated as either 100~$\!\%$ correlated (1.0) or  
uncorrelated (0.0).}
\begin{center}
\begin{tabular}{l c c c}
\hline 
\hline
Category                   & Electron       & Muon       & Correlation       \\
\hline
Statistical Uncertainty    & 14.3           & 15.3       & 0.0               \\
\hline
Acceptance:                &                &            &                   \\
Simulation Statistics      & 3.6            & 3.9        & 0.0               \\
Boson $\pt$ Model          & 1.2            & 1.0        & 1.0               \\
PDF Model                  & 36.9           & 35.4       & 1.0               \\
$\pt$ Scale and Resolution & 0.8            & 5.6        & 1.0               \\
$\et$ Scale and Resolution & 9.5            & 0.0        & 0.0               \\
Material Model             & 20.2           & 0.0        & 0.0               \\
Recoil Energy Model        & 6.8            & 9.4        & 1.0               \\
\hline
Efficiencies:              &                &            &                   \\
Vertex $z_0$ Cut           & 11.7           & 11.5       & 1.0               \\
Track Reconstruction       & 11.1           & 10.9       & 1.0               \\
Trigger                    & 2.9            & 32.7       & 0.0               \\
Lepton Reconstruction      & 11.1           & 19.7       & 0.0               \\
Lepton Identification      & 24.1           & 23.8       & 0.0               \\
Lepton Isolation           & 9.4            & 9.7        & 0.0               \\
$\Z$-rejection Cut         & 0.0            & 4.6        & 0.0               \\
Cosmic Ray Algorithm       & 0.0            & 0.3        & 0.0               \\
\hline
Backgrounds:               &                &            &                   \\
Hadronic                   & 23.1           & 10.8       & 1.0               \\
$\zll$                     & 1.5            & 9.2        & 1.0               \\
$\wtnu$                    & 1.3            & 2.3        & 1.0               \\
Cosmic Ray                 & 0.0            & 2.2        & 0.0               \\
\hline
\hline
\end{tabular}
\label{tab:Werrcat}
\end{center}
\end{table}

We use the Best Linear Unbiased Estimate (BLUE)~\cite{res:blue,res:blue2} 
method to combine measurements in the electron and muon channels.  For 
the $\wlnu$ measurements, we identify twenty categories of uncertainties, 
several of which are correlated in the electron and muon channels.  
Table~\ref{tab:Werrcat} lists these categories and summarizes the raw 
contribution of each (in $\pb$) to the $\W$ cross section measurements 
in the electron and muon channels.  Based on the information in this 
table, we combine the measurements in the two lepton channels and obtain
\begin{eqnarray}
\sigmabrwln =
  2.749 &\pm& 0.010({\it stat.}) \nonumber\\
        &\pm& 0.053({\it syst.})\nonumber\\ 
        &\pm& 0.165({\it lum.})~\nb, \nonumber\\
\end{eqnarray}
which has a precision of 2.0~$\!\%$, not including the uncertainty
associated with the measured integrated luminosity of our samples.
The uncertainty on luminosity is not included in the calculation of
the combined value.

The combination of the $\zll$ cross section measurements in 
the electron and muon channels is based on the same procedure.  
In this case, we identify seventeen categories of 
uncertainties, some of which are correlated between channels.  
Table~\ref{tab:Zerrcat} provides a list of these categories 
and summarizes the raw contribution of each (in $\pb$) to 
the $\Z$ cross section measurements in the electron and muon 
channels.  The additional acceptance for forward electrons 
in the plug calorimeter modules reduces the statistical
uncertainty associated with the $\Z$ cross section measurement 
in the electron channel, which thus has a larger weight in the 
final combination.  The combined result is       
\begin{eqnarray}
\sigmabrzgll =
  254.9 &\pm& 3.3({\it stat.}) \nonumber\\
        &\pm&  4.6({\it syst.}) \nonumber\\
        &\pm& 15.2({\it lum.})~\pb, \nonumber\\
\end{eqnarray}
which has a precision of 2.2~$\!\%$, not including the uncertainty
associated with the measured integrated luminosity of our samples.  As
discussed previously, the combined cross section given here is the
cross section for dileptons in the mass range 66~$\GeVCSq <
M_{\ell\ell} <$~116~$\GeVCSq$ including contributions from both
$\gamma^{*}$ and $\Z$ boson exchange.  In order to convert the
measured cross section into a cross section for pure $\Z$ boson
exchange over the entire mass range, one must multiply the measured
value by the correction factor presented earlier, $F
=$~1.004~$\pm$~0.001.

\begin{figure*}[t]
\begin{center}
\includegraphics[width=5.in]{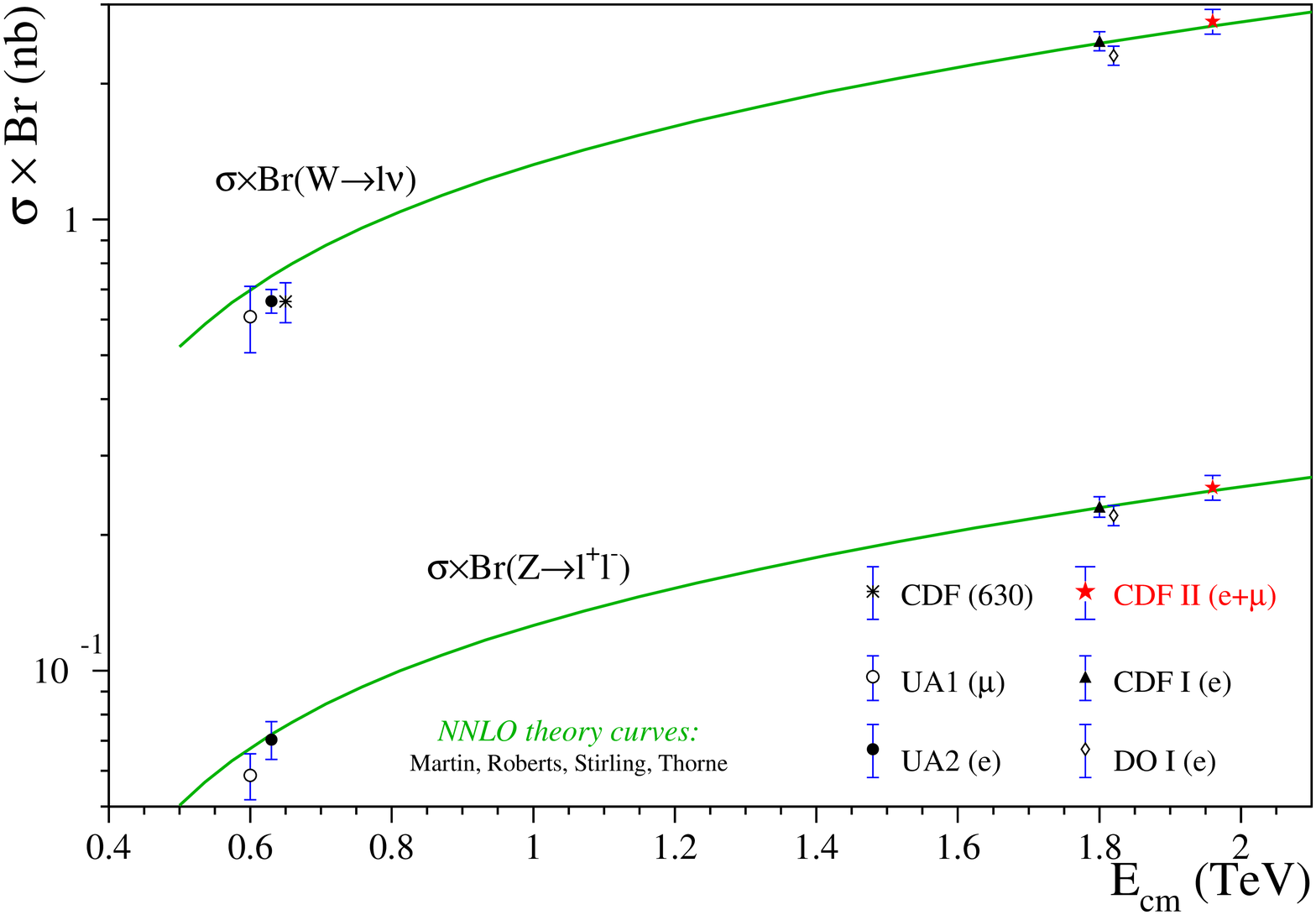}
\caption{$\wlnu$ and $\zll$ cross section measurements as 
a function of the $\ppbar$ center-of-mass energy, $E_{\mathrm{CM}}$.
The solid lines correspond to the theoretical NNLO Standard Model
calculations from
\cite{int:nnlo0,int:nnlo4,int:nnlo1,int:nnlo2,int:nnlo3}.}
\label{fig:sigma_th}
\end{center}
\end{figure*}

A comparison of the predictions
from~\cite{int:nnlo0,int:nnlo4,int:nnlo1,int:nnlo2,int:nnlo3} for
$\sigmabrwln$ and $\sigmabrzll$ as a function of the $\ppbar$ center
of mass energy, $E_{\mathrm{CM}}$, with our measured values and other
experimental results~\cite{int:ua1,int:cdf_sigmas,int:d0_B} are shown
in Fig.~\ref{fig:sigma_th}.

\begin{table}
\caption{Uncertainty categories for the inclusive $\Z$ 
cross section measurements.  These values are absolute 
contributions to $\sigmazee$ in $\pb$.  The uncertainties
in the electron and muon channels for each category 
are treated as either 100~$\!\%$ correlated (1.0) or 
uncorrelated (0.0).}
\begin{center}
\begin{tabular}{l c c c}
\hline 
\hline
Category                   & Electron       & Muon       & Correlation       \\
\hline
Statistical Uncertainty    & 3.93           & 5.87       & 0.0               \\
\hline
Acceptance:                &                &            &                   \\
Simulation Statistics      & 0.61           & 1.01       & 0.0               \\
Boson $\pt$ Model          & 0.16           & 0.19       & 1.0               \\
PDF Model                  & 1.96           & 4.94       & 1.0               \\
$\pt$ Scale and Resolution & 0.10           & 0.13       & 1.0               \\
$\et$ Scale and Resolution & 0.67           & 0.00       & 0.0               \\
Material Model             & 2.45           & 0.00       & 0.0               \\
Recoil Energy Model        & 0.00           & 0.00       & 0.0               \\
\hline
Efficiency:                &                &            &                   \\
Vertex $z_0$ Cut           & 1.08           & 1.04       & 1.0               \\
Track Reconstruction       & 1.42           & 1.98       & 1.0               \\
Trigger                    & 0.17           & 2.05       & 0.0               \\
Lepton Reconstruction      & 1.43           & 1.24       & 0.0               \\
Lepton Identification      & 3.39           & 3.48       & 0.0               \\
Lepton Isolation           & 1.21           & 1.77       & 0.0               \\
Cosmic Ray Algorithm       & 0.00           & 0.15       & 0.0               \\
\hline
Backgrounds:               &                &            &                   \\
Hadronic                   & 1.10           & 0.08       & 1.0               \\
$\ztt$                     & 0.02           & 0.04       & 1.0               \\
$\wlnu$                    & 0.17           & 0.00       & 1.0               \\
Cosmic Ray                 & 0.00           & 1.76       & 0.0               \\
\hline
\hline
\end{tabular}
\label{tab:Zerrcat}
\end{center}
\end{table}

\subsubsection{Combination of the $R$ measurements}
\label{subsec:res_combr}

The same BLUE method is also used to combine our measurements 
of $R_{e}$ and $R_{\mu}$.  For our cross section ratio 
measurements we identify fifteen categories of uncertainties, 
some of which are correlated between our measurements in the 
electron and muon channels.  Table~\ref{tab:Rerrcat} lists 
these categories and summarizes the raw contribution of each 
to the $R_{e}$ and $R_{\mu}$ measurements.  Since most of the
uncertainties related to efficiency factors are uncorrelated 
in the electron and muon channels, the corresponding 
uncertainties are combined into a single net uncertainty for 
uncorrelated efficiencies.  The exception is the uncertainty 
on COT track reconstruction efficiency which is 100~$\!\%$ 
correlated between the two channels. The combined result is
\begin{equation}
R = 10.84 \pm 0.15({\it stat.}) \pm 0.14({\it syst.}) 
\end{equation}
which is precise to 1.9~$\!\%$.  

\begin{table}
\caption{Uncertainty categories for the $R$ measurements.
The uncertainties in the electron and muon channels for each 
category are treated as either 100~$\!\%$ correlated (1.0) or 
uncorrelated (0.0).}
\begin{center}
\begin{tabular}{l c c c}
\hline 
\hline
Category                   & Electron       & Muon       & Correlation       \\
\hline
Statistical Uncertainty    & 0.1748         & 0.2659     & 0.0               \\
\hline 
Acceptance Ratio:          &                &            &                   \\
Simulation Statistics      & 0.0293         & 0.0472     & 0.0               \\
Boson $\pt$ Model          & 0.0020         & 0.0044     & 1.0               \\
PDF Model                  & 0.0701         & 0.0836     & 1.0               \\
$\pt$ Scale and Resolution & 0.0012         & 0.0167     & 1.0               \\
$\et$ Scale and Resolution & 0.0184         & 0.0000     & 0.0               \\
Material Model             & 0.0322         & 0.0000     & 0.0               \\
Recoil Energy Model        & 0.0267         & 0.0377     & 1.0               \\
\hline
Efficiency Ratio:          &                &            &                   \\
Uncorrelated               & 0.1204         & 0.0999     & 0.0               \\
Track Reconstruction       & 0.0169         & 0.0437     & 1.0               \\
\hline
Backgrounds:               &                &            &                   \\
Hadronic                   & 0.0437         & 0.0399     & 1.0               \\
Uncorrelated Electroweak   & 0.0089         & 0.0094     & 0.0               \\
Correlated Electroweak     & 0.0057         & 0.0369     & 1.0               \\
Cosmic Ray                 & 0.0000         & 0.0689     & 0.0               \\
\hline
Correction Factor, $F$     & 0.0107         & 0.0109     & 1.0               \\
\hline
\hline
\end{tabular}
\label{tab:Rerrcat}
\end{center}
\end{table}

\subsection{Extraction of Standard Model Parameters}

As previously discussed, the precision value for $R$ 
obtained from the combination of our measurements in 
the electron and muon channels can be used to measure
various Standard Model parameters and in the process
test the predictions of the model.  The ratio of cross 
sections can be expressed as
\begin{equation}
R = 
\frac{\sigmappw}{\sigmappz} 
\frac{\Gamma(\wlnu)}{\Gamma(\zll)}
\frac{\Gamma(Z)}{\Gamma(W)} 
\label{eq:rth}.
\end{equation}
Using the precision LEP measurements for $\Gamma(\zll)/
\Gamma(Z)$ at the $\Z$ pole mass and the NNLO calculation 
of $\sigmappw/\sigmappz$ by
\cite{int:nnlo0,int:nnlo4,int:nnlo1,int:nnlo2,int:nnlo3}, we extract
the Standard Model parameter $Br(\wlnu) = \Gamma(\wlnu)/\gam$ from
Eq.~\ref{eq:rth} using our measured value of $R$.  Using the Standard
Model prediction for $\Gamma(\wlnu)$, we also make an indirect
measurement of $\gam$ and based on this value place a constraint on
the CKM matrix element $\vcs$.

\subsubsection{Extraction of $Br(\wlnu)$}  
\label{subsec:res_br}

\begin{figure}[t!]
\begin{center}
\includegraphics[width=3.5in]{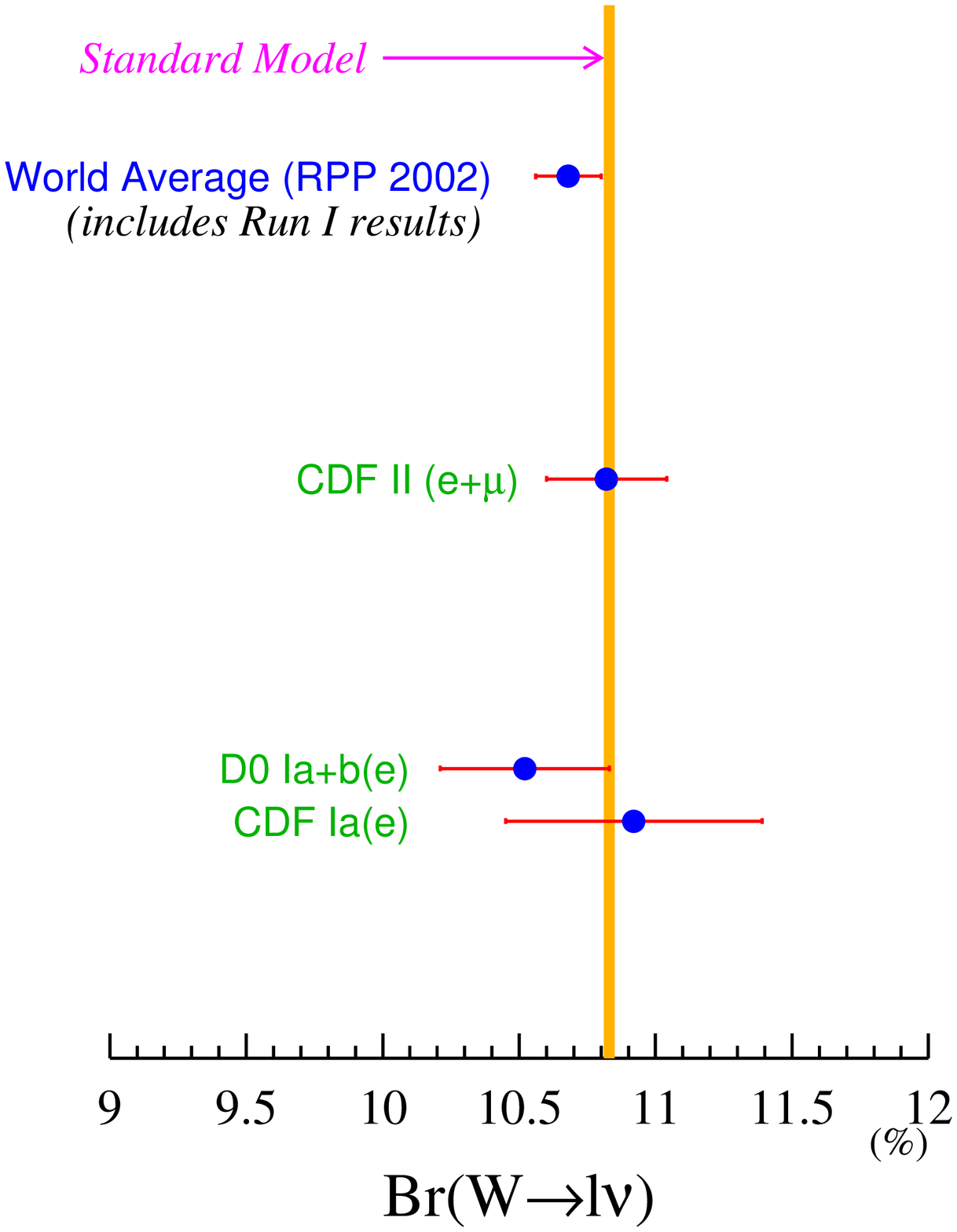}
\caption{Comparison of our measured value of $Br(\wlnu)$ with previous 
hadron collider measurements~\cite{int:cdf_ratio1,int:cdf_ratio2,int:d0_B},   
the current world average of experimental results~\cite{int:pdg}, and the 
Standard Model expectation~\cite{int:pdg}.}
\label{fig:brleptsummary}
\end{center}
\end{figure}

\begin{figure}[t!]
\begin{center}
\includegraphics[width=3.5in]{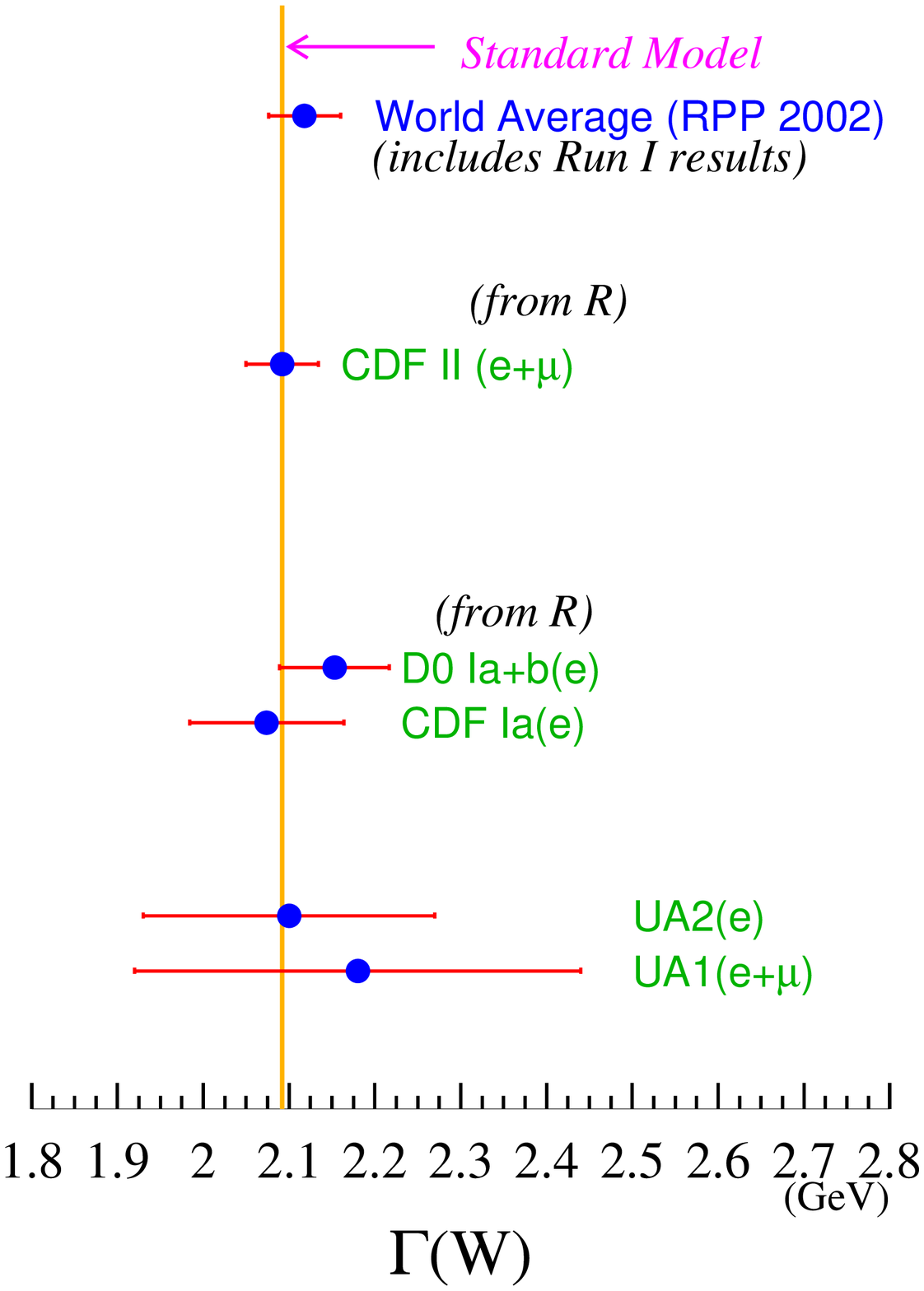}
\caption{Comparison of our measured value of $\gam$ with previous hadron 
collider measurements~\cite{int:ua1,int:ua2,int:cdf_ratio1,int:cdf_ratio2,
int:d0_B}, the current world average of experimental results~\cite{int:pdg}, 
and the Standard Model expectation~\cite{int:pdg}.}
\label{fig:gammasummary}
\end{center}
\end{figure}

The required parameters to extract $Br(\wlnu)$ from our measured 
$R$ value using Eq.~\ref{eq:rth} are the predicted ratio of $\W$ 
and $\Z$ production cross sections and the measured value of 
$Br(\zll) = \Gamma(\zll)/\Gamma(Z)$.  The value of $\sigmawen/
\sigmazee$ obtained from the NNLO calculations provided by 
\cite{int:nnlo0,int:nnlo4,int:nnlo1,int:nnlo2,int:nnlo3} is 3.3696 
with associated relative uncertainties of 0.0056 coming from the PDF
model and 0.0043 coming from electroweak and CKM matrix parameters
used in the calculations (see Sec.~\ref{subsec:res_eratio}).  The
experimental value of $Br(\zll) =$ 0.033658~$\pm$~0.000023 as measured
at LEP is taken from~\cite{int:pdg}.

When extracting $Br(\wlnu)$ from $R$, it is important to 
consider correlated uncertainties in the ratio of predicted  
cross sections and the ratio of acceptances, $A_{Z}/A_{W}$, used
in the measurement of $R$.  In a sense, we measure $Br(\wlnu)$
by equating $R_{\mathrm{phys}}$ to $R_{\mathrm{meas}}$, where
\begin{equation}
 R_{\mathrm{phys}} \equiv
      \frac{\sigmawen}{\sigmazee} \, 
      \frac{Br(\wlnu)}{Br(\zll)} 
\end{equation}
and
\begin{equation}
 R_{\mathrm{meas}} \equiv 
      \frac{N_{W}^{\mathrm{obs}}-N_{W}^{\mathrm{bck}}}{A_W\epsilon_W} \,
      \frac{A_Z\epsilon_Z}{N_{Z}^{\mathrm{obs}}-N_{Z}^{\mathrm{bck}}}~{\rm .} 
\end{equation}
Then, 
\begin{eqnarray}
   Br(\wlnu) = &&
   \frac{N_{W}^{\mathrm{obs}}-N_{W}^{\mathrm{bck}}}{N_{Z}^{\mathrm{obs}}-N_{Z}^{\mathrm{bck}}} \,
   \frac{\epsilon_Z}{\epsilon_W} \, \nonumber\\
  && \times \left( \frac{A_Z\sigma_Z}{A_W\sigma_W} \right)
    \, Br(\zll)~{\rm .} \nonumber\\
\label{eq:wlvbr}
\end{eqnarray}
The ratio of the acceptance times the cross section for $\Z$
and $\W$ bosons on the right-hand side of Eq.~\ref{eq:wlvbr} is 
affected by uncertainties in the PDF model.  To account properly
for correlations between the PDF uncertainties associated with 
each of these four quantities, we independently calculate a 
PDF model uncertainty for the quantity contained within the 
parentheses using the method described in Sec.~\ref{sec:acc}.  
The measured PDF model uncertainties on this quantity are 
found to be slightly larger than for those on $A_Z/A_W$ alone 
(0.9~$\!\%$ versus 0.6~$\!\%$ in the electron channel and 
1.0~$\!\%$ versus 0.8~$\!\%$ in the muon channel).  These 
correlated uncertainties are separately accounted for in our 
extraction of $Br(\wlnu)$ from the measured value of $R$.  We 
obtain 
\begin{equation}
  Br(\wlnu) = 0.1082~\pm~0.0022
\end{equation}
where the uncertainty contributions are from $R$ ($\pm$~0.00212), 
the predicted ratio of cross sections ($\pm$~0.00047), and the $\zll$ 
branching ratio ($\pm$~0.00007).  The Standard Model value for 
this parameter is 0.1082~$\pm$~0.0002, and the world average of 
experimental results is 0.1068~$\pm$~0.0012~\cite{int:pdg}, both of 
which are in good agreement with our measured value.  A summary of 
$Br(\wlnu)$ measurements is shown in Fig.~\ref{fig:brleptsummary}.

\subsubsection{Extraction of $\gam$}  
\label{subsec:res_emugamma}

An indirect measurement of $\gam$ can be made from our measured value
of $Br(\wlnu)$ using the Standard Model value for the leptonic partial
width, $\Gamma(\wlnu)$.  We use the fitted value for $\Gamma(\wlnu)$
of 226.4 $\pm$ 0.4~$\MeV$~\cite{int:pdg}.  Based on this value, we
obtain
\begin{equation}
  \gam = 2092 \pm 42 ~\MeV 
\end{equation}
which can be compared to Standard Model prediction of
2092~$\pm$~3~$\MeV$~\cite{int:pdg} and the world average of
experimental results, 2118~$\pm$~42~$\MeV$~\cite{int:pdg}.  A summary
of $\gam$ experimental measurements is shown in
Fig.~\ref{fig:gammasummary}.  Our indirect measurement is in good
agreement with the fit~\cite{int:pdg} and the theoretical prediction
as well as other measurements in literature.

An alternative approach for obtaining $\gam$ is to first use the 
predicted values for both $\Gamma(\wlnu)$ and $\Gamma(\zll)$ to 
extract a ratio of the total widths, $\gam/\Gamma(Z)$, from the 
measured value of $R$.  The precisely measured value of $\Gamma(Z)$ 
from the LEP experiments (2495.2 $\pm$ 2.3~$\MeV$~\cite{int:pdg}) is 
then used to extract a value for $\gam$.  Using this approach we 
obtain    
\begin{equation}
  \frac{\gam}{\Gamma(Z)} = 0.838 \pm 0.017
\end{equation}
for the ratio of total widths, which can be compared to the Standard 
Model prediction of 0.8382 $\pm$ 0.0011~\cite{int:pdg}.  Based on the 
measured value of $\Gamma(Z)$ we obtain  
\begin{equation}
  \gam = 2091 \pm 42 ~\MeV~{\rm ,} 
\end{equation}
where the uncertainty on the measured value for $\Gamma(Z)$ makes 
a negligible contribution to the total uncertainty.  Since the 
measurement of $\Gamma(Z)$ is independent of the measurement of 
the branching ratio $Br(\zll)$, both extracted values of $\gam$ are 
independent to some degree.

\subsubsection{Extraction of $\vcs$}  
\label{subsec:res_vcs}

In the Standard Model the total $\W$ width is a sum over partial 
widths for leptons and quarks where the latter subset involves a 
sum over certain CKM matrix elements \cite{int:pdg}:
\begin{eqnarray}
\label{vcseq}
\Gamma_W \simeq && 3\Gamma_W^0 +
  3\left( 1 + \frac{\alpha_{\mathrm{s}}}{\pi} + 1.409(\frac{\alpha_{\mathrm{s}}}{\pi})^2
            - 12.77(\frac{\alpha_{\mathrm{s}}}{\pi})^3 \right) \,\nonumber\\
&& \times \sum_{\mathrm{[no~top]}}
   | V_{qq^\prime} |^2 \, \Gamma_W^0~{\rm .}
\end{eqnarray}
Only the first two rows of the CKM matrix contribute as 
decays to the top quark are kinematically forbidden.
Thus the relevant CKM matrix elements are $V_{\mathrm{ud}}$, 
$V_{\mathrm{us}}$, $V_{\mathrm{cd}}$, $V_{\mathrm{cs}}$, 
$V_{\mathrm{ub}}$, and $V_{\mathrm{cb}}$.  Of these, $\vcs$ 
contributes the largest uncertainty.  We use the indirect 
measurement of $\gam$ from our measured value of $Br(\wlnu)$
as a constraint on $\vcs$ based on world average measurements 
of all the other CKM matrix elements and find
\begin{equation}
     |\vcs| = 0.976 \pm 0.030~{\rm ,}
\end{equation}
using $\alpha_{\mathrm{s}} = 0.120$ and $\Gamma_W^0 = 226.4$~$\MeV$
\cite{int:pdg}.  Our measured value is more precise than the direct
measurement at LEP, $|\vcs| =$~0.97~$\pm$~0.11~\cite{res:vcs1,res:vcs2}, but
not as precise as the combined value from LEP and Run~I at the
Tevatron, $|\vcs| =$~0.996~$\pm$~0.013~\cite{res:pete}.

\subsection{Summary}
\label{subsec:res_sum}

\begin{table*}[t]
\caption{Standard Model parameters extracted from the measured 
ratio of $\W$ and $\Z$ production cross sections, $R$.}
\begin{tabular}{l c c c}
\hline
\hline
Quantity         & Our Measurement       & World Average       & SM Value            \\
\hline
$Br(\wlnu)$      & 0.1082 $\pm$ 0.0022   & 0.1068 $\pm$ 0.0012 & 0.1082 $\pm$ 0.0002 \\
$\gam$ in $\MeV$ & 2092 $\pm$ 42         & 2118 $\pm$ 42       & 2092  $\pm$ 3    \\
$\gam/\Gamma(Z)$ & 0.838 $\pm$ 0.017     & 0.849 $\pm$ 0.017   & 0.838 $\pm$ 0.001   \\
$\vcs$           & 0.976 $\pm$ 0.030     & 0.996 $\pm$ 0.013   & N/A                 \\
$g_{\mu}/g_{e}$  & 0.991 $\pm$ 0.012     & 0.993 $\pm$ 0.013   & 1                   \\
\hline
\hline
\end{tabular}
\label{tab:extracted}
\end{table*}

We have performed measurements for the $\W$ and $\Z$ boson
production cross sections in the electron and muon decay 
channels based on 72~$\pbinv$ of $\ppbar$ collision data 
at $\sqrt{s} =$~1.96~$\TeV$.  We calculate the ratio of the 
$\W$ and $\Z$ cross sections, $R$, in each lepton channel 
and combine them to obtain a value which is precise to 
1.9~$\!\%$.  The precision will improve when more data are 
analyzed.  From this ratio we extract the leptonic $\W$ 
branching ratio, the $\W$ width, the ratio of the $\W$ 
and $\Z$ widths, and constrain the CKM matrix element 
$\vcs$.  A summary of extracted quantities is given in 
Table~\ref{tab:extracted}.

\begin{acknowledgments}
We thank the Fermilab staff and the technical staffs of the
participating institutions for their vital contributions. This work
was supported by the U.S. Department of Energy and National Science
Foundation; the Italian Istituto Nazionale di Fisica Nucleare; the
Ministry of Education, Culture, Sports, Science and Technology of
Japan; the Natural Sciences and Engineering Research Council of
Canada; the National Science Council of the Republic of China; the
Swiss National Science Foundation; the A.P. Sloan Foundation; the
Bundesministerium fuer Bildung und Forschung, Germany; the Korean
Science and Engineering Foundation and the Korean Research Foundation;
the Particle Physics and Astronomy Research Council and the Royal
Society, UK; the Russian Foundation for Basic Research; the Comision
Interministerial de Ciencia y Tecnologia, Spain; in part by the
European Community's Human Potential Programme under contract
HPRN-CT-2002-00292; and the Academy of Finland.
\end{acknowledgments}


\clearpage

\bibliography{wzxsec_jphysg}

\clearpage

\end{document}